\documentclass[a4paper,12pt]{article}
\pdfoutput=1
\usepackage{jheppub}
\usepackage[utf8]{inputenc}
\usepackage{url}
\usepackage{mathrsfs}
\usepackage{pdflscape}
\usepackage{refcount}
\usepackage{booktabs}
\usepackage{todonotes}
\usepackage{enumitem}
\usepackage{adjustbox}
\usepackage{afterpage}
\usepackage{float}
\usepackage{dsfont}
\usepackage{epigraph}
\usepackage{scalerel}
\usepackage[all,2cell]{xy}

\usepackage{bbm}
\usepackage[bbgreekl]{mathbbol}
\usepackage[T1]{fontenc}
\usepackage{kpfonts,baskervald}

\usepackage{import}
\usepackage{physics}
\usepackage{slashed}
\usepackage{IEEEtrantools}
\usepackage{cleveref}

\usepackage{amsthm}
\theoremstyle{definition}
\newtheorem{theorem}{Theorem}[section]
\newtheorem{definition}[theorem]{Definition}
\newtheorem{lemma}[theorem]{Lemma}
\newtheorem{exercise}[theorem]{Exercise}
\newtheorem{example}[theorem]{Example}
\newtheorem{remark}[theorem]{Remark}

\usepackage{caption}
\usepackage{subcaption}

\usepackage{tikz}
\usepackage{tikz-feynman}
\usetikzlibrary{positioning}
\usetikzlibrary{arrows}
\usetikzlibrary{decorations.pathreplacing}
\usetikzlibrary{decorations.markings}
\usetikzlibrary{shapes.multipart}
\usetikzlibrary{shapes.geometric}
\usetikzlibrary{calc}
\usetikzlibrary{decorations.pathmorphing}
\usetikzlibrary{shapes.misc}
\usetikzlibrary{cd}
\usetikzlibrary{angles}
\usetikzlibrary{tqft}

\newcounter{course}
\counterwithin*{section}{course}
\counterwithin{equation}{section}
\counterwithin{figure}{section}
\counterwithin{theorem}{section}
\renewcommand*{\theHsection}{\the\value{course}.\the\value{section}}
\newcommand{\newcourse}[3][]{
    \stepcounter{course}

    \phantomsection
    \addcontentsline{toc}{section}{{\large \Roman{course}. #2}\\
    \texorpdfstring{\textnormal{\textit{#3}}\vspace{-0.2cm}}{}}

    \begin{center}
        {\LARGE {\bf \Roman{course}. #2}}\\

        \vspace{0.2cm}
        \bigskip
        L{\small ECTURED} B{\small Y} #3\\
        \ifthenelse { \equal {#1} {} }  %if no note takers, no second line
    { \mbox{} }   % if #1 == blank
        {\small N{\footnotesize OTES} B{\footnotesize Y} #1}\\
        ---
    \end{center}
}

\DeclareMathOperator{\Aut}{Aut}

\DeclareMathOperator{\Diff}{Diff}
\DeclareMathOperator{\End}{End}

\DeclareMathOperator{\Hom}{Hom}
\DeclareMathOperator{\id}{id}

\DeclareMathOperator{\Map}{Map}

\DeclareMathOperator{\pt}{pt}
\DeclareMathOperator{\Spin}{Spin}

\DeclareMathOperator{\coker}{coker}

\DeclareRobustCommand{\mstrut}{^{\vphantom{1*\prime y\vee M}}}

\DeclareMathOperator{\SO}{SO}

\let\O\relax
\DeclareMathOperator{\O}{O}
\DeclareMathOperator{\U}{U}
\DeclareMathOperator{\SU}{SU}
\DeclareMathOperator{\GL}{GL}

\DeclareMathOperator{\Alg}{Alg}
\DeclareMathOperator{\Bord}{Bord}
\DeclareMathOperator{\Bun}{Bun}
\DeclareMathOperator{\Cat}{Cat}
\DeclareMathOperator{\Edges}{Edge}
\DeclareMathOperator{\Fun}{Fun}

\DeclareMathOperator{\Man}{Man}
\DeclareMathOperator{\Rep}{Rep}
\DeclareMathOperator{\Set}{Set}
\DeclareMathOperator{\Sq}{Sq}
\DeclareMathOperator{\Vect}{Vect}
\DeclareMathOperator{\Vertices}{Vert}
\DeclareMathOperator{\codim}{codim}

\makeatletter

\DeclareSymbolFont{symbolsC}{U}{txsyc}{m}{n}
\def\re@DeclareMathSymbol#1#2#3#4{%
    \let#1=\undefined
    \DeclareMathSymbol{#1}{#2}{#3}{#4}}
\re@DeclareMathSymbol{\varparallel}{\mathop}{symbolsC}{9}

\makeatother

\newtheorem{problem}{\color{blue}Problem}
\newtheorem{corollary}{Corollary}
\newtheorem{proposition}{Proposition}

\abstract{XXX }

\begin{document}
%\maketitle

\thispagestyle{empty}

\begin{center}

{\Huge Simons Lectures on Categorical Symmetries}

\vspace{5cm}

{\Large VOLUME 1}

\vspace{2cm}

\medskip

%{\Large Perimeter Institute Summer School}

%\medskip

%{\Large June 13--17, 2022}

%\vspace{1cm}

{\large Lecturers:}

\medskip

\textit{Clay C{\'o}rdova, Michele Del Zotto, Dan Freed, David Jordan, and Kantaro Ohmori}

\vspace{1cm}

{\large Note Takers:}

\medskip

\textit{Davi Costa, Jonte G\"odicke, Aaron Hofer, Davide Morgante, Robert Moscrop, Elias Riedel Gårding, and Anja \v Svraka}

% Leon Liu, Luuk Stehouwer  :: Constantin
% Arun Debray:: partially for Mike

%\vspace{2cm}

%\medskip

%{\Large Les Diablerets Summer School}

%\medskip

%{\Large September 3--8, 2023}

%\vspace{1cm}

%{\large Lecturers:}

%\textit{, Shu-Heng Shao, and Constantin Teleman}

%\vspace{1cm}
%\medskip
%{\large Note Takers:}

\vspace{1cm}
\medskip

Editors:

\textit{Michele Del Zotto and Claudia Scheimbauer}

\end{center}

\newpage
\thispagestyle{empty}

\phantom{\textit{This page was intentionally left blank}}

\newpage 
\mbox{}
\vfill
\begin{center}
\begin{minipage}{0.8\textwidth}
\begin{center}
We dedicate this volume of lectures to the memory of Jim Simons.  His famous
dialogs with C.~N.~Yang laid the groundwork for the vigorous interaction
between mathematicians and physicists that continues to this day.  The present
volume of summer school lectures is part of this vibrant enterprise, the work
of one of many Simons Collaborations.  We offer these lectures as a tribute to
Jim's influential ideas, his generosity in all forms, and his enduring legacy.
\end{center}
\end{minipage}
\end{center}
\vfill
\mbox{}
\newpage
\thispagestyle{empty}

\phantom{\textit{This page was intentionally left blank}}

\newpage 

\mbox{}
\medskip
\vspace{2.5cm}
\begin{flushright}
%\begin{quote}
There once was a Simons Foundation,\\
that craved categorification,\\
the symmetries grew\\
from effective to blue\\
and escaped to a higher dimension.\\
%\end{quote}
\rule{4cm}{0.4pt}\\
%{\em Simons Collaboration Members}
\end{flushright}

\bigskip
\bigskip
\bigskip
\bigskip
\begin{center}
\includegraphics[scale=0.55, angle=90]{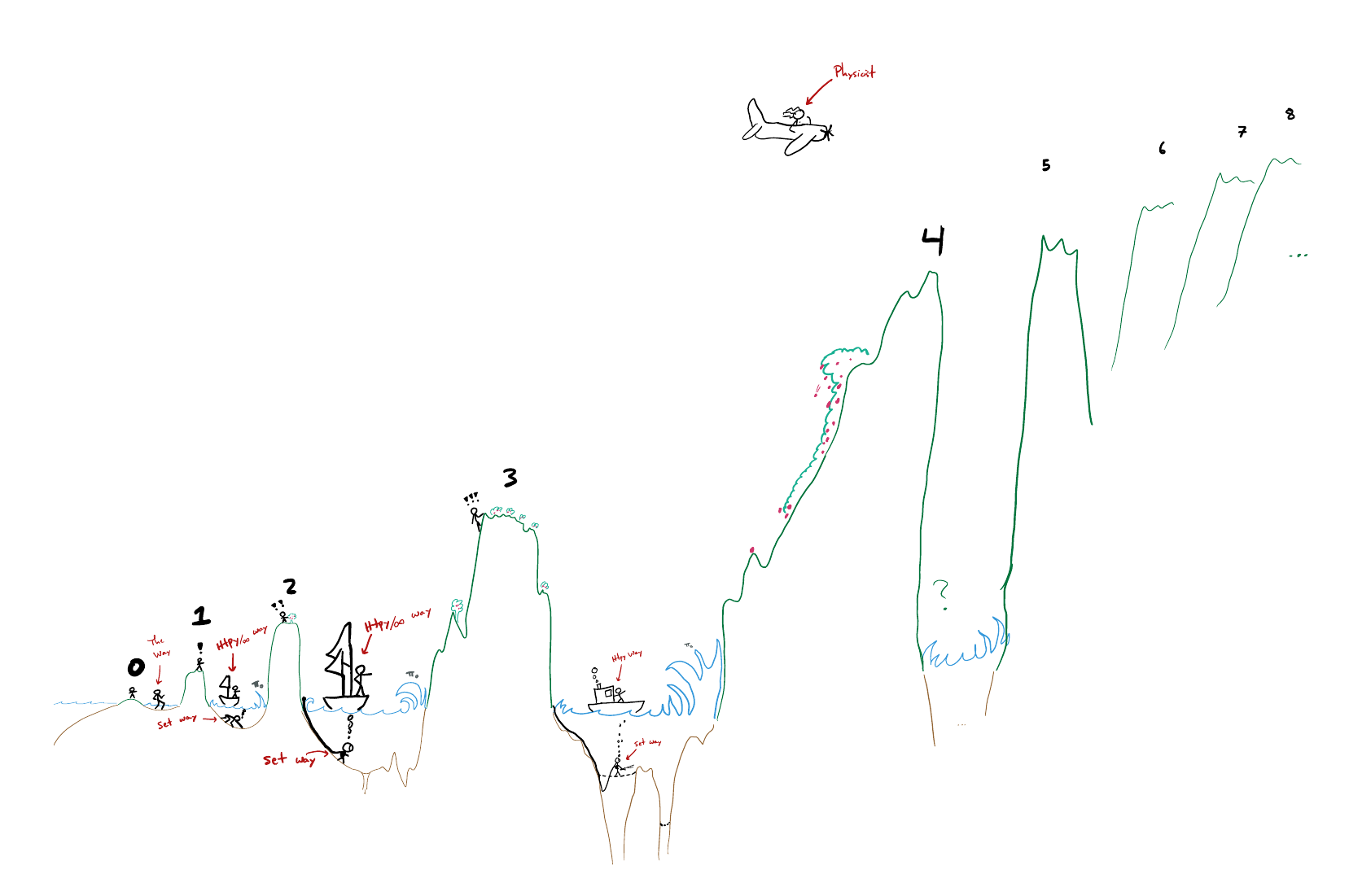}
\end{center}
\begin{flushright}
\tiny illustration by David Ayala
\end{flushright}

%\epigraph{
%There once was a Simons Foundation,\\
%it sought out categorification,\\
%the symmetries grew\\
%from effective to blue\\
%but it escaped to a higher dimension.
%}

\newpage
\thispagestyle{empty}

\phantom{\textit{This page was intentionally left blank}}

%If we need a foreword/opening
%
\phantomsection
\addcontentsline{toc}{section}{{\large Foreword}}

\noindent {\Huge\textbf{Foreword}}

\vspace{5cm}

This is the first volume collecting notes from the Global Categorical Symmetries (GCS) Summer Schools. The first school was organized by Ibrahima Bah, Michele Del Zotto, Theo Johnson-Freyd, Julia Plavnik, and Constantin Teleman at Perimeter Institute in June 13-17, 2022, while the second one was organized by Alberto Cattaneo, Lennart Döppenschmitt, Lukas M\"uller, and Claudia Scheimbauer at the Swiss Map Research Station in Les Diableret in September 3-8, 2023 as part of the Simons Collaboration on Global Categorical Symmetries. Each school consisted of four series of lectures, most of which have been collected in the chapters of this volume. With the exception of the lectures of Dan Freed, each lecture is a patchwork of notes taken by several note-takers that we have merged together in the Sisyphean effort of introducing global categorical symmetries of quantum fields to the masses. Shu-Heng Shao's lectures on non-invertible symmetries at the second school were based on the notes of his lectures at the TASI summer school, which are available at \url{https://arxiv.org/abs/2308.00747}.

\medskip

These volumes are devoted to interested newcomers: we only assume (basic) knowledge of quantum field theory (QFT) and some relevant maths. We try to give appropriate references for non-standard materials that are not covered. 
%We plan to continue publishing proceedings of these schools in the future.
Our aim in this first volume is to illustrate some of the main questions and ideas together with some of the methods and the techniques necessary to begin exploring global categorical symmetries of QFTs.

\medskip

We thank all note-takers for doing a fantastic job in translating the lectures to paper and filling in details. In particular, special thanks goes to Elias Riedel Gårding for in addition adapting Dan Freed's tex code to the given layout.
We thank Arun Debray for providing additional notes and Theo Johnson-Freyd for supporting the project.
Finally, we would like to thank all participants of the schools for their interest, attendance, many questions and comments, and for making the schools a success.

\begin{flushright}
\today

Michele Del Zotto and Claudia Scheimbauer
\end{flushright}

\newpage

\thispagestyle{empty}

\phantom{\textit{This page was intentionally left blank}}

\newpage 

\tableofcontents

%\newpage
%\newyear{\Large Perimeter Institute Summer School}{\Large June 13--17, 2022}

\newpage

{\clearpage%Nicer colours
% \definecolor{airforceblue}{rgb}{0.36, 0.54, 0.66}
% \definecolor{charcoal}{rgb}{0.21, 0.27, 0.31}
% \definecolor{darkslategray}{rgb}{0.18, 0.31, 0.31}
% \definecolor{scgreen}{rgb}{0.0, 0.34, 0.25}
% \definecolor{sgreen}{rgb}{0.2, 0.8, 0.6}
% \definecolor{coolblack}{rgb}{0.0, 0.18, 0.39}
% \definecolor{internationalkleinblue}{rgb}{0.0, 0.18, 0.65}

%%%%%%%%%%%%%%%%%%%%%%%%%%%%%%%%%%%%%%%%%%%%%%%%%%%%%%%%%%%%%%%%%%%
%%%%%%%%%%%%%%%%%%%%%%%%Handy macros%%%%%%%%%%%%%%%%%%%%%%%%%%%%%%%
%%%%%%%%%%%%%%%%%%%%%%%%%%%%%%%%%%%%%%%%%%%%%%%%%%%%%%%%%%%%%%%%%%%
\renewcommand*\rm[1]{\mathrm{#1}}

%For cohomology group macros
\newcommand\HHm{\mathrm{H}}
\newcommand\cHH{\check{\mathrm{H}}}

%For correct infinitesimals.
\newcommand*\diff{\mathop{}\!\mathrm{d}}
\renewcommand*\Diff{\mathop{}\!\mathrm{D}}

\newcommand*\I{\mathrm{i}}

%Calligraphic letters
\newcommand*\cA{\mathcal{A}}
\newcommand*\cB{\mathcal{B}}
\newcommand*\cC{\mathcal{C}}
\newcommand*\cD{\mathcal{D}}
\newcommand*\cE{\mathcal{E}}
\newcommand*\cF{\mathcal{F}}
\newcommand*\cG{\mathcal{G}}
\newcommand*\cH{\mathcal{H}}
\newcommand*\cI{\mathcal{I}}
\newcommand*\cJ{\mathcal{J}}
\newcommand*\cK{\mathcal{K}}
\newcommand*\cL{\mathcal{L}}
\newcommand*\cM{\mathcal{M}}
\newcommand*\cN{\mathcal{N}}
\newcommand*\cO{\mathcal{O}}
\newcommand*\cP{\mathcal{P}}
\newcommand*\cQ{\mathcal{Q}}
\newcommand*\cR{\mathcal{R}}
\newcommand*\cS{\mathcal{S}}
\newcommand*\cT{\mathcal{T}}
\newcommand*\cU{\mathcal{U}}
\newcommand*\cV{\mathcal{V}}
\newcommand*\cW{\mathcal{W}}
\newcommand*\cX{\mathcal{X}}
\newcommand*\cY{\mathcal{Y}}
\newcommand*\cZ{\mathcal{Z}}

\newcommand*\bbC{\mathbb{C}}
\newcommand*\bbN{\mathbb{N}}
\newcommand*\bbR{\mathbb{R}}
\newcommand*\bbZ{\mathbb{Z}}
\newcommand*\Frf{\mathfrak{f}}
\newcommand*\Frg{\mathfrak{g}}
\newcommand*\scrt{\mathscr{T}}
\newcommand*\bord{\mathbf{Bord}_S^{\langle d,d-1\rangle}}
\newcommand*\bordn{\mathbf{Bord}_S^{\langle d,0\rangle}}

%%%%%%%%%%%%%%%%%%%%%%%%%%%%%%%%%%%%%%%%%%%%%%%%%%%%%%%%%%%%%%%%%%
\newcourse[Robert Moscrop]{An introduction to symmetries in quantum field theory}{Kantaro Ohmori}

\section{Introduction}

It is well known that quantum field theories have global symmetries which can be exploited to organise operators according to their representations. Global symmetries are very useful tools to constrain the dynamics of quantum fields. Well-known examples of applications are Ward identities constraining correlation functions with selection rules, or anomalies, that can be exploited to constrain the structure of RG flows and hence the phase diagrams of the quantum fields.\footnote{\ Unless otherwise stated by anomalies here we will always mean 't Hooft anomalies for global symmetries. We refer our readers to the lectures by Clay C{\'o}rdova to learn more about anomalies.}  During the past decade, building upon well-known structures in supergravity and string theory, we have understood how to generalise symmetries to provide conserved quantum numbers for extended operators. The main aim of this series of lectures is to explain such generalization and some of its ramifications. A slogan for this whole program is the following\medskip

\begin{center}
{\bf Main message:} \textit{Generalised symmetries of quantum fields are topological operators.}
\end{center}

\medskip

\noindent Here a few remarks are in order: 
\begin{enumerate}
\item In these lectures we will be working with Euclidean spacetimes, hence we do not pay attention to the distinction between operators and defects. In a Lorentzian quantum field this distinction is meaningful: a defect is an operator extended along the time direction. In presence of a defect, the Hilbert space of the theory obtained from a Hamiltonian quantization changes. In a Lorentzian quantum field one can speak about topological symmetry defects and topological symmetry operators. The two notions become equivalent upon Wick rotation.
\item An operator of a quantum field is topological if its dependence on its support is a function only of the homotopy class of the latter up to (crucial) contact terms. This last statement is what ensures the conservation of the corresponding charge: topological operators commute with insertions of the stress-energy tensor by construction.
\end{enumerate}

\medskip

\noindent In this course we are not going to be systematically presenting a final version of a theory of generalised symmetry, as the latter is currently being developed. Rather, we will first give a context which allows discussing generalizations of symmetries in a straightforward manner, and then give few examples of generalized symmetries with applications. The rough plan of the course is as follows: In section \ref{sec:qftdefn} we will be introducing the notions of QFT as a functor, extended operators, and defects; In section \ref{sec:topsym} we will explain the main message relating topological operators and symmetries; In section \ref{sec:confinement} we will discuss one-form symmetries in gauge theories and their relation to confinement; in section \ref{sec:noninv} we will discuss a simple example of non-invertible symmetries.

\section{Lecture 1: Defining QFT categorically}\label{sec:qftdefn}

\paragraph{References} Readers should also consult other lectures, particularly David Jordan's and references therein. Kapustin's ICM proceeding\footnote{A.~Kapustin, \emph{Topological Field Theory, Higher Categories, and Their Application},
  International Congress of Mathematicians, 4, 2010. \href{https://arxiv.org/abs/1004.2307}{{\tt
   	arXiv:1004.2307}}}  is also informative to locate the context in physics.

\subsection{QFT as a functor}
In this lecture we will only talk about relativistic Euclidean quantum field theory $\rm{QFT}_d$ in $d$ spacetime dimensions. We define what we mean by $\rm{QFT}$ as a functor
\begin{gather}
	\cZ:\bord \rightarrow \mathbf{Vect}.
\end{gather}
That is, a functor from category of bordisms with structures defined by $S$ to the category of vector spaces. For example, $S$ could include differentiable structure, Riemannian metric, spin-structure or possession of a $G$-bundle. More specifically, $\bord$ is the category whose objects are the $(d-1)$-dimensional closed manifolds with $S$-structure and the morphisms are $S$-bordisms between two objects $M$ and $\overline{N}$ where $M,N\in \rm{Obj}(\bord)$ and $\overline{N}$ denotes the orientation reversal on $N$. We present a typical bordism in \cref{fig:bord}

\begin{figure}[h]
	\centering
    \begin{tikzpicture}        
        \draw[rounded corners =8pt,gray] (0.05,1,-0.2) -- (1.5,1.2,0) -- (3,1.5,0)-- (4.5,2.4,0) -- (6,3,-0.2);
        
        \draw[rounded corners=8pt, gray] (-0.05,-1,0.2) -- (1.5,-1.2,0) -- (3,-1.5,0) -- (4.5,-2.4,0) -- (5.9,-3,0.2);
        
        \draw[rounded corners=10pt, gray] (6,-1,-0.2) -- (5,-0.7,0) --(4,0,0)-- (5,0.7,0) --(6,1,0.2);
        
        \draw[rounded corners=8pt, gray] (1.5,0,0) -- (2.25,-0.3,0) -- (3,0,0);
        \draw[rounded corners=8pt, gray] (1.8,-0.1,0) -- (2.3,0.1,0) -- (2.7,-0.1,0);
        \node[left] at (-0.4,0,0){\textcolor{teal}{$M$}};
        \node[right] at (6.2,0,0){\textcolor{red}{$\overline{N}$}};
        
        \draw[teal] (0,1,0) 
        \foreach \t in {5,10,...,355}
            {--(0,{cos(\t)},{sin(\t)})}
        -- (0,1,0);
        
        \draw[red] (6,2,1)
        \foreach \t in {5,10,...,355}
            {--(6,{2+cos(90+\t)},{sin(90+\t)})}
        -- (6,2,1);
        \draw[red] (6,-2,1)
        \foreach \t in {5,10,...,355}
            {--(6,{-2+cos(90+\t)},{sin(90+\t)})}
        -- (6,-2,1);
        \node[above] at (3,1.6,0){\textcolor{gray}{$W$}};
    \end{tikzpicture}
	\caption{Picture of a bordism with $\partial W=M\sqcup \overline{N}$.}\label{fig:bord}
\end{figure}

In summary, $\cZ$ takes a $(d-1)$-dimensional manifold $M$ and returns a vector space $\cZ(M)$ which we call the state space. Additionally, $\cZ$ takes an $S$-bordism $W$ and returns a linear map 
\begin{gather}
	\cZ(W):\cZ(M)\rightarrow \cZ(N).
\end{gather}
Furthermore, $\cZ$ must be monoidal; meaning that it must preserve the `tensor-product' structure of both categories
\begin{gather}
	\cZ(M_1\sqcup M_2) = \cZ(M_1)\otimes \cZ(M_2).
\end{gather}
\begin{example}
	For $M\in\rm{Obj}(\bord)$, then we have $\cZ(M\times [0,\tau])=U_M(\tau)$ is just the time evolution operator. The functoriality of $\cZ$ then gives us
	\begin{gather}
		U_M(\tau_2)\circ U_M(\tau_1)=U_M(\tau_1+\tau_2).
	\end{gather}
	This allows us to write $U_M(\tau)$ as $e^{-\tau H_M}$ where $H_M$ is the Hamiltonian.  
	
	Alternatively, one can bend the cylinder to obtain a bordism from $M\sqcup \overline{M}$ to the empty set. $\cZ$ then gives us a pairing
	\begin{gather}
		\cZ(\tilde{W}):\cZ(M)\otimes \cZ(\overline{M})\rightarrow \bbC.
	\end{gather} 
	This gives us the analogue of the inner product on the state space. Both of these bordisms are drawn in \cref{fig:cyclin}.
		\begin{figure}
	\centering
    \begin{tikzpicture}
    		\draw[teal] (-3,0,1)
        \foreach \t in {5,10,...,355}
            {--(-3,{cos(90+\t)},{sin(90+\t})}
        -- (-3,0,1);
        \draw[red] (0,0,1)
        \foreach \t in {5,10,...,355}
            {--(-0,{cos(90+\t)},{sin(90+\t})}
        -- (0,0,1);
        \draw[gray] (-2.95,1,-0.2) -- (0.05,1,-0.2);
        \draw[gray] (-3.05,-1,0.2) -- (-0.05,-1,0.2);
        \node[above] at (-1.45,1.05,0){\textcolor{gray}{$M\times[0,\tau]$}};
        \node[left] at (-3.4, 0, 0){\textcolor{teal}{$M$}};
        \node[right] at (0.4, 0, 0){\textcolor{red}{$\overline{M}$}};
        
      	\draw[teal] (3,2,0.5)
        	\foreach \t in {5,10,...,355}
            {--(3,{2+0.5*cos(90+\t)},{0.5*sin(90+\t)})}
        -- (3,2,0.5);
        
        \draw [teal] (3,-2,0.5)
        \foreach \t in {5,10,...,355}
            {--(3,{-2+0.5*cos(90+\t)},{0.5*sin(90+\t)})}
        -- (3,-2,0.5);
            
        \draw[gray] (3.06,2.54,0) 
        \foreach \t in {5,10,...,175}
            {--({2.95+2.55*sin(\t)},{2.55*cos(\t)},0)}
        -- (2.94, -2.54,0);
        
        \draw[gray] (2.92,1.46,0) 
        \foreach \t in {5,10,...,175}
            {--({2.95+1.47*sin(\t)},{1.47*cos(\t)},0)}
        -- (3.06, -1.47,0);
        
        \node[left] at (2.8, 2,0){$\textcolor{teal}{M}$};
        \node[left] at (2.8, -2,0){$\textcolor{teal}{\overline{M}}$};
        \node[right] at (5.6,0,0){\textcolor{gray}{$\tilde{W}$}};
    \end{tikzpicture}
    \caption{L{\small EFT}: A simple cylindrical bordism corresponding to time evolution. R{\small IGHT}: A bent cylindrical bordism which can be seen as an inner product.}\label{fig:cyclin}
	\end{figure}
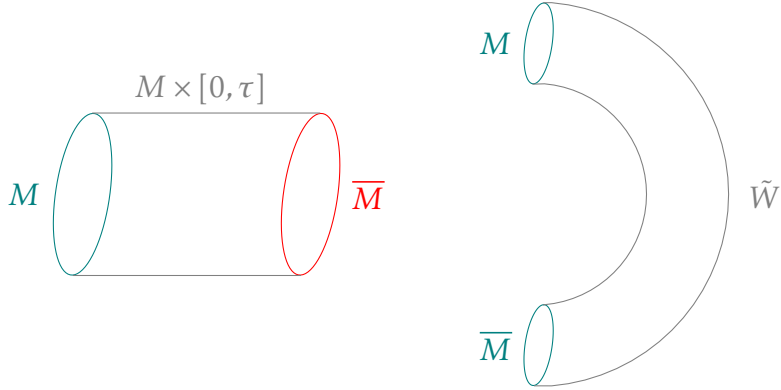
\end{example}

\subsection{Finite gauge theory}\label{sec:finite}
Now consider untwisted finite gauge theory. We first fix the structure $S$ to the orientation, that ensure the objects of $\bord$ are oriented manifolds without metric, and fix $p\in \bbZ$ with $-1 \leq p \leq d-1$. Then $G$ will be a finite abelian group. Then we have
\begin{gather}
	\cZ(M) = \langle \HHm^{p+1}(M,G)\rangle_\bbC,
\end{gather}
which is the $\bbC$-span of a finite set. In other words, for each element of $A_i\in\HHm^{p+1}(M,G)$ we get a basis state $|A_i\rangle$ of the state space. Then, if we denote the inclusion map to a submanifold $\cM$ as $i_\cM$, we view the action on the morphism as
\begin{gather}
	\cZ(W)|A_i\rangle = c(W)\sum_{A\in\HHm^{p+1}(W,G)} |i_N^* \overline{A}\rangle,
 \label{eq: KO finite gauge theory partition function}
\end{gather}
where we only sum over the states such that $i_M^* A=A_i$ so we restrict to states with the given initial condition $A_i$. This provides us with the analogue of the path integral for a finite gauge theory with trivial action. 

In \eqref{eq: KO finite gauge theory partition function}, $c(W)$ is the constant it is
\begin{equation}
    c(W) = \prod_{i=0}^p \lvert H^{p-i}(W,G)\rvert^{(-1)^i}.
    \label{eq: KO partition function constant}
\end{equation}
This prefactor is needed for the functoriality of $Z$. 
For example, take $p=0$, $W = M \times S^1$ with $M$ closed $d-1$-dim manifold. We want $Z(W) = \mathrm{Tr}Z(M) = \lvert H^{p+1}(M,G)\rvert$, which is realized by the above choice. 

\medskip
\noindent {\bf Problem 1.} Take $d=2$, $p=0$ and $G=\bbZ_2$. Now calculate $\cZ(S^1)$ and the action of $\cZ$ on the bordism given by a `pair of pants' or `co-pair of pants'.

\section{Lecture 2: Extended operators and defects: Extended QFT$_d$}\label{sec:topsym}
\subsection{Extension to higher categories}
As formulated above, we cannot easily define extended operators or defects. In order to do so, we have to enhance $\bord$ to something more intricate that can account for these extended objects. This is achieved by using a {\it higher category}.

\paragraph{References} Readers are strongly encouraged to read the original paper.\footnote{D.~Gaiotto, A.~Kapustin, N.~Seiberg and B.~Willett, \emph{Generalized Global Symmetries},  JHEP 02
(2015) 172, \href{https://arxiv.org/abs/1412.5148}{{\tt
   	 	arXiv:1412.5148}}}

To motivate the use of a higher category, let us consider what we would want from $\cZ$ on closed manifolds of increasingly high codimension. 
\begin{itemize}
	\item {\bf Codimension 0.} We already covered this case, it is simply a bordism from the $d$-dimensional empty manifold to itself. As such, we have that $\cZ(M_d)\in \bbC$. 
	\item {\bf Codimension 1.} Again this case is familiar, we want $\cZ(M_{d-1})$ to give a vector space.
	\item {\bf Codimension 2.} Extrapolating the above two cases, here we now want something that is to vectors spaces as vector spaces are to their ground field. This is an object called a $2$-vector space, which is in itself a category.
	\item {\bf Codimension $\mathbf{n}$.} Continuing the above process will tell us that at codimension-$n$ we are looking for an $n$-vector space. For $n>2$, this is a higher category.
\end{itemize}
With these considerations in mind, we see that the correct structure to enhance $\bord$ to is a $d$-category whose objects are points, morphisms are bordisms of points, 2-morphisms are bordisms between the $1$-dimensional bordisms and so on. We will denote this as $\bordn$. Now we can state that an extended QFT in $d$-dimensions is given by a functor
\begin{gather}
	\cZ:\bordn \rightarrow \mathscr{C}_d,
\end{gather}
where $\mathscr{C}_d$ is an appropriate $d$-category. Furthermore, $\cZ$ should still preserve the monoidal structure given by disjoint unions.

In particular, we can think of $a\in\rm{Obj}(\cZ(S^{q-1}))$ as a codimension-$q$ defect/extended operator. In the non-TQFT setting, we should take the radius of the sphere to zero in order to maintain this identification. An equivalent formulation, which we do not delve deeply into here, is that of decorated bordisms.

Now consider a closed manifold $M$ with two defects of codimension-$q$ inserted on $M$. Then we can imagine cutting these out of the manifold by using a $(q-1)$-sphere. Then evaluating $\cZ$ on this set-up, we obtain the correlation function of the two defects.

Similarly, we can consider a junction of defects within $M$. We can then obtain a more intricate correlation function by wrapping the junction with an appropriate bordism. For example, a junction of three lines can be wrapped by a pair of pants, then the evaluation of $\cZ$ gives us the correlation function of the three line defects in such a configuration.

More generally, between two $(d-1)$-dimensional manifolds $M$ and $N$ we can insert defects not only inside the bordism itself, but also extend such defects to the boundaries; that is, to $M$ and $\overline{N}$ themselves. In this case, we have that $\cZ(W):\cZ(M)\rightarrow\cZ(N)$ also pushes forward the defects from the boundaries in accordance to their extension in the bulk.

We say that $a$ is topological if the evaluation of $\cZ$ on a closed manifold $M$ with an insertion of $a$ depends only topologically on the  submanifold the defect is inserted on.

\medskip 
\noindent {\bf Finite gauge theory.} Returning to the example in \cref{sec:finite}, here we take $A\in \HHm^{p+1}(M,G)=[M,B^pG]$. In this set-up, he defects arrange in to two types. The first of which are defects of Wilson type, where we have
\begin{gather}
	w_D=\phi\left(\I\int_D A\right),
\end{gather} 
where $\phi\in G^\vee=\rm{Hom}(G,\rm{U}(1))$. The second type are called disordered defects. In this case, we excise the tubular neighborhood $ND$ of $D^{p+2}$ from the total space $M$. Locally it is $D^{p+2} \times S^{p+1}$, and along $S^{p+1}$ direction we enforce
\begin{gather}
	\int_{S^{p+1}} A = g \in G
\end{gather}
when summing over background fields in the evaluation of the partition function.\footnote{
When evaluating $Z$, the cohomology in equation \eqref{eq: KO partition function constant} has to be replaced by the cohomology relative to the boundary of $DN$.
}

\medskip
\noindent {\bf Problem 2.} Evaluate $\cZ$ on $S^{d+1}$ with a Wilson line inserted upon an $S^{p+1}$ and a `t Hooft line inserted upon an $S^{d-p-2}$ which are Hopf linked. If need be, take $d=3$ and $p=0$ so this is a genuine Hopf link of two $S^1$.
\subsection{Topological operators and symmetry}
{\bf Theme.} The generalised symmetries in QFT are equal to, and defined by, topological operators and defects.

Assume we have a codimension-1 topological operator $U$. Then we can insert this operator along any cylinder without extension to the boundaries. Now we define an operator $\hat{U}$ to be the operator in the limit where the length of the cylinder shrinks to zero. Then $\hat{U}$ is a map from $\cZ(M)$ to itself. However, since $U$ is topological, we could move it to any given time slice in the cyclinder before shrinking it. This tells us that $[\hat{U},H]=0$. 

This $\hat{U}$ isn't random, however. In fact, there exist operators that are consistent with time evolution and commute with the Hamiltonian, but do not arise from a symmetry. An example of this would be an operator which projects to a specific eigenspace of $H$.

\begin{example}
	In perturbative QFT, continuous symmetries of $S$ are captured by Noether's theorem from which we obtain a $(d-1)$-form current $j$ that satisfies $\diff j=0$. We then build a defect operator as
	\begin{gather}
		U_\alpha [D] = e^{\I\alpha \int_D j},
	\end{gather}
	where $\alpha\in \bbR/2\pi\bbZ$. This is topological since $j$ is conserved. Generally, when there is a symmetry group $G$, then we can build defects $U_g[D]$ labelled by $g\in G$ of codimension-1 by enforcing a twist on the Hilbert space when crossing over $D$. These topological operators have a fusion rule given by
	\begin{gather}
		U_{g_1} U_{g_2} = U_{g_1\cdot g_2}.
	\end{gather}
In particular, this is an invertible symmetry precisely because the fusion respects the group multiplication law. All in all, we have that the conventional symmetries of a system are captured by codimension-1 invertible topological operators. As such, we call these operators symmetry operators.
\end{example}

\noindent {\bf Backgrounds.} General insertion of conventional discrete symmetry operators can be understood as a network of junctions on a manifold that respects the group multiplication law. We can conveniently describe such a network by a cocycle $B\in Z^1(M,G)$ which we call a background field.

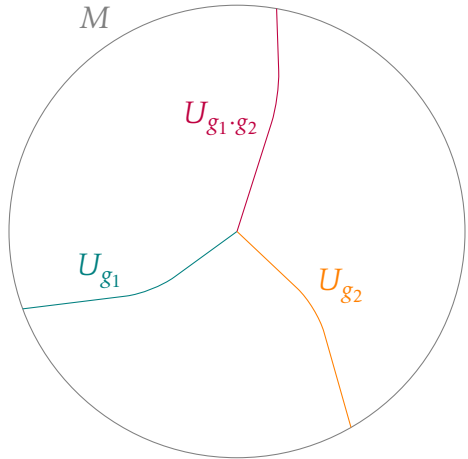
\begin{figure}
    \centering
    \begin{tikzpicture}[scale=3]
            \draw[gray] (0,0) circle (1cm);
            \draw[teal,rounded corners=10pt] ({cos(200)},{sin(200)}) -- ({cos(200)/2+0.1},{sin(200)/2-0.1}) -- (0,0);
            \draw[purple,rounded corners=10pt] ({cos(80)},{sin(80)}) -- ({cos(80)/2+0.1},{sin(80)/2+0.1}) -- (0,0);
            \draw[orange,rounded corners=10pt] ({cos(300)},{sin(300)}) -- ({cos(300)/2+0.1},{sin(300)/2+0.1}) -- (0,0);
            \node[above] at (-0.6,-0.3){\textcolor{teal}{$U_{g_1}$}};
            \node[right] at (0.3, -0.24){\textcolor{orange}{$U_{g_2}$}};
            \node[left] at (0.15,0.5){\textcolor{purple}{$U_{g_1\cdot g_2}$}};
            \node[left] at ({cos(120)},{sin(120)+0.08}){\textcolor{gray}{$M$}};
    \end{tikzpicture}
    \caption{A junction of codimension-1 symmetry operators. Such a configuration is captured by a cocycle $B\in Z^1(M,G)$.}
    \label{fig:junc}
\end{figure}

On the other hand, when $G$ is continuous, one can have a non-flat $G$-connection describing the general set-up instead. The action is then shifted as
\begin{gather}
	S\mapsto S+\I\int B\wedge j+\ldots.
\end{gather}
The key takeaway is that QFTs come with an `intrinsic' symmetry defined by topological operators.

\section{Lecture 3: One-form symmetries and confinement in gauge theory}\label{sec:confinement}
Previously, we've talked about conventional symmetries mostly. By this we mean that we've talked about codimension one invertible topological operators mainly. But what if we start working at higher codimension?

\paragraph{References} The original reference is again\footnote{D.~Gaiotto, A.~Kapustin, N.~Seiberg and B.~Willett, \emph{Generalized Global Symmetries},  JHEP 02
(2015) 172, \href{https://arxiv.org/abs/1412.5148}{{\tt
   	 	arXiv:1412.5148}}}.

Consider a set of codimension-$(p+1)$ invertible topological operators. The symmetry associated with these operators are called $p$-form symmetries. In this section we will aim to understand the $p=1$ case which is important in gauge theory.
\subsection{Topological operators in Yang-Mills}
In the framework of continuous gauge theory, we have a gauge field $A$ which we understand as a $G$-connection. The associated field strength is then given by
\begin{gather}
	F=\diff A + A \wedge A,
\end{gather}
from which we can define the Yang-Mills action
\begin{gather}
	S_{\rm{YM}} = \int \rm{Tr}\left(\tfrac{1}{g}^2 F\wedge *F + \theta F\wedge F\right).
\end{gather}
Similarly, one can write down the partition function as
\begin{gather}
	\cZ[X] = \int_{\cC_X} [\diff A] e^{-S_{\rm{YM}}[A]},
\end{gather}
where $\cC_X$ is the space of $G$-connections on $X$ modulo gauge equivalence. We now ask about the possible topological operators in this set-up.  

When $G$ is abelian, the connection $A$ gives rise to a topological operator classified by $F/2\pi \in \HHm^2(X,\bbZ)$. More generally, for $G$ non-abelian, we have a topological operator classified by $w_2(A)\in\HHm^2(X,\pi_1(G))$. Focussing on this case,
we can write a dimension-$2$ topological operator as
\begin{gather}
	\omega[D] = \phi\left(\int_D w_2(A)\right),
\end{gather}
with $\phi\in\pi_1(G)^\vee$. This $\phi$ defines a one-form symmetry of the form $B\pi_1(G)^\vee$ called the magnetic one-form symmetry.

The second type of topological operator we can form is of disordered type, called the Gukov-Witten operator. We take $A$ to be singular along a manifold of dimension two which we denote by $D^2$. Then we have to demand that
\begin{gather}
	\rm{PExp}\left( \oint A\right) = g \in Z(G),
\end{gather}
where we have integrated over a surface wrapping $D^2$. We therefore have a background field $B\in\HHm^2(X,Z(G))$ for which we can write a partition function
\begin{gather}
	\cZ[X,B] = \int_{\overline{\cC}_X, w_2(A)=B} [\diff A]e^{-S_{\rm{YM}}[A]}
\end{gather}
where $\overline{\cC}_X$ is the space of $G/Z(G)$-connections on $X$ modulo gauge equivalence. This is called the electric one-form symmetry.

In the case of usual symmetries, the operators charged under the symmetry are simply local operators. But for a one-form symmetry, the charged operators are lines. In our case we have a non-topological line operator given by a Wilson loop
\begin{gather}
	W_R(\gamma) = \rm{Tr}\; \rm{PExp}\left[ \oint_\gamma A\right],
\end{gather}
where $R$ is a representation and $\gamma$ a closed loop. We also recognise the holonomy of $A$ around $\gamma$ inside the trace. Given such a Wilson line, we can link it with the topological operator defining the electric one-form symmetry. By deforming this topological operator and allowing to cross the Wilson line results in picking up an element $g$ in the centre of $G$ within the Wilson loop. Explicitly, we have
\begin{gather}
	W_R'(\gamma) = \rm{Tr}\;\left( \rm{PExp}\left[ \int_x^{x_0} A\right] \cdot g \cdot \rm{PExp}\left[ \int^x_{x_0} A\right]\right),
\end{gather}
where $x_0$ is the base point of integration and $x$ is the point at which we have crossed the loop and the topological operator. Since $g$ is central, we end up with
\begin{gather}
	W_R'(\gamma) =\left(\frac{\rm{Tr}_R\; g}{\rm{Tr}_R\rm{\; id.}}\right) W_R(\gamma).
\end{gather}
By taking $R$ to be an irreducible representation of $G$, the prefactor we pickup is just a phase.
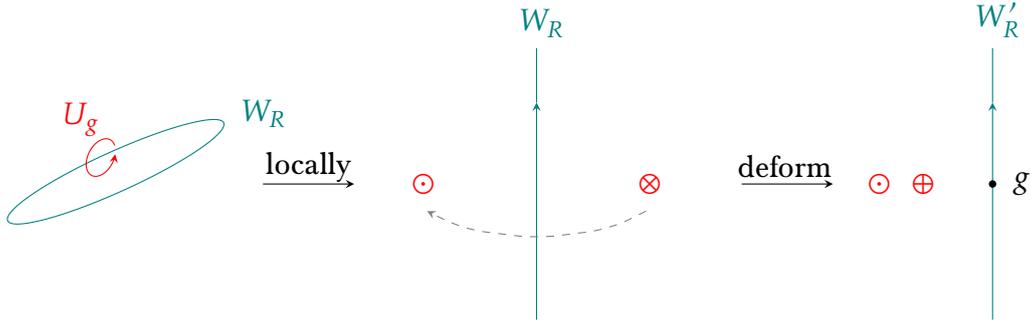
\begin{figure}
    \centering
    \begin{tikzpicture}[scale=1.2]
            \draw[teal,-stealth] (0,0) -- (0,2.4);
            \draw[teal] (0,2.4) -- (0,3);
            \node[right,red] at (1,1.5){$\otimes$};
            \node[left,red] at (-1,1.5){$\odot$};
            \draw[gray,-stealth, dashed, rounded corners=8pt] (1.2,1.2) -- (0.8,1) -- (0,0.9)-- (-0.8,1) -- (-1.2,1.2);

            \draw[teal,-stealth] (5,0) -- (5,2.4);
            \draw[teal] (5,2.4) -- (5,3);
            \draw[black, fill=black] (5,1.5) circle (1pt);
            \node[right] at (5.1,1.5){$g$}; 
            \node[left,red] at (4,1.5){$\odot$};
            \node[right,red] at (4,1.5){$\oplus$};
            
            \node[teal,above] at (0.05,3){$W_R$};
            \node[teal,above] at (5.05,3){$W_R'$};
            
            \draw[teal] (-4.5,1.25,-1) 
            \foreach \t in {5,10,...,355}
                {--({-5+0.5*cos(\t)},{1.25+0.5*sin(\t)},{-1+2.8*sin(\t)})}
            -- (-4.5,1.25,-1);
            \draw[red, stealth-] ({-4.72+0.15*sin(80)},{1.85+0.15*cos(80)},{0.15-0.15*cos(80)})
            \foreach \t in {5,10,...,330}
                {--({-4.72+0.15*sin(80+\t)},{1.85+0.15*cos(80+\t)},{0.15-0.15*cos(80+\t)})};
            \node[above,teal] at (-3,2,0){$W_R$};
            \node[above,red] at (-5,1.9,0){$U_g$};
            
            \draw[-stealth] (-3,1.5,0) -- (-2,1.5,0);
            \draw[-stealth] (2.25,1.5,0) -- (3.25,1.5,0);
            \node[above] at (-2.5,1.4,0){locally};
            \node[above] at (2.72,1.45,0){deform};
    \end{tikzpicture}
    \caption{Deformation of a topological symmetry operator wrapping a Wilson loop. The condition that $\mathrm{PExp}\oint_\gamma A=g$ causes an insertion of $g$ on the Wilson loop at the point of crossing.}
    \label{fig:crossing}
\end{figure}

\begin{example}
	For $G=\rm{U}(1)$, the irreducible representations are classified by $q\in\bbZ$. These integers are then interpreted as the electric charge of the Wilson line $W_q$ which, in turn, can be interpretted as the worldline of an infinitely heavy particle of charge $q$. Since $\rm{U}(1)$ is abelian, we can take any $g\in G$ as our central element, say $g=e^{\I \alpha}$. Then the phase we pickup is simple $\rm{Tr}_q\, g = e^{\I \alpha q}$, where $q$ is now seen as the one-form symmetry charge.
\end{example}

\noindent {\bf Problem 3.} Take $G=\rm{SU}(2)$, $g\in  Z(\rm{SU}(2))=\bbZ_2$ and $R_s$ the spin-$s$ representation. What is the phase obtained for $W_{R_s}$?
\subsection{Confinement in Yang-Mills theory}
Physically, confinement roughly means that we cannot find the colour of individual quarks. In particular, if we have a quark ant-quark pair, confinement characterises the potential $V(L)$ that the pair feel at distance $L$ as $V(L)\propto L$ as $L$ gets large. 

However, it would be better to have a symmetry-based description of the above. We now aim to describe this more mathematically by exploiting the one-form symmetry of the system. To do so, we model a quark anti-quark pair as a Wilson line as in \cref{fig:meson}. The expectation of such a Wilson line is then
\begin{gather}
	\langle W_R \rangle \sim e^{-T V(L)}.
\end{gather}
If the quarks are confined, then we must have
\begin{gather}
	\langle W_R \rangle \sim e^{-TL},
\end{gather}
as $T\rightarrow \infty$ and $L\rightarrow \infty$; this is an {\it area law}. On the other hand, if the quarks are de-confined we must have that $V(L)$ tends to a constant $v$ as $L\rightarrow \infty$, from which we get a {\it perimeter law} for $\langle W_R \rangle$.
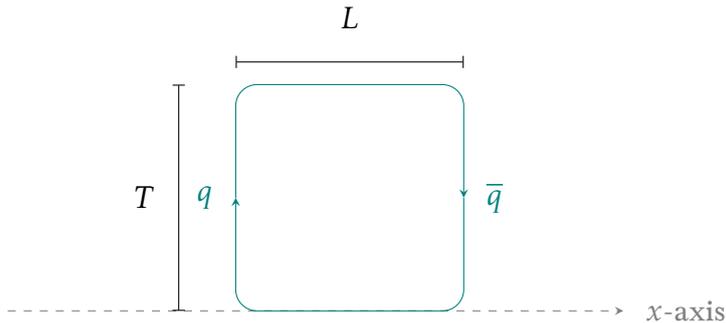
\begin{figure}
	\centering
	\begin{tikzpicture}[scale=1.5]
		\draw[gray,dashed, -stealth] (-3,-1) -- (2.4,-1);
		%\draw[gray, dashed] (0,-1,0) -- (0,-1,-5);
		%\draw[gray, dashed] (0,-1) -- (0,2);
		\node[right] at (2.5,-1){\textcolor{gray}{$x$-axis}};
		\draw[teal, stealth-, rounded corners =8pt] (-1,0) -- (-1,-1) -- (0,-1) -- (1,-1) -- (1,0);
		\draw[teal, -stealth, rounded corners =8pt] (-1,0) -- (-1,1) -- (0,1) -- (1,1) -- (1,0);
		\node[left] at (-1.1,0){\textcolor{teal}{$q$}};
		\node[right] at (1.1,0){\textcolor{teal}{$\overline{q}$}};
		\draw[|-|] (-1,1.2) -- (1,1.2);
		\draw[|-|] (-1.5,-1) -- (-1.5, 1);
		\node[above] at (0,1.4){$L$};
		\node[left] at (-1.6,0){$T$};

	\end{tikzpicture}
	\caption{The Wilson line model of a meson.}\label{fig:meson}
\end{figure}

It is useful to understand the limit of this expectation value as $TL\rightarrow \infty$ while taking $L \gg T$. We do this by taking the vertical segments in \cref{fig:meson} out to infinity and focus on the horizontal segments. This leads to a factorisation of the form
\begin{gather}\label{eq:xax}
	\lim_{TL\rightarrow \infty} \langle W_R(L,T)\rangle = \langle W_R (\ell_{t=0})\rangle \langle W_R (\ell_{t=T})\rangle = |W_R(x-\rm{axis})|^2,
\end{gather}
where $\ell$ refers to a horizontal segment of the loop. If deconfined, the Wilson line causes only the local effect around itself, resulting in a perimeter law of the form $\langle W_R\rangle = e^{-c\, \rm{per}(\gamma)}$ where $\rm{per}(\gamma)$ denotes the perimeter of the support of the loop and $c$ is a constant. With a generic value of $c$, the term on the right hand side is zero. However, we can circumvent this by introducing a fine-tuned local counterterm to define another Wilson loop by
\begin{gather}
	\tilde{W}_R(\gamma) = W_R \, \rm{exp}\left( c\oint_\gamma \diff l\right)= W_R \, \rm{exp}\left( c\, \rm{per}(\gamma)\right).
\end{gather}
With this, the right hand side of \cref{eq:xax} is now a non-zero constant. Now consider the set up in \cref{fig:xax}. By carrying out the same limit we see that
\begin{gather}
	 \langle 0 | U_g^{\dagger} \tilde{W}_R(x-\text{axis}) U_g|0\rangle = e^{\I \alpha} \langle 0|\tilde{W}_R|0\rangle \neq 0
\end{gather}
where we've used the crossing rule between the Wilson loop and the topological symmetry operator. This differs from $\langle 0 | \tilde{W}_R |0\rangle$ by a phase, thus telling us that the action of $U_g$ away from the $x$-axis on states does not preserve the vacuum. This is an example of one-form spontaneous symmetry breaking.
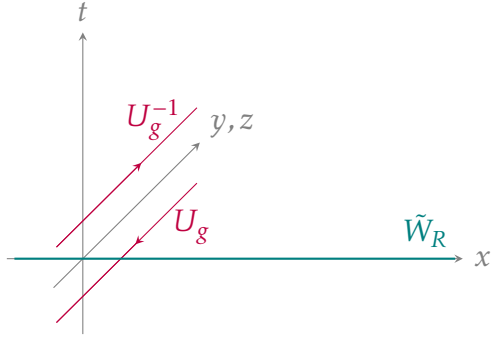
\begin{figure}[ht]
	\centering
	\begin{tikzpicture}
	\draw[gray, -stealth] (0,-1,0) -- (0,3,0);
		\draw[purple] (0,-0.5,0.9) -- (0,-0.5,-3.9);
		\draw[purple, -stealth reversed] (0,-0.5,0.9) -- (0,-0.5,-2); 
		\draw[gray, -stealth] (-1,0,0) -- (5,0,0);	
		\draw[gray, -stealth] (0,0,1) -- (0,0,-4);
		\draw[purple] (0,0.5,0.9) -- (0,0.5,-3.9);
		\draw[purple, -stealth] (0,0.5,0.9) -- (0,0.5,-2); 
		\draw[teal, thick] (-0.9,0,0) -- (4.9,0,0);
		\node[above] at (4.5,0,0){\textcolor{teal}{$\tilde{W}_R$}};
		\node[right] at (5,0,0){\textcolor{gray}{$x$}};
		\node[above] at (0.5,0,-3.8){\textcolor{gray}{$y,z$}};
		\node[above] at (0,3,0){\textcolor{gray}{$t$}};
		\node[above] at (0,0.5,-2.4){\textcolor{purple}{$U_g^{-1}$}};
		\node[right] at (0.1,-0.5,-2.4){\textcolor{purple}{$U_g$}};
	\end{tikzpicture}
	\caption{The Wilson loop and topological symmetry operator set up after taking $T,L\rightarrow \infty$ with $L\gg T$.}\label{fig:xax}
\end{figure}

As an example, in Maxwell theory we have the Coulomb potential $V(L)\propto L^{-1}$. As such, this gives a perimeter law and we see that the theory spontaneously breaks the $\rm{U}(1)$ electric one-form symmetry due to the deconfinement. This also implies that the photon can be understood as a Nambu-Goldstone boson! However, we need to be careful when considering the magnetic one-form symmetry of the Maxwell theory. Na\"ive application of the Nambu-Goldstone theorem would lead you to believe that there are two photons, but in reality a single photon actually takes this into account. The theorem relating the number of broken generators and NG-bosons for 0-form symmetry (in a relativistic system) does not hold for 1-form symmetry. 

Now instead consider the same meson model in confinement. Then we have that area law does not allow a local counterterm to cancel its contribution. The same calculation then tells us that
\begin{gather}
	| W_R(x-\rm{axis})|^2 = 0.
\end{gather} 
In contrast to the deconfined case, we cannot remedy this behavior by a local counterterm dressing $W_R$. Thus, we conclude that the topological symmetry operators associated to the one-form electric symmetry preserves the vacuum. As such, the one-form symmetry is not spontaneously broken during confinement.

In pure Yang-Mills theory we typically expect that confinement (where the symmetry is preserved) occurs at low temperatures while deconfinement (where the symmetry is broken) occurs at high temperature. At some temperature $\Lambda^*$ there should be some kind of sharp phase transition between the two regimes. This is the exact opposite of the situation for a 0-form symmetry! 

\medskip
\noindent {\it Remark.} If we consider gauge theory with some additional matter $\psi$, then the electric one-form symmetry is explicitly broken from $Z(G)$ to the matter preserving subgroup $\{ g\in Z(G): g\cdot \psi= \psi\}$. The intuition here is that we can now end Wilson lines on matter states in the same representation as the line. If we try to probe this line by a topological operator, we see that we can simply unlink the operator and the link and shrink it to zero. 

\medskip
\noindent {\bf Problem 4.} For a 0-form $\bbZ_n$ symmetry that is spontaneously broken and gapped, we see that there are $n$ distinct vacua. What is the analogue of this statement for the $\bbZ_n$ one-form symmetry $B\bbZ_n$? As a hint, consider $\rm{U}(1)$ gauge theory with a charge $q$ scalar field and find a phase where it is gapped and the electric one-form symmetry is broken. 

\section{Lecture 4: Non-invertible chiral symmetry}\label{sec:noninv}

\paragraph{References} Original references are \footnote{Y.~Choi, H.~T.~Lam and S.-H.~Shao, \emph{Noninvertible Global Symmetries in the Standard Model},  Phys. Rev. Lett. 129 (2022) 161601, \href{https://arxiv.org/abs/2205.05086}{{\tt
   	 	arXiv:2205.05086}}}\footnote{C.~Cordova and K.~Ohmori, \emph{Noninvertible Chiral Symmetry and Exponential Hierarchies},  Phys.
Rev. X 13 (2023) 011034, \href{https://arxiv.org/abs/2205.06243}{{\tt
   	 	 	arXiv:2205.06243}}}.

Consider only four dimensional massless QED. That is, $\rm{U}(1)$ gauge theory with a massless charge 1 electron $\Psi=(\psi_L,\psi_R)$. This theory classically has a chiral symmetry given by
\begin{gather}
	\psi_R \mapsto e^{\I\alpha}\psi_R, \quad \psi_L\mapsto \psi_L.
\end{gather}
The ABJ anomaly captures the fact that the chiral symmetry does not persist in the quantum theory. In particular, we have that the corresponding (three-form) Noether current $j$ obeys
\begin{gather}
	\diff j = \tfrac{1}{8\pi^2} F\wedge F. 
\end{gather}
This can be computed from the triangle famous triangle diagram:
\begin{center}
	\begin{tikzpicture}
		\begin{feynman}
			\vertex (a1) {\(\diff j\)};
			\vertex [right=of a1] (a);
			\vertex [above right=of a] (b);
			\vertex [below right=of a] (c);
			\vertex [right=of b] (d) {\(A\)};
			\vertex [right=of c] (e) {\(A\)};
			
			\diagram*{
				(a) -- [fermion] (c) -- [fermion] (b) -- [fermion] (a);			
				(d) -- [photon] (b);
				(e) -- [photon] (c);
				(a1) -- [scalar] (a);
			};
		\end{feynman}
	\end{tikzpicture}
\end{center}
The textbook explanation for this is that the chiral symmetry is quantum mechanically broken. Is this actually the case?

Locally we can write $F\wedge F = \diff(A\wedge\diff A)$. We can therefore locally modify the current to 
\begin{gather}
	j\rightarrow j' = j-\tfrac{1}{8\pi^2} A\wedge \diff A,
\end{gather}
which is locally conserved $\diff j'=0$. Globally, a general $\rm{U}(1)$ bundle has an instanton number given by
\begin{gather}
	I=\int_M F\wedge F \neq 0.
\end{gather}

\noindent {\bf Problem 5.} Find such a non-zero configuration.

\medskip
\noindent Now we imagine writing a general correlator involving $\diff j$ as a summation over various instanton sectors. Explicitly, we write
\begin{gather}
	\langle \diff j(x)\, \cO(y)\, \cO(z)\rangle = \sum_{\{I\}} \langle \diff j(x)\, \cO(y)\, \cO(z)\rangle_I.
\end{gather}
In particular, the correlator for the $I=0$ sector vanishes while the rest do not. However, on $\bbR^4$ or $S^4$ there is no $\rm{U}(1)$-bundle with a non-zero instanton number. This tells us that a selection rule on $\bbR^4$ is satisfied. However, if the gauge group is non-abelian, then we can have instantons even on $\bbR^4$.

Now consider the non-topological operator
\begin{gather}
	C_\alpha = e^{\I \alpha \int_\Sigma j}.
\end{gather}
What we want is some form of corrected operator heuristically of the form
\begin{gather}
	D_\alpha = e^{\I \alpha \int_\Sigma (j-A\wedge \diff A)},
\end{gather}
but this is not globally gauge invariant. But if we have $\alpha \in 2\pi\bbZ$, we have a defect of the form
\begin{gather}
	\tilde{D}_\alpha = e^{\I \alpha\int_\Sigma A\wedge \diff A},
\end{gather}
which is invertible and gives rise to the integer quantum Hall effect. This story can be generalised for $\alpha=2\pi p/q$ where we get the fractionalised quantum Hall effect. This helps us arrive at
\begin{gather}
	D_\alpha[\Sigma] = C_\alpha[\Sigma] \times \cA^{q,p}[\Sigma],
\end{gather}
where the multiplication by $\cA^{q,p}[\Sigma]$ means we stack an appropriate TQFT on top of the defect. In particular, $\cA^{q,p}[\Sigma]$ describes the fractional quantum Hall effect. This hints at the fact that if the gauge group is abelian, the ABJ-anomalous chiral symmetry is, in fact, a non-invertible symmetry!

}

\newpage
{\clearpage\newcourse[Elias Riedel Gårding]{Introduction to anomalies in quantum field theory}{Clay Córdova}
\label{cordova}
% --- Local command definitions ---

\newcommand{\eqbox}[3]{\begin{IEEEeqnarraybox}[][#1]{#2}#3\end{IEEEeqnarraybox}}
\newcommand{\eps}{\varepsilon}
\newcommand{\e}{\mathrm{e}}
\renewcommand{\i}{\mathrm{i}}
\newcommand{\Z}{\mathbb{Z}}
\newcommand{\R}{\mathbb{R}}
\newcommand{\C}{\mathbb{C}}
\renewcommand{\O}{\mathscr{O}} % TODO: Conflict with Freed?
\newcommand{\OO}{\operatorname{O}}
\newcommand{\Pin}{\operatorname{Pin}}
\newcommand{\T}{\mathsf{T}}

% Shorter overlines (long bars) from
% https://tex.stackexchange.com/a/537290
\newcommand{\lbari}[1]{\,\overline{\!{#1}}} % overline short italic
\newcommand{\lbar}[1]{\mskip.5\thinmuskip\overline{\mskip-.5\thinmuskip {#1} \mskip-.5\thinmuskip}\mskip.5\thinmuskip} % overline short

% ---

These lectures are based on the works by Gaiotto, Komargodski, Kapustin, Seiberg as well as on work by Córdova, Lam, Freed, Seiberg.\footnote{D. Gaiotto, A. Kapustin, Z. Komargodski, N. Seiberg, \textit{Theta, Time Reversal, and Temperature}, \href{https://doi.org/10.1007/JHEP05(2017)091}{JHEP 05 (2017) 091}, \href{https://arxiv.org/abs/1703.00501}{arXiv:1703.00501} and C. Córdova, D. S. Freed, H. T. Lam, N. Seiberg, \textit{Anomalies in the Space of Coupling Constants and Their Dynamical Applications I}, \href{https://doi.org/10.21468/SciPostPhys.8.1.001}{SciPost Phys. 8 (2020) 1}, 001,\href{https://arxiv.org/abs/1905.09315}{\textt{arXiv:1905.09315}}, \textit{Anomalies in the Space of Coupling Constants and Their Dynamical Applications II}, \href{https://doi.org/10.21468/SciPostPhys.8.1.002}{SciPost Phys. 8 (2020) 1}, 002,\href{https://arxiv.org/abs/1905.13361}{\textt{arXiv:1905.09315}}} We refer our readers to those papers for more detailed references.

{\section{Lecture 1. Invitation: Degenerate ground states in quantum mechanics}\label{cordova}
The setup of quantum mechanics is a Hilbert space $W$ (for example $W =
L^2(\R)$), where quantum states are unit vectors in $W$ modulo the equivalence
\begin{equation}
  w_1 \sim w_2 \qq{if} w_1 = \e^{\i \alpha} w_2
\end{equation}
for some $\alpha \in \R$. Time evolution is generated by a self adjoint positive
hamiltonian operator $H = H^*$.
We work in the Heisenberg picture, where operators evolve by
\begin{equation}
  \O(t) = \e^{\i H t} \O(0) \e^{-\i H t}.
\end{equation}
The eigenvalues of $H$, $E_i$, are energies. Their multiplicity is called the
degeneracy. We are most interested in the \emph{ground state}, the state of
smallest $E$.

We will focus on the degeneracy of the ground state. Our two main questions will
be
\begin{enumerate}
\item When is the ground state degenerate?
\item Is the degeneracy stable under deforming $H$?
\end{enumerate}

\begin{example}
  Consider $n$ particles moving on $\R$. The Hilbert space and Hamiltonian are
  \begin{equation}
    W = L^2(\R^n) \qquad H = -\sum_{i=1}^n \pdv[2]{x_i}
    + V(x_1, \dots, x_n).
  \end{equation}
\end{example}
\begin{theorem}
  Let $V \geq 0$ and $V \in L^2_\mathrm{loc}(\R^n)$ (meaning that $V$ is
  square-integrable on compact sets). Further assume that $V \to \infty$ as
  $\abs{x} \to \infty$. Then the above $H$ has a non-degenerate ground state.
\end{theorem}
\begin{exercise}
  Prove this (see Reed, Simon IV XIII).
\end{exercise}
As a consequence of this, for example a double well potential
(\cref{fig:double-well}) has a unique ground state.

\begin{figure}[h]
  \centering
  \includegraphics[width=.5\textwidth]
  {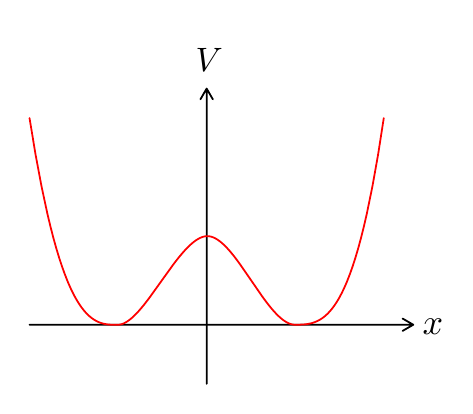}
  \caption{Double well potential.}
  \label{fig:double-well}
\end{figure}

We see that making degenerate ground states is not so easy. The goal for these
lectures will be to develop a theory of invariants---\emph{anomalies}---that
imply robust, degenerate ground states in QFT (this can also be applied in QFT).

A key role in our analysis is played by \emph{symmetry}. In QM there are two
varieties of symmetry:
\begin{enumerate}
\item Unitary transformations $U\colon W \to W$ with $\comm{H}{U} = 0$
\item Antiunitary transformations $T\colon W \to W$ with $\comm{H}{T} = 0$.
  Antiunitarity means that $T(\lambda w) = \lbar{\lambda} T(w)$.
  Any such operator is related to time-reversal symmetry.
\end{enumerate}
The symmetry acts on operators by conjugation:
\begin{equation}
  \O \mapsto U \O U^\dagger
\end{equation}

\emph{Answer to a question:} The above theorem says that it is hard to break
symmetries spontaneously in quantum mechanics. We will develop a theory of how
to circumvent this difficulty.

\begin{example}
  Consider a particle on a circle.
  The variable $x$ is periodic: $x \sim x + 2\pi$.
  We take the Lagrangian
  \begin{equation}
    L = \frac{1}{2} \dot{x}^2 + \frac{\theta}{2\pi} \dot{x}
  \end{equation}
  where $\theta \sim \theta + 2\pi$ is a parameter\footnote{As $\theta \to
    \theta + 2\pi$, $H$ changes by conjugation by a unitary operator. The
    spectrum remains the same.}. We may calculate the canonical Hamiltonian to
  be
  \begin{equation}
    H = \frac{1}{2}\qty(-\i \dv{x} - \frac{\theta}{2\pi})^2.
  \end{equation}
  It is easy to solve this example explicitly. The eigenfunctions of $H$ are
  Fourier modes
  \begin{equation}
    \exp(\i n x), \qquad n \in \Z
  \end{equation}
  with energies
  \begin{equation}
    E_n = \frac{1}{2}\qty(n - \frac{\theta}{2\pi})^2.
  \end{equation}

  It is useful to make a plot of how the energies change as $\theta$ varies
  (\cref{fig:energy-change-plot}).
  \begin{figure}[h]
    \centering
    \includegraphics[width=.7\textwidth]
    {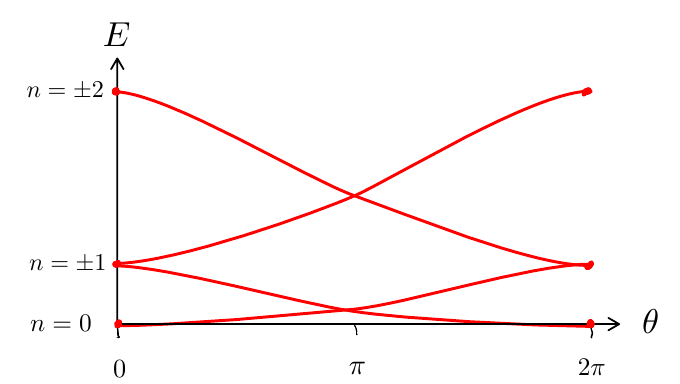}
    \caption{The change in energy levels as $\theta$ varies.}
    \label{fig:energy-change-plot}
  \end{figure}

  At $\theta = 0$, there is a unique ground state given by $n = 0$ and doubly
  degenerate excited states $\pm n$ for $n > 0$. But at $\theta = \pi$,
  something magic has happened: All energy eigenvalues are doubly degenerate.
  We would like to find out why.

  \emph{Wrong idea}: ``Maybe the level crossing is generic in any one-parameter
  family?'' This is wrong:
  \begin{exercise}
    Determine the codimension of the level crossing loci in a multiparameter
    $H$.
  \end{exercise}

  The right idea is to look at the symmetries of the problem:
  \begin{enumerate}
  \item The shift $x \mapsto x + a$ with $a$ a time independent constant.
    This defines unitaries $U_a$, which form the group $U(1)$ under composition.
    This is a symmetry for any value of $\theta$.
  \item Reflections $C\colon x \mapsto -x$. This is a symmetry only at $\theta =
    0$ and $\theta = \pi$.
    That it is a symmetry at $\theta = 0$ is manifest from the Lagrangian. That
    it is a symmetry at $\theta = \pi$ requires a little more thought. The term
    $\frac{\theta}{2\pi} \dot{x}$ seems to change sign, but since $\pi = -\pi
    \pmod{2\pi}$, this is in fact a symmetry.
  \end{enumerate}

  On the eigenfunctions, the shift acts by a phase:
  \begin{equation}
    U_a(w_n) = \exp(\i n a) w_n.
  \end{equation}
  The reflection is more subtle. At $\theta = 0$, it acts as
  $C(w_n) = w_{-n}$. At $\theta = \pi$, it instead acts as
  $C(w_n) = w_{-n + 1}$.

  \begin{exercise}
    Convince yourself that this is how $C$ acts.
  \end{exercise}

  Now let us ask a seemingly benign question: What group is generated by $U_a$
  and $C$? This seems like a simple question, but it in fact has two different
  answers:
  \begin{enumerate}
  \item By construction, when acting on operators (which are generated by
    $\pdv{x}$), we realise $\mathrm{O}(2)$.
  \item Acting on states at $\theta = \pi$, we find an additional phase:
    \begin{IEEEeqnarray*}{rCl}
      C U_a C^{-1}(w_n)
      &=& C U_a(w_{-n+1}) = C\qty(\e^{\i a (-n+1)} w_{-n+1}) \\
      &=& \e^{\i a} U_{-a}(w_n) \IEEEyesnumber
    \end{IEEEeqnarray*}
    rather than the $U_{-a}(w_n)$ we would naively expect. This is a projective
    representation associated to a double cover
    \begin{equation}
      1 \to \Z_2 \to \Pin^-(2) \to \OO(2) \to 1.
    \end{equation}
  \end{enumerate}
  Comments:
  \begin{enumerate}
  \item The discrepancy between states and operators is possible because states
    are \emph{rays} rather than vectors. It is possible to have a cocycle $\mu
    \in H^2(\OO(2), \U(1))$ in multiplication on states. Operators, which
    transform by conjugation, are blind to $\mu$.
  \item All states have the same such $\mu$ ($\mu$ is not a property of an
    individual energy level, but of the whole Hilbert space). Operators act
    transitively on states. $\theta = 0$ has $\mu = 0$ while $\theta = \pi$ has
    $\mu \neq 0$.
  \item Each eigenspace of $H$ forms a projective representation of $\OO(2)$
    with a given $\mu$. In particular, if $\mu \neq 0$, then the dimension of
    each eigenspace must be larger than $1$. This is the advertised
    non-degeneracy. It comes from the basic fact that there is no such thing as
    a one-dimensional projective representation.
  \end{enumerate}
\end{example}

Projective representations, which we have just seen, are the easiest examples of
anomalies, both in QM and in QFT. They are robust under symmetry preserving
deformations.

\begin{exercise}
  Modify the Hamiltonian above by
  \begin{equation}
    H \mapsto H + \frac{\lambda}{2\pi} \cos(2x).
  \end{equation}
  \begin{itemize}
  \item Show that, for small $\lambda$ at $\theta = 0$,
    \begin{equation}
      \abs{E_{+1} - E_{-1}} = \lambda + \order{\lambda^2}
    \end{equation}
    (the degeneracy of the excited states is lifted).
  \item Prove that, for all $\lambda$ at $\theta = \pi$, the degeneracy
    persists.
  \end{itemize}

  (Here, the additional term breaks the shift symmetry except for $a = \pi$, but
  this is all that is needed to ensure degeneracy.)
\end{exercise}

\begin{example}
  Consider real fermions with $\T$ symmetry. $\psi^i(t)$ for $i = 1, \dots, N$
  ($N$ even for simplicity) are classically grassmann, and quantised as
  \begin{equation}
    \{\psi^i, \psi^j\} = 2\delta^{ij}
  \end{equation}
  and the time-reversal symmetry acts as
  \begin{equation}
    \T \psi^i (t) \T^{-1} = -\psi^i(-t).
  \end{equation}
  The Hilbert space has $\dim(W) = 2^{N/2}$. Take the Hamiltonian $H = 0$ so
  that all states are degenerate. What possible $\T$-invariant deformations can
  we add?
  \begin{enumerate}
  \item Quadratic deformations? These would have the form $\Delta H = \i m
    \psi^1 \psi^2$ ($m \in \R$), but then $\T \Delta H \T^{-1} = -\Delta H$, so
    this is prohibited by $\T$-invariance.
  \item Quartic deformations: Here it is helpful to ground the fermions into
    complex pairs:
    \begin{equation}
      a_n = \frac{1}{\sqrt{2}} \qty(\psi^{2m-1} + \i \psi^{2n}).
    \end{equation}
    Then
    \begin{equation}
      \{a_n, a_m^\dagger\} = \delta_{mn}
    \end{equation}
    and each pair of $\psi$s generates a two-component space
    $W_\pm$:
    \begin{IEEEeqnarray*}{rClCrCl}
      a(w_-) &=& 0, &\qquad& a^\dagger(w_-) &=& w_+. \\
      a(w_+) &=& 0, &\qquad& a^\dagger(w_+) &=& w_-.
    \end{IEEEeqnarray*}
    A general state is $w_{\pm\pm\dots\pm}$ with $N/2$ labels.

    For the ``magic'' case of $N = 8$,\footnote{This is the same $8$ as in Bott
      periodicity.} a quartic deformation looks like
    \begin{equation}
      \delta H = 4q \psi^1 \psi^2 \psi^3 \psi^4
    \end{equation}
    ($q \in \R$, $q > 0$), or rewritten (by some algebra) as
    \begin{equation}
      \Delta H = -q \qty(a_1 a_1^\dagger - \frac{1}{2})
      \qty(a_2 a_2^\dagger - \frac{1}{2}).
    \end{equation}

    We find a two-fold degeneracy of aligned and anti-aligned states:
    \begin{equation}
      \eqbox{c}{rLl}{
        w_{++}, w_{--} &\colon\quad& \Delta E = -q \\
        w_{+-}, w_{-+} &\colon& \Delta E = q.
      }
    \end{equation}
  \end{enumerate}
\end{example}

\begin{exercise}
  Show that for $N = 8$ there exists a $\T$-invariant quartic deformation
  leading to a unique ground state.
\end{exercise}

Comments:
\begin{enumerate}
\item We have seen that $8$ is the first value of $N$ such that there is a
  $\T$-invariant deformation breaking the ground state degeneracy.
  $N$ mod $8$ is protected by $\T$.
\item This is another example of an anomaly.
\end{enumerate}
}
{\section{Lecture 2: Anomalies and inflow}
\subsection{Background fields}
Consider the setting of $d$-dimensional QFT (for $d = 1$, this is QM).
The symmetry structure defines the classical background fields ``$A$''.
These define the arguments of the partition function $Z[A]$.
Physicists usually divide symmetries into
\begin{enumerate}
\item ``Internal symmetries'': Finite group, compact Lie group, higher group
    \dots
    For a finite group, $A$ is a connection on an associated bundle, or using
    operators
    \begin{equation}
        \eqbox{c}{s}{finite group \\ ($\pi$-finite space)}
        \to \text{flat connection $A$} \leftrightarrow
        \text{network of symmetry defects}
    \end{equation}
    (where the second arrow is essentially Poincaré duality),
    \begin{equation}
      \eqbox{c}{s}{compact Lie group \\ ($\pi$-finite space)}
      \to \text{general connection $A$} \leftrightarrow
      \eqbox{c}{s}{local data: current correlators, \\
        global data: bundle topology.}
    \end{equation}
    and so on.
\item Spacetime symmetries. These have the background fields
    \begin{itemize}
    \item Lorentz symmetry: metric
    \item Fermion number: spin structure
    \item T-reversal symmetry: Stiefel--Whitney class $w_1$
        (unorientable spacetime)
    \end{itemize}
    We use the letter $A$ to collectively denote all such backgrounds.
\end{enumerate}

\subsection{Gauge transformation of background fields}
The data of $A$ is often subject to gauge redundancies
\begin{equation}
    A \mapsto A^\lambda
\end{equation}
where $\lambda$ is a gauge parameter. For connections, these are standard gauge
transformations, for example
\begin{equation}
    A \mapsto A + \dd{\lambda}
\end{equation}
for $\U(1)$.
(For a finite group, we correspondingly shift the cocycle by a coboundary).

\emph{Key point}: In defining $Z[A]$, we typically pick explicit
representatives (explicit connections). Then we need to ask whether the
partition function is gauge invariant:
\begin{equation}
  Z[A] \overset{?}{=} Z[A^\lambda].
\end{equation}

Gauge invariance of $Z[A]$ is closely related to topological invariance of
symmetry defects.

\begin{example}
  Symmetry group $U(1) \ni \exp(i\varphi)$. The symmetry operator is a
  codimension 1 defect (see \cref{fig:defect-dual-wilson-line}). The associated
  background field is $A = \varphi \delta(\tau - \tau_0) \dd{\tau}$, which we
  find by requiring
  \begin{equation}
    \exp(\oint A) = \exp(i\varphi)
  \end{equation}
  along the dual Wilson line.

  The defect at a specific location $\tau = \tau_0$ gives rise to an explicit
  choice of connection. Moving the defect from $\tau_0$ to $\tau_1$ requires the
  gauge transformation $\lambda = \varphi \qty(\theta(\tau - \tau_1) -
  \theta(\tau - \tau_0))$.

  \begin{figure}[h]
    \centering
    \includegraphics[width=.5\textwidth]
    {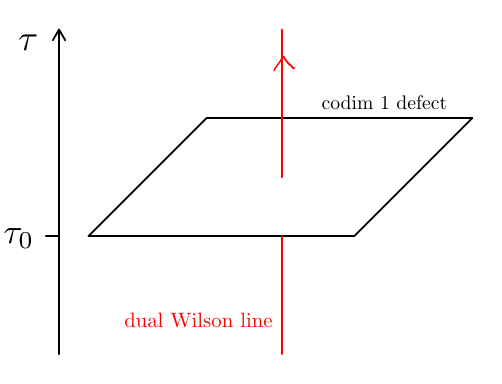}
    \caption{A symmetry defect and an associated dual Wilson line.}
    \label{fig:defect-dual-wilson-line}
  \end{figure}
\end{example}

When we contemplate the possibility that $Z[A] \neq Z[A^\lambda]$, we are
contemplating the breakdown of topological invariance of symmetry defects.
What kind of failure is possible?

\emph{Key idea}: As long as the defects are separated (their supports are not
coincident), we demand topological invariance holds exactly.
Failure can happen at \emph{coincident points}.

For example, in quantum mechanics, two symmetry defects $g$ and $h$ can be moved
back and forth along the time line as long as the do not touch. But it is
possible that when they coincide (fuse), the result is different from the single
symmetry defect $gh$; see \cref{fig:qm-fusion}. This happens exactly when the
representation is projective.
\begin{figure}[h]
  \centering
  \includegraphics[width=.4\textwidth]
  {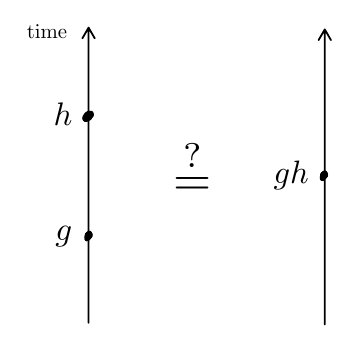}
  \caption{Possibility of an anomaly in quantum mechanics.}
  \label{fig:qm-fusion}
\end{figure}
In a 2d QFT, an analogous diagram is the one shown in \cref{fig:2d-fusion}.
This type of diagram captures \emph{higher group cohomology}, in this case
$H^3$.
\begin{figure}[h]
  \centering
  \includegraphics[width=.6\textwidth]
  {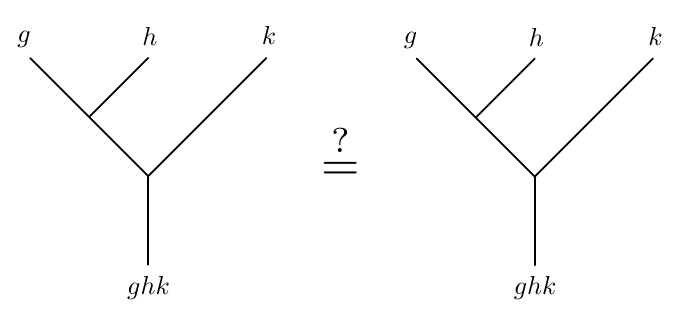}
  \caption{Possibility of an anomaly in a 2d QFT.}
  \label{fig:2d-fusion}
\end{figure}

Make an ansatz allowing $Z$ to change by a phase:
\begin{equation}
  Z[A^\lambda] = Z[A] \exp(-2\pi\i \int_X a(\lambda, A))
\end{equation}
where $X$ is spacetime.
\begin{itemize}
\item We demand that $\alpha(\lambda, A)$ is a \emph{local} functional (meaning
  that it satisfies some cutting and glueing rules). This is where we put in the
  fact that non-invariance cannot come from separated points.

  It is helpful to look again at the $U(1)$ example. Coupling to $A$ means
  including a term
  \begin{equation}
    S \supseteq \int_X A \wedge *J
  \end{equation}
  where $J$ is the 1-form conserved current. A gauge transformation $A \to A +
  \dd{\lambda}$ induces a change in the action
  \begin{equation}
    \delta S = \int_X \dd{\lambda} \wedge * J
    \sim \int_X \lambda \wedge \dd{* J}.
  \end{equation}
  Here $\dd{*J}$ vanishes at separated points; this means that
  \begin{equation}
    \eval{\ev{\dd{*J}(x) \O(y_1) \dots \O(y_n)}}_{A = 0} = 0
  \end{equation}
  if $x \neq y_i$ for all $i$. But we allow contact terms, nonzero at coincident
  points. That $\alpha$ is local means that ambiguity or failure in the
  topological property only happens at coincident points.

\item Why have we enforced that the ambiguity is only in the phase, that is,
  $\abs{Z[A^\lambda]} = \abs{Z[A]}$? In general, the modulus $\abs{Z[A]}$ of the
  partition function has a probability interpretation, which we would like to
  maintain. This is analogous to the situation of rays in quantum mechanics.

  The partition function $Z[A]$ is subject to an ambiguity
  \begin{equation}
    Z[A] \to Z[A] \exp(2\pi\i\int_X \beta(A))
  \end{equation}
  where, again, $\beta$ is local. This is what physicists call the freedom to
  change ``scheme''.
  
  This leads to a cohomology problem. A classic definition of the set of possible
  anomalies is
  \begin{equation}
    \{\alpha\} / (\alpha \sim \alpha + \delta \beta),
  \end{equation}
  the set of $\alpha$s modulo the freedom of changing scheme.
\end{itemize}

\subsection{Inflow}
\textbf{Hypethesis/definition/theorem}: For any anomaly $[\alpha]$ there is a
$(d+1)$-dimensional \emph{invertible} theory (sometimes called a ``classical''
theory, to indicate that nothing fluctuates, or an SPT) with partition function
\begin{equation}
  \exp(2\pi\i\int_Y \omega(A))
\end{equation}
such that, if $\partial Y = X$, then
\begin{equation}
  \exp(2\pi\i\int_Y \omega(A^\lambda) - 2\pi\i\int_Y \omega(A))
  = \exp(2\pi\i\int_X\alpha(\lambda,A)).
\end{equation}

Let us illustrate how this works with a geometric picture: Given $X$, construct
a $(d+1)$-manifold $X \times_\lambda S^1$, the \emph{mapping torus} for
$\lambda$, by propagating $X$ along a new time direction $t$ as in
\cref{fig:inflow-mapping-torus}. Call the fields on the new manifold
$\tilde{A}$, and require that $\eval{\tilde{A}}_{t_i} = A$ and
$\eval{\tilde{A}}_{t_f} = A^\lambda$. Then glue the two endpoints $t_i$ and
$t_f$ together.
\begin{figure}[h]
  \centering
  \includegraphics[width=.8\textwidth]
  {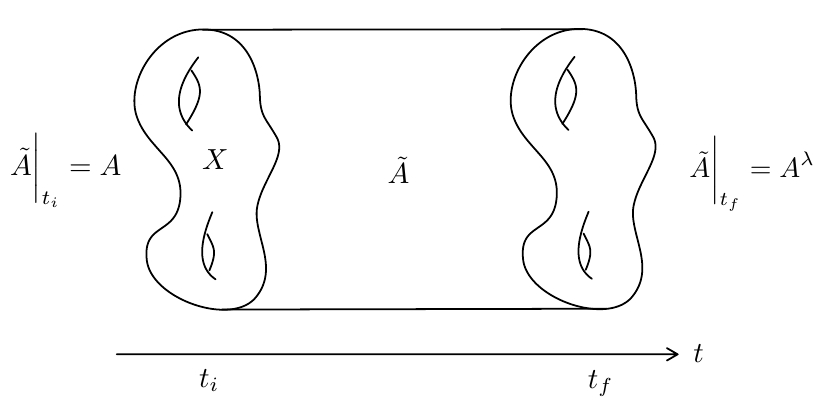}
  \caption{Construction of the mapping torus.}
  \label{fig:inflow-mapping-torus}
\end{figure}
Then the anomalous phase can be expressed as the partition function of the
invertible theory:
\begin{equation}
  \text{anomalous phase}
  = \exp(2\pi\i\int_X\alpha(\lambda,A))
  = \exp(2\pi\i\int_{X \times_\lambda S^1}\omega(\tilde{A})).
\end{equation}

The inflow point of view allows us to define a modified but fully gauge
invariant partition function
\begin{equation}
  \tilde{Z}[A] = Z[A] \exp(2\pi\i\int_Y \omega(A))
\end{equation}
where $\partial Y = X$. This satisfies
\begin{equation}
  \tilde{Z}[A^\lambda] = \tilde{Z}[A],
\end{equation}
and the price we have paid for this is that our original theory is now the
boundary of an invertible field theory.

We have arrived at the following paradigm: An anomalous field theory is the
boundary of an invertible field theory. Anomalies could be defined by invertible
theories.
}
{\section{Lecture 3: Examples and consequences of inflow}

\subsection{Examples}
Last time we constructed the fully gauge invariant partition function
\begin{equation}
  \tilde{Z}[A] = Z[A] \exp(2\pi\i\int_Y \omega(A))
\end{equation}
where $A$ are background fields, $Z[A]$ is the anomalous partition function, and
the added factor describes an invertible theory on a space $Y$ such that
$\partial Y = X$.

\begin{example}[Particle on a circle, revisited]
  For the Lagrangian
  \begin{equation}
    L = \frac{1}{2} \dot{x}^2 + \frac{\theta}{2\pi} \dot{x}
  \end{equation}
  at $\theta = 0, \pi$, where the symmetry is $\OO(2)$,
  consider first backgrounds in $\U(1) \subset \OO(2)$. Let $A = A_0 \dd{t}$.
  Then the modified Lagrangian
  \begin{equation}
    L \to \frac{1}{2} \qty(\dot{x} + A_0)^2
    + \frac{\theta}{2\pi} \qty(\dot{x} + A_0) + k A_0,
    \qquad k \in \Z
  \end{equation}
  is manifestly invariant under
  \begin{equation}
    x \to x - \lambda, \qquad A \to A_0 + \dv{\lambda}{t}.
  \end{equation}
  Now we consider also reflections. Is $\e^{-S}$ invariant under
  \begin{equation}
    C\colon x \mapsto -x, \quad A_0 \mapsto -A_0?
  \end{equation}
  \begin{enumerate}
  \item $\theta = 0$: $L = \frac{1}{2} \qty(\dot{x} + A_0)^2 + kA_0$ is
    invariant if $k = 0$.
  \item $\theta = \pi$: Here
    \begin{equation}
      L = \frac{1}{2} \qty(\dot{x} + A_0)^2
      + \frac{\pi}{2\pi} \qty(\dot{x} + A_0) + k A_0,
      \qquad k \in \Z
    \end{equation}
    \begin{exercise}
      Show that under $C$,
      \begin{equation}
        \e^{-S} \to \e^{-S} \exp(\i(2 k + 1) \oint A_0 \dd{t})
      \end{equation}
      (here $S$ is the Euclidean action and the time coordinate $t$ is compact).
    \end{exercise}
    $C$-invariance would require $2 k + 1$, which is impossible as $k$ is an
    integer.

    \emph{Comment}: Failure of $C$-invariance in the presence of $A$ is the
    anomaly. $\delta S = (2k + 1) \i \oint A \dd{t}$ is the anomalous phase.
  \end{enumerate}
\end{example}

Now let us see the same thing from the inflow viewpoint. Extend $S^1 = \partial
Y$ and extend $A$ into $Y$. The invariant partition function is
\begin{equation}
  \tilde{Z}[A] = Z[A] \exp(\frac{\i}{2} \int_Y F).
\end{equation}
The second factor describes an invertible theory. If $\partial Y$ had been
$\emptyset$, it would simply have been $(-1)^{c_1}$ where $c_1$ is the first
Chern class.
\begin{exercise}
  Verify that $\tilde{Z}[A]$ is $C$-invariant.
\end{exercise}

More generally, a finite symmetry $G$ acts linearly on operators, but
projectively on states, described by $\mu \in H^2(G, \R/\Z)$.
We can view $[A] \in H^1(X, H)$ as a map $X \to BG$. Then
\begin{equation}
  A^*(\mu) \in H^2(X, \R/\Z)
\end{equation}
gives an invertible theory
\begin{equation}
  \label{eq:general-invertible-theory}
  \exp(2\pi\i\int_Y A^*(\mu)).
\end{equation}
\begin{exercise}
  Let $\Z_2^\times$ denote the shift $x \mapsto x + \pi$ (preserved by
  $\cos(2x)$). Unpack \eqref{eq:general-invertible-theory} for
  $G = \Z_2^\times \times \Z_2^C$.
\end{exercise}

\begin{example}[$\T$-invariant fermions]
  Consider again real fermions $\{\psi^i, \psi^j\} = 2\delta^{ij}$; $i = 1,
  \dots N$. We saw in lecture 1 that $\T$-invariance led to degenerate ground
  states unless $N = 0 \pmod{8}$.

  $\T^2 = 1$ requires that the Euclidean spacetime manifold has a $\Pin^-$
  structure. (Why the minus sign? After Wick rotation; this leads to
  $(\text{reflection})^2 = (-1)^F$.

  A useful table:
  $w_1 \in H^1(\cdot, \Z_2)$ and $w_2 \in H^2(\cdot, \Z_2)$ are the first and
  second Stiefel--Whitney classes of the manifold.
  \begin{center}
    \begin{tabular}{lll}
      Possible structures & $w_1$ & $w_2$ \\
      \hline
      $\SO$ & 0 & any \\
      $\OO$ & any & any \\
      $\Spin$ & 0 & 0 \\
      $\Pin^+$ & any & 0 \\
      $\Pin^-$ & any & $w_1 \cup w_1$
    \end{tabular}
  \end{center}

  If $Y$ is a closed $2$-manifold, then
  \begin{equation}
    \label{eq:w2}
    w_2 = \mathrm{euler} \bmod{2} = w_1 \cup w_1.
  \end{equation}
  Therefore any such $Y$ has a $\Pin^-$ structure.
  \begin{exercise}
    Prove \eqref{eq:w2}. Which $2$-manifolds admit a $\Pin^+$ structure?
  \end{exercise}

  \emph{Abstract desctiption of an invertible theory:}
  The $2$-dimensional $\Pin^-$ bordism group is $\Omega_2^{\Pin^-} \cong \Z_8$.
  An invertible theory is given by $Z[Y] \in \U(1)$.
  Compatibility with bordisms:
  \begin{equation}
    Z[Y] \in \Hom(\Omega_2^{\Pin^-}, \U(1)) \cong \Z_8
  \end{equation}
  (the generator is the \emph{ABK invariant of the associated quadratic form}).

  A result, that we do not prove, is that for $N$ real fermions with $\T^2 = 1$,
  the anomaly theory is $N \cdot \text{generator} \in \Hom(\Omega^{\Pin^-},
  \U(1))$.

  \emph{Physical picture}: Consider $N$ 2d real (majorana) fermions $\chi^i$
  with Lagrangian
  \begin{equation}
    L = \i \chi \slashed\partial \chi + \i m \underbrace{\chi_L \chi_R}_{
      \clap{\footnotesize\text{2 chiralities}}}.
  \end{equation}
  \begin{exercise}
    Check that this is $\T$-invariant with mass.
  \end{exercise}
  On a manifold with boundary, $\eval{\chi_L^i}_{\partial Y} =
  \eval{\chi_R^i}_{\partial Y} = \psi^i$. So we have $N$ real fermions on the
  boundary. For large $m$, or equivalently low energies, the $\chi$ are very
  massive, so
  \begin{equation}
    \lim_{\abs{m} \to \infty} Z_\chi(Y)
  \end{equation}
  describes an invertible theory, a ``suitable eta invariant''.
  (In fact, it is generally true that the effective action of the long-distance
  limit of a trivially gapped theory is invertible; we are using this here.)
  In summary, this is a direct construction of an invertible theory whose
  boundary is the theory of $N$ fermions.
\end{example}

\subsection{General consequences of anomalies}
The general theme is that anomalies protect ``non-triviality'' of families of
QFTs related by continuous deformations that preserve the symmetry type
(possible background fields that we can turn on). For example,
\begin{itemize}
\item Dialing a coupling constant (a potential, \dots).
\item Adding massive fields (called spectators in condensed matter theory).
  Think of this as ``bringing down fields from infinitely high mass''.
\item RG flows, triggered by a symmetry preserving operators.
  (This is, in a sense, covered by the previous two points and their inverses.
  RG flow removes massive degrees of freedom and changes coupling constants.)
\end{itemize}

Denote by $\lambda$ a parameter on such a family. Then everything in our inflow
paradigm depends on $\lambda$:
\begin{equation}
  \tilde{Z}_\lambda[A] = Z_\lambda[A] \exp(2\pi\i\int_Y \omega_\lambda(A))
\end{equation}
In particular, $\omega_\lambda$ is the Lagrangian of a family of invertible
theories. Let $[\omega_\lambda]$ denote its deformation class (anything
connected to $\omega_{\lambda = 0}$); this is an invariant of the family.
(Sometimes the word ``anomaly'' is used for $\omega_\lambda$, sometimes for
$[\omega_\lambda]$.)

This is powerful. Often $\omega$ is characterised by discrete data (a level, an
element of a torsional cobordism group). Then $[\omega_\lambda] \neq 0$.
\begin{exercise}
  Show that $[\omega_\lambda] \neq 0$ for the QM examples we have discussed
  earlier.
\end{exercise}

\emph{Key point (anomaly matching)}: For $Z_\lambda[A]$ the anomalous theory,
that is, the boundary of the invertible theory $\omega_\lambda$, if
$[\omega_\lambda] \neq 0$, the the boundary theory cannot be invertible anywhere
in the family (because an invertible theory does not need a nontrivial extension
to one dimension higher).

Let us unpack this for RG flow. The parameter $\lambda$ is the distance scale:
$\lambda = 0$ is the UV and $\lambda = \infty$ is the IR. We can characterise
possible IR behaviours of relativistic QFTs by their mass gap:
\begin{enumerate}
\item Gapless: Massless fields, or more generally an interacting conformal field
  theory
\item Gapped: All particles have mass.
  \begin{enumerate}
  \item Topological QFT
  \item Topological invertible QFT. This is impossible if $[\omega_\lambda]
    \neq 0$
  \end{enumerate}
\end{enumerate}
In summary, if $[\omega_\lambda] \neq 0$, the IR theory must be nontrivial in
some way: either there are massless degrees of freedom, or it is a nontrivial TQFT.
}
{\section{Lecture 4: Anomalies in 4d Yang--Mills theory}

Consider the setting of 4d Yang--Mills theory. For simplicity, take the gauge
group to be $\SU(N)$. The theory is defined by the Euclidean action
\begin{equation}
  S = \frac{1}{g^2} \int \Tr(f \wedge *f)
  - \frac{\i\theta}{8\pi^2} \int \Tr(f \wedge f).
\end{equation}
Here $g$ is the coupling ``constant'' and $\theta \sim \theta + 2\pi$ is an
angle.

The \emph{instanton number} is given by the Chern--Weil formula
\begin{equation}
  I = \frac{1}{8\pi^2} \int_X \Tr(f \wedge f) = c_2 \in \Z,
\end{equation}
the second Chern number.

Physicists believe that the partition function is defined by a functional
integral
\begin{equation}
  Z = \sum_{I \in \Z} \int Da \exp(-\frac{1}{g^2}
  \int \Tr(f \wedge *f) + \i\theta I).
\end{equation}
where $\int Da$ is the integral over the space of connections with a fixed
bundle topology given by $I$. Note that $\theta$ controls how different bundle
topologies contribute to the sum.

The lore (supported by considerable evidence) is that at $\theta = 0$, the IR is
\emph{trivially gapped}, or \emph{invertible}.

\emph{Question}: Can the IR be invertible for all $\theta$?

\emph{Answer}: \textbf{No.} For at least one value of $\theta$, the IR is
noninvertible.

We will prove\footnote{At a physicists' level of rigour} this by finding an
anomaly. For this we first need to discuss the symmetries.

\begin{enumerate}
\item The centre of $\SU(N)$ is isomorphic to $\Z_N$, so there is a one-form
  symmetry $\Z_N^{(1)}$. The charged objects are Wilson lines. (We can use their
  expectation values to detect confinement.) The charge of an irrep $R$ is the
  number of boxes in a Young tableau, modulo $N$.

  A background field for his symmetry is abstractly $A \in H^2(X, \Z_N)$. We
  access it concretely via $\SU(N)/\Z_N$-bundles. For these we have an
  additional characteristic class $w_2 \in H^2(X, \Z_N)$. This is the
  obstruction to lifting the bundle to an $\SU(N)$-bundle.

  \begin{exercise}
    Consider a bundle with structure group $\U(1) \subseteq \SO(3) = \SU(2) /
    \Z_2$. Show that $w_2 = c_1 \bmod 2$, where $c_1 = \frac{F}{2\pi}$.
  \end{exercise}

  The partition function coupled to $A$ is
  \begin{equation}
    Z[A] = \sum_{\substack{I \\ w_2 = A \text{ fixed}}} \int Da\, \e^{-S}
  \end{equation}
  where the sum is over $\SU(N)/\Z_N$-bundles. By summing over $w_2$, we recover
  YM.
\item Time-reversal ($\T$) symmetry. The $\theta$ term is odd under reflection
  (if we change orientation of $X$, $c_2[X]$ changes sign). This means that for
  general $\theta$, $\T$ is not a symmetry. However, at two special points,
  $\theta = 0$ and $\theta = \pi$, it is.

  Note: $\exp(-S) = \exp(-\frac{1}{g^2} \int \Tr(f \wedge *f) + i\theta I)$, so
  at $\theta = 0$ or $\pi$, $\e^{-S}$ is real. This is expected from
  reflection positivity (the Euclidean version of unitarity) and $\T$
  (reflection symmetry).
\end{enumerate}

We will examine the interplay between $\T$ and $\Z_N^{(1)}$. For this we have to
learn about an interesting topic, \emph{fractional instantons}. The quantisation
of $I = \frac{1}{8\pi^2} \int_X \Tr(f \wedge f)$ is different for $\SU(N)$ and
$\SU(N) / \Z_N$. We use the relation
\begin{equation}
  \U(N) \cong \frac{\U(1) \times \SU(N)}{\Z_N}.
\end{equation}
to understand $\SU(N) / \Z_N$-bundles via $\U(N)$-bundles.
For $\U(N)$-bundles we have
\begin{equation}
  H^2(X, \Z_Z) \ni w_2 = c_1 \mod{N}.
\end{equation}
(Here $c_1 = \frac{\Tr(f)}{2\pi}$.) It is a fact that, for a $\U(N)$-bundle,
\begin{equation}
  \Z \ni c_2 = \frac{1}{8\pi^2} \int_X \Tr(f \wedge f) - \Tr(f) \wedge \Tr(f).
\end{equation}
This leads to, for an $\SU(N) / \Z^2$, bundle,
\begin{equation}
  I = c_2 - \frac{N - 1}{2N} \int \widehat{w_2} \cup \widehat{w_2}
\end{equation}
where $\widehat{w_2}$ is the integer lift.
What is the integer lift? The \emph{Pontryagin square} is an operation
\begin{equation}
  P\colon H^2(X, \Z_N) \to H^4(X, \Z_{\gcd(2,N)N})
\end{equation}
It has properties:
\begin{itemize}
\item $\bmod{N}$, $P(w) = w \cup w$.
\item If lifts exist, $P$ is the cup of lifts.
\item If $N$ is even and $X$ spin, $P$ is divisible by two.
\end{itemize}

This gives us a formula for $\SU(N) / \Z_N$-bundles:
\begin{equation}
  I = \text{integer} - \frac{N - 1}{2N} \int_X P(w_2).
\end{equation}
Configurations with non-integer $I$ are sometimes called ``fractional
instantons''.
\begin{exercise}
  Construct an $\SU(3)$-bundle on $T^4$ with connection such that $I$ is
  fractional. (If you feel ambitious, try to make it self-dual. This is one of
  the few setups where this can be explicitly done.)
\end{exercise}

Let us zoom in at the interesting point: $\theta = \pi$ and compute the anomaly.
We couple the theory to a background $\Z_N^{(1)}$ field and check
$\T$-invariance (just like we did for the particle on a circle). The coupling
now is
\begin{equation}
  \Im(-S) = \i\pi I + \frac{2\pi\i k}{2N} \int_X P(A).
\end{equation}
The latter is a new counterterm that is available, subject to $k \in \Z$ and $k
\sim k + \gcd(2,N)N$. We evaluate:
\begin{equation}
  \T(e^S) = \e^S \exp(2\pi\i\qty(I + \frac{2k}{2N} \int_X P(A))).
\end{equation}
\begin{exercise}
  Derive this formula.
\end{exercise}
In the absence of background fields, $I$ is an integer and the last term
vanishes, so $\e^S$ is $\T$-invariant. In the presence of background fields,
$\T$-invariance requires $2k = N - 1$. If $N$ is even, there is no solution for
$k$! This is an anomaly between $\T$ and $\Z_N^{(1)}$.

\emph{Note}: How bad is this failure? The failure of $\T$-invariance is only by
background fields. When $A$ is off, the theory has the symmetry.

We can try to write an associated 5d invertible theory. It is defined on
manifolds, in general without orientation (because of $\T$-symmetry we are in an
$\OO$-type cobordism category) and $\Z_N^{(1)}$ background fields $A$. Looking
at the $\Z_2^{(1)}$, the action is
\begin{equation}
  \exp(-\pi\i\int_Y \widehat{w_1}(TY) \cup P(A))
\end{equation}
where $\widehat{w_1}(TY)$ is an integral uplift of $w_1$ with twisted integral
coefficients.

Yang--Mills is the boundary of this invertible theory, which characterises a
mixed anomaly between time-reversal and the $1$-form symmetry. We have reached a
striking conclusion: Yang--Mills theory has an anomaly that was not known until
about five years ago!

\emph{Dynamical implications:}
At $\theta = \pi$, $\SU(N)$ YM (for even $N$) cannot be trivially gapped.
What does the IR look like? (Not that the answer could be different for
different $N$.)
\begin{itemize}
\item Gapless: Possible (but sounds exotic)
\item Gapped? Must be a nontrivial (i.e.~non-invertible) TQFT.
  \begin{enumerate}
  \item Deconfined: $\Z_N^{(1)}$ is spontaneously broken. This would be
    something like a discrete gauge theory.
  \item $\T$-symmetry spontaneously broken: Then $\dim(\mathcal{H}_{S^3}) \geq
    2$ and there is a domain wall relating the degenerate vacua.
    Lore (with good evidence) is that this is the possibility that is realised.
    There are believed to be exactly two ground states.
  \item A TQFT with both $\T$ and $\Z_N^{(1)}$ not spontaneously broken.
    A result (Córdova, Ohmori, from the suggested readings): This is
    mathematically impossible.
  \end{enumerate}
  (These possibilities are not mutually exclusive. For example both symmetries
  could be spontaneously broken.)
\end{itemize}
}
% {\input{freed-4.tex}}
}

\newpage

{\clearpage%%%%%%%%%%%%%%%%%%%%%%%%
%%%%%Local commands%%%%%
%%%%%%%%%%%%%%%%%%%%%%%%
\newcommand{\cat}[1]{\ensuremath{\text{\normalfont\sffamily{#1}}}}
\renewcommand*\rm[1]{\mathrm{#1}}
\newcommand\cN{\mathcal{N}}
\newcommand\cL{\mathcal{L}}
\newcommand{\cZ}{\mathcal{Z}}
\newcommand{\bbC}{\mathbb{C}}
\newcommand{\bbZ}{\mathbb{Z}}
\newcommand{\HHm}{\operatorname{H}}
\newcommand*\I{\mathrm{i}}

\newcourse[Davide Morgante]{Symmetry Categories 101}{Michele Del Zotto}

The main aim of the course is to begin an exploration of the structure of symmetries in quantum field theory (QFT) and their dynamical consequences in light of the mantra that \textit{symmetries in QFT are described by topological operators/defects}. Our purpose here is to discuss a very simple example of QFT in detail. Our choice is massless quantum electrodynamics in four spacetime dimensions. We will study the symmetries that this theory has and we will demonstrate in particular that chiral symmetry is non-invertible, as a consequence of the ABJ anomaly,\footnote{S. L. Adler, \textit{Axial vector vertex in spinor electrodynamics}, \href{https://doi.org/10.1103/PhysRev.177.2426}{Phys.Rev. 177 (1969)}, 2426-2438 and J.S. Bell(CERN) and R. Jackiw, \textit{A PCAC puzzle:$\pi_0 \to \gamma\gamma$ in the $\sigma$-model} \href{https://doi.org/10.1007/BF02823296}{Nuovo Cim.A 60 (1969)}, 47-61, as well as lectures by K. Ohmori and C. Córdova in this volume.} and it has a higher structure. Our aim is to analyze the latter explicitly using field theory methods. This topic will be the core of this course (lectures 2 and 3), building on an ongoing collaboration with Christian Copetti, Kantaro Ohmori and Yifan Wang.\footnote{C. Copetti, M. Del Zotto, K. Ohmori, Y. Wang, \textit{Higher Structure of Chiral Symmetry}, \href{https://arxiv.org/abs/2305.18282}{arXiv:2305.18282}} Before diving in this example, in the first lecture we will discuss a heuristic argument based on decoupling from gravity for which we expect symmetries of any relativistic quantum field must be organized by higher categories. In the fourth lecture we will discuss generalized Takahashi-Ward identities and further applications of these generalized symmetries.

\section{Lecture 1: Heuristic derivation of symmetry categories}

Dan Freed lectures at PI in 2022 (later in this volume) started by remarking that thinking about global categorical symmetries both words global and categorical can be dropped. Moreover, he continued, since no global symmetries should be present in a quantum theory of gravity, also the word symmetry should be dropped when writing the actual fundamental theory of Nature... but then... what are we doing here??

Funnily enough, precisely this remark, namely the lack of symmetries in a fundamental theory, can be used to argue that symmetries in QFT \textit{must} be described by higher categories.  This will be the topic of today's lecture.

\subsection{Global symmetries are emergent}

The main purpose of this section is to argue the following

\medskip

\noindent \textbf{Claim: }\textit{In any fundamental theory of Nature global symmetries must be emergent, meaning that they hold only below a certain energy scale $\Lambda$. $\Lambda$ is moreover smaller than the Planck scale.}

\medskip

There are many different ways to convince ourselves this must be the case. A 
compelling argument was presented by Harlow and Ooguri using the 
holographic principle.\footnote{D. Harlow, and H. Ooguri, \textit{Symmetries in quantum field theory and quantum gravity}, \href{https://doi.org/10.1007/s00220-021-04040-y}{Comm. Math. Phys. 383 (2021) 3}, 1669-1804, \href{https://arxiv.org/abs/1810.05338}{\textt{arXiv:1810.05338}}} 
Another argument uses the relation between string compactification 
and geometric engineering limits. Here we present a more classic  heuristic  
argument based on some assumptions about the physics of black holes.\footnote{See eg. T. Banks and N. Seiberg, \textit{Symmetries and Strings in Field Theory and Gravity}, \href{https://doi.org/10.1103/PhysRevD.83.084019}{Phys. Rev. D 83 (2011) 084019}, \href{https://arxiv.org/abs/1011.5120}{\textt{arXiv:1011.5120}} and C. Cordova, K. Ohmori and T. Rudelius, \textit{Generalized symmetry breaking scales and weak gravity conjectures}, \href{https://doi.org/10.1007/JHEP11(2022)154}{JHEP 11 (2022) 154}, \href{https://arxiv.org/abs/2202.05866}{\textt{arXiv:2202.05866}}}

Consider a simple gravitational theory with a global conserved $\rm{U}(1)$ charge and a free scalar field $\phi$ with global charge $q$ under such symmetry. Condensing a large number of $\phi$ particles, say $N\gg 1$, in a very small region of space the density of the system is enough to make it collapse to form a black hole. By charge conservation this black hole has a global conserved charge $Nq$ (the same of the system of $\phi$ particles we started from). Due to Hawking radiation, our black hole evaporates, emitting Hawking quanta of thermal radiation. Black hole evaporation is a well-known semiclassical statistical process due to the pair production of particles and antiparticles near the horizon of the black hole. It can happen that only one of the two pair produced conjugate states ends up falling back and crossing the horizon, while the other escapes. Hawking argued that this process causes the BH to evaporate.\footnote{S.W. Hawking, \textit{Particle creation by black holes}, \href{https://projecteuclid.org/journals/communications-in-mathematical-physics/volume-43/issue-3/Particle-creation-by-black-holes/cmp/1103899181.full}{Comm. Math. Phys. 43
(1975) 199}} Due to the statistical nature of this process we expect that charge is, on average, conserved. By evaporation the black hole shrinks, becoming smaller and smaller. The semiclassical analysis by Hawking, however, holds as long as quantum gravitational effects become relevant. These are expected to kick in at length scales that are smaller than the Plank length $\ell_P\sim10^{-35}m$, which is one of the fundamental constants of Nature.

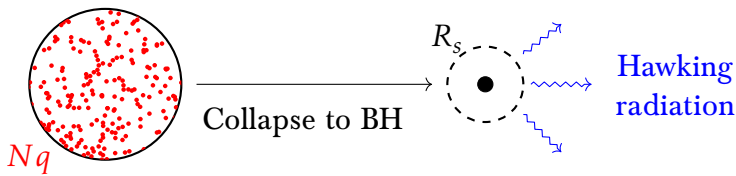
\begin{figure}[H]
\centering
\begin{tikzpicture}
  \begin{scope}
    \draw[thick] (0,0) circle (1);
    \clip (0,0) circle (1);
    \pgfmathsetseed{24122015}
    \foreach \p in {1,...,1050}
    { \fill[red] (2*rand,2*rand) circle (.03);}
  \end{scope}
  \node[] (charge) at (-1,-1) {$\color{red}Nq$};
  %%%%%%%%%%%%%%%%%%%%%%%
  \draw[thick, dashed] (5,0) circle (0.5);
  \node[] (A) at (4.5,0.6) {$R_s$};
  \draw[fill=black] (5,0) circle (0.1);
  \draw[->] (1.2,0) -- (4.3,0);
  \node[] (txt) at (2.6,-.5) {Collapse to BH};
  %%%%%%%%%%%%%%%%%%%%%%
  \foreach \in / \out in {(5.5,.4)/{(6,.8)}, (5.6,0)/{(6.4,0)}, (5.5,-.4)/{(6,-.9)}}
  \draw [->,line join=round, decorate, decoration={
    zigzag,
    segment length=4,
    amplitude=.9,post=lineto,
    post length=2pt
  }, blue]  \in -- \out;
  \node[blue] (txt2) at (7.5,0) {\shortstack{Hawking\\ radiation}};
\end{tikzpicture}
\caption{Visual sketch of the black hole argument}
\end{figure}

At the end of the evaporation process we end up with a very tiny black hole, of size comparable to the Planck length, that has a large global charge $Nq$. This Plank-size charged black hole is stable: it cannot radiate no-longer and because of its large global charge it cannot decay. Now, we can repeat this same process \textit{ad libitum}, thus producing infinitely many such small Planck-sized black holes with large global symmetry charges. Notice that our collection of tiny black holes have no other charge other than their conserved global charge by construction, and therefore they do not interact with each other.\footnote{\ Global charges are not gauged and therefore are not associated to any form of interaction.} Moreover, they are indistinguishable one another (for a far lying observer). Bringing many of these tiny black-holes together, therefore, one obtains a single microstate from the perspective of statistical physics. This black hole microstate however, is in facts a ``macrostate'' that has an area as large as we please (just by producing and bringing in more and more Plank size black holes, we can increase the area of this black hole microstate as much as we desire). This results in a violation of one of the most important laws of black hole thermodynamics, the Bekenstein-Hawking  entropy formula,\footnote{A. Bekenstein, \textit{Black holes and the second law}, \href{https://doi.org/10.1007/BF02757029}{Lettere al Nuovo Cimento. 4 (1972) 15}, 99–104.} namely the statement that 
%\vspace{5pt}
\begin{equation}
  \left(\begin{gathered}\text{the entropy} \\ \text{of the black hole}\end{gathered}\right) \,\, \sim \,\, {1 \over 4} \,\left(\begin{gathered}
  \text{the area} \\ \text{of the black hole}\end{gathered}\right)\,
\end{equation}
in suitable units (as a first order approximation). On the LHS of this equation we have a quantum feature of the black hole (after Boltzmann the entropy of a statistical system is proportional to the logarithm of the number of quantum microstates of the system). The RHS of this equation is a classical property of the black hole (the area of its event horizon). This law of physics is violated by our gedanken scalar field: we can make the area of the black hole as large as we please while still having a single ``macrostate'', and this is a contradiction.\footnote{\, Morally, variations on this theme would allow to rule out any sort of conserved charge in a fundamental theory where gravity is dynamical... The devil is in the details, however, and to date the lack of global symmetries in theories of quantum gravity is among the most celebrated conjectures about the quantum behavior of gravity.}  How can one save the day? The key idea is that there should be another fundamental energy scale $\Lambda$ such that all conservation laws from global symmetries are violated at energies above it. In particular, when the evaporating black hole radius reaches a size comparable with $1/\Lambda$ symmetry violations start occurring, and the charge conservation laws preventing the black hole to completely disappear no longer hold.\footnote{\ Recall we are using a collection of units for which $\hbar = c = 1$. In these units, lengths are inversely proportional to energies.} As long as $\Lambda$ is below the Plank energy scale $M_P = 1 / \ell_P$, the symmetry violating effects will kick-in before the Hawking evaporation breaks down, allowing the BH to disappear and preventing the existence of stable Plank-size black holes with large conserved global charges. To avoid this contradiction, our claim better holds.

%%%%%%%%%%%%%%%%%%%%%%%%%%%%%%%%%%%%%%%%%%%%%%%%%%%%%%%%%%%%%%%%%%%%
%%%%%%%%%%%%%%%%%%%%%%%%%%%%%%%%%%%%%%%%%%%%%%%%%%%%%%%%%%%%%%%%%%%%
%%%%%%%%%%%%%%%%%%%%%%%%%%%%%%%%%%%%%%%%%%%%%%%%%%%%%%%%%%%%%%%%%%%%
%%%%%%%%%%%%%%%%%%%%%%%%%%%%%%%%%%%%%%%%%%%%%%%%%%%%%%%%%%%%%%%%%%%%
%%%%%%%%%%%%%%%%%%%%%%%%%%%%%%%%%%%%%%%%%%%%%%%%%%%%%%%%%%%%%%%%%%%%

\subsection{Emergent symmetries and higher categories: a heuristic derivation}

In the previous section we have heuristically established that global symmetries must be emergent in any fundamental theory, namely there is an energy scale $\Lambda$ which is smaller than the Planck scale, above which all symmetries are violated. In the rest of this lecture, we will use this point of view to argue that global symmetries are indeed organized by higher categories and to learn some more about the expected structures involved. We will proceed inductively on $d$, the number of dimensions of spacetime, starting from $d=1+1$.

\medskip

In a $(1+1)$-dimensional theory all excitations are particles, 
extended along their worldlines, ie. (geodesic) curves describing 
their trajectories in spacetime. The physics along each such line 
is governed by a one-dimensional gravitational path integral. 
Consider the system at low energy scales, well below $M_P$, such 
that gravitational interactions can be neglected and we obtain an 
effective field theory description. It is in this limit that 
quantum field theory emerges as a low energy effective theory 
describing a system decoupled from gravity. In particular, we expect to be able 
to take this limit in such a way that most excitations of the original gravitational 
system decouple completely, while retaining a smaller collection 
of excitations that are still interacting among each other. We can  characterize these using the physics of the 
corresponding quantum mechanical worldline theories: some will be gapped and
flow to their ground states at energies below $\Lambda$, while 
other will still have a non-trivial dynamics along their 
worldlines at those scales. Moreover, these lines 
can be organized according to the energies required 
to displace their positions in spacetime. When the 
energy cost to displace the worldline is zero and the corresponding worldline is gapped, we 
call the lines ``topological''. When the energy cost 
to displace a worldline is infinite, it becomes a 
relic (a defect or a domain wall). When the energy 
cost is finite, these are worldlines of dynamical 
excitations of the quantum field. In practice one 
needs to explicitly check this for each line by 
computing the energy cost of a displacement, but for 
the sake of our heuristic argument this is not going 
to be necessary. These collections of lines have an 
interplay arising by crossing each other or merging 
one another upon stacking, in particular these 
operations lead to an action of the collection of 
topological lines on the 
non-topological ones. Since topological lines can be deformed at no energy cost, this gives an interesting way of establishing equivalences (Ward-Takahashi identities) between inequivalent field configurations -- we will see an example momentarily. For this reason it is natural to interpret the topological lines as the global symmetry of the quantum field emerging below the scale $\Lambda$.

\medskip

It is natural to interpret the topological lines as  objects of 
a \textit{symmetry category} $\cat{C}$. Given two lines $\mathcal{D}_1, 
\mathcal{D}_2\in\cat{C}$ they can join at a point, where an 
operator is located. If this operator is topological as well, we 
can slide it freely along the line. The collection of such 
operators form a $\bbC$-linear vector space,\footnote{\ Indeed, 
consider a pointlike operator $\mathcal O_{12}(p)$ giving a 
topological interface between two line operators $\mathcal D_1$ 
and $\mathcal D_2$ supported on two lines $\gamma_1$ and $
\gamma_2$ such that $\partial \gamma_i = p$. Multiplying such an 
operator by a complex constant give another topological pointlike 
operator. Adding two such operators, gives another operator $
\mathcal O_{12}(p) + \mathcal{O}'_{12}(p) = (\mathcal O + \mathcal 
O')_{12}(p)$.} the vector space of topological interfaces between 
the corresponding worldline quantum mechanics, which we denote $
\Hom(\mathcal{D}_1,\mathcal{D}_2)$. In this notation the vector 
space $\Hom(\mathcal{D},\mathcal{D}) = \End(\mathcal{D})$ is the 
vector space of local operators of the quantum mechanics. These 
requirements gives the category a $\bbC$-linear structure since 
these $\Hom$ spaces are $\bbC$-linear vector spaces.

The topological lines can fuse with each other, by bringing them 
together. This process does not depend on the detail of the 
merging, since the dependence of the symmetry lines is topological 
on their support. This operation at the level of the corresponding quantum mechanical 
worldlines, correspond to the $\otimes$-product of the associated Hilbert spaces. 
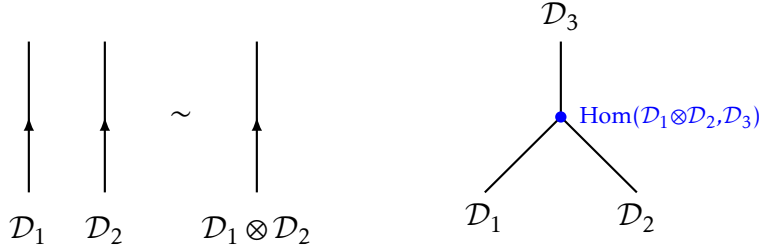
\begin{figure}[H]
  \centering
  \begin{tikzpicture}[baseline={(0,1)},scale=1]
    \foreach \in /\obj in {1/{\mathcal{D}_1},2/{\mathcal{D}_2},4/{\mathcal{D}_1\otimes \mathcal{D}_2}} {
    \draw[postaction={decorate},decoration={
      markings,
      mark=at position .5 with {\arrow{latex}}}, thick] (\in,0) -- (\in,2);
    \node[] (A\in) at (\in,-.5) {$\obj$};}
    \node[] (sim) at (3,1) {$\sim$};
    %%%%%%%%%%%%%%%%
    \draw[thick] (7,0) -- (8,1)  -- (9,0);
    \draw[thick] (8,1) node[circle,fill=blue,inner sep=1.5pt,minimum size=.1pt] {} -- (8,2);
    \node[right,blue] (hom) at (8.1,1) {$\scriptstyle\Hom(\mathcal{D}_1\otimes \mathcal{D}_2,\mathcal{D}_3)$};
   
    \node[below] (B1) at (7,0) {$\mathcal{D}_1$};
    \node[below] (B2) at (9,0) {$\mathcal{D}_2$};
    \node[above] (B3) at (8,2) {$\mathcal{D}_3$};
  \end{tikzpicture}
  \caption{Parallel and partial fusion of two defects}\label{fig:fusion}
\end{figure}

This implies that the symmetry category has a monoidal structure, with a fusion 
product -- see Figure \ref{fig:fusion}. It is interesting to remark that this category has also simple objects, corresponding to the topological lines which 
are the IR-fixed points of some quantum mechanical system with a single ground state. In this case 
\begin{equation}
  \Hom(\mathcal D,\mathcal D)\simeq \bbC\,.
\end{equation}
The fusion product of lines is associative, but because of their topological nature, the associativity of the product is captured by an interface, the \textit{associator morphism}
\begin{equation}
  F_{\mathcal{D}_1, \mathcal{D}_2, \mathcal{D}_3}\in\Hom\big((\mathcal{D}_1\otimes \mathcal{D}_2)\otimes \mathcal{D}_3, \mathcal{D}_1\otimes (\mathcal{D}_2\otimes \mathcal{D}_3)\big)
\end{equation}
whose construction is encoded in the diagram depicted in Figure \ref{fig:associators}.
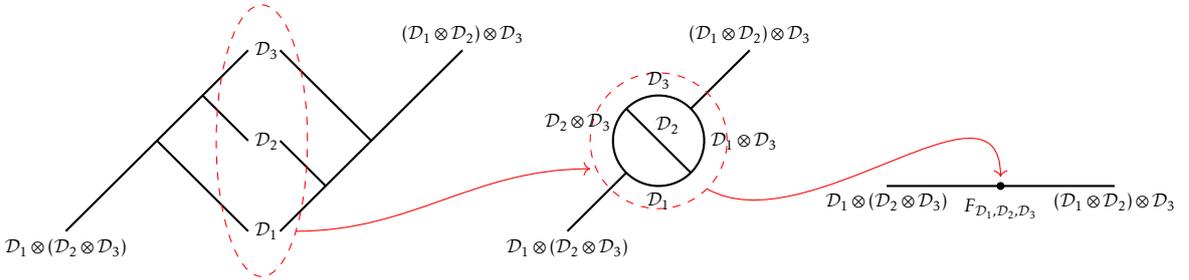
\begin{figure}[H]
  \centering
  \begin{tikzpicture}[scale=.6,every node/.append style={transform shape}]
    %\begin{scope}[transform canvas={scale=1,every node/.append style={transform shape}}]
    \foreach \coord/\name in {(4,0)/1,(4,2)/2,(4,4)/3}
    \node[right] (D\name) at \coord {$\mathcal{D}_\name$};
    \node[below] (prod1) at (0,0) {$\mathcal{D}_1\otimes(\mathcal{D}_2\otimes \mathcal{D}_3)$};
    \draw[-, thick] (4,4) -- (0,0);
    \draw[-, thick] (4,2) -- (3,3);
    \draw[-, thick] (4,0) -- (2,2);

    %%%%%%%%%%%%%%%%%%%
    \foreach \coord/\name in {(4.7,0)/D_11,(4.7,2)/D_22,(4.7,4)/D_33}
    \node[] (\name) at \coord {};
    \node[above] (prod2) at (8.7,4) {$(\mathcal{D}_1\otimes \mathcal{D}_2)\otimes \mathcal{D}_3$};
    \draw[-, thick] (4.7,4) -- (6.7,2);
    \draw[-, thick] (4.7,2) -- (5.7,1);
    \draw[-, thick] (4.7,0) -- (8.7,4);
    %\end{scope}
    %%%%%%%%%%%%%%%%%%%
    \draw[->,red] (5.045,0) to[out=0, in=180] (11.5,1.38);
    \draw[dashed,red] (4.3,2) ellipse (1 and 3);
    %%%%%%%%%%%%%%%%%%%
    \node[below] (C) at (11,0) {$\mathcal{D}_1\otimes(\mathcal{D}_2\otimes \mathcal{D}_3)$};
    \node[above] (D) at (15,4) {$(\mathcal{D}_1\otimes \mathcal{D}_2)\otimes \mathcal{D}_3$};
    \node[circle split, draw, thick, rotate=-45,  minimum size=2cm] (circ) at (13,2) {};
    \draw[thick,-]  (11,0) to  (12.293,1.293);
    \draw[thick,-]  (15,4) to  (13.707,2.707);
    \draw[dashed,red] (circ) circle (1.5);

    \node[above] (D22) at (13.2,2) {$\mathcal{D}_2$};
    \node[below] (D11) at (13,1) {$\mathcal{D}_1$};
    \node[left] (D2D3) at (12.1,2.4) {$\mathcal{D}_2\otimes \mathcal{D}_3$};
    \node[above] (D33) at (13,3) {$\mathcal{D}_3$};
    \node[right] (D1D3) at (14,2) {$\mathcal{D}_1\otimes \mathcal{D}_3$};
    %%%%%%%%%%%%%%%%%%%
    \draw[->,red] (14.061,0.939) to[out=-30,in=90] (20.5,1.2); 
    %%%%%%%%%%%%%%%%%%%
    \node[below] (start) at (18,1) {$\mathcal{D}_1\otimes(\mathcal{D}_2\otimes \mathcal{D}_3)$};
    \node[below] (end) at (23,1)  {$(\mathcal{D}_1\otimes \mathcal{D}_2)\otimes \mathcal{D}_3$}; 

    \draw[-,thick] (start.north) -- (end.north);
    \node[circle, fill=black, inner sep=.2pt, minimum size=5pt] (E) at (20.5,1) {};
    \node[] (F) at (20.5,0.5) {$F_{\mathcal{D}_1,\mathcal{D}_2,\mathcal{D}_3}$};
  \end{tikzpicture}
  \caption{Fusing the two networks which encode different ordering for the fusion of three defects. Shrinking down the defect bubble, one obtains a morphism which encodes associativity.}
  \label{fig:associators}
\end{figure}
Assuming we can fix a basis of simple topological lines, the 
associator can be written as a matrix whose elements are known as 
\textit{F-symbols}. Because of the topological nature of the junctions of topological lines, the associator morphisms satisfy a \textit{pentagon identity}, which is 
expressed as the commutativity of the following diagram
\begin{figure}[H]
  \centering
  \begin{tikzpicture}[commutative diagrams/every diagram]
    \node (P0) at (90:3cm) {$\mathcal{D}_1\otimes (\mathcal{D}_2\otimes (\mathcal{D}_3\otimes D_4))$};
    \node (P1) at (90+72:3cm) {$\mathcal{D}_1\otimes ((\mathcal{D}_2\otimes \mathcal{D}_3)\otimes D_4))$} ;
    \node (P2) at (90+2*72:3cm) {\makebox[5ex][r]{$(\mathcal{D}_1\otimes (\mathcal{D}_2\otimes \mathcal{D}_3))\otimes D_4$}};
    \node (P3) at (90+3*72:3cm) {\makebox[5ex][l]{$((\mathcal{D}_1\otimes \mathcal{D}_2)\otimes \mathcal{D}_3)\otimes D_4$}};
    \node (P4) at (90+4*72:3cm) {$(\mathcal{D}_1\otimes \mathcal{D}_2)\otimes (\mathcal{D}_3\otimes D_4)$};
    \path[commutative diagrams/.cd, every arrow, every label]
    (P0) edge node[swap] {$F_{\mathcal{D}_2,\mathcal{D}_3,D_4}$} (P1)
    (P1) edge node[swap] {$F_{\mathcal{D}_1,\mathcal{D}_2\otimes \mathcal{D}_3,D_4}$} (P2)
    (P2) edge node {$F_{\mathcal{D}_1,\mathcal{D}_2,\mathcal{D}_3}$} (P3)
    (P4) edge node {$F_{\mathcal{D}_1\otimes \mathcal{D}_2,\mathcal{D}_3,D_4}$} (P3)
    (P0) edge node {$F_{\mathcal{D}_1,\mathcal{D}_2,\mathcal{D}_3\otimes D_4}$} (P4);
    \draw[->,blue,thick] (.5,0) arc (0:340:.5);
    \node[blue] (txt) at (.1,-1) {\shortstack{Associativity\\constraint}};
    \end{tikzpicture}
    \caption{The \textit{pentagonator} gives a constraint for the F-symbols}
\end{figure}
As a consequence we expect that topological lines form a $\mathbb C$-linear monoidal category. Indeed the latter consists of a category $\cat{C}$ endowed with an $\otimes$ operation that satisfies the above requirements, plus the additional requirement that there is a unitor $\mathbf{1}_\cat{C}$ such that $\mathbf{1}_\cat{C}\otimes L \simeq L$ and $L\otimes  \mathbf{1}_\cat{C} \simeq L$. In our context the latter is given by an empty worldline, which always exists. The existence of evaluation and coevaluation maps follows from the CPT theorem: given $\mathcal D$ we expect there is a conjugate line $\overline{\mathcal D}$ and the two behave as a pair of particle/antiparticle. At sufficiently high energies the existence of an evaluation morphism $\mathcal D \otimes \overline{\mathcal D} \to \mathbf{1}_\cat{C}$ maps to the annihilation process, while the coevaluation morphism $\mathbf{1}_\cat{C} \to \mathcal D \otimes \overline{\mathcal D}$ maps to the creation of a particle/antiparticle pair from the vacuum (see David Jordan's lecture in this volume). Since the identity is simple by construction, the resulting structure is that of a \textit{tensor category}.\footnote{\, For systems with decomposition this can be relaxed to a multitensor category -- the identity is no longer simple and splits into as many copies as there are universes. Our readers that are interested in decomposition of 2d theories can consult E. Sharpe, \textit{An introduction to decomposition}, \href{https://arxiv.org/abs/2204.09117}{\textt{arXiv:2204.09117}}.} Finally, if the the system has finitely many topological lines, the resulting symmetry category is a fusion category.\footnote{\ By definition a finite tensor category is a fusion category.}

\medskip

\noindent \textbf{Remark/Exercise:} Consider a class of topological defect lines labeled by elements $g$ of a finite group $G$, with one-dimensional morphism spaces dictated by the group multiplication law: $L_g \otimes L_{g'} \simeq L_{gg'}$. Convince yourselves that in this case the datum of $F$ is equivalent to the datum of a class $\alpha(g,h,k)\in \HHm^3(G,\U(1))$:\footnote{\, Notice that this class encodes an anomaly. Interested readers can consult e.g. C.-M.Chang, Y.-H. Lin, S.-H. Shao, Y. Wang, X. Yin, \textit{Topological Defect Lines and Renormalization Group Flows in Two Dimensions}, \href{https://doi.org/10.1007/JHEP01(2019)026}{JHEP 01 (2019) 026}, \href{https://arxiv.org/pdf/1802.04445}{\textt{arXiv:1802.04445}} and R. Thorngren, Y. Wang, \textit{Anomalous symmetries end at the boundary}, \href{https://link.springer.com/article/10.1007/JHEP09(2021)017}{JHEP 09 (2021) 017}, \href{https://arxiv.org/pdf/2012.15861}{\textt{arXiv:2012.15861}}}
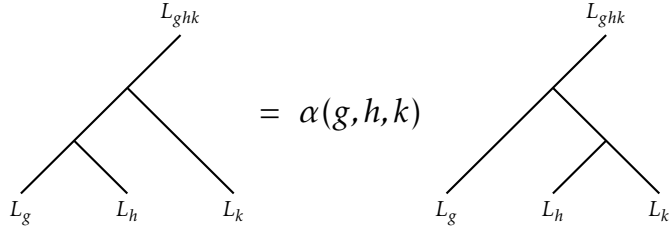
\begin{figure}[H]
  \centering
  \begin{tikzpicture}[scale=.7,every node/.append style={transform shape}]
    \foreach \coord/\name in {(0,0)/L_g,(2,0)/L_h,(4,0)/L_k}
    \node[below] (\name) at \coord {$\name$};
    \node[above] (prod1) at (3,3) {$L_{ghk}$};
    \draw[-, thick] (0,0) -- (3,3);
    \draw[-, thick] (2,0) -- (1,1);
    \draw[-, thick] (4,0) -- (2,2);
    %%%%%%%%%%%%%%%%
    \node[] (eq) at (6,1.5) {\scalebox{1.5}{$=\ \alpha(g,h,k)$}};
    %%%%%%%%%%%%%%%%
    \foreach \coord/\name in {(8,0)/L_g,(10,0)/L_h,(12,0)/L_k}
    \node[below] (\name) at \coord {$\name$};
    \node[above] (prod1) at (11,3) {$L_{ghk}$};
    \draw[-, thick] (8,0) -- (11,3);
    \draw[-, thick] (10,0) -- (11,1);
    \draw[-, thick] (12,0) -- (10,2);
    \end{tikzpicture}
    \caption{Deforming the network could result in an overall phase which will encode the anomaly of the associated symmetry}
    \label{fig:group_class}
\end{figure}

\noindent \textbf{Remark 2:} In our analysis above we have neglected the distinction between bosonic and fermionic excitations. Our analysis can be extended to include fermions by considering super-Hilbert spaces.

\bigskip

Before proceeding with the inductive argument and going to higher dimension, we first give an example of a categorical symmetry which is relevant, for example, in the Ising model as described in the lectures by Shu-Heng Shao at this school.\footnote{\ The latter were based on the TASI lecture notes: S.-H. Shao, \textit{What's done cannot be undone: TASI lecture on non-invertible symmetries}, \href{https://arxiv.org/abs/2308.00747}{\textt{arXiv:2308.00747}}, that the interested readers can consult for more details.}

\subsection{Example: Tambara-Yamagami symmetry}
Given a finite group $G$, we can construct the corresponding Yambara-Tamagami category \cat{TY}$(G)$ as follows. The objects are a collection of topological lines labelled by elements of the group $G$, $L_g$, and a duality topological line $\cN$. The fusion algebra of these objects is given by
\begin{equation}
    L_g \otimes L_h \simeq L_{gh},\qquad \cN\otimes\overline{\cN} \simeq \sum_{g\in G}L_g,\qquad \cN\otimes L_g =\cN = L_g \otimes \cN
    \label{eq:TYfusion}
\end{equation}
This is one of the simplest symmetry categories that are not group-like: the line $\cN$ is an example of a non-invertible symmetry. What is the consequence of such symmetry for a 2d quantum field theory? Let us assume that our system has a $\cat{TY}(G)$ symmetry and see what are the consequences for one of the simplest observables, the torus partition function. Consider placing our field theory on a torus $T^2$, and notice that at any point on the torus we can ``bubble out'' an insertion of a topological defect line on a loop, $S^1$ (due to the topological nature of such insertion we can shrink the $S^1$ it back to zero-size at no cost). The only price to pay for this operation is to divide by the value of the parition function of the topological line on $S^1$ (which computes its quantum dimension), thus:
\begin{eqnarray*}
        \begin{tikzpicture}[baseline={(1,0)}]
            \draw[fill=blue!50,opacity=.5,even odd rule] (0,0) ellipse (1 and .75) 
		(-0.5,0) arc(120:60:1 and 1.25)  arc(-60:-120:1 and 1.25) coordinate[pos=0.25] (xt);
	    \draw (-0.5,0) arc(-120:-130:1 and 1.25) (0.5,0) arc(-60:-50:1 and 1.25);
	    \draw[] (-65:1 and .75) to[out=40,in=10] 
		node[pos=0.2,right]{} (xt);
	    \draw[dashed] (xt) to[out=-170,in=-140] (-65:1 and .75);
	    \draw[] (0.8,0.05) arc(0:360:0.8 and .5) {};
     \end{tikzpicture}\quad&\sim\qquad
     \begin{tikzpicture}[baseline={(1,1)}]
            \draw[fill=blue!50, opacity=.5] (0,0) rectangle (2,2);
            \draw[postaction={decorate},decoration={
              markings,
              mark=at position .145 with {\arrow{latex}},
              mark=at position .375 with {\arrow{latex}},
              mark=at position .620 with {\arrowreversed{latex}},
              mark=at position .875 with {\arrowreversed{latex}},
            }
            ]
            (0,0) -- +(2,0) -- +(2,2) -- +(0,2) -- cycle; 
        \end{tikzpicture}&=
         \frac{1}{\cN(S^1)}\ \begin{tikzpicture}[baseline={(1,1)}]
            \draw[fill=blue!50, opacity=.5] (0,0) rectangle (2,2);
            \draw[postaction={decorate},decoration={
              markings,
              mark=at position .145 with {\arrow{latex}},
              mark=at position .375 with {\arrow{latex}},
              mark=at position .620 with {\arrowreversed{latex}},
              mark=at position .875 with {\arrowreversed{latex}},
            }
            ]
            (0,0) -- +(2,0) -- +(2,2) -- +(0,2) -- cycle; 
            \draw[postaction={decorate},decoration={
              markings,
              mark=at position .100 with {\arrow{latex}}}, draw=red, thick] (1,1) circle (.5);
            \node[] (line) at (1.5,1.5) {$\color{red}\scriptstyle\cN$};
        \end{tikzpicture}
\end{eqnarray*}
Now we can exploit the topology of the torus to make the line fuse against itself and exploit the fusion algebra:
\begin{eqnarray*}
=\frac{1}{\cN(S^1)}\ \begin{tikzpicture}[baseline={(1,1)}]
            \draw[fill=blue!50, opacity=.5] (0,0) rectangle (2,2);
            \draw[postaction={decorate},decoration={
              markings,
              mark=at position .145 with {\arrow{latex}},
              mark=at position .375 with {\arrow{latex}},
              mark=at position .620 with {\arrowreversed{latex}},
              mark=at position .875 with {\arrowreversed{latex}},
            }
            ]
            (0,0) -- +(2,0) -- +(2,2) -- +(0,2) -- cycle; 
            \foreach \in / \out in {(.7,0)/{(0:90:.7)},(0,1.3)/{(270:360:.7)},(2,.7)/{(90:180:.7)},(1.3,2)/{(180:270:.7)}}
            \draw[postaction={decorate},decoration={
              markings,
              mark=at position .5 with {\arrow{latex}}}, draw=red, thick] \in arc \out;
            \node[] (line) at (1,1) {$ \color{red}\scriptstyle\cN$};
            \draw[blue, dashed] (1,0) ellipse (.7cm and .3cm);
            \node[blue] (glue) at (1.8, -.5) {Fuse!};
        \end{tikzpicture}&\hspace{-12pt}=
        \dfrac{1}{\cN(S^1)}\ \begin{tikzpicture}[baseline={(1,1)}]
            \draw[fill=blue!50, opacity=.5] (0,0) rectangle (2,2);
            \draw[postaction={decorate},decoration={
              markings,
              mark=at position .145 with {\arrow{latex}},
              mark=at position .375 with {\arrow{latex}},
              mark=at position .620 with {\arrowreversed{latex}},
              mark=at position .875 with {\arrowreversed{latex}},
            }
            ]
            (0,0) -- +(2,0) -- +(2,2) -- +(0,2) -- cycle; 
            \draw[postaction={decorate},decoration={
              markings,
              mark=at position .100 with {\arrow{latex}}}, draw=red, thick] (1,1) circle (.4);
              \foreach \in / \out in {(1,0)/{(1,.6)},(1,1.4)/{(1,2)},(.6,1)/{(0,1)},(2,1)/{(1.4,1)}}
              \draw[draw=blue, thick] \in--\out;
              \node[] (line) at (1.5,1.5) {$\color{red}\scriptstyle\cN$};
              \node[] (line) at (1.5,.3) {$\color{blue}\scriptstyle{\sum L_g}$};
        \end{tikzpicture}&= \hspace{3.8pt}\sum_{g,h\in G} \hspace{8.5pt}
        \begin{tikzpicture}[baseline={(1,1)}]
          \draw[fill=blue!50, opacity=.5] (0,0) rectangle (2,2);
          \draw[postaction={decorate},decoration={
            markings,
            mark=at position .145 with {\arrow{latex}},
            mark=at position .375 with {\arrow{latex}},
            mark=at position .620 with {\arrowreversed{latex}},
            mark=at position .875 with {\arrowreversed{latex}},
          }
          ]
          (0,0) -- +(2,0) -- +(2,2) -- +(0,2) -- cycle;
          \foreach \in / \out in {(.5,0)/{(.5,2)},(0,1.5)/{(2,1.5)}}
          \draw[draw=blue,thick,postaction={decorate},decoration={
            markings,
            mark=at position .5 with {\arrow{latex}}}] \in--\out;
          \node[] (line1) at (.8,.3) {$\color{blue}\scriptstyle{L_g}$};
          \node[] (line2) at (1,1.2) {$\color{blue}\scriptstyle{L_h}$};
      \end{tikzpicture}
\end{eqnarray*}
In the final step we have shrunk the line back to a point, leaving behind the two sums over $L_g$ lines. The latter is equivalent to the partition function of the theory with the $G$ symmetry gauged. Therefore the theory $\mathcal{T}$ on the torus $T^2$ is isomorphic to the theory $\mathcal{T}/G$ on $T^2$, namely
\begin{equation}
  \cZ_{\mathcal{T}}\left(
  \begin{tikzpicture}[scale=.5, baseline={(0,-.1)}]
		\draw[fill=blue!50,opacity=.5,even odd rule] (0,0) ellipse (1 and .75) 
		(-0.5,0) arc(120:60:1 and 1.25)  arc(-60:-120:1 and 1.25) coordinate[pos=0.25] (xt);
	   \draw (-0.5,0) arc(-120:-130:1 and 1.25) (0.5,0) arc(-60:-50:1 and 1.25);
	   \draw[] (-65:1 and .75) to[out=40,in=10] 
		node[pos=0.2,right]{} (xt);
	   \draw[dashed] (xt) to[out=-170,in=-140] (-65:1 and .75);
	   \draw[] (0.8,0.05) arc(0:360:0.8 and .5) {};
	\end{tikzpicture}\right)=\cZ_{\mathcal{T}/G}\left(
        \begin{tikzpicture}[scale=.5, baseline={(0,-.1)}]
		\draw[fill=blue!50,opacity=.5,even odd rule] (0,0) ellipse (1 and .75) 
		(-0.5,0) arc(120:60:1 and 1.25)  arc(-60:-120:1 and 1.25) coordinate[pos=0.25] (xt);
	   \draw (-0.5,0) arc(-120:-130:1 and 1.25) (0.5,0) arc(-60:-50:1 and 1.25);
	   \draw[] (-65:1 and .75) to[out=40,in=10] 
		node[pos=0.2,right]{} (xt);
	   \draw[dashed] (xt) to[out=-170,in=-140] (-65:1 and .75);
	   \draw[] (0.8,0.05) arc(0:360:0.8 and .5) {};
	\end{tikzpicture}\right)
\end{equation}
This is an example of a duality symmetry obtained as a consequence of the non-trivial fusion algebra of the topological defect lines in the $\cat{TY}(G)$ category: if a 2d theory $\mathcal T$ has such symmetry category, then it is equivalent to the theory $\mathcal T/G$.  As discussed in detail in Shao's lectures, for the Ising model, we have an explicit realization of this type of symmetry with $G = \mathbb Z_2$ and $\cN$ the Kramers-Wannier duality defect.

%%%%%%%%%%%%%%%%%%%%%%%%%%%%%%%%%%%%%%%%%%%%%%%%%%%%%%%%%%%%%%%%%%%%
%%%%%%%%%%%%%%%%%%%%%%%%%%%%%%%%%%%%%%%%%%%%%%%%%%%%%%%%%%%%%%%%%%%%
%%%%%%%%%%%%%%%%%%%%%%%%%%%%%%%%%%%%%%%%%%%%%%%%%%%%%%%%%%%%%%%%%%%%
%%%%%%%%%%%%%%%%%%%%%%%%%%%%%%%%%%%%%%%%%%%%%%%%%%%%%%%%%%%%%%%%%%%%
%%%%%%%%%%%%%%%%%%%%%%%%%%%%%%%%%%%%%%%%%%%%%%%%%%%%%%%%%%%%%%%%%%%%

\subsection{Heuristic derivation: the inductive step}
We can now proceed with our heuristic argument considering the higher dimensional cases. In the higher dimensional case $d$ is greater than two, and in general we expect that the UV theory will not only have particle-like excitations with worldlines, rather it will have also excitations supported in dimension $1< k < d$, such as strings that have non-trivial \textit{worldsheets} ($k=2$), or even more generally $k$-dimensional membranes that have their own \textit{worldvolumes} ($k>2$). These higher dimensional configurations are governed by $k$-dimensional quantum field theories (QFTs) along their worldvolumes, which play the same role as the quantum mechanics in the $d=2$ case. We expect that these worldvolumes are governed by the dynamics of the corresponding $k$-dimensional QFTs below the $\Lambda$ energy scale. In $k>1$ case, the flow becomes more complex and we can now, roughly, distinguish between two possibilities for the IR dynamics of the theory
\begin{itemize}
  \item Gapped, meaning that there is a non-zero energy gap between the ground state and the first excited state. In this case, the IR theory is some $k$-dimensional TFT coupled to the bulk $d$-dimensional system;
  \item Gapless, meaning that the IR theory is not gapped. In this case, the IR theory is some $k$-dimensional CFT coupled to the bulk $d$-dimensional system
\end{itemize} 
The gapless case is extremely interesting. This leads to coupled bulk-boundary systems which have very interesting dynamics and also nice mathematical structures and applications. In these lectures however we are going to focus on the gapped case, since our focus are symmetries of $d$-dimensional QFTs. Extending these construction to the case of coupled bulk-boundary systems is an interesting open question. Notice that it is relatively simple to give examples of $k$-dimensional QFT that flowing in the IR can become topological. As an interesting exercise to our readers, they can try to formulate examples of gauge theories in the UV that flow to topological BF theories in the IR starting from Abelian Higgs models.\footnote{For a discussion see eg. T. Banks and N. Seiberg, \textit{Symmetries and Strings in Field Theory and Gravity}, \href{https://doi.org/10.1103/PhysRevD.83.084019}{Phys. Rev. D 83 (2011) 084019}, \href{https://arxiv.org/abs/1011.5120}{\textt{arXiv:1011.5120}}.} 

Also in $d$ dimensions we can organize the $k$-dimensional membranes according to the energy cost of their displacement in spacetime. If it is infinite, we will have some relics (or heavy defects), if it is finite we have some dynamical excitations, if it is zero and the worldvolume QFT is gapped we have \textit{topological membranes}. It is clear that the case of dimension grater than $2$ is much richer. Mimicking what we did for the $d=1+1$ case, the topological membranes will act on the non-topological ones by merging one another or crossing. Now however the collection of topological membranes is much richer. The symmetry category can be organized according to the codimension of the topological membranes. It is natural to identify the objects of the symmetry category with the topological membranes of smallest codimension. In a $d$-dimensional QFT, therefore, we expect to have objects that correspond to $(d-1)$-dimensional topological membranes which are coupled to the bulk. Let us call one such object $\mathcal D^{(0)}$.\footnote{\ The convention is that an topological membrane $\mathcal D^{(p)}$ is of codimension $p+1$, and can hence act by linking on $p$-dimensional field configurations and by crossing on configurations of dimension $>p$.} The idea is that by assumption, the endomorphism of one such object
\begin{equation}
    \text{End}_\cat{C}(\mathcal D^{(0)},\mathcal D^{(0)})
\end{equation}
has to be the symmetry of a $(d-1)$ dimensional topological theory which is coupled to the bulk $d$-dimensional theory. By inductive hypothesis, we know that the symmetries of a $(d-1)$ dimensional theory are indeed encoded by a (multi)-fusion $(d-2)$-categories -- compare to the lectures by David Jordan in this volume (it is indeed useful to think of $n$-categories as categories enriched over $(n-1)$-categories, which in particular implies that the hom-sets between the objects are $(n-1)$-categories themselves).\footnote{\, For more about this perspective on multifusion higher $n$-categories see eg. T. Johnson-Freyd, \textit{On the Classification of Topological Orders}, \href{https://doi.org/10.1007/s00220-022-04380-3}{Commun.Math.Phys. 393 (2022) 2}, 989-1033, \href{https://arxiv.org/pdf/2003.06663}{\textt{arXiv:2003.06663}}.} Thus we find that our symmetry category is indeed a category enriched over a (multi)-fusion $(d-2)$-category. This concludes our inductive argument: we expect that symmetries, described by the topological membranes of our QFT, are organized by a $(d-1)$ higher category \cat{C}, such that morphisms correspond to a $(d-2)$-dimensional topological interfaces and higher morphisms corresponds to topological interfaces of lower codimensions. Of course it would be nice to have a better axiomatic characterization of these structures, but, as discussed in David Jordan's lectures, it is much better to approach the question in the homotopy way. 
\begin{quote}
  \textbf{Moral:} \textit{The symmetries of a $d$-dimensional QFT are encoded by its topological membranes which are organized in a $(d-1)$-higher tensor category.}
\end{quote}
%In this sense, the homs in the $d$-category are $(d-1)$-categories themselves. 
In particular, notice that we have a grading of the category based on the co-dimension of the topological defect they host. This definition resonates with the physical intuition, realizing a network of topological defects as a system of TFTs and interfaces between them across various dimensions. We refer to these networks of topological membranes as the higher structure of the symmetry category.

Let us remark here that there is already an interesting structure for each single simple object. Well, firstly we need to clarify what is meant by simple object in this context. We propose to characterise simple objects as the ones that cannot host point-like topological operators on their support. This is because the presence of a non-trivial point-like operator call it $\mathcal O_{\mathcal D}$ on $\mathcal D$ implies that the corresponding defect Hilbert space must have multiple vacua on any closed spatial manifold.\footnote{\ These can indeed be constructed as Euclidean path integrals on semi-infinite cylinders with insertions of $\mathcal O_{\mathcal D}^n$.} Now this implies that there are multiple selection sectors in the limit of infinite volume, and hence that one such defect can be split in simpler components. Let us stress that even if $\mathcal D$ is simple according to our definition, the corresponding $\End(\mathcal D,\mathcal D)$ is not going to be necessarily trivial since as per our discussion above it encodes the symmetries of the associated TFT. In particular there could be a gaugeable part $\mathbb{A}\subseteq\End(\mathcal D,\mathcal D)$ giving rise to a new simple object $\mathcal{D}/\mathbb{A}$ by gauging $\mathbb{A}$ on the world-volume of $\mathcal{D}$. Moreover, gauging $\mathbb{A}$ on half-space will result in a non-trivial interface, and hence an element $\Hom(\mathcal{D},\mathcal{D}/\mathbb{A})$
\begin{figure}[H]
  \begin{center}
    \begin{tikzpicture}
      \draw (0,0) -- (4,-1) -- (5,0);
      \draw[dashed] (0,0)-- (1,1)-- (5,0);

      \draw (0,2) -- (4,1) -- (5,2) -- (1,3)-- (0,2);
      \draw (0,0) -- (0,2);
      \draw (4,-1)--(4,1);
      \draw (5,0) -- (5,2);
      \draw[dashed] (1,1) -- (1,3);

      \draw[fill=blue!50,opacity=.5] (2,-0.5) -- (2,1.5) -- (3,2.5) -- (3,0.5) -- (2,-0.5);

      \draw[fill=red!50,opacity=.2] (0,0) -- (2,-.5) -- (2,1.5) -- (3,2.5) -- (1,3) -- (0,2) -- (0,0);
      \draw[fill=red!50,opacity=.2] (2,-.5) -- (4,-1) -- (5,0) -- (5,2) -- (3,2.5) -- (2,1.5) -- (2,0.5);

      \node[red,rectangle,draw] (A) at (0,4) {$\mathcal D$}; 
      \node[red,rectangle,draw] (B) at (5,4) {$\mathcal{D}/\mathbb{A}$};
      \node[blue,rectangle,draw] (C) at (2.5,4) {$\Hom(\mathcal{D},\mathcal{D}/\mathbb{A})$};
      
      \draw[->,red] (A.south) to[out=270, in=90] (1.5,1.5);
      \draw[->,red] (B.south) to[out=270, in=90] (3.5,1.5);
      \draw[->,blue] (C.south) to[out=270,in=90] (2.5,1.5);
    \end{tikzpicture}
  \end{center}
\end{figure}
This implies that not only we do not expect to have a conventional Schur lemma, we moreover expect also to have possible morphisms between simple objects--a fact well-known to experts in higher categories.

One might even consider stacking a TFT decoupled from the bulk on the defect $\mathcal D$ and gauge/condense part of the symmetry of the resulting system, giving rise to even further simple objects with non-trivial morphisms. However, physics requires the action of $\cat{C}$ on the QFT operators to be faithful somehow, meaning that for each topological membrane there should be a non-topological one that transforms non-trivially with respect to it. This operation induces a sort of equivalence among topological defects, a sort of \textit{categorical gauge principle}: when characterizing the symmetries, one can always stack the topological membranes with some decoupled factor, if that helps in performing some computation, and retain only the faithful part eventually. We denote the equivalence class of a defect with respect to this operation as $[[\mathcal D]]$. This equivalence plays an important role when determining the higher structure of a symmetry category.

Our focus in the next lecutures will be to study explicitly the higher structure of a symmetry category for a four-dimensional quantum field theory, the massless limit of quantum electrodynamics.

\section{Lecture 2: Chiral symmetry of massless QED}
In this lecture we begin our excursion in the symmetry categories of field theories. Our example of choice is the chiral symmetry in $(3+1)d$ quantum electrodynamics (QED).\footnote{As revisited in: C. Cordova, K. Ohmori, \textit{Noninvertible Chiral Symmetry and Exponential Hierarchies}, \href{https://doi.org/10.1103/PhysRevX.13.011034}{Phys.Rev.X 13 (2023) 1}, 011034,\href{https://arxiv.org/pdf/2205.06243}{\textt{arXiv:2205.06243}} and Y. Choi, H.T. Lam, S.-H. Shao, \textit{Noninvertible Global Symmetries in the Standard Model}, \href{https://doi.org/10.1103/PhysRevLett.129.161601}{Phys.Rev.Lett. 129 (2022) 16}, 161601, \href{https://arxiv.org/abs/2205.05086}{\textt{arXiv:2205.05086}}.} The Lagrangian for the theory is
\begin{equation}
  \cL_{\text{QED}}=-\frac{1}{4e^2}F\wedge\star F+\I\bar{\psi}\slashed{D}_A\psi
\end{equation}
where $\psi$ is a Dirac fermion that couples to the Maxwell field through the covariant derivative $\slashed{D}_A=i\gamma^\mu(\partial_\mu-iA_\mu)$, $\mu=0,1,2,3$ is a spacetime index, $\gamma^\mu$ are the gamma matrices of the corresponding Clifford algebra $\{\gamma^\mu,\gamma^\nu\} = 2\eta^{\mu\nu}$ in conventions where $\eta^{00} = -1$, and $F = \dd A$ is the curvature of the gauge connection. This Lagrangian has a chiral $\U(1)_\chi$ symmetry which acts on the Dirac fermion as follows
\begin{equation}
  \psi\rightarrow e^{\I\frac{\alpha}{2}\gamma_5}\psi,\qquad \alpha\in[0,2\pi)
\end{equation}
where $\gamma^5 = i \gamma^0 \gamma^1 \gamma^2 \gamma^3$. The Nöther theorem associates to such a one-parameter family of transformation a corresponding Nöther current, such that $d \star j_\chi = 0$. 
As reviewed in the lectures by Ohmori and Cordova in this volume, this equation does not hold at the quantum level. Within quantum field correlators the following ABJ anomalous conservation equation holds instead:
\begin{equation}\label{eq:chiralviol}
  \langle \dd \star j_\chi \cdots \rangle = \langle\frac{1}{8\pi^2}F\wedge F \cdots \rangle
\end{equation}
Such an anomaly can be interpreted by the descent procedure: the anomaly inflow action $\mathcal{A}_5$ obeys $\dd{\mathcal{A}}_5=2\pi I_6$, where
\begin{equation}
  I_6 = \frac{1}{2(2\pi)^3}F\wedge F\wedge F_\chi\,,
\end{equation}
and $F_\chi$ is the curvature of a classical background for the chiral $\U(1)_\chi$. This effect is what forces this symmetry to become non-invertible. Let us proceed by showing how to construct corresponding codimension-one topological defects.

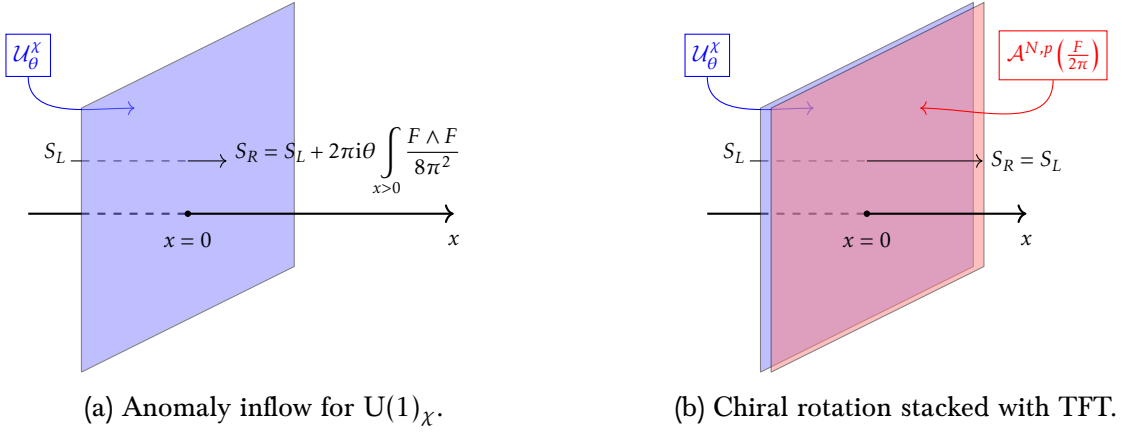
\begin{figure}[t]
  \centering
  \begin{subfigure}[b]{0.45\textwidth}
    \begin{tikzpicture}[scale=.7,every node/.append style={transform shape}]   
      \node[] (A) at (10,-.5) {$x$};
      \node[blue,draw] (chi) at (2,3) {$\mathcal{U}^\chi_\theta$};
      %\node[red,draw] (tau) at (8,3) {$\mathcal{T}$};
      \draw[-,dashed] (3,1) -- (5,1);
  
      \draw[->, blue] (chi.south) to[out=270, in=180] (4,2);
      \draw[-, thick,dashed] (3,0) -- (5,0);
      \draw[fill=blue!50, opacity=.5] (3,-3) -- (7,-1) -- (7,4) -- (3,2) -- (3,-3);
      %\draw[fill=red!50, opacity=.5] (3.2,-3) -- (7.2,-1) -- (7.2,4) -- (3.2,2) -- (3.2,-3);
      \draw[-, thick] (2,0) -- (3,0);
      \draw[->, thick] (5,0) -- (10,0);
      \draw[fill=black] (5,0) circle (0.05);
      %\draw[->, red] (tau.south) to[out=270,in=0] (6,2); 
  
      \node[] (zero) at (5,-.5) {$x=0$};
      \node[] (SL) at (2.5,1.15) {$S_L$};
      \node[] (SR) at (8,1) {$S_R=S_L+2\pi \I \theta\displaystyle{\int\limits_{x>0}}\dfrac{F\wedge F}{8\pi^2}$};
      \draw[-] (2.8,1) -- (3,1);
      \draw[->] (5,1) -- (SR);
    \end{tikzpicture}
    \caption{Anomaly inflow for $\U(1)_\chi$.}
  \end{subfigure}
  \hfill
  \begin{subfigure}[b]{0.45\textwidth}
      \centering
      \begin{tikzpicture}[scale=.7,every node/.append style={transform shape}]   
        \node[] (A) at (8,-.5) {$x$};
        \node[blue,draw] (chi) at (2,3) {$\mathcal{U}^\chi_\theta$};
        \node[red,draw] (tau) at (8.5,3) {$\mathcal{A}^{N,p}\left(\frac{F}{2\pi}\right)$};
        \draw[-,dashed] (3,1) -- (5,1);
    
        \draw[->, blue] (chi.south) to[out=270, in=180] (4,2);
        \draw[-, thick,dashed] (3,0) -- (5,0);
        \draw[fill=blue!50, opacity=.5] (3,-3) -- (7,-1) -- (7,4) -- (3,2) -- (3,-3);
        \draw[fill=red!50, opacity=.5] (3.2,-3) -- (7.2,-1) -- (7.2,4) -- (3.2,2) -- (3.2,-3);
        \draw[-, thick] (2,0) -- (3,0);
        \draw[->, thick] (5,0) -- (8,0);
        \draw[fill=black] (5,0) circle (0.05);
        \draw[->, red] (tau.south) to[out=270,in=0] (6,2);
    
        \node[] (zero) at (5,-.5) {$x=0$};
        \node[] (SL) at (2.5,1.15) {$S_L$};
        \node[] (SR) at (8,1) {$S_R=S_L$};
        \draw[-] (2.8,1) -- (3,1);
        \draw[->] (5,1) -- (SR);
      \end{tikzpicture}
      \caption{Chiral rotation stacked with TFT.}
  \end{subfigure}
     \caption{Building the non-invertible chiral symmetry defect}
\end{figure}
Consider inserting a codimension-one domain wall at $x=0$ which implements a chiral rotation with angle $\theta$. Here the chiral background field is taken to be

\begin{equation}
  A_\chi = \theta\delta(x)\dd{x}
\end{equation}
By anomaly inflow, the bulk action changes across the domain wall 
\begin{equation}
  S_L \rightarrow S_R=S_L+2\pi i\theta\int_{x>0}\frac{F\wedge F}{8\pi^2}
\end{equation}
This jump implies that the chiral rotation defect, while being topological, does not define a symmetry of our theory, rather it produces a shift in the $\theta$ angle, and hence defines a topological interface to a different theory. Can one undo such shift? If we can find a 3d TFT that has an anomaly which is equal and opposite to the above shift, we can stack it along the worldvolume of the chiral rotation interface resulting in a genuine topological defect of codimension 1. If we can find one such theory, this procedure will generate a topological defect implementing the chiral rotation. As we shall see below, the price to pay for this is that the resulting defect is non-invertible. In these lectures we will solve this problem for values of the angle $\theta=\frac{p}{N}\in\mathbb{Q}/\bbZ$. For further simplicity let us assume we are working only on Spin 4-manifolds. What do we want from this TFT? Let us fix $\theta = p/N$ so that the offending contibution is $2\pi i \frac{p}{N}\int_{x>0}\frac{F\wedge F}{8\pi^2}$. We are after a 3d TFT that has an equal and opposity anomaly, hence the 3d TFT we are after
\begin{itemize} 
  \item must have a $\bbZ_N^{(1)}$ $1$-form symmetry, and a corresponding collection of topological lines, with a 2-form $\mathbb Z_N$ background field we denote $B^{(2)}$ in what follows;
  \item must have the right 't Hooft anomaly, meaning that upon turning the $B^{(2)}$ background field, the 3d TFT is a boundary of the SPT with anomaly
  \begin{equation}
    \mathcal{A}_4=-\frac{\I 2\pi p}{2N}\int \mathcal{P}\left(B^{(2)}\right)
  \end{equation}
  where $\mathcal{P}$ is the Pontryagin square operation. In particular this implies that the topological lines of the 3d TFT have a non-trivial braiding. Upon identifying $B^{(2)}={F}/{2\pi}$ modulo $N$, this cancels the offending contribution.
\end{itemize} 
It is interseting to remark that there are actually many choices of TFTs which would satisfy these conditions. But as proven by Hsin, Lam and Seiberg\footnote{P.-S. Hsin, H. T. Lam Lam, N. Seiberg, \textit{Comments on One-Form Global Symmetries and Their Gauging in 3d and 4d},\href{https://doi.org/10.21468/SciPostPhys.6.3.039}{SciPost Phys. 6 (2019) 3, 039}, \href{https://arxiv.org/abs/1812.04716}{\textt{arXiv:1812.04716}}} any such TFT $\mathcal{T}$ can be decomposed into a product of a minimal TFT $\mathcal{A}_{N,p}(B^{(2)})$ and a \textit{decoupled} TFT $\mathcal{T}^\prime$
\begin{equation}
    \mathcal{T}=\mathcal{A}_{N,p}\otimes\mathcal{T}^\prime
\end{equation}
One can find $\mathcal{T}^\prime$ from $\mathcal{T}$ by computing 
\begin{equation}
    \frac{\mathcal{A}_{N,-p}\otimes \mathcal{A}_{N,p}}{\bbZ_N}=\operatorname{id}
\end{equation}
All the lines needed for the symmetry are in $\mathcal{A}_{N,p}$. This is consistent with what we discussed in the first lecture: there are several options to define the 3d topological defects that implement chiral rotations in massless QED, each differing by a choice of decoupled TFT. Since the latter is decoupled, these choices are immaterial and give rise to an example of the categorical gauge equivalence principle we mentioned in the first lecture. Of course once a choice is made, we need to stick to it, to work in a consistent set of conventions. Here, we are going to choose as representatives the minimal 3d TFT by Hsin, Lam, Seiberg. Therefore our symmetry defect is going to be given by 
\begin{equation}
  D_{\frac{p}{N}}^{(0)}(M^{(3)})=\mathcal{A}^{N,p}\qty(\frac{F}{2\pi})\otimes \mathcal{U}_{\frac{p}{N}}^\chi(M^{(3)})
\end{equation}
We have one very explicit example, given in Shu-Heng Shao's lectures, for the case $p=1$
\begin{equation}
    \mathcal{D}^{(0)}_{\frac{1}{N}}(M^{(3)})=\int\left[D a\right]_{M^{(3)}}\exp\left(\frac{\I N}{4\pi}\oint_{M^{(3)}} a\dd{a}+\I\oint_{M^{(3)}}a\wedge\frac{F}{2\pi}+\I\frac{2\pi}{N}\oint_{M^{(3)}}\star j_\chi\right)
\end{equation}
In this case we can clearly see the various pieces we discussed 
\begin{itemize}
    \item The topological Chern-Simons action on the defect world-volume
    \begin{equation}
        \exp \I S_{\mathcal{T}}[a]=\exp\left(\frac{\I N}{4\pi}\oint_{M^{(3)}}a\dd{a}\right)
    \end{equation}
    \item The chiral symmetry defect
    \begin{equation}
        \mathcal{U}^\chi_{\frac{1}{N}}=\exp\left(\I \frac{2\pi}{N}\oint_{M^{(3)}}\star j_\chi\right)
    \end{equation}
    \item The coupling of the TFT to the background 
    \begin{equation}
        \I\oint_{M^{(3)}}a\wedge\frac{F}{2\pi}
    \end{equation}
\end{itemize}
The full minimal TFT is then just
\begin{equation}
    \mathcal{A}_{N,1}\left[\frac{F}{2\pi}\right]=\frac{\I N}{4\pi}\oint_{M^{(3)}}a\dd{a}+\I \oint_{M^{(3)}}a\wedge\frac{F}{2\pi}
\end{equation}
For generic values of $N,p$ co-prime, we have that the spectrum of abelian anyons hosted by the TFT have spin and $\bbZ_N$ charge given by
\begin{equation}
    \begin{cases}
        \Theta(L^k)=\exp\left(\I \pi \frac{pk^2}{N}\right)&\text{Spin}\\
        q(L)=p&\text{Charge}
    \end{cases}
\end{equation}
Some remarks are now in order
\begin{enumerate}
    \item Wrapping $\mathcal{D}^{(0)}_{\frac{p}{N}}(S^3)$ without operator insertions is going to give 
    \begin{equation}
        \mathcal{D}^{(0)}_{\frac{p}{N}}(S^3)=\mathcal{A}_{N,p}(S^3)=\frac{1}{\sqrt{N}}
    \end{equation}
    since there is no chiral charge inside the $S^3$, the chiral part acts trivially. This tells us that the operator has a quantum dimension and therefore is non-invertible.
    \item Stacking the non-invertible operator with its orientation reversal gives a condensate 
    \begin{equation}
        \mathcal{D}^{(0)}_{\frac{p}{N}}(M^{(3)})\overline{\mathcal{D}^{(0)}_{\frac{p}{N}}}(M^{(3)})=\sum\limits_{\Sigma^{(2)}\in\HHm_2(M^{(3)},\bbZ_N)}\eta_{\frac{1}{N}}^{(1)}e^{\frac{2\pi \I}{N}p Q(\Sigma^{(2)})}
    \end{equation}
    where 
    \begin{equation}
        \eta_{\frac{1}{N}}^{(1)}(\Sigma^{(2)})=\exp\left(\frac{\I}{N}\int_{\Sigma^{(2)}}\frac{F}{2\pi}\right)
    \end{equation}
    is the generator of the $\bbZ_N^{(1)}\subset \U(1)^{(1)}_m$ magnetic $1$-form symmetry and $Q(\Sigma^{(2)})$ is the triple self-intersection number. The sum is taken over $2$-cycles in $M^{(3)}$.
    \item The non-invertibility of the defect can also be probed by its action on the Hilbert space of the theory. The QED Hilbert space on the compact $3$-manifold $M^{(3)}=S^1\times S^2$ decomposes into flux sectors graded by the magnetic flux along $S^2$
    \begin{equation}
        \mathcal{H}_\text{QED}(S^1\times S^2)=\bigoplus_m\mathcal{H}_m,\qquad m=\int_{S^2}\frac{F}{2\pi}
    \end{equation}
    We want to understand the action of $\mathcal{D}^{(0)}_{\frac{p}{N}}$ on this Hilbert space. To do so, we need to know the partition function of the minimal TFT $\mathcal{A}_{N,1}$ on this spatial manifold. Notice first that the TFT couples to the magnetic $1$-form symmetry so that the flux sector $m$ "talks" with the defect. This magnetic flux induces an anyon line along the $S^1$ which is just the $m$-th power of the generating line $L^m$. One can compute then 
    \begin{equation}
        \mathcal{A}_{N,1}\left(S^1\times S^2,\  L^m
        \begin{tikzcd}
             S^1\arrow[loop left]
        \end{tikzcd} \right)=\begin{cases}
            0&m\neq0 \mod N\\
            1&m=0 \mod N
        \end{cases}
    \end{equation}
\end{enumerate}
So on the Hilbert space, the $D^{(0)}_{\frac{p}{N}}$ defect acts as a projector onto sectors $\mathcal{H}_m$ with $m=0 \mod N$, 
\begin{equation}
    \mathcal{D}^{(0)}_{\frac{p}{N}}(S^1\times S^2)=\mathcal{U}^\chi_{\frac{p}{N}}\otimes\mathcal{P}_{m,N}
\end{equation}
and it is therefore  manifestly non-invertible.

\section{Lecture 3: Higher structure of Chiral symmetry}

\subsection{Morphisms}
Now that we have the most general structure of the defect, we can study various morphisms to explore the higher-structure of the symmetry. Of course, the $1$-form symmetry defects $\eta_\alpha^{(1)}(\Sigma^{(2)})$ should fuse according to their group law meaning that
\begin{equation}
    \Hom_{(2)}\qty(\eta_\alpha^{(1)}\otimes\eta_\beta^{(1)},\eta_\gamma^{(1)})
\end{equation}
is non-trivial only for $\alpha+\beta=\gamma$. These defects can also end on the chiral symmetry defect $\mathcal{D}^{(0)}_{\frac{p}{N}}$ but only restricting to the $\bbZ_N^{(1)}$ subgroup of $\U(1)^{(1)}$ to ensures the coupling to the minimal theory $\mathcal{A}_{N,p}$. This junction defines a $2$-morphism which is identified with the generating line $L$ of $\mathcal{A}_{N,p}$
\begin{equation}
    \Hom_{(2)}\qty(\eta_{\frac{pk}{N}}^{(1)},\mathds{1}_{\mathcal{D}_{\frac{p}{N}}}^{(1)})=L^k
\end{equation}
If we want to fuse together the chiral defects, they should inherit the a group-like fusion from the $\mathcal{U}^\chi$. In particular 
\begin{equation}
    \Hom\qty(\mathcal{D}_{\frac{p}{N}}^{(0)},\mathcal{D}_{\frac{p^\prime}{N^\prime}}^{(0)})=\begin{cases}
        \End(\mathcal{A}_{N,p})&\text{if }\frac{p}{N}=\frac{p^\prime}{N^\prime}\mod 1\\
        0
    \end{cases}
\end{equation}
Let us consider the following fusion
\begin{equation}
    \mathcal{D}^{(0)}_{\frac{1}{4}}\otimes\mathcal{D}^{(0)}_{\frac{1}{4}}\rightarrow\mathcal{D}^{(0)}_{\frac{1}{2}}
\end{equation}
Let us call $L$ and $L^\prime$ the two generators for the $\mathcal{A}_{4,1}$. These can come from a $\eta^{(1)}_{\frac{1}{4}}$ ending on the defects.
\begin{figure}
    \centering
    \begin{tikzpicture}[scale=.9,every node/.append style={transform shape}]
      \begin{scope}[]
        \node[below] (L) at (0,0) {$\mathcal{D}^{(0)}_{\frac{1}{4}}$};
        \node[below] (R) at (4,0) {$\mathcal{D}^{(0)}_{\frac{1}{4}}$};
        \node[circle,fill=orange,inner sep=1.5pt,minimum size=.1pt] (M) at (2,2) {};
        \node[above] (U) at (2,4) {$\mathcal{D}^{(0)}_{\frac{1}{2}}$};

        \node[circle,fill=blue,inner sep=1.5pt,minimum size=.1pt] (LL) at (.4,1) {};
        \node[above left, blue] (LL1) at (.4,1) {$L$};

        \draw[thick] (L) to[out=90, in=210] (M) to[out=-30, in=90] (R);
        \draw[thick] (M) to (U);

        \draw[-,thick, blue] (-1.5,1) -- (LL.center);
        \node[above,blue] (U) at (-1.5,1) {$\eta^{(1)}_{\frac{1}{4}}$};

        \node[right,orange,draw] (Z) at (3,5) {$\bbZ_2$ gauging interface};
        \draw[<-, orange] (M) to[out=20, in=270] (Z);
        
        \node[] (sim) at (5,2) {\scalebox{1.5}{$\simeq$}};
      \begin{scope}[xshift=6cm]
        \node[below] (L) at (0,0) {$\mathcal{D}^{(0)}_{\frac{1}{4}}$};
        \node[below] (R) at (4,0) {$\mathcal{D}^{(0)}_{\frac{1}{4}}$};
        \node[circle,fill=blue,inner sep=1.5pt,minimum size=.1pt] (M) at (2,2) {};
        \node[above] (U) at (2,4) {$\mathcal{D}^{(0)}_{\frac{1}{2}}$};

        %\node[circle,fill=blue,inner sep=1.5pt,minimum size=.1pt] (LL) at (.4,1) {};
        \node[above left, blue] (LL1) at (2,2) {$L$};

        \draw[thick] (L) to[out=90, in=210] (M) to[out=-30, in=90] (R);
        \draw[thick] (M) to (U);

        \draw[-,thick, blue] (0,2) -- (M);
        \node[above,blue] (U) at (0,2) {$\eta^{(1)}_{\frac{1}{4}}$};
      \end{scope}
    \end{scope}
    \end{tikzpicture}
    \caption{The line $L$ in $\mathcal{A}_{2,1}$ generated by a $\bbZ_4^{(1)}$ $1$-form symmetry defects remains stuck on the gauging interface since it is not neutral under the symmetry generated by $L^2\otimes L^{\prime 2}$.}
    \label{eq:stucklines}
\end{figure}
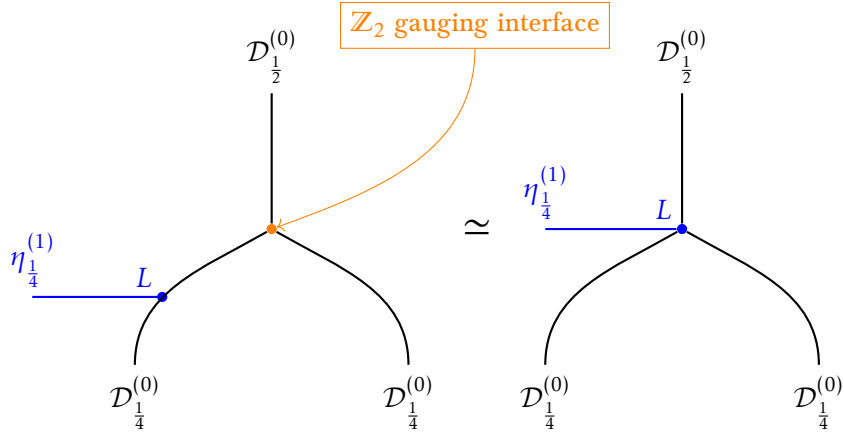
On the other side, we have a line $\tilde{L}$ coming from the $\mathcal{A}_{2,1}$ theory. This has charge $q(\Tilde{L})=\frac{1}{2}$ and therefore its square is transparent. But before fusion $q(L^2\otimes L^{\prime 2})\neq 0$. Therefore on one side we have more lines than the ones after the fusion. However, the line $L^2\otimes L^{\prime 2}$ is actually transparent to the bulk $1$-form symmetry and can be therefore gauged. After gauging this $\bbZ_2$ subgroup, we are left with a set of lines, modulo fusion with $\mathds{1}+ L^2\otimes L^{\prime 2}$, generated by
\begin{eqnarray}
    L\otimes L^\prime, &q(L\otimes L^\prime)=2,&\Theta(L\otimes L^\prime)=i\\
    L \otimes L^{\prime 3}, &q(L\otimes L^{\prime 3})=0,&\Theta(L\otimes L^{\prime 3})=-i
\end{eqnarray}
As we can see, the second lines are decoupled, which means they appear only in the spectrum of a decoupled TFT coefficient in the fusion product of the symmetry defects 
\begin{equation}
  \mathcal{D}^{(0)}_{\frac{1}{4}}(M^{(3)})\otimes \mathcal{D}^{(0)}_{\frac{1}{4}}(M^{(3)})=\mathcal{A}_{2,1}(M^{(3)})\mathcal{D}^{(0)}_{\frac{1}{2}}(M^{(3)})
\end{equation}
With this gauging interface, we can construct the homs that follow the group law of $\mathbb{Q}/\bbZ$ in full generality
\begin{equation}
  \Hom\qty(\mathcal{D}^{(0)}_{\frac{p}{N}}\otimes \mathcal{D}^{(0)}_{\frac{p^\prime}{N^\prime}},\mathcal{D}^{(0)}_{\frac{p}{N}+\frac{p^\prime}{N^\prime}})\qquad\text{by gauging }\bbZ_{\gcd(p+q,N)}
\end{equation}
The decoupled TFT is really "useless" since it is completely decoupled from the bulk. 
\subsection{Associators}
We now would like to understand the associator morphism in the category of chiral defects. These are going to be constructed in a manner similar to the one in Figure \ref{fig:associators} but they are also going to depend on the gauging interfaces. Let us call the gauging interfaces $A_{ijk}$ and the various symmetry defects just by their $\mathbb{Q}/\bbZ$ charge as not to clutter too much the notation. Graphically, the associator $F_{\mathcal{D}_{p_1/N_1},\mathcal{D}_{p_2/N_2},\mathcal{D}_{p_3/N_3}}$ will take the diagrammatic form shown in Figure (\ref{fig:associator2}).
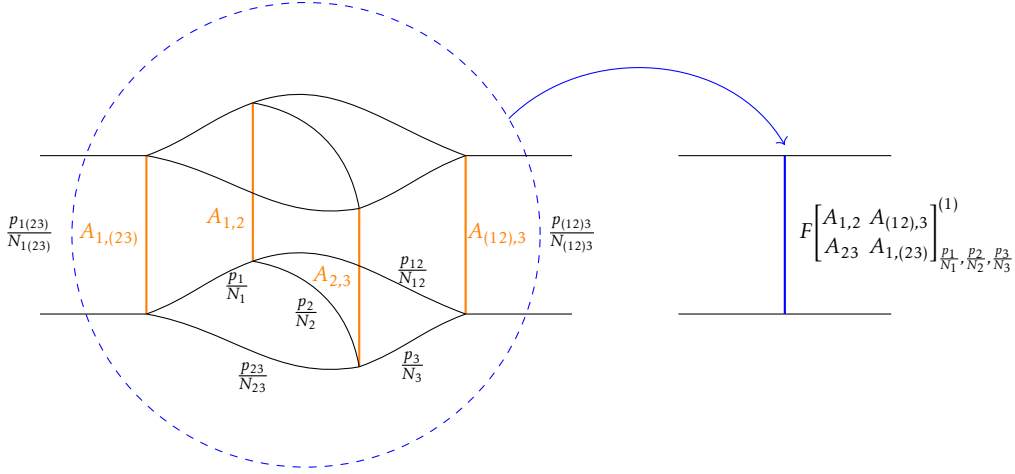
\begin{figure}
  \centering
  \begin{tikzpicture}[scale=.7, every node/.style={transform shape}]
    \begin{scope}
      %%%%%%%lower coord
      \coordinate (LD) at (0,0); %left down
      \coordinate (CLD) at (2,1); %center left down
      \coordinate (CRD) at (4,-1); %center right down
      \coordinate (RD) at (6,0); %right down
      %%%%%%%upper coord
      \coordinate (LU) at (0,3); %left up
      \coordinate (CLU) at (2,4); %center left up
      \coordinate (CRU) at (4,2); %center right up
      \coordinate (RU) at (6,3); %right up

      \draw[dashed, blue] (3,1.5) circle (4.4);
      \foreach \name in {L,CL,CR,R}
        \draw[-, thick, orange] (\name D) -- (\name U);

      \draw[-] (-2,0) -- (LD) to[out=-10, in=190] (CRD);
      \draw[-] (LD) to[out=20, in=200] (CLD);
      \draw[-] (8,0) -- (RD) to[out=200, in=20] (CRD);
      \draw[-] (RD) to[out=160, in=20] (CLD);

      \draw[-] (-2,3) -- (LU) to[out=-10, in=190] (CRU);
      \draw[-] (LU) to[out=20, in=200] (CLU);
      \draw[-] (8,3) -- (RU) to[out=200, in=20] (CRU);
      \draw[-] (RU) to[out=160, in=20] (CLU);
      
      \foreach \name in {U, D}
        \draw[-] (CL\name) to[out=-10, in=100] (CR\name);

      %\fill[blue!30, opacity=.5] (-2,0) -- (LD) -- (LU) -- (-2,3) -- cycle;
      %\fill[blue!30, opacity=.5] (8,0) -- (RD) -- (RU) -- (8,3) -- cycle;

      %\fill[blue!30, opacity=.5] (LD) to[out=-10, in=190] (CRD) -- (CRU) to[out=190,in=-10] (LU) -- cycle;
      %\fill[blue!20, opacity=.2] (LD) to[out=20, in=200] (CLD) -- (CLU) to[out=200,in=20] (LU) -- cycle;
      %\fill[blue!20, opacity=.2] (CLD) to[out=20, in=160] (RD) -- (RU) to[out=160,in=20] (CLU) -- cycle;
      %\fill[blue!30, opacity=.5] (CRD) to[out=20, in=200] (RD) -- (RU) to[out=200,in=20] (CRU) -- cycle;
      %\fill[red!50, opacity=.2] (CRD) -- (CRU) to[out=100, in=-10] (CLU) -- (CLD) to[out=-10, in=100] cycle;

      \node[] at (-2.2,1.5) {$\frac{p_{1(23)}}{N_{1(23)}}$};
      \node[] at (8,1.5) {$\frac{p_{(12)3}}{N_{(12)3}}$};
      \node[] at (2,-1.2) {$\frac{p_{23}}{N_{23}}$};
      \node[] at (1.7,.5) {$\frac{p_{1}}{N_{1}}$};
      \node[] at (5,.8) {$\frac{p_{12}}{N_{12}}$};
      \node[] at (5,-1) {$\frac{p_{3}}{N_{3}}$};
      \node[] at (3,0) {$\frac{p_{2}}{N_{2}}$};
      \node[orange] at (1.5,1.8) {$A_{1,2}$};
      \node[orange] at (-.7,1.5) {$A_{1,(23)}$};
      \node[orange] at (3.5,.7) {$A_{2,3}$};
      \node[orange] at (6.6,1.5) {$A_{(12),3}$};
    \end{scope}
    \draw[->,blue] ({4.4*cos(30)+3},{4.4*sin(30)+1.5}) to[out=45, in=120] (12,3.2);
    \begin{scope}[xshift=10cm]
      \draw[-] (0,0) -- (4,0);
      \draw[-] (0,3)-- (4,3);
      \draw[-,blue, thick] (2,0) -- (2,3);
      \node at (4.3,1.5) {$F\mqty[A_{1,2}&A_{(12),3}\\A_{23}&A_{1,(23)}]^{(1)}_{\frac{p_1}{N_1},\frac{p_2}{N_2},\frac{p_3}{N_3}}$};
    \end{scope}
  \end{tikzpicture}
  \caption{Higher associator $F$ for the chiral symmetry defect}
  \label{fig:associator2}
\end{figure}

The associator is "morally" a $2d$ TFT, but contrary to the defect we started with that didn't have local operators, this has point like topological operators arising from the lines in $\mathcal{A}_{N,p}$ that remain stuck to the gauging interface (Figure \ref{eq:stucklines}). This $2d$ TFT can therefore be decomposed over selection sectors 
\begin{equation}
    \mathcal{T}_{2d}=\bigoplus_i\mathcal{T}_{\pi_i}
\end{equation}
labelled by the local operators in the corresponding fusion category satisfying 
\begin{equation}
    \pi_i\otimes \pi_j =\delta_{ij}\pi_i
\end{equation}
Because of the coupling with the bulk $1$-form symmetry $\U(1)^{(1)}_m$, these vertices are also charged under it. The expansion then reads
\begin{equation}
    F^{(1)}=\bigoplus_{q=0}^{N-1}\qty(\bigoplus_{i:q_i=q}\mathcal{T}_{\pi_i})\otimes \eta^{(1)}_{\frac{q}{N}}
\end{equation}
where $q_i$ is the $1$-form symmetry charge of $\pi_i$.

Now we would like to find a way to detect the associator. Based on the idea of the action on Wilson lines in the Ising model discussed by Shu-Heng Shao, we could try to throw a 't Hooft line at the associator and see what happens\footnote{Being pedantic, it is not really the monopole line that we move through the associator but vice versa being the latter topological and the former not.}. Since the monopole line is charged under the $\U(1)^{(1)}_m$ we expect that when passing through the defect $\mathcal{D}^{(0)}_{\frac{p}{N}}$ will be mapped to a non-genuine line operator constructed attached to the $L$ generator of $\mathcal{D}^{(0)}_{\frac{p}{N}}$ through a $\eta^{(1)}_{\frac{p}{N}}$ surface.
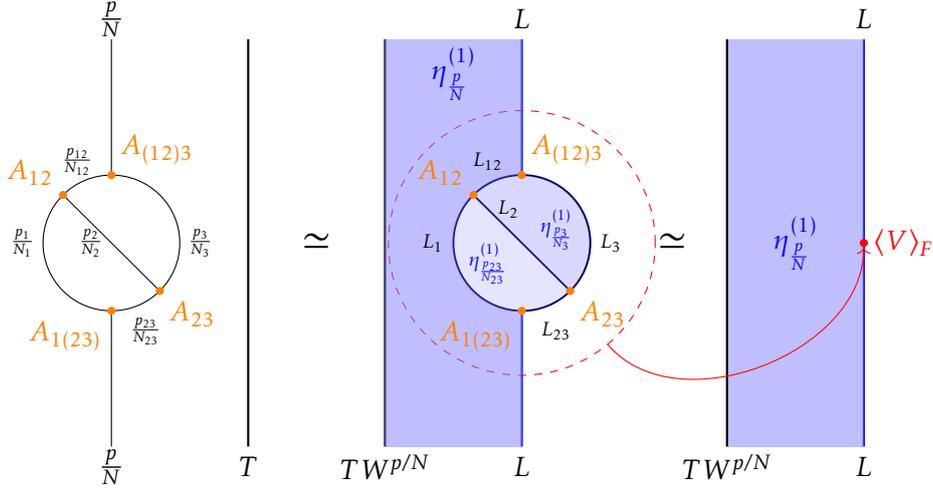
\begin{figure}
  \centering
  \begin{tikzpicture}[scale=.9,every node/.style={transform shape}]
    \tikzset{
      clip even odd rule/.code={\pgfseteorule}, % Credit to Andrew Stacey 
      invclip/.style={
          clip,insert path=
              [clip even odd rule]{
                  [reset cm](-\maxdimen,-\maxdimen)rectangle(\maxdimen,\maxdimen)
              }
      }
    } 

    %\draw[help lines, step=1] (0,0) grid (14,6);
    \draw[-, thick] (4,0) -- (4,6);

    \node[circle split, draw, rotate=-45,  minimum size=2cm] (circ) at (2,3) {};
    \draw[-] (2,0) -- (2,2);
    \draw[-] (2,4) -- (2,6);

    \node at (5,3) {\scalebox{1.5}{$\simeq$}};

    \begin{scope}[xshift=-2cm]
      \draw[-, thick] (8,0) -- (8,6);
      \node[circle split, draw, rotate=-45,  minimum size=2cm, blue, thick] (circ) at (10,3) {};
      \draw[red, dashed] (10,3) circle (1.95);
      \draw[-, blue, thick] (10,0) -- (10,2);
      \draw[-, blue, thick] (10,4) -- (10,6);

      \begin{scope}
        \clip[rotate around={-45:(11,2)}] (8,2) rectangle ++(3,2);
        \node[circle split, rotate=-45,  minimum size=2cm, fill=blue!30, draw, fill opacity=.5] (circ) at (10,3) {};
      \end{scope}
      \begin{scope}
        \clip[rotate around={45:(11,2)}] (8,2) rectangle ++(3,3);
        \node[circle split, rotate=-45,  minimum size=2cm, fill=blue!20, draw, fill opacity=.5] (circ) at (10,3) {};
      \end{scope}

      \begin{scope}
        \begin{pgfinterruptboundingbox}
        \clip[invclip] (10,3) circle (1);
        \fill[blue!50, opacity=.5] (8,0) -- (10,0) -- (10,6) -- (8,6) -- cycle;
        \end{pgfinterruptboundingbox}
      \end{scope}
    \end{scope}

    \foreach \s in {2,8}{
      \foreach \angle/\pos/\name in {90/above right/(12)3,135/above left/12,270/below left/1(23),315/below right/23}{
        \node[circle,fill=orange,inner sep=1.2pt,minimum size=.04pt] (P\s\angle) at ({\s+cos(\angle)},{3+sin(\angle)}) {};
        \node[\pos,orange] at (P\s\angle)  {$A_{\name}$};
      }
    }

    \begin{scope}[xshift=-2cm]
      \node[blue] at (9,5.5) {$\eta^{(1)}_{\frac{p}{N}}$};
      \node[blue] at (9.5,2.7) {\scalebox{.7}{$\eta^{(1)}_{\frac{p_{23}}{N_{23}}}$}};
      \node[blue] at (10.5,3.2) {\scalebox{.7}{$\eta^{(1)}_{\frac{p_3}{N_3}}$}};

      \node[] at (8,-.3) {$TW^{p/N}$};
      \node[] at (10,-.3) {$L$};
      \node[] at (10,6.3) {$L$};
      \node[] at (8.7,3) {\scalebox{.7}{$L_1$}};
      \node[] at (11.3,3) {\scalebox{.7}{$L_3$}};
      \node[] at (10.5,1.7) {\scalebox{.7}{$L_{23}$}};
      \node[] at (9.5, 4.2) {\scalebox{.7}{$L_{12}$}};
      \node[] at (9.8,3.5) {\scalebox{.7}{$L_2$}};
    \end{scope}

    \node[] at (4,-.3) {$T$};
    \node[] at (2,-.3) {$\frac{p}{N}$};
    \node[] at (2,6.3) {$\frac{p}{N}$};
    \node[] at (.7,3) {\scalebox{.7}{$\frac{p_1}{N_1}$}};
    \node[] at (3.3,3) {\scalebox{.7}{$\frac{p_3}{N_3}$}};
    \node[] at (2.5,1.7) {\scalebox{.7}{$\frac{p_{23}}{N_{23}}$}};
    \node[] at (1.5, 4.2) {\scalebox{.7}{$\frac{p_{12}}{N_{12}}$}};
    \node[] at (1.7,3) {\scalebox{.7}{$\frac{p_2}{N_2}$}};

    \node at (10.2,3) {\scalebox{1.5}{$\simeq$}};
    \draw[blue,thick] (13,0) -- (13,6);
    \fill[blue!50, opacity=.5] (11,0)--(13,0)--(13,6)--(11,6) -- cycle;
    \draw[-,thick] (11,0) -- (11,6);
    \node[circle,fill=red,inner sep=1.2pt,minimum size=.04pt] (V) at (13,3) {};

    \node[red,right] at (V) {$\langle V\rangle_F$};
    \node[blue] at (12,3) {$\eta^{(1)}_{\frac{p}{N}}$};
    \node[] at (11,-.3) {$TW^{p/N}$};
    \node[] at (13,-.3) {$L$};
    \node[] at (13,6.3) {$L$};

    \draw[->, red] ({8+1.95*cos(-50)},{3+1.95*sin(-50)}) to[out=-50, in=-90] (V);
  \end{tikzpicture}
  \caption{A 't Hooft line passing through the associator diagram}
\end{figure}
As a simple example, consider the definition of the associator morphism in Figure (\ref{fig:associator2}) setting $N_1=N_2=N$, $N_3=0$ and $p_1=-p_2=p$. Once we pass the 't Hooft line through the associator diagram, we crate various $1$-form symmetry defects as well as lines in $\mathcal{A}_{N,p}$. After closing the bubble, we can see that the associator computes a one-point function in the associator TFT. From this construction we can easily see that this one-point function is just the one for the vertex operator $V(L\otimes \overline{L})$ which vanishes identically. In fact
\begin{equation}
  F[A\ A]_{\frac{p}{N},\frac{p}{N},-\frac{p}{N}}=\bigoplus_{k=0}^{N-1}\eta_{\frac{k}{N}}^{(1)}(\Sigma^{(2)})=\mathcal{C}_N^{(1)}
\end{equation}
is just a condensate of the $1$-form symmetry defect. When we pass a 't Hooft line through it
\begin{equation}
  F[A\ A]_{\frac{p}{N},\frac{p}{N},-\frac{p}{N}}\otimes T=T\otimes \sum_{k=0}^{N-1}e^{2\pi \I \frac{k}{N}}\eta_{\frac{k}{N}}^{(1)}(\Sigma^{(2)})=T\otimes \sum_{k=0}^{N-1}e^{2\pi \I \frac{k}{N}}=0
\end{equation}
where in the last identity we just shrink the $\eta^{(1)}$ defect. This gives us a selection rule
\begin{equation}
  \left\langle V(L\otimes\overline{L})\right\rangle=0
\end{equation}
which leads us to the conclusion that $F$-symbols are actually physical.

\section{Lecture 4: Takahashi-Ward identities}

\subsection{Takahashi-Ward Identities}
We can now explore the consequences of the higher structure from the generalized symmetry in the form of Takahashi-Ward identities which are satisfied by the partition function of a QFT with the insertion of the these symmetry operators. Consider putting our $4d$ theory on a four-sphere $S^4$. On this we can insert local operators which carry some charge
\begin{equation}
    q(O_i)=q_i,\qquad \text{under }\U(1)_\chi^{(0)}
\end{equation}
As we have shown before, the non-invertible chiral symmetry defect carries a quantum dimension
\begin{equation}
    \mathcal{D}_{\frac{p}{N}}^{(0)}\qty(
        \begin{tikzpicture}[scale=.3, baseline={(0,-.1)}]
        \shade[ball color = blue!50, opacity = 0.5] (0,0) circle (2cm);
        \draw (0,0) circle (2cm);
        \draw (-2,0) arc (180:360:2 and 0.6);
        \draw[dashed] (2,0) arc (0:180:2 and 0.6);
        \fill[fill=blue!50] (0,0) circle (1pt);
\end{tikzpicture})=\frac{1}{\sqrt{N}}
\end{equation}
But, whenever there is some local insertion of an operator charged under the $\U(1)_\chi^{(0)}$, this will measure its charge
\begin{equation}
    \mathcal{D}_{\frac{p}{N}}^{(0)}\qty(
        \begin{tikzpicture}[scale=.3, baseline={(0,-.1)}]
        \shade[ball color = blue!50, opacity = 0.5] (0,0) circle (2cm);
        \draw (0,0) circle (2cm);
        \draw (-2,0) arc (180:360:2 and 0.6);
        \draw[dashed] (2,0) arc (0:180:2 and 0.6);
        \fill[fill=blue!50] (0,0) circle (1pt);
        \node[circle,fill=red,inner sep=1.5pt,minimum size=.1pt] (M) at (1,1) {};
        \node[red, above] at (2,1) {$O_1$};
\end{tikzpicture})=\frac{1}{\sqrt{N}}e^{2\pi \I\frac{p}{N}q_1}
\end{equation}
This equation is meant to be thought to be in a $S^4$-correlator between the the local operator and the symmetry defect: when we exchange the local operator with the symmetry defect we gain a phase. Then by shrinking the symmetry defect we get its quantum dimension. So in general we put some local operators on $S^4$ and bubble up a $\mathcal{D}_{p/N}^{(0)}$ defect on $S^3\subset S^4$, move it around the $S^4$ touching all the charged operators, and then close it back again
\begin{equation}
    \left\langle
        \begin{tikzpicture}[scale=.5, baseline={(0,-.1)}]
        \shade[ball color = blue!50, opacity = 0.5] (0,0) circle (2cm);
        \draw (0,0) circle (2cm);
        \draw (-2,0) arc (180:360:2 and 0.6);
        \draw[dashed] (2,0) arc (0:180:2 and 0.6);
        \fill[fill=blue!50] (0,0) circle (1pt);
        \node[circle,fill=red,inner sep=1.5pt,minimum size=.1pt] (M) at (-1,-1) {};
        \node[circle,fill=red,inner sep=1.5pt,minimum size=.1pt] (M) at (-1,.7) {};
        \node[red, above] at (2,1) {$O_i$};
        \draw[dashed, blue, thick] (.5,-1.2) circle (.4 and .4); 
        \foreach \x in {1,...,8}{\draw[blue, -latex] ({.5+.4*cos(\x*360/8)},{-1.2+.4*sin(\x*360/8)} )-- ++({.5*cos(\x*360/8)},{.5*sin(\x*360/8)});}
        \node[blue] at (1.5,-3) {$D_{p/N}^{(0)}(S^3)$};
\end{tikzpicture}\right\rangle=\prod_i e^{2\pi \I \frac{p}{N}q_i}\ \left\langle 
\begin{tikzpicture}[scale=.3, baseline={(0,-.1)}]
        \shade[ball color = blue!50, opacity = 0.5] (0,0) circle (2cm);
        \draw (0,0) circle (2cm);
        \draw (-2,0) arc (180:360:2 and 0.6);
        \draw[dashed] (2,0) arc (0:180:2 and 0.6);
        \fill[fill=blue!50] (0,0) circle (1pt);
        \node[circle,fill=red,inner sep=1.5pt,minimum size=.1pt] (M) at (1,1) {};
        \node[circle,fill=red,inner sep=1.5pt,minimum size=.1pt] (M) at (-1,-1) {};
        \node[circle,fill=red,inner sep=1.5pt,minimum size=.1pt] (M) at (-1,.7) {};
        \node[red, above] at (2,1) {$O_i$};
\end{tikzpicture}\right\rangle
\end{equation}
This last equality then implies that the correlator is non-zero only when
\begin{equation}
    \prod_i e^{2\pi \I \frac{p}{N}q_i}=1
\end{equation}
We can do more. Let us consider the theory on $S^2\times S^2$ and evaluate how the chiral symmetry defect can give us a constraint on the partition function of the theory. There are various way of visualizing $S^2\times S^2$: first of all we consider one of the $S^2$ as an $S^1$ fibration over an interval with size shrinking at the ends. Then, this gives a fibration of $S^2\times S^2$ over $S^2\times S^1$. We proceed to visualize the fiber as $S^2$ times an interval with the two endpoints identified as in Figure (\ref{fig:S2S2}).
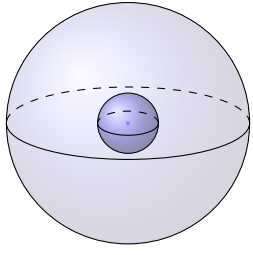
\begin{figure}
    \centering
    \begin{tikzpicture}[scale=.8, baseline={(0,-.1)}]
        \shade[ball color = blue!50, opacity = 0.2] (0,0) circle (2cm);
        \draw (0,0) circle (2cm);
        \draw (-2,0) arc (180:360:2 and 0.6);
        \draw[dashed] (2,0) arc (0:180:2 and 0.6);
        \fill[fill=blue!50] (0,0) circle (1pt);

        \shade[ball color = blue!50, opacity = 0.5] (0,0) circle (.5cm);
        \draw (0,0) circle (.5cm);
        \draw (-.5,0) arc (180:360:.5 and 0.2);
        \draw[dashed] (.5,0) arc (0:180:.5 and 0.2);
        \fill[fill=blue!50] (0,0) circle (.25pt);    
    \end{tikzpicture}
    \caption{In the picture the inner and outer spheres are identified.}
    \label{fig:S2S2}
\end{figure}
For our analysis, a transverse slice of the manifold will be sufficient. To start let us bubble a $\mathcal{D}_{p/N}^{(0)}$ defect on a contractible $S^3\subset S^2\times S^2$. In the $S^2\times S^1$ fiber this will look like an $S^2$.

\begin{eqnarray*}
    \begin{tikzpicture}[baseline={(0,0)}]
        \draw[] (0,0) circle (.5);
        \draw[] (0,0) circle (1.5);
    \end{tikzpicture}&=\sqrt{N}\ 
    \begin{tikzpicture}[baseline={(0,0)}]
        \draw[] (0,0) circle (.5);
        \draw[] (0,0) circle (1.5);
        \draw[thick, red] (-1,0) circle (.4);
    \end{tikzpicture}&=\sqrt{N}\ 
    \begin{tikzpicture}[baseline={(0,0)}]
        \tikzset{
          clip even odd rule/.code={\pgfseteorule},
          invclip/.style={
              clip,insert path=
                  [clip even odd rule]{
                      [reset cm](-\maxdimen,-\maxdimen)rectangle(\maxdimen,\maxdimen)
                  }
          }
        } 
        \draw[] (0,0) circle (.5);
        \draw[] (0,0) circle (1.5);
        \begin{scope}
            \clip[] (0,0) circle (1.5);
            \draw[thick, red] (-1.5,0) circle (.4 and .2);
        \end{scope}
        \begin{scope}
            \begin{pgfinterruptboundingbox}
                \clip[invclip] (0,0) circle (.5);
                \draw[thick, red] (-.5,0) circle (.4 and .2);
            \end{pgfinterruptboundingbox}
        \end{scope}
        \draw[dashed, blue] (-1,0) circle (.3);
        \node[draw, blue] (txt) at (2,1.5) {Fuse};
        \draw[<-, blue] (-1,.3) to[out=90, in=-90] (txt);
    \end{tikzpicture}\\[5pt]
    %%%%%%%%%%%%%%%%%%%%%%%%%%%%%%%
   \sqrt{N}\ \begin{tikzpicture}[baseline={(0,0)}]
        \tikzset{
          clip even odd rule/.code={\pgfseteorule},
          invclip/.style={
              clip,insert path=
                  [clip even odd rule]{
                      [reset cm](-\maxdimen,-\maxdimen)rectangle(\maxdimen,\maxdimen)
                  }
          }
        } 
        \draw[] (0,0) circle (.5);
        \draw[] (0,0) circle (1.5);
        \begin{scope}
            \begin{pgfinterruptboundingbox}
                \clip[invclip] (0,0) circle (.5);
                \clip[] (0,0) circle (1.5);
                \draw[thick, red,-] (0,0) -- (-2,1) -- (-2,-1) -- cycle;
            \end{pgfinterruptboundingbox}
        \end{scope}
        \begin{scope}
            \clip[] (0,0) -- (-2,1) -- (-2,-1) -- cycle;
            \draw[-, blue, thick] (-1,1) -- (-1,-1);
            \node[blue] at (-1.2,0) {$\mathcal{C}_N$};
        \end{scope}
        \node[circle,fill=green,inner sep=1.2pt,minimum size=.04pt] (L1) at (-1,0.5) {};
        \node[circle,fill=green,inner sep=1.2pt,minimum size=.04pt] (L2) at (-1,-0.5) {};
        \node[green, above right] at (L1) {$L$};
        \node[green, below right] at (L2) {$L$};
    \end{tikzpicture}&=\sqrt{N}\ 
    \begin{tikzpicture}[baseline={(0,0)}]
    \tikzset{
          clip even odd rule/.code={\pgfseteorule},
          invclip/.style={
              clip,insert path=
                  [clip even odd rule]{
                      [reset cm](-\maxdimen,-\maxdimen)rectangle(\maxdimen,\maxdimen)
                  }
          }
        } 
        \draw[] (0,0) circle (.5);
        \draw[] (0,0) circle (1.5);
        \begin{scope}
            \begin{pgfinterruptboundingbox}
                \clip[invclip] (-1.5,-.2) rectangle ++(3,.4);
                \draw[red, thick] (0,0) circle (.7);
                \draw[red, thick] (0,0) circle (1.2);
            \end{pgfinterruptboundingbox}
        \end{scope}
        \draw[red, thick] (-0.671,.2) -- (-0.458,.2);
        \draw[red, thick] (-0.671,-.2) -- (-0.458,-.2);
        \draw[red, thick] (-1.183,.2) -- (-1.487,.2);
        \draw[red, thick] (-1.183,-.2) -- (-1.487,-.2);

        \draw[red, thick] (0.671,.2) arc (180:360:.256 and .1);
        \draw[red, thick] (0.671,-.2) arc (180:0:.256 and .1);

        \draw[blue, thick, -] (-0.849,.849) to[out=-80, in=80, tension=2] (-0.849, -0.849);
        \node[circle,fill=green,inner sep=1.2pt,minimum size=.04pt] (L1) at (-0.849,.849) {};
        \node[circle,fill=green,inner sep=1.2pt,minimum size=.04pt] (L2) at (-0.849, -0.849) {};
    \end{tikzpicture}&=\sqrt{N}\ \begin{tikzpicture}[baseline={(0,0)}]
    \tikzset{
          clip even odd rule/.code={\pgfseteorule},
          invclip/.style={
              clip,insert path=
                  [clip even odd rule]{
                      [reset cm](-\maxdimen,-\maxdimen)rectangle(\maxdimen,\maxdimen)
                  }
          }
        } 
        \draw[] (0,0) circle (.5);
        \draw[] (0,0) circle (1.5);
        \begin{scope}
            \begin{pgfinterruptboundingbox}
                \clip[invclip] (-1.5,-.2) rectangle ++(1,.4);
                \draw[red, thick] (0,0) circle (.7);
                \draw[red, thick] (0,0) circle (1.2);
            \end{pgfinterruptboundingbox}
        \end{scope}
        \draw[red, thick] (-0.671,.2) -- (-0.458,.2);
        \draw[red, thick] (-0.671,-.2) -- (-0.458,-.2);
        \draw[red, thick] (-1.183,.2) -- (-1.487,.2);
        \draw[red, thick] (-1.183,-.2) -- (-1.487,-.2);

        \draw[blue, thick, -] (-0.849,.849) to[out=-80, in=80, tension=2] (-0.849, -0.849);
        \node[circle,fill=green,inner sep=1.2pt,minimum size=.04pt] (L1) at (-0.849,.849) {};
        \node[circle,fill=green,inner sep=1.2pt,minimum size=.04pt] (L2) at (-0.849, -0.849) {};

        \draw[-, thick, orange] (.7,0)--(1.2,0);
        \node[circle,fill=pink,inner sep=1.2pt,minimum size=.04pt] (L1) at (.7,0) {};
        \node[circle,fill=pink,inner sep=1.2pt,minimum size=.04pt] (L2) at (1.2, 0) {};
    \end{tikzpicture}\\[5pt]
    %%%%%%%%%%%%%%%%%%%%%%%%%%%%%%%%%%%%%%%%%%%%
    \sqrt{N}\ \begin{tikzpicture}[baseline={(0,0)}]
    \tikzset{
          clip even odd rule/.code={\pgfseteorule},
          invclip/.style={
              clip,insert path=
                  [clip even odd rule]{
                      [reset cm](-\maxdimen,-\maxdimen)rectangle(\maxdimen,\maxdimen)
                  }
          }
        } 
        \draw[] (0,0) circle (.5);
        \draw[] (0,0) circle (1.5);
        \begin{scope}
            \begin{pgfinterruptboundingbox}
                \clip[invclip] (-1.5,-.2) rectangle ++(1,.4);
                \draw[red, thick] (0,0) circle (.7);
                \draw[red, thick] (0,0) circle (1.2);
            \end{pgfinterruptboundingbox}
        \end{scope}
        \draw[red, thick] (-0.671,.2) -- (-1.183,.2);
        \draw[red, thick] (-0.671,-.2) -- (-1.183,-.2);
        \draw[blue, thick, -] (-0.9,.2) -- (-0.9, -0.2);
        \node[circle,fill=green,inner sep=1.2pt,minimum size=.04pt] (L1) at (-0.9,.2) {};
        \node[circle,fill=green,inner sep=1.2pt,minimum size=.04pt] (L2) at (-0.9,-.2) {};

        \draw[-, thick, orange] (.7,0)--(0.5,0);
        \draw[-, thick, orange] (1.2,0)--(1.5,0);
        \node[circle,fill=pink,inner sep=1.2pt,minimum size=.04pt] (L3) at (.7,0) {};
        \node[circle,fill=pink,inner sep=1.2pt,minimum size=.04pt] (L4) at (1.2, 0) {};
    \end{tikzpicture}&=\sqrt{N}\ 
    \begin{tikzpicture}[baseline={(0,0)}]
    \tikzset{
          clip even odd rule/.code={\pgfseteorule},
          invclip/.style={
              clip,insert path=
                  [clip even odd rule]{
                      [reset cm](-\maxdimen,-\maxdimen)rectangle(\maxdimen,\maxdimen)
                  }
          }
        } 
        \draw[] (0,0) circle (.5);
        \draw[] (0,0) circle (1.5);
        \begin{scope}
            \begin{pgfinterruptboundingbox}
                \clip[invclip] (1,0) circle (.3);
                \draw[blue, thick] (0,0) circle (1.05);
            \end{pgfinterruptboundingbox}
        \end{scope}
        \draw[red, thick] (1,0) circle (.3);
        \draw[orange, thick] (0.7,0) -- (0.5,0);
        \draw[orange, thick] (1.3,0) -- (1.5,0);
        \node[circle,fill=green,inner sep=1.2pt,minimum size=.04pt] (L1) at (1,0.3) {};
        \node[circle,fill=green,inner sep=1.2pt,minimum size=.04pt] (L2) at (1,-0.3) {};
        \node[circle,fill=pink,inner sep=1.2pt,minimum size=.04pt] (L3) at (0.7,0) {};
        \node[circle,fill=pink,inner sep=1.2pt,minimum size=.04pt] (L4) at (1.3,0) {};
    \end{tikzpicture}&=\frac{1}{N}\sum_{b,c\in \bbZ_N}e^{\frac{2\pi \I }{N}(p^{-1})_N b\cup c}\; 
    \begin{tikzpicture}[baseline={(0,0)}]
    \tikzset{
          clip even odd rule/.code={\pgfseteorule},
          invclip/.style={
              clip,insert path=
                  [clip even odd rule]{
                      [reset cm](-\maxdimen,-\maxdimen)rectangle(\maxdimen,\maxdimen)
                  }
          }
        } 
        \draw[] (0,0) circle (.5);
        \draw[] (0,0) circle (1.5);
        \draw[blue, thick] (0,0) circle (1.05);
        \draw[orange, thick] (0.5,0) -- (1.5,0);      
    \end{tikzpicture}
\end{eqnarray*}
where the phase comes from the unlinking of the Hopf link between the two lines $L^{p^{-1}b}$ and $(L^\prime)^{p^{-1}c}$.

And so
\begin{equation}
    \mathcal{Z}_\mathrm{QED}(S^2\times S^2)=\sum_{b,c\in\bbZ_N}\exp\qty(\frac{2\pi \I}{N}(p^{-1})_N b\cup c)\,\mathcal{Z}_\mathrm{QED}(S^2\times S^2)
\end{equation}
where in the second partition function we have magnetic fluxes for $b,c$ in $S^2\times S^2$. This gives the well known self-duality of QED where the theory on $S^2\times S^2$ is equivalent to it's gauging of the $(\bbZ_N^{(1)})_m$ with discrete torsion.

The people more mathematically inclined might have noticed that the Hopf link formed by the two lines $L, L^\prime$ is precisely the surgery diagram for the handlebody decomposition of $S^2\times S^2$. Indeed we can view this manifold as the following null-bordism 
\begin{equation}
    *\rightarrow S^3\xrightarrow{H^1_2}S^1\times S^2\xrightarrow{H_2^2}S^3\rightarrow*
\end{equation}
where the maps $H^i_2$ are $4$-dimensional $2$-handle gluing operations. We can associate the Ward identity to this bordism by nucleating a symmetry defect $\mathcal{D}_{p/N}$ on the left side of the bordism and shrink it on the other side after passing it through the surgery maps.

\begin{center}
    \textbf{Main message:} We can use topology to probe the higher structure of the symmetry
\end{center}
\subsection{Dynamical constraints: a lightning review}
The symmetries of a theory can be encoded in a bulk/boundary system. We refer to Freed's lectures in this volume for more details and references about this construction. Schematically here the idea is to find an isomorphism between the $d$-dimensional theory of interest and a bulk/boundary system consisting of a $d+1$-dimensional slab hositing at $d+1$-dimensional TQFT, delimited by two boundaries: one boundary is a topological boundary condition for the TQFT, while the other host a relative field theory, responsible for the dynamics. Since the construction is topological, we can bring the topological boundary on top of the relative theory and obtain the desired isomorphism with the absolute $d$-dimenisonal theory of interest. The pair bulk-TQFT/topological boundary encodes the structure and properties of all topological defects surviving the isomorphism, hence gives an alternative approach to characterize the topological defects. A particularly powerful feature of this construction is that it has a nice interplay with symmetry preserving RG flows: when going from the UV to the IR, only the relative theory changes since the RG-flow is a property of the dynamical boundary:

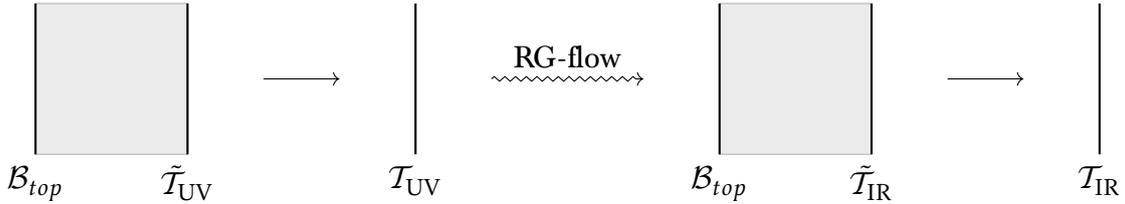
\begin{figure}[H]
    \centering
    \begin{tikzpicture}
        \draw[fill=gray!50, opacity=.3] (0,0) rectangle (2,2);
        \draw[-, thick] (0,0) -- (0,2);
        \draw[-, thick] (2,0) -- (2,2);
        \node[below] at (0,0) {$\mathcal{B}_{top}$};
        \node[below] at (2,0) {$\Tilde{\mathcal{T}}_{\mathrm{UV}}$};

        \draw[->] (3,1) -- (4,1);
        \draw[-, thick] (5,0)--(5,2);
        \node[below] at (5,0) {$\mathcal{T}_\mathrm{UV}$};
        \draw [->,
        line join=round,
        decorate, decoration={
            zigzag,
            segment length=4,
            amplitude=.9,post=lineto,
            post length=2pt
        }]  (6,1) -- (8,1);
        \node[above] at (7,1) {RG-flow};
        \begin{scope}[xshift=9 cm]
            \draw[fill=gray!50, opacity=.3] (0,0) rectangle (2,2);
            \draw[-, thick] (0,0) -- (0,2);
            \draw[-, thick] (2,0) -- (2,2);
            \node[below] at (0,0) {$\mathcal{B}_{top}$};
            \node[below] at (2,0) {$\Tilde{\mathcal{T}}_{\mathrm{IR}}$};

            \draw[->] (3,1) -- (4,1);
            \draw[-, thick] (5,0)--(5,2);
            \node[below] at (5,0) {$\mathcal{T}_\mathrm{IR}$};
        \end{scope}
    \end{tikzpicture}
    \caption{Topological symmetry theory and RG flow}
\end{figure}
Consider than the case of a model with a $\bbZ_N^{(1)}$ duality symmetry. In the $\mathcal{T}_\rm{UV}$ side we know that we can gauge on half spacetime and get an interface between the initial theory and the $\mathcal{T}_\rm{UV}/\bbZ_N^{(1)}$ one. When the latter is dual to the former, one can compose the half-spacetime gauging interface with the duality and obtain a topological duality defect, which is in particular non-invertible.\footnote{See e.g. J. Kaidi, K. Ohmori, and Y. Zheng, \textit{Kramers-Wannier-like Duality Defects in (3+1)D Gauge Theories}, Phys.Rev.Lett. 128 (2022) 11, 111601, \href{https://arxiv.org/abs/2111.01141}{arXiv:2111.01141} and Y. Choi, C. Cordova, P.-S. Hsin, H. T. Lam, S.-H. Shao \textit{Non-invertible duality defects in 3+1 dimensions}, Phys. Rev. D (105) (2022), \href{https://arxiv.org/abs/2111.01139}{arXiv:2111.01139}} As discussed in Freed's lectures, we can realize this gauging in terms of operations in the topological boundary (symmetry) side (from Dirichlet to Neumann). This operation then must survive along a symmetry preserving RG-flow. So we can predict that in the IR phase there is must be a topological duality defect. This is a strong constraint on the IR properties of the quantum field theory of interest.\footnote{Y. Choi, C. Cordova, P.-S. Hsin, H. T. Lam, S.-H. Shao \textit{Non-invertible Condensation, Duality, and Triality Defects in 3+1 Dimensions}, Commun.Math.Phys. 402 (2023) 1, 489-542, \href{https://arxiv.org/abs/2204.09025}{arXiv:2204.09025}} 

Let's consider a bosonic $4d$ theory that is trivially gapped in the IR, for instance. Then one can think of this gauging as a property of a $4d$ SPT phase (since the theory is trivially gapped). But bosonic $4d$ SPTs are classified, and have action of the form
\begin{equation}
    \mathcal{Z}_\mathrm{SPT}[A^{(2)}]=\exp \frac{2\pi \I }{N}p\int_X A^{(2)}\cup A^{(2)}
\end{equation}
where $N$ is odd and $p\in\bbZ_N$. The self duality under the gauging of the $\bbZ_N$ acts like a Fourier transform on the action, and the equality with the initial one is just given by numerology on the $p,N$ coefficients. With this one can infer that a UV theory with a $\bbZ_N$ self-duality defect is going to be trivially gapped in the IR iff $N$ is odd and its decomposition over prime powers $N=y_1^{\ell_1}\cdots y_k^{\ell_k}$ has $y_i\equiv 1\mod 4$. When $N$ is even the theory does not admit a trivially gapped IR phase.}

\newpage

{\clearpage% newcommands can go here
\newcommand{\N}{\mathbb N}
\newcommand{\Z}{\mathbb Z}
\newcommand{\Q}{\mathbb Q}
\newcommand{\R}{\mathbb R}
\newcommand{\C}{\mathbb C}

\newcommand{\Disk}{\mathrm{Disk}}
\makeatletter
\newcommand*\bigcdot{\mathpalette\bigcdot@{.5}}
\newcommand*\bigcdot@[2]{\mathbin{\vcenter{\hbox{\scalebox{#2}{$\m@th#1\bullet$}}}}}
\makeatother

\newcommand{\picA}{\includegraphics[scale=0.1]{figure3.jpeg}}
\newcommand{\picB}{\includegraphics[scale=0.09]{figure4.jpeg}}

\theoremstyle{definition}
\newtheorem{expectation}[theorem]{Expectation}
\newtheorem{conjecture}[theorem]{Conjecture}
\newtheorem{idea}[theorem]{Idea}

%%%%%%%%%%%%%%%%%%%%%%%%%%%%%%%%%%%%%%%%%%%%%%%%%%%%%%%%%%%%%%%%%%
\newcourse[Jonte G\"odicke, Aaron Hofer, and Anja \v Svraka]{Applied Cobordism Hypothesis}{David Jordan}

%Aaron
\section{Lecture 1: Introduction}

Before getting into the abstract world of $(\infty,n)$-categories, let us briefly go over a bit of the motivation coming from physics.

\subsection{Motivation}
For an (Euclidean) $D$-dimensional QFT $Z$ we expect the following assignments: 
\begin{itemize}
    \item For any $D$-dimensional manifold $M^D$ without boundary, a number $Z(M^D) \in \mathbb{C}$, called the \emph{partition function};
    \item For any $(D-1)$-dimensional manifold $M^{(D-1)}$, a vector space $Z(M^{(D-1)})$ of \emph{boundary-conditions};
    \item For any cylinder $M^{(D-1)} \times [0,t]$ of length $t \in \R_{>0}$, a \emph{time evolution operator} 
    \begin{align}
        Z(M^{(D-1)}) \stackrel{e^{-tH}}{\to} Z(M^{(D-1)}) \,.
    \end{align}
\end{itemize}
More generally, for any $D$-dimensional manifold $M$ with boundary decomposed as $\partial M = M_{\mathrm{in}} \sqcup M_{\mathrm{out}}$, we want to be able to assign a ``matrix element"
\begin{align}
    Z(M)(F_{\mathrm{in}},F_{\mathrm{out}}) = \int_{F|_{M_{\mathrm{in}}} = F_{\mathrm{in}}}^{F|_{M_{\mathrm{out}}} = F_{\mathrm{out}}} e^{- S[F]} \in \mathbb{C}
\end{align}
 called the \emph{path integral}, for any pair of boundary conditions $F_{\mathrm{in}} \in Z(M_{\mathrm{in}})$ and $F_{\mathrm{out}} \in Z(M_{\mathrm{out}})$. Here the integral means we want to sum over all possible field configurations $F$ which restrict to $F_{\mathrm{in}}$ and $F_{\mathrm{out}}$ on the respective boundaries. The picture behind this heuristic description is the following:
\begin{align}
    \begin{tikzpicture}[baseline=1.5cm]
        %manifold
        \draw[black,thick] (0,0) -- (0,0.75);
        \draw[black,thick] (1,0) -- (1,0.6);
        \draw[black,thick] (2,0) -- (2,0.6);
        \draw[black,thick] (3,0) -- (3,0.75);
        \draw[black,thick] (1,2.25) -- (1,3);
        \draw[black,thick] (2,2.25) -- (2,3);
        \draw[black,thick] (2,0.6) arc (0:180:0.5 and 0.4);
        \draw[black,thick] (0,0.75) .. controls (0,1.5) and (1,1.5) .. (1,2.25);
        \draw[black,thick] (3,0.75) .. controls (3,1.5) and (2,1.5) .. (2,2.25);
        %genus
        \draw[black,thick] (1.8,1.6) arc (-10:-170:0.3 and 0.15);
        \draw[black,thick] (1.7,1.52) arc (5:175:0.2 and 0.1);
        %boundaries
        \draw[black!30!red,thick] (2,3) arc (0:180:0.5 and 0.2);
        \draw[black!30!red,thick,dashed] (2,3) arc (0:-180:0.5 and 0.2);
        \draw[black!30!green,thick] (0.5,0) ellipse (0.5 and 0.2);
        \draw[black!30!green,thick] (2.5,0) ellipse (0.5 and 0.2);
        %decoration 
        \node at (1.5,-0.5) [black!30!green] {$M_{\mathrm{in}}$};
        \node at (1.3,3.5) [black!30!red] {$M_{\mathrm{out}}$};
        \node at (-0.4,1.5) {$M$};
        \node at (4,-0.5) [black!30!green] {$F_{\mathrm{in}}$};
        \node at (4,3.5) [black!30!red] {$F_{\mathrm{out}}$};
        \node at (4.6,1.5) {$Z(M)$};
        \draw[->] (4,0) -- (4,3);
        \draw[->,dashed,black!30!green] (3.8,-0.7) .. controls (2.25,-1.5) and (0.5,-1).. (0.5,-0.3);
        \draw[->,dashed,black!30!green] (3.8,-0.7) .. controls (3.25,-1) and (2.5,-0.7) .. (2.5,-0.3);
        \draw[->,dashed,black!30!red] (3.9,3.8) .. controls (3.8,4.1) and (2,4) .. (1.85,3.3);
    \end{tikzpicture}
\end{align}
Here the dashed arrows signify that the boundary conditions $F_{\mathrm{in}}$ and $F_{\mathrm{out}}$ are localised on $M_{\mathrm{in}}$ and $M_{\mathrm{out}}$, respectively. Unfortunately, there exists no measure on the space of fields to make this a rigorous mathematical entity. Moreover, in the presence of gauge symmetries the situation becomes even worse as we need to be even more careful in order to only account for physically relevant information.\footnote{In mathematical terms, gauge symmetry tells us that we are not interested in the fields themselves, but only their equivalence class w.r.t.\ some equivalence relation dictated by the physical system.}

The Atiyah-Segal axioms allow us to formalize this assignment for topological QFTs (TFTs or TQFTs) as a \emph{functor}
\begin{align}
    \mathrm{Cob^{D,D-1}_{\mathrm{fr}/\mathrm{or}}} \to \mathrm{Vect},\, \mathrm{Hilb},\,\dots
\end{align}
which sends a disjoint union to the tensor product (i.e.~a symmetric monoidal functor).
Functoriality corresponds to matrix multiplication in the sense that the gluing of two $D$-dimensional manifolds $M_1$ and $M_2$ with $(M_1)_{\mathrm{out}} \cong (M_2)_{\mathrm{in}}$
\begin{align}
    \begin{tikzpicture}[baseline=2.4cm]
        %collar
        \draw[black!30!green,thick] (2,2.4) arc (0:180:0.5 and 0.2);
        \draw[black!30!green,thick,dashed] (2,2.4) arc (0:-180:0.5 and 0.2);
        %manifold
        \draw[black,thick] (0,0) -- (0,0.6);
        \draw[black,thick] (1,0) -- (1,0.6);
        \draw[black,thick] (2,0) -- (2,0.6);
        \draw[black,thick] (3,0) -- (3,0.6);
        \draw[black!30!green,thick] (2,0.6) arc (0:180:0.5 and 0.4);
        \draw[black,thick] (0,0.6) .. controls (0,1.35) and (1,1.35) .. (1,2.1);
        \draw[black,thick] (3,0.6) .. controls (3,1.35) and (2,1.35) .. (2,2.1);
        \draw[black,thick] (2,0.6) arc (0:180:0.5 and 0.4);
        \draw[black,thick] (0,0.6) .. controls (0,1.35) and (1,1.35) .. (1,2.1);
        \draw[black,thick] (3,0.6) .. controls (3,1.35) and (2,1.35) .. (2,2.1);
        \draw[black,thick] (0,4) -- (0,4.6);
        \draw[black,thick] (1,4) -- (1,4.6);
        \draw[black,thick] (2,4) -- (2,4.6);
        \draw[black,thick] (3,4) -- (3,4.6);
        \draw[black,thick] (1,2.1) -- (1,2.7);
        \draw[black,thick] (2,2.1) -- (2,2.7);
        \draw[black,thick] (2,4) arc (0:-180:0.5 and 0.4);
        \draw[black,thick] (0,4) .. controls (0,3.45) and (1,3.45) .. (1,2.7);
        \draw[black,thick] (3,4) .. controls (3,3.45) and (2,3.45) .. (2,2.7);
        \draw[black,thick] (2,0.6) arc (0:180:0.5 and 0.4);
        \draw[black,thick] (0,0.6) .. controls (0,1.35) and (1,1.35) .. (1,2.1);
        \draw[black,thick] (3,0.6) .. controls (3,1.35) and (2,1.35) .. (2,2.1);
        \draw[black,thick] (1.8,1.6) arc (-10:-170:0.3 and 0.15);
        \draw[black,thick] (1.7,1.52) arc (5:175:0.2 and 0.1);
        %boundaries
        \draw[black,thick] (1,4.6) arc (0:180:0.5 and 0.2);
        \draw[black,thick,dashed] (1,4.6) arc (0:-180:0.5 and 0.2);
        \draw[black,thick] (3,4.6) arc (0:180:0.5 and 0.2);
        \draw[black,thick,dashed] (3,4.6) arc (0:-180:0.5 and 0.2);
        \draw[black,thick] (2.5,0) ellipse (0.5 and 0.2);
        \draw[black,thick] (2.5,0) ellipse (0.5 and 0.2);
        \draw[black,thick] (0.5,0) ellipse (0.5 and 0.2);
        \draw[black,thick] (0.5,0) ellipse (0.5 and 0.2);
        %decoration 
        \node at (-1,0) {$(M_1)_{\mathrm{in}}$};
        \node at (-1,2.4) {$(M_1)_{\mathrm{out}}$};
        \node at (-1,4.7) {$(M_2)_{\mathrm{out}}$};
        \node at (-1.4,1.1) {$M_1$};
        \draw[->] (-1,0.3) -- (-1,2.1);
        \node at (-1.4,3.8) {$M_2$};
        \draw[->] (-1,2.7) -- (-1,4.5);
    \end{tikzpicture}
\end{align}
can be written as a ``sum" over the intermediate fields
\begin{align}
    \int e^{- S[F]} = \int_{F|_{(M_1)_{\mathrm{out}}}} \int e^{- S[F]} \int e^{- S[F]}.
\end{align}
To be a bit more precise, the symmetric monoidal category of framed/oriented \textbf{cobordisms} consists of:
\begin{align}
    \mathrm{Cob^{D,D-1}_{\mathrm{fr}/\mathrm{or}}} = \begin{cases}
        \textrm{objects } = (D-1) \textrm{ manifolds}\\
        \textrm{arrows } = D \textrm{ manifolds}
    \end{cases}
\end{align}
The subscripts $\mathrm{fr}$ and $\mathrm{or}$ correspond to framings and orientations, respectively.\footnote{More generally, there is a cobordism category for any other type of tangential structure such as e.g.\ spin. In particular, different physical theories require different tangential structures to be defined.} Recall here that a $D$-\emph{framed} $D$-dimensional manifold is a manifold $M$ together with a trivialization of its tangent bundle $TM \cong M \times \mathbb{R}^{D}$. For a $(D-1)$-dimensional manifold $M'$ we require a trivialization of $TM \oplus \mathbb{R}$. Equivalently, there is a consistent way to assign a basis of $T_pM$ for every point $p \in M$: 
\begin{figure}
\centering
    \begin{tikzpicture}[baseline=1cm]
        %manifold
        \draw[black,very thick] (2.5,0) ellipse (0.5 and 0.2);
        \draw[black,very thick] (0.5,0) ellipse (0.5 and 0.2);
        \draw[black,very thick] (2,2) arc (0:180:0.5 and 0.2);
        \draw[black,very thick,dashed] (1,2) arc (-180:0:0.5 and 0.2);
        \draw[black,thick] (2,0) arc (0:180:0.5 and 0.6);
        \draw[black,thick] (0,0) .. controls (0.1,1.2) and (0.9,0.8) .. (1,2);
        \draw[black,thick] (3,0) .. controls (2.9,1.2) and (2.1,0.8) .. (2,2);
        %framing
        \draw[black!30!green,thick,->] (0.5,0.5) -- (0.5,0.8);
        \draw[black!30!green,thick,->] (0.5,0.5) -- (0.8,0.5);
        \draw[black!30!green,thick,->] (2.4,0.5) -- (2.4,0.8);
        \draw[black!30!green,thick,->] (2.4,0.5) -- (2.7,0.5);
        \draw[black!30!green,thick,->] (1.5,1.2) -- (1.5,1.5);
        \draw[black!30!green,thick,->] (1.5,1.2) -- (1.8,1.2);
    \end{tikzpicture}
\end{figure}

In the definition of $ \mathrm{Cob^{D,D-1}_{\mathrm{fr}/\mathrm{or}}}$ we further demand that the framing/orientation on the boundaries are compatible with the ones in the bulk. Note here that the category $\mathrm{Cob^{D,D-1}_{\mathrm{fr}/\mathrm{or}}}$ only carries topological information, in particular the numbers $Z(M^D)$ are topological invariants for any closed $D$-dimensional manifold $M^D$.
\subsection{TQFT vs TFT in physics}
In physics, the path integral
$Z(M^D) = \int_{\mathrm{ fields}} e^{- S[F]}$ is often given in terms of an \emph{action functional} 
\begin{align}
    S[F] = \int_{M^D} \mathcal{L}(F)
\end{align} 
where $\mathcal{L}$ is an object called the \emph{Lagrangian} of the theory which maps fields $F$ to top forms $\Omega^D(M^D)$. Usually the Lagrangian $\mathcal{L}$ is given as the sum of a \emph{kinetic} $\mathcal{L}_{\mathrm{kin}}$ and a \emph{potential} $\mathcal{L}_{\mathrm{pot}}$ term. 

Now, if we want to make the path integral mathematically precise in the Atiyah-Segal axioms, we run into the following problem: The Lagrangian $\mathcal{L}$ is typically metric dependent. For instance, in gauge theory, one considers as fields connections $A$ on some principle bundle $P$ over $M^D$. For this type of theory the kinetic part of the Lagrangian is given by
\begin{align}
    \mathcal{L}_{\mathrm{kin}} = F_A \wedge *F_A
\end{align}
where $F_A$ is the curvature of the connection $A$ and $*$ is the metric dependent Hodge star operator. Thus, for these theories, we have no chance of interpreting $Z(M^D) = \int_{\mathrm{ fields}} e^{- S[F]}$ as a topological invariant of $M^D$.

From the physics perspective, a \emph{topological} quantum field theory is a QFT that is not metric dependent. Let us go through a few examples:
\begin{enumerate}
    \item Topological twists of SUSY QFTs: Here, the fields live in some kind of graded vector space. We can then find some kind of differential $Q$ acting on the fields (i.e.\ $F \in \mathrm{Ch}_\C$) so that we can take the cohomology $\frac{\mathrm{ker}(Q)}{\mathrm{im}(Q)}$. If our differential is nice enough\footnote{We only consider topological twists, there are more general twists that can lead to different theories. So called holomorphic twists for example lead to conformal field theories.} the path integral becomes well-defined and might loose its metric dependence. These types of theories have a lot of applications in mathematics (Floer theory, Donaldson theory, Mirror Symmetry, Khovanov homology, ...) 

    \item Finite gauge theories: Instead of working with a Lie group, we can just work with a finite group. In this case the path integral reduces to a finite sum and everything can be well defined.

    \item Finite homotopy types: These are a generalization of finite gauge theories and were discussed in detail in Constantin Teleman's lecture. 
\end{enumerate}

We will be most interested in the application to symmetries. The basic idea here is to implement the notion of topological symmetry of a QFT via coupling the given QFT to a TFT of one dimension higher, often called the symmetry TFT of the QFT. 

This idea can be used as a powerful tool to analyze general QFTs. To formulate this recall that the infrared (IR) limit of a QFT is the theory which results from ``tuning down" the metric dependence. More precisely, it is the fixed point our given theory flows towards under the renormalization group flow. Roughly speaking the renormalization group tells us how the QFT we are considering behaves under changing the overall energy scale. In particular, the fixed points under this are scale invariant QFTs.\footnote{These QFTs often tend to be invariant under the bigger group of conformal transformations.} 
\begin{figure}[h!]
    \centering
    \begin{tikzpicture}
        %UV-manifold
        \draw[thick] (0,0) -- (2,0) -- (3,1) -- (1,1) -- cycle;
        \filldraw[fill=white,draw=black,thick] (0.7,0.5) .. controls (0.8,0.6) .. (1,0.5) -- (1,0.5) .. controls (1.15,0.75) .. (1.3,0.5) -- (1.3,0.5) .. controls (1.5,1.5) .. (1.7,0.5) -- (1.7,0.5) .. controls (1.83,0.85) .. (2,0.5)  -- (2,0.5) .. controls (2.2,0.7) .. (2.3,0.5);
        %IR-manifold
        \draw[thick] (5,0) -- (7,0) -- (8,1) -- (6,1) -- cycle;
        %arrow
        \draw[thick,->] (3.2,0.5) -- (4.8,0.5);
        \node at (4,0.8) {RG};
    \end{tikzpicture}
    \label{fig:rgflowdavid}
\end{figure}

The key insight now is that the topological symmetries of the theory in the high energy or ultraviolet (UV) regime will be the same as the topological symmetries of the theory in the IR regime, since they were already metric independent from the start! So in particular two theories can only flow to the same IR limit if the IR limit can consistently support both symmetry TFTs.

\section{Lecture 2: Extended TFT}
The $1$-categorical formulation of a TFT $Z$ as a functor allows us to compute the number $Z(M^D)$ for a closed $D$-dimensional manifold $M^D$ by cutting $M^D$ into ``simpler" pieces:

However, this is often still too complicated! Instead, we want to be able to cut our manifolds up further so that we only need to understand $Z$ on disks $\mathbb{D}^D$.
This lecture aims to set up the algebraic machinery that allows us to make this plausible. For this, we need to go into the world of higher categories.

There exists an $n$-category (actually an $(\infty,n)$-category) of framed cobordisms $\mathbf{Cob}^{n}_{\mathrm{fr}}$ consisting of: 
\begin{itemize}
    \item Objects: $n$-framed points;
    \item $1$-morphisms: $n$-framed $1$-manifolds (with boundary);
    \item $2$-morphisms: $n$-framed $2$-manifolds (with boundary \& corners);
    \item $\dots$
    \item $n$-morphisms: the \emph{space} of framed $n$-manifolds;
\end{itemize}
It is useful to think of an $n$-framed point as a ``small" $n$-cube.

As mentioned in the first lecture, we can also consider different tangential structures such as orientations or spin structures on our manifolds leading to $n$-categories of cobordisms with tangential structure. In this lecture we will mainly focus on the framed cobordism $n$-category and occasionally consider the oriented one, denoted by $\mathbf{Cob}^{n}_{\mathrm{or}}$, as well.

Let us go through some examples of objects and $k$-morphisms for the case $n=2$ in the oriented setting. 
There are two non trivial objects, the positively oriented point $\{ \bigcdot _{+}\}$ and the negatively oriented point $\{ \bigcdot _{-}\}$. A simple example of a $1$-morphism from $\{ \bigcdot _{+}\} \sqcup \{ \bigcdot _{-}\}$ to $\varnothing$ is given by the interval and can be visualized as: 
\begin{figure}[H] \label{evcoev}
\centering
\begin{tikzpicture}[scale=0.8]
    \draw[black,thick] plot[smooth, tension=1.7] coordinates{ (0,0) (1,1) (2,0) };
    \fill (0,0) circle[radius=2pt];
    \fill (2,0) circle[radius=2pt];
    %decoration 
    \node at (0,-0.5) [black] {$\{\bigcdot_{+}\}$};
    \node at (2,-0.5) [black] {$\{\bigcdot_{-}\}$};
    \node at (1,1.5) [black] {$\varnothing$};
    \node at (1,1) [black, thick] {$\succ$};
\end{tikzpicture} 
\end{figure}

An example of a $2$-morphism between $1$-morphisms from $\{ \bigcdot _{+}\} \sqcup \{ \bigcdot _{-}\}$ to itself is given by:

\begin{figure}[H]
\centering
\begin{tikzpicture}[scale=0.5]
%\node (0,7) {$\circlearrowleft$};
\draw[thick] (-1.25, 0.875) node [anchor=south] {$\circlearrowleft$} -- (-2, -1) arc (-90:90:1.5 and 1) -- (-1, 3.5) -- (3, 3.5) -- (2,1) arc (90:270:1.5 and 1) -- (3,1.5) -- (2.2, 1.5);
\draw[thick] (-0.52,-0.2) .. controls (0, 1.8) and (1, 1.8) .. (0.52,0.2);
\end{tikzpicture}
\end{figure}

When defining a higher category, it does not suffice to only prescribe the $k$-morphisms; we also need to know how to compose them! This should come as no surprise. Suppose we are given some QFT; then we can organize all its topological defects into a higher category with $k$-morphisms given by defects in codimension $k$. In this picture, the composition of $k$-morphisms is nothing else then the fusion of the corresponding defects:
\begin{figure}[h!]
    \centering
    \begin{tikzpicture}
        \draw[thick] (0,0) -- (1,1) -- (1,3) -- (0,2) -- cycle;
        \filldraw[fill=white,thick] (0.5,0) -- (1.5,1) -- (1.5,3) -- (0.5,2) -- cycle;
        \draw[ultra thick] (3,0) -- (4,1) -- (4,3) -- (3,2) -- cycle;
        \draw[thick,->] (1.6,1.5) -- (2.7,1.5);
    \end{tikzpicture}
\end{figure}

For $\mathbf{Cob}^{n}_{\mathrm{fr}}$, the composition is defined by gluing the corresponding manifolds. However, there are some subtleties to make this precise, as we will now discuss.

We know how to compose morphisms in a $1$-category: there is an algebraic operation $\circ$ which satisfies the axiom of associativity. To compose $1$-morphisms in a $2$-category (e.g.\ a tensor category) we again need an algebraic operation $\circ$, but now we also need an associativity \emph{isomorphism} $(-\circ -) \circ - \stackrel{\cong}{\Rightarrow} - \circ  (- \circ -)$ which itself needs to fulfil some coherence axioms, the pentagon axiom in this case. In particular there can be multiple solutions to these coherence axioms, i.e.\ the associativity isomorphism is the extra structure we need to choose! 

The appearance of these coherence isomorphisms can also be seen in the defect picture. Instead of the naive idea of fusing the defects together, it can actually matter in which way we fuse them precisely; this can be implemented by an invertible defect in one codimension higher:
\begin{figure}[h!]
    \centering
    \begin{tikzpicture}
        %lhs
        \draw[thick] (0,0) -- (1,1) -- (1,3) -- (0,2) -- cycle;
        \filldraw[fill=white,thick] (0.5,0) -- (1.5,1) -- (1.5,3) -- (0.5,2) -- cycle;
        %rhs
        \draw[ultra thick] (4,2) -- (4,3) -- (3,2) -- (3,1);
        \draw[thick] (3,1) -- (2.5,0) -- (3.5,1) -- (4,2);
        \filldraw[fill=white,thick] (3,1) -- (3.5,0) -- (4.5,1) -- (4,2);
        \draw[ultra thick,black!30!blue] (3,1) -- (4,2);
        %arrow
        \draw[thick,->] (1.6,1.5) -- (2.7,1.5);
    \end{tikzpicture}
\end{figure}

We might expect, and sometimes even assume, that we can keep going up the dimensional ladder like this; however, really, we can't and we shouldn't.

One example where we can see the problem already happens in linear algebra: given two fixed vector spaces $V$ and $W$, the tensor product $V \otimes W$ is not unique! In fact, any vector space $X$ that satisfies a certain property will be isomorphic to our chosen tensor product $V \otimes W$ and is, in this regard, just as good, our just as canonical.

\subsection{$\boldsymbol{\infty}$-categories}
To circumvent these kinds of issues, we will now finally step into the realm of $\infty$-categories.

Unfortunately, we can't give a full, proper introduction to the topic, as this would take an entire lecture\footnote{See Scheimbauer's lectures at the GCS School in 2024 for an introduction to this topic.} in its own right. Instead, we are going to focus on the basic ideas. The relevant construction of an $(\infty,n)$-category we have in mind is the one of so called \emph{iterated complete Segal spaces}.
In an $(\infty,1)$-category, morphisms no longer form just a set but a whole space.
Moreover, if we have two morphisms $f \colon X \to Y$ and $g \colon Y \to Z$, then we don't expect to get a unique composition $g \circ f$; instead, we get a whole family of equivalent possible composite morphisms. To make this a bit more precise let us consider two morphisms $f, f' \in \Hom(X,Y)$, in a higher category it makes sense to ask if $f$ and $f'$ are \emph{isomorphic}. We can then further ask: if we have two different morphisms between morphisms $\varphi,\psi \colon f \to f'$, so-called higher morphisms, witnessing the isomorphism $f \cong f'$, are they also isomorphic, and if they are, what about even higher morphisms between them?  
This basic idea allows us to think of $\Hom(X,Y)$ as a space with the morphisms, such as $f$ and $f'$ as points, the higher morphisms, e.g.\ $\varphi$ and $\psi$, as paths between these points, and the even higher morphisms as paths between the paths, i.e., \ as homotopies. We can picture this idea as:
\begin{figure}[h!]
    \centering
    \begin{tikzpicture}
        %space
        \draw[thick] (0,0) .. controls (2,1) .. (4,0);
        \draw[thick] (0,0) .. controls (2,-1) .. (4,0);
        \filldraw[black] (0,0) circle (0.05);
        \filldraw[black] (4,0) circle (0.05);
        %labels
        \node at (0,0) [left] {$f$};
        \node at (4,0) [right] {$f'$};
        \node at (2,1) [above] {$\varphi$};
        \node at (2,-1) [below] {$\psi$};
        \node at (2,-0.1) [rotate=90] {$\Longleftarrow$};
        \node at (2,0.1) [rotate=90] {$\Longrightarrow$};
    \end{tikzpicture}
\end{figure}
What this really tells us is that higher categories contain spaces if we restrict ourselves to isomorphisms. In practice, this means that we can use tools and ideas from topology to study categories with a different origin. In the following sections, we will see this for more algebraic categories. 

Coming back to the issues regarding composition, one might expect that we have composition maps
\begin{equation}
    \Hom(X,Y) \times \Hom(Y,Z) \to \Hom(X,Z)
\end{equation}
between the morphism spaces, and indeed, this is how it works in the model of $(\infty,1)$-categories given by simplicially or topologically enriched categories.
However, for many examples this is too strong  and what we have is a \emph{space of compositions} $C$, which comes with a map
\begin{equation}
    C \to \Hom(X,Y) \times \Hom(Y,Z)
\end{equation}
which tells us where the composable morphisms live. The key point now is that this composition space $C$ needs to be contractible (in a relative sense).

In the example of the tensor product of two vector spaces $V$ and $W$, the contractability of the space of possible tensor products amounts to the statement that for any two possible realizations $(V\otimes W)_1$ and $(V \otimes W)_2$ of the tensor product there is an isomorphism $(V\otimes W)_1 \cong (V\otimes W)_2$, and for any two of such isomorphism there is a higher isomorphism. However, in this case, there are no higher morphisms, thus the possible isomorphisms need to be the same!

With this, we can now understand why composition in $\mathbf{Cob}^n_{\mathrm{fr}}$ is subtle; if we glue two cylinders to a torus, what we really do is cutting up the torus and remember where we made the cuts. However, all of these possible cuts are just as good up to homotopy, i.e., \ the space of possible compositions is contractible:

\begin{figure}[h!]
    \centering
    \begin{tikzpicture}
        %cut1
        \draw[blue, thick, dashed] (0,0.55) arc (90:270:0.1 and 0.24);
        \draw[blue, dashed] (0,0.55) arc (90:270:-0.1 and 0.24);
         %cut2
        \draw[blue, thick, dashed] (0,-0.65) arc (270:90:0.1 and 0.24);
        \draw[blue, dashed] (0,-0.65) arc (270:90:-0.1 and 0.24);
        %arrows
        \node at (-0.07,0.33) [blue,left,scale=0.4] {$\leftarrow$};
        \node at (0.07,0.33) [blue,right,scale=0.4] {$\rightarrow$};
        \node at (-0.07,-0.43) [blue,left,scale=0.4] {$\leftarrow$};
        \node at (0.07,-0.43) [blue,right,scale=0.4] {$\rightarrow$};
        %surface
        \draw[thick] (0,-0.05) ellipse (1.2 and 0.6);
        %handle1
        \filldraw[fill=white,thick] (0.375,0) arc (0:-180:0.4 and 0.2);
        \filldraw[fill=white,thick] (0.325,-0.1) arc (0:180:0.35 and 0.15);
    \end{tikzpicture}
\end{figure}

The upshot of this crash course is that the whole machinery of $\infty$-categories is set up so that we can "forget" about all of these subtleties in practice.

\subsection{Fully extended TFTs}
\begin{definition}
    A \textbf{fully extended framed/oriented} $n$-dimensional \textbf{topological field theory (TFT)} is a symmetric monoidal functor
    \begin{align}
        (\mathbf{Cob}^n_{\mathrm{fr/or}},\sqcup) \to (\mathbf{S},\otimes)
    \end{align}
    where $(\mathbf{S},\otimes)$ is some symmetric monoidal ``algebraic" $n$-category.
\end{definition}
Some examples of what we mean by an ``algebraic" $n$-category, and which will be discussed in more detail later on, are:
\begin{itemize}
    \item The category of vector spaces with the tensor product $(\Vect_\mathbb{C},\otimes_\mathbb{C})$;
    \item The $2$-category of linear categories with the cartesian product $(\mathrm{Cat}_\mathbb{C},\times)$
    \item The $3$-category of (finite) tensor categories with the so-called Deligne tensor product $(\mathbf{Tens},\boxtimes)$;
    \item The $4$-category of braided tensor categories again with the Deligne tensor product $(\mathbf{BrTens},\boxtimes)$;
\end{itemize}

Using this terminology, we can make the dream of understanding a TFT by understanding what it does on disks, mentioned in the beginning of this lecture, precise in terms of the so called \emph{cobordism hypothesis}\footnote{In mathematics a hypothesis is a conjectural statement where the correct framework is still undefined.} , first formulated by Baez-Dolan\footnote{J.~C.~Baez,  and J.~Dolan \emph{Higher‐dimensional algebra and topological quantum field theory}, Journal of Mathematical Physics. 36 (11): 6073–6105. 1995. \href{https://arxiv.org/abs/q-alg/9503002}{{\tt arXiv:9503002}}} and brought into its current form by Lurie\footnote{J.~Lurie, \emph{On the classification of
topological field theories},
  Current developments in mathematics, 2008, Int. Press, Somerville, MA, 2009,
  pp.~129--280. \href{http://arxiv.org/abs/arXiv:0905.0465}{{\tt
  arXiv:0905.0465}}}.

\begin{conjecture}[Cobordism hypothesis]
    There is an equivalence (of $\infty$-groupoids) between $\{n\text{-dim extended TQFTs}\} = \mathrm{Fun}^{\otimes}(\mathbf{Cob}_\mathrm{fr}^n,\mathbf{S})$ and the fully-dualizable subcategory $\mathbf{S}^{\mathrm{fd}}$ given by
    \begin{align}
        Z \in \mathrm{Fun}^{\otimes}(\mathbf{Cob}_\mathrm{fr}^n,\mathbf{S}) \mapsto Z(\{\bigcdot_{+}\}).
    \end{align}
\end{conjecture}

We can interpret this statement algebraically as saying that 
$\mathbf{Cob}_\mathrm{fr}^n$ is the symmetric monoidal $n$-category with duals freely generated by $\{\bigcdot_{+}\}$.

\begin{definition}
    An object $X \in \mathbf{S}$ is \textbf{dualizable} if there exists an object ${}^\vee \! X$ in $\mathbf{S}$ together with $1$-morphisms
\begin{align}
    \begin{tikzpicture}[baseline]
        \draw[black,thick] (0.5,0) arc (0:180:0.5cm);
        \node at (-0.5,0) [black,thick,below] {${}^\vee \! X $};
        \node at (0.5,0) [black,thick,below] {$X$};
        \node at (92:0.5cm) [black,very thick, scale=0.65] {$\prec$};
    \end{tikzpicture}
    &= \mathrm{ev}_X \colon {}^\vee \! X \otimes X \to \id,\\
     \begin{tikzpicture}[baseline]
        \draw[black,thick] (0.5,0) arc (0:-180:0.5cm);
        \node at (0.5,0) [black,thick,above] {${}^\vee \! X $};
        \node at (-0.5,0) [black,thick,above] {$X$};
        \node at (-92:0.5cm) [black,very thick, scale=0.65] {$\prec$};
    \end{tikzpicture}
    &= \mathrm{coev}_X \colon \id \to X \otimes {}^\vee \! X,
\end{align}
called the \textbf{left evaluation} and \textbf{coevaluation} morphisms fulfilling the so-called \textbf{Zorro identities} (or \textbf{Zorro moves}): 
\begin{align}
        \begin{tikzpicture}[baseline=-2mm]
            \draw[black,thick] (0,0)  arc (360:180:0.5);
            \draw[black,thick] (1,0)  arc (0:180:0.5);
            \node at (-0.485,-0.495) [black,very thick, scale=0.65] {$\prec$};
            \node at (0.51,0.5) [black,very thick, scale=0.65] {$\prec$};
            \draw[black,thick] (1,0) -- (1,-0.8) node[black,below] {$X$};
            \draw[black,thick] (-1,0) -- (-1,0.8) node[black,above] {$X$};
        \end{tikzpicture}
        =
        \begin{tikzpicture}[baseline=-2mm]
            \draw[black,thick] (0,-0.8) node [black,below] {$X$} -- (0,0.8) node[black,above] {$X$};
            \node at (0,0) [black, very thick, scale=0.65,rotate=90] {$\succ$};
        \end{tikzpicture},
        \qquad
        \begin{tikzpicture}[baseline=-2mm]
            \draw[black,thick] (0,0)  arc (0:180:0.5);
            \draw[black,thick] (1,0)  arc (360:180:0.5);
            \draw[black,thick] (1,0) -- (1,0.8) node[black,above] {${}^\vee \!X$};
            \draw[black,thick] (-1,0) -- (-1,-0.8) node[black,below] {${}^\vee \!X$};
            \node at (-0.485,0.495) [black,very thick, scale=0.65] {$\prec$};
            \node at (0.51,-0.5) [black,very thick, scale=0.65] {$\prec$};
        \end{tikzpicture}
        =
        \begin{tikzpicture}[baseline=-2mm]
            \draw[black,thick] (0,-0.8) node[black,below] {${}^\vee \!X$} -- (0,0.8) node[black,above] {${}^\vee \!X$};
            \node at (0,0) [black, very thick,scale=0.65,rotate=90] {$\prec$};
        \end{tikzpicture}.
\end{align}
In traditional notation, the first diagram reads as the following equation of 
\begin{align}
    \mathrm{id}_X = (\mathrm{id_X} \otimes \mathrm{ev}_X ) \circ (\mathrm{coev}_X \otimes \mathrm{id_X}).
\end{align}
\end{definition}
A prototypical example of dualizable objects are finite dimensional vector spaces where the Zorro identities state that the pairing between a finite dimensional vector space and its dual is non-degenerate.

We can also consider something similar called adjointability for morphisms. The prototypical examples are adjoint functors in the $2$-category $\mathrm{Cat}$. 

\begin{definition}
    A subcategory $\mathbf{S}' \subset \mathbf{S}$ is called  \textbf{fully-dualizable} if every object has duals and for $k<n$ every $k$-morphism has adjoints.
    An object $X \in \mathbf{S}$ is called \textbf{fully-dualizable} if it lies in a fully-dualizable subcategory.
\end{definition}

%Anja
%\newpage
\section{Lecture 3: Duals and Adjoints in Higher Morita Categories}

Let us provide some motivation by considering duals and adjoints in the cobordism category. 

\begin{example}\label{example:handles}
   In the oriented cobordism category, the dual object of a positively oriented point is the negatively oriented point. The evaluation and coevaluation maps are determined by the following oriented cobordisms.
    
\begin{figure}[H] \label{evcoev}
\centering
\begin{tikzpicture}[scale=0.8]
    \draw[black,thick] plot[smooth, tension=1.7] coordinates{ (0,0) (1,1) (2,0) };
    \draw[black,thick] plot[smooth, tension=1.7] coordinates{ (4,1) (5,0) (6,1) };]
    \fill (0,0) circle[radius=2pt];
    \fill (2,0) circle[radius=2pt];
    \fill (4,1) circle[radius=2pt];
    \fill (6,1) circle[radius=2pt];
    %decoration 
    \node at (0,-0.5) [black!30!blue] {$\{\bigcdot_{-}\}$};
    \node at (2,-0.5) [black!30!red] {$\{\bigcdot_{+}\}$};
    \node at (6,1.5) [black!30!blue] {$\{\bigcdot_{-}\}$};
    \node at (1,1) [black, thick] {$\prec$};
    \node at (5,0) [black, thick] {$\prec$};
    \node at (4,1.5) [black!30!red] {$\{\bigcdot_{+}\}$};
    \node at (1,-1.5) {$\mathrm{ev}_{\{\bigcdot_{+}\}}$};
    \node at (5,-1.5) {$\mathrm{coev}_{\{\bigcdot_{+}\}}$};
\end{tikzpicture}.
\end{figure}

By the definition of a dual, the compositions
$$\{ \bigcdot _{+}\} \xrightarrow[]{\mathrm{coev}_{\{ \bigcdot _{+}\}} \otimes \mathrm{id}_{\{ \bigcdot _{+}\}}} \{ \bigcdot _{+}\} \otimes {\{ \bigcdot _{-}\}} \otimes \{ \bigcdot _{+}\} \xrightarrow[]{\id_{\{ \bigcdot _{+}\}} \otimes \mathrm{ev}_{\{ \bigcdot _{+}\}}} \{ \bigcdot _{+}\}
\text{ and } {\{ \bigcdot _{-}\}} \xrightarrow[]{ \mathrm{id}_{{\{ \bigcdot _{-}\}}} \otimes \mathrm{coev}_{\{ \bigcdot _{+}\}}  } {\{ \bigcdot _{-}\}} \otimes \{ \bigcdot _{+}\} \otimes {\{ \bigcdot _{-}\}} \xrightarrow[]{ \mathrm{ev}_{\{ \bigcdot _{+}\}} \otimes \mathrm{id}_{{\{ \bigcdot _{-}\}}} } \{ \bigcdot _{-} \}$$ should be equal to identities on objects, respectively $\id_{\{ \bigcdot _{+}\}}$ and $\id_{\{ \bigcdot _{-}\}},$

These Zorro identities are topologically obvious, as seen in the figure below. 

\begin{figure}[H] \label{snake1}
\centering
        \begin{tikzpicture}[baseline=-2mm]
            \draw[black,thick] (0,0)  arc (360:180:0.5);
            \draw[black,thick] (1,0)  arc (0:180:0.5);
            \node at (-0.3,-0.9) {$\mathrm{coev}_{\{\bigcdot_{+}\}}$};
            \node at (0.4,0.9) {$\mathrm{ev}_{\{\bigcdot_{+}\}}$};
            \node at (-0.485,-0.495) [black,very thick, scale=0.65] {$\prec$};
            \node at (0.51,0.5) [black,very thick, scale=0.65] {$\prec$};
            \draw[black,thick] (1,0) -- (1,-0.8) node[red,below] {$\{ \bigcdot _{+}\}$};
            \draw[black,thick] (-1,0) -- (-1,0.8) node[red,above] {$\{ \bigcdot _{+}\}$};
        \end{tikzpicture}
        =
        \begin{tikzpicture}[baseline=-2mm]
            \draw[black,thick] (0,-0.8) node [red,below] {$\{ \bigcdot _{+}\}$} -- (0,0.8) node[red,above] {$\{ \bigcdot _{+}\}$};
            \node at (0,0) [black, very thick, scale=0.65,rotate=90] {$\succ$};
            \node at (0.6,0) {$\id_{\{\bigcdot_{+}\}}$};
        \end{tikzpicture}
        \qquad
        \begin{tikzpicture}[baseline=-2mm]
            \draw[black,thick] (0,0)  arc (0:180:0.5);
            \draw[black,thick] (1,0)  arc (360:180:0.5);
            \draw[black,thick] (1,0) -- (1,0.8) node[blue,above] {$\{ \bigcdot _{-}\}$};
            \draw[black,thick] (-1,0) -- (-1,-0.8) node[blue,below] {$\{ \bigcdot _{-}\}$};
            \node at (-0.485,0.495) [black,very thick, scale=0.65] {$\prec$};
            \node at (0.51,-0.5) [black,very thick, scale=0.65] {$\prec$};
            \node at (-0.3,0.9) {$\mathrm{ev}_{\{\bigcdot_{+}\}}$};
            \node at (0.4,-0.9) {$\mathrm{coev}_{\{\bigcdot_{+}\}}$};
        \end{tikzpicture}
        =
        \begin{tikzpicture}[baseline=-2mm]
            \draw[black,thick] (0,-0.8) node[blue,below] {$\{ \bigcdot _{-}\}$} -- (0,0.8) node[blue,above] {$\{ \bigcdot _{-}\}$};
            \node at (0,0) [black, very thick,scale=0.65,rotate=90] {$\prec$};
            \node at (0.6,0) {$\id_{\{\bigcdot_{-}\}}$};
        \end{tikzpicture}
\end{figure}

If we start climbing on our homotopy ladder further, the first step would be extending the cobordism category to a 2-category by allowing manifolds with corners to play the role of 2-morphisms.
Now, we can discuss the higher dualizability of objects. Namely, we call for evaluation and coevaluation maps to have adjoints. In the subsequent discussion, we will temporarily set aside considerations related to the plus and minus orientations. 

Let's examine an illustrative case involving $ \mathrm{ev}$, the evaluation on the object $\{ \bigcdot \}$, and its left adjoint $ \mathrm{ev}^{L}$, depicted below. We need the data of a unit and a counit for the adjunction $ \mathrm{ev}^{L} \dashv \mathrm{ev}$. We will denote these with $\eta:\id_{\emptyset} \xrightarrow[]{} \mathrm{ev}\circ \mathrm{ev}^{L} $ and $\epsilon: \mathrm{ev}^{L} \circ \mathrm{ev} \xrightarrow[]{}  \id_{ \{ \bigcdot \} ^ \vee \otimes \{ \bigcdot \} }$. These maps will manifest the adjunction between the morphisms $ \mathrm{ev}$ and $ \mathrm{ev}^{L}$, provided that the compositions 

$$ \mathrm{ev} \xrightarrow[]{\eta \circ \id_{\mathrm{ev}}} \mathrm{ev} \circ \mathrm{ev}^{L} \circ \mathrm{ev} \xrightarrow[]{\id_{\mathrm{ev}} \circ  \epsilon } \mathrm{ev} \quad \text{ and } \quad \mathrm{ev}^{L} \xrightarrow[]{\id_{\mathrm{ev}^{L}} \circ \eta} \mathrm{ev}^{L} \circ \mathrm{ev} \circ \mathrm{ev}^{L} \xrightarrow[]{\epsilon \circ \id_{\mathrm{ev}^{L}}} \mathrm{ev}^{L}$$ yield identity transformations on $ \mathrm{ev}$ and $ \mathrm{ev}^{L}$.

Let us fix the convention of reading all the diagrams from left to right and bottom to top. Now, we can think of the counit map as tracing the rainbow; by this, we think of the manifold

\begin{figure}[H] \label{rainbowcounit}
\centering
\begin{tikzpicture}[scale=0.6]
\draw[thick,xshift=2cm]  plot[smooth, tension=1] coordinates{ (1,1) (0,0) (1,-1) } node[black,below] {$ \mathrm{ev}^{L}$};
\draw[red,thick] plot[smooth, tension=1] coordinates{ (0,-1) (1.5,1) (3,-1) };
\draw[thick] plot[smooth, tension=1] coordinates{ (0,1) (1,0) (0,-1) } node[black,below] {$ \mathrm{ev}$};
\draw[red,thick] plot[smooth, tension=1] coordinates{ (1,0) (1.5,0.5) (2,0) };
\draw[red,thick] plot[smooth, tension=1] coordinates{ (0,1) (1.5,2) (3,1) };
\end{tikzpicture}.
\end{figure}

 Observe that topologically, this manifold corresponds to the saddle. The next step is to pinch the corners to be able to interpret it as a manifold with corners and consequently as a 2-morphism in the cobordism category. 
 
\begin{figure}[H] \label{saddle}
\centering
\begin{tikzpicture}[scale=0.5]
\node at (-3.5,1.5) [black!50!green] {$\epsilon := $};
\node at (1.3, 4.5) [red] {$\id_{\{ \bigcdot \} ^ \vee \otimes \{ \bigcdot \} }$};
\node at (-2.5, -1.5) [black!30!blue] {$ \mathrm{ev}$};
\node at (2.5, -1.5) [black!30!blue] {$ \mathrm{ev}^{L}$};
\draw[thick] (-1.25, 0.875) -- (-2, -1) arc (-90:90:1.5 and 1) -- (-1, 3.5) -- (3, 3.5) -- (2,1) arc (90:270:1.5 and 1) -- (3,1.5) -- (2.2, 1.5);
\draw[thick] (-0.52,-0.2) .. controls (0, 1.8) and (1, 1.8) .. (0.52,0.2);
\draw[thick,black!30!blue] (-2, -1) arc (-90:90:1.5 and 1);
\draw[thick,black!30!blue]  (2,1) arc (90:270:1.5 and 1);
\draw[thick,red]  (-1, 3.5) -- (3, 3.5);
\draw[thick,red] (3,1.5) -- (2.2, 1.5);
\end{tikzpicture}
\end{figure}

The unit of the adjunction will be given by 
 
 \begin{figure}[H] \label{snake}
    \centering
    \begin{tikzpicture}[scale=0.6]
    \draw[thick,black!30!blue] (0,0) ellipse (1.5 and 1);
    \node[black!30!blue] at (0,2) {$ \mathrm{ev} \circ \mathrm{ev}^{L}$};
    \node[red!50!blue] at (-2.5,-1) {$\eta :=$};
    \draw[thick]  plot[smooth, tension=2] coordinates{ (-1.5,0) (0,-2) (1.5,0)};
    \node at (0,-3) [thick] {$\id_{\emptyset}$};
\end{tikzpicture} .
\end{figure}

Finally, the second Zorro identity can now be visualized as follows.

\begin{figure}[H] \label{snake}
    \centering
    \begin{tikzpicture}[scale=0.6]
    \draw[thick,red!50!blue] (0,0) ellipse (1.5 and 1);
    \node[red!50!blue] at (-2.5,0) {$\eta$};
    \draw[thick,black!30!green] plot[smooth, tension=2] coordinates{ (1.5,0.1) (2.25,0.8) (2.98,0.1) };
    \draw[thick]  plot[smooth, tension=1.5] coordinates{ (-0.5,3.5) (-1.5,2.5)  (0,1.5) };
    \draw[thick,black!30!green,xshift=3cm]  plot[smooth, tension=1.5] coordinates{ (1,1) (0,0) (1.5,-1) };
    \draw[thick,yshift=-4cm, xshift=3cm]  plot[smooth, tension=1.5] coordinates{ (1,1) (0,0) (1.5,-1) };
    \draw[thick,red!50!blue]  plot[smooth, tension=2] coordinates{ (-1.5,0) (0,-2) (1.5,0)};
    \draw[thick,black!30!green] (0,-1) -- (0,1.5) -- (4.5,1.5) -- (4.5,-1) ;
    \draw[thick,black!30!green] (-0.5,0.95) -- (-0.5,3.5) -- (4,3.5) -- (4,1) ;
    \node[black!30!green] at (4.5,3.7) {$\epsilon$};
    \draw[thick,yshift=1cm,xshift=-1.5cm] (0,-1) -- (0,1.5);
    \draw[thick] (4.5,-1) -- (4.5,-5);
    \draw[thick] (3,0) -- (3,-3.65);
    \draw[thick] (4,1) -- (4,-3);
    \node at (-2.5,3.7) {$\id_{\mathrm{ev}^{L}}$};
    \node at (5.5,-4) {$\id_{\mathrm{ev}^{L}}$};
    
    %=
    \draw[black,thick] (5.5,0.15) -- (6,0.15);
    \draw[black,thick] (5.5,-0.15) -- (6,-0.15);
    
    %RHS
    \draw[thick]  plot[smooth, tension=1.5] coordinates{ (8,3.5) (7,2.5) (8.5,1.5) };
    \draw[thick]  plot[smooth, tension=1.5] coordinates{ (8,-2.5) (7,-3.5) (8.5,-4.5) };
    \draw[thick] (6.98,2.6) -- (6.98,-3.2);
    \draw[thick] (8,3.5) -- (8,-2.5);
    \draw[thick] (8.5,1.5) -- (8.5,-4.5);
     \node at (9.5,0) {$\id_{\mathrm{ev}^{L}}$};
\end{tikzpicture} 
\end{figure}

 Topologically, it is evident that the relation holds. Similarly, one can check that the first Zorro identity holds. We have now demonstrated that the evaluation possesses a left adjoint. Following a similar line of reasoning for the adjoint of the coevaluation, one can deduce that the positively oriented point is 2-dualizable.

 Extending our inquiry into higher dimensions and inspecting adjoints of the units and counits, we observe a consistent pattern throughout.
 As a consequence, we can assert that $\mathbf{Cob}^{n}_{\mathrm{or}}$ is fully dualizable. Following a similar line of arguments, one can show that $\mathbf{Cob}^{n}_{\mathrm{fr}}$ also is  fully dualizable. Our focus is on the functors mapping from the fully dualizable category $\mathbf{Cob}^{n}_{\mathrm{fr}}$ into a category $\mathbf{S}$. It is crucial to emphasize that the image of such a functor must necessarily reside within the fully dualizable part of $\mathbf{S}$. In other words, a TFT ensures that the dualizablity data in the $\mathbf{Cob}^{n}_{\mathrm{fr}}$ will be pushed through to $\mathbf{S}$.

The discussion above concludes why, to every $n$-dimensional framed TFT $\mathrm{Z}$ in $\mathbf{S}$, we can assign a fully dualizable object in $\mathbf{S}$ by evaluating it on $\{\bigcdot _{+}\}$. This is the easy direction of the cobordism hypothesis. The deep and very intricate part of the statement demonstrates that any fully dualizable object can be lifted (in an essentially unique way) to a framed TFT.
\end{example}

\begin{remark}
    Even though the visual representations suggest so, the coevaluation and the left adjoint of the evaluation are not the same morphisms. The codomain of the coevaluation map is $X \otimes X^ \vee$, while the left adjoint of the evaluation maps into $X^\vee \otimes X$. In a symmetric monoidal category, the relation between the two is given by the equation $\mathrm{ev}_{L} = \beta \, \circ \, (q^{-1}  \otimes \mathrm{id}_{X^\vee}) \, \circ \, \mathrm{coev}$, where $\beta$ is the braiding and $q^{-1}$ pseudo-inverse of the Serre functor\footnote{Proposition 4.2.3 in J.~Lurie, \emph{On the classification of
topological field theories},
  Current developments in mathematics, 2008, Int. Press, Somerville, MA, 2009,
  pp.~129--280. \href{http://arxiv.org/abs/arXiv:0905.0465}{{\tt
  arXiv:0905.0465}}}.
\end{remark}

\begin{example}\label{example:vect}
    Let's go back to an example of 1-categories and 1-dualizability. The first case we inspect is $\mathbf{S}= \mathrm{Vect}^{\otimes}_{\mathbb{C}}$.
    Let $\mathrm{Z}$ be an oriented topological field theory $\mathrm{Z}\colon \mathrm{Cob}^1_{\mathrm{or}} \xrightarrow{}  \mathrm{Vect}^{\otimes}_{\mathbb{C}}$ such that $\mathrm{Z}$ evaluated at a point is a vector space $V$, $\mathrm{Z}(\{\bigcdot _{+}\})=V$. Since $\mathrm{Z}(\{\bigcdot _{+}\})$ has to be dualizable in $\mathrm{Vect}^{\otimes}_{\mathbb{C}}$, we are interested in identifying the dualizable objects in this category. Let $W$ be any vector space. One can show that $W$ has an evaluation and coevaluation satisfying the Zorro identities if and only if $W$ is a finite-dimensional vector space. Since $V$ has to be finite-dimensional, we can choose a finite basis $\{v_i\}_{1\leq i \leq n}$. Denote the dual space with  $V^{\vee}=\mathrm{Hom}(V,\mathbb{C})$ and the dual basis with $\{v^i\}_{1\leq i \leq n}$. Define the evaluation map $\mathrm{ev}_{V}: V^{\vee} \otimes V \xrightarrow[]{} \mathbb{C}$ as a map that evaluates every $w \in V^{\vee}$ at $v \in V$. The coevaluation map $\mathbb{C} \xrightarrow[]{} V \otimes V^{\vee}$ is given by $1 \mapsto \sum_{1\leq i \leq n} v_i \otimes v^i$. The compatibility between the evaluation and coevaluation essentially encodes matrix multiplication. 
\end{example}

\begin{exercise}
Show that under the notation above, the following equation holds
\begin{align*}
		\mathrm{Z} \Big(\
            \begin{tikzpicture}[baseline={([yshift=-0.8ex, xshift=-0.2ex]current bounding box.center)}]
	       \draw[black, thick] (0,0) circle (0.4cm);
            \end{tikzpicture} 
            \, \Big) (1) = \underset{1\leq i \leq n}{\sum} \text{ev}_{V} (v^{i} \otimes v_{i})= \text{dim}_{\mathbb{C}} V.
\end{align*}
\end{exercise}

\begin{exercise}
Show that to define a 1-dimensional TFT, it suffices to give a finite-dimensional vector space.
\end{exercise}

    The first step in climbing our homotopy ladder was extending the cobordism category to a 2-category by allowing manifolds with corners. In this case, we also have to extend our target category, previously a category of vector spaces over $\mathbb{C}$. We still want to be able to retrieve the $\mathrm{Vect}_{\mathbb{C}}$ as the 1-category of endomorphisms of the unit object of our new 2-category. The natural choice is the Morita category, which we discuss in the following examples.

\begin{definition}
    Let $\Alg_{1} (\mathrm{Vect}_{\mathbb{C}})$ be a \textbf{2-category of associative algebras} over $\mathbb{C}$. Let $A$ and $B$ be associative algebras. Then the 1-morphisms $M$ and $N$ from $B$ to $A$ are defined to be the $(A, B)$-bimodules. The 2-morphisms between $M$ and $N$ are bimodule homomorphisms. The vertical composition is given by the standard composition of bimodule homomorphisms. The horizontal composition is given by the relative tensor product of bimodules.
    The monoidal structure comes from the tensor product of algebras, and the monoidal unit is $\mathbb{C}$.   
\end{definition}

%CHECK:
\begin{definition}
    For every algebra $A$ over $\mathbb{C}$, we define the \textbf{opposite algebra} $A^{op}$, which has the same underlying vector space, but algebra multiplication is given by $a * b =  b a$. 
\end{definition}

\begin{remark}
    Notice that an $A$-$B$-bimodule structure is precisely the same data as an $A \otimes B^{op}$-module structure. 
\end{remark}

Now, let us undertake a more comprehensive examination of dualizability within the $\Alg_{1}(\mathrm{Vect}_{\C})$.

\begin{theorem}
    Every algebra in $\Alg_{1}(\mathrm{Vect}_{\C})$ is 1-dualizable. 
\end{theorem}

\begin{remark}
    The dual of the algebra $A$ is given by the opposite algebra $A^{op}$. The evaluation map is given by $A$ seen as a $(A^{op} \otimes A, \C)$-bimodule. The Zorro axioms will hold trivially.
\end{remark}

    Since we're diving into higher dualizability, the next question to tackle is whether the evaluation map has adjoints. The following results will be beneficial for our further analysis. 
    
\begin{lemma}
   If $M$ is an $(A, B)$-bimodule, then $M$ admits a right adjoint $N$ if and only if $M$ is finitely generated and projective as an $A$-module.
\end{lemma}

   We conclude that the evaluation map has a right adjoint if and only if $A$ is finitely generated and projective $A^{op} \otimes A$-module. Unraveling the definitions shows that the condition of being projective over $A^{op} \otimes A$ amounts to separability of the algebra, which over $\C$ is equivalent to $A$ being semisimple. Thus, $A$ has to be a finite direct sum of matrix algebras over $\C$ by the Artin-Wedderburn theorem. 
   Similarly, the evaluation map has a left adjoint if and only if $A$ is finitely generated and projective $\C$-module. Any algebra's projectivity condition is fulfilled since $\C$ is a field.
   Finally, we conclude that an algebra is 2-dualizable if and only if it is finite-dimensional and semisimple.

\begin{exercise}
    The following is an exercise that unravels the parallel situation of the discussions above for the category of $\mathbb{K}$-linear categories, $(\mathbf{Cat}_{\mathbb{K}}, \times)$, and a category of $\mathbb{K}$-linear presentable categories, $(\mathbf{Pr}_{\mathbb{K}}, \otimes)$. 
\begin{enumerate}[label=(\alph*)]
    \item Let $\mathcal{C}$ be a $\mathbb{K}$-linear small category. Show that such a category is always 1-dualizable and that the opposite category gives the dual.
    \item Let $\mathcal{C}$ be a $\mathbb{K}$-linear presentable category. Show that $\mathcal{C}$ is 1-dualizable if and only if $\mathcal{C}$ is a co-completion of some small category.
    \item In both cases (a) and (b), category $\mathcal{C}$ is 2-dualizable if and only if it is finite semisimple.
\end{enumerate}
\end{exercise}

\subsection{Orientations}

 Since we are considering manifolds with $n$-framings, one can globally change the framings and induce a group action of the orthogonal group $\mathrm{O}(n)$, and thus also of the special orthogonal group $\mathrm{SO}(n)$, on the $\mathbf{Cob}_{\mathrm{fr}}^{n}$ category. Therefore, we have an $\mathrm{SO}(n)$-action on the category of $n$-dimensional framed TFTs $\mathrm{Fun}^{\otimes}(\mathbf{Cob}_{\mathrm{fr}}^{n},\mathbf{S})$. By the cobordism hypothesis, this induces an action on the category of fully dualizable objects~$\mathbf{S}^{\mathrm{fd}}$.

    Given that we are doing homotopy theory, the notion of $\mathrm{SO}(n)$-action is more subtle. The data of the action is given by the map between the classifying spaces $B\mathrm{SO}(n) \xrightarrow[]{} B\mathrm{Aut}(\mathbf{S})$.
The oriented TFTs can now be seen as fixed points of the framed TFTs with respect to the $\mathrm{SO}(n)$-action as above. 
$$[\mathrm{Fun}^{\otimes}(\mathbf{Cob}_{\mathrm{fr}}^{n},\mathbf{S}) ]^{\mathrm{SO}(n)} \simeq \mathrm{Fun}^{\otimes}(\mathbf{Cob}_{\mathrm{or}}^{n},\mathbf{S})  $$

\begin{lemma}\label{lemma:orientedcob} [\textbf{Corollary of Cobordism Hypothesis}]
    There is an equivalence between the fully extended n-dimensional oriented topological field theories and homotopy fixed points of the fully-dualizable objects in $\mathbf{S}$ with respect to the action of $\mathrm{SO}(n)$, namely 
    $$\mathrm{Fun}^{\otimes}(\mathbf{Cob}_{\mathrm{or}}^{n},\mathbf{S}) \simeq \big[\mathbf{S}^{\mathrm{fd}}\big]^{h\mathrm{SO}(n)}.$$
\end{lemma}

    Heuristically speaking, a framed TFT involves the extra choice of framing, while the oriented TFT forgets the framing. We think of this as quotienting out by $\mathrm{SO}(n)$, but this is not an exact statement as it is not true that every $n$-manifold can be framed. However, every manifold can be locally framed, and the homotopy fixed point structure gives the data we need to patch the local information. The homotopy theory deals with this issue.

\begin{definition}\label{homotopyfixedpoint}
    A \textbf{homotopy fixed point} is a null homotopy between the action $B\mathrm{SO}(n) \xrightarrow[]{} B\text{Aut}(\mathbf{S})$ and a point.
\end{definition}

\begin{example}\label{example:so2}
    Let us now work through an example where $n=2$. The group $\mathrm{SO}(2)$ is isomorphic to the circle group $S^{1}$, therefore $B\mathrm{SO}(2) \cong BS^{1} \cong \C \mathbb{P}^{\infty}$. This example is, in particular, nice to work with because $\C \mathbb{P}^{\infty}$ has a nice iterative structure as a CW complex, namely $e_{0} \cup e_{2} \cup e_{4} \cup ...$, where in every even degree there is exactly one cell.  

    In our case, the target category $\mathbf{S}$ is a 2-category. Hence, all the homotopy groups $\pi_{\geq 3} \cong 0$ are trivial. The consequence of this is that to give a map from the $B\mathrm{SO}(2)$ into $B \mathrm{Aut} (\mathbf{S})$, one only has to specify what happens on the 0-cell and the 2-cell. The data of the map from $\C \mathbb{P}^{\infty}$ into a 2-type is oblivious of the images of higher cells. In other words, if we know the images of $e_0$ and $e_2$, the map can be uniquely extended to the map from the entire $B\mathrm{SO}(2)$.
    
    Let us return to the example  $\mathbf{S}^{\mathrm{fd}}=\Alg_{1} (\mathrm{Vect}_{\mathbb{C}})^{\mathrm{fd}}$. Every 2-dualizable algebra $A$ is finite dimensional as a vector space, and therefore there is a natural isomorphism between $A$ and a double dual $A^{**}$, $A \cong A^{**}$. Even though this isomorphism is natural, it doesn't have to commute with the action of $A$. Using this natural isomorphism, we can get a new bimodule where we twist the action of $A$ on $A^{**}$. Let us denote that bimodule by $_{A} {A^{**}}_{A}$. Now, we look at the 2-morphism between the identity 1-morphism $_{A} A_{A}$ and $_{A} {A^{**}}_{A}$. This bimodule morphism $_{A} A_{A} \xrightarrow[]{} _{A} {A^{**}}_{A}$ is the image of the $e_2$ cell under the action map. The data of the homotopy fixed point is a natural isomorphism which identifies the morphism $_{A} A_{A} \xrightarrow[]{} _{A} {A^{**}}_{A}$ with the trivial action $_{A} A_{A} \xrightarrow[]{} _{A} A_{A}$. In other words, this is the natural isomorphism between $_{A} A_{A}$ and $_{A} {A^{**}}_{A}$. This data corresponds exactly to the additional structure of symmetric Frobenius algebra on $A$. This structure has a nice interpretation in terms of pictures. 
    If we start with a TFT $\mathrm{Z}$ attached to the algebra $A$, the value of $\mathrm{Z}$ on a circle $S^1$ is the 0-Hochschild homology of $A$, i.e., a trace of $A$.
\begin{align*}
		\mathrm{Z}	\Big(\,
            \begin{tikzpicture}[baseline={([yshift=-0.8ex, xshift=-0.2ex]current bounding box.center)}]
	       \draw[black, thick] (0,0) circle (0.4cm);
            \end{tikzpicture}  
            \Big)\ = \text{HH}_0(A)= \text{Tr}(A)
\end{align*}

\begin{align*}
		\mathrm{Z}	\Big(\,  \begin{tikzpicture}[baseline={([yshift=-0.8ex, xshift=-0.2ex]current bounding box.center)}]
	       \draw[black, thick] (0,0) circle (0.4cm);
            \end{tikzpicture} 
            \Big) = \mathrm{Z}	\Big(\,  \begin{tikzpicture}[baseline={([yshift=-0.8ex, xshift=-0.2ex]current bounding box.center)}]
	    \draw[thick]  plot[smooth, tension=1.5] coordinates{ (0.5,0.7) (0,0.35) (0.5,0) };
            \fill (0.5,0.7) circle[radius=1.5pt];
            \fill (0.5,0) circle[radius=1.5pt];
            \end{tikzpicture} 
            \Big) \underset{\mathrm{Z}( \, \begin{tikzpicture}[baseline={([yshift=-0.8ex, xshift=-0.2ex]current bounding box.center)}]
            \fill (0.5,0.1) circle[radius=1.5pt];
            \fill (0.5,-0.1) circle[radius=1.5pt];
            \end{tikzpicture}  \, )}{\otimes} \mathrm{Z} \Big( \, \begin{tikzpicture}[baseline={([yshift=-0.8ex, xshift=-0.2ex]current bounding box.center)}]
	    \draw[thick]  plot[smooth, tension=1.5] coordinates{ (-0.5,0.7) (0,0.35) (-0.5,0) };
            \fill (-0.5,0.7) circle[radius=1.5pt];
            \fill (-0.5,0) circle[radius=1.5pt];
            \end{tikzpicture} \, \Big)=
            \mathrm{Z} (ev) \underset{\mathrm{Z}(\{ \bigcdot \} \sqcup \{ \bigcdot \} )}{\otimes} \mathrm{Z} (ev^L) = A \underset{A \otimes A^{op}}{\otimes} A = A_{/{[A,A]}}
\end{align*}
Here $ \mathrm{ev}^L = coev \circ \beta$ is the left (and right) adjoint of the evaluation.
The Frobenius structure is the data of the trace

    \begin{align*}
		\mathrm{Z}	\Big(\,	\begin{tikzpicture}[baseline={([yshift=-0.8ex]current bounding box.center)}] \node[rotate=0] at (0,0) {	
		\resizebox{0.07\textwidth}{!}{%
			\begin{tikzpicture}[ tqft/cobordism/.style={draw},tqft/every boundary component/.style={draw}] 
			\pic [tqft/cap];
			\end{tikzpicture}	} }; \end{tikzpicture} \Big)= \text{tr}\colon A_{/{[A,A]}} \xrightarrow[]{} \mathbb{C}.
    \end{align*}
The trace is non-degenerate and symmetric, which can be seen as a consequence of certain topological diagrams.
\end{example}

\begin{example}
    Let $A$ be a group algebra of a finite group $G$ with identity element $e$, $A= \mathbb{C}[G]$. This has a Frobenius structure where the trace is given for $h\in G \subset A$ by $\text{tr}(h)=\delta_{h,e}$. Let $\mathrm{Z}$ be the oriented TFT assigned to $A$. Then the value of $\mathrm{Z}$ on the surfaces with genus $g$ is
    $$Z_{G}(\Sigma_{g})= \sum_{[G \xrightarrow[\underset{G}{\circlearrowleft}]{} E \, \xrightarrow[]{} \Sigma_g]} \frac{1}{|\mathrm{Aut}(E)|} =\frac{1}{|G|} \sum_{G \xrightarrow[\underset{G}{\circlearrowleft}]{} E \, \xrightarrow{} \Sigma_g} 1,$$
    where the first sum goes over isomorphism classes of principal $G$-bundles over $\Sigma_g$. Finally, this sum equals the number of principle $G$-bundles.
\end{example}

\subsection{Higher Morita Categories}

 This section discusses the target higher categories we intend to use to understand the calculus of defects. We start by giving an alternative definition of the tensor categories using ideas of algebras. Usually, to define a tensor category, one has to specify a category with tensor product, unit, associators, and unitors such that the pentagon axiom is satisfied. Instead, we use the topological category of disks denoted by $\Disk_{\mathrm{fr}}^{1}$ that has as objects finite disjoint unions of $\mathbb{R}$ and whose spaces of morphisms are spaces of framed embeddings of finite disjoint unions. Namely, the 1-morphisms are framed embeddings; the 2-morphisms are isotopies between the embeddings, and the 3-morphisms are isotopies of isotopies between embeddings, and so on.

 \begin{definition}
    Let $\mathbf{Pr}_{\C}$ be a 2-category of locally presentable $\C$-linear categories, co-continuous functors, and natural transformations.
\end{definition}

 \begin{definition}
     A \textbf{tensor category} $\mathcal{A}$ is a $\Disk_{\mathrm{fr}}^1$-algebra in the category of categories $\mathbf{Pr}_{\C}$, that is a symmetric monoidal functor $\mathcal{A}: \Disk_{\mathrm{fr}}^1 \xrightarrow[]{} \mathbf{Pr}_{\mathbb{C}}.$
 \end{definition}
\begin{remark}
\label{rem1}
    If we have such a symmetric monoidal functor $\mathcal{A}$, it gives us the data of a category $ \mathcal{C}\coloneqq\mathcal{A}(\R)$. Furthermore, for every embedding $\iota\colon \overset{n}{\underset{k=1}{\sqcup}} \R \hookrightarrow \R$ of finite disjoint union of open intervals into $\R$, we get a map $\mathcal{A}(\iota\colon \overset{n}{\underset{k=1}{\sqcup}} \R \hookrightarrow \R): \overset{n}{\underset{k=1}{\otimes}} \mathcal{C} \rightarrow \mathcal{C}$. This way, we obtain a monoidal structure on the category $\mathcal{C}$. The data of the associator comes from isotopies between the appropriate inclusions of finite disjoint unions of disks. Formally, one should still prove that the pentagon axiom holds. Let us provide some visual interpretations. The symmetric monoidal functor $\mathcal{A}$ can be pictured living on $\R$ where every interval is labeled with the monoidal category~$\mathcal{C}$.
    
\begin{figure}[H]
    \centering
    \begin{tikzpicture}
        \draw[thick,black!40!blue] (0,0) -- (4,0);
        %Labels
        \node[thick,black!40!blue] at (0.4,0) {$\Big($};
        \node[thick,black!40!blue] at (1,0) {$\Big)$};
        \node[thick,black!40!blue] at (1.5,0) {$\Big($};
        \node[thick,black!40!blue] at (2.5,0) {$\Big)$};
        \node[thick,black!40!blue] at (3,0) {$\Big($};
        \node[thick,black!40!blue] at (3.8,0) {$\Big)$};
        \node at (0.7,0.4) {$\mathcal{C}$};
        \node at (2,0.4) {$\mathcal{C}$};
        \node at (3.4,0.4) {$\mathcal{C}$};

    \end{tikzpicture}
        \caption{a $\Disk_{\mathrm{fr}}^1$-algebra}
\end{figure}

\end{remark}

%Lecture 4
    Having explored an example of an algebra in dimension one within the category of categories, we will now focus on the two prominent examples of algebras in dimensions two and three. We have already seen the one-dimensional topological category of disks, so we proceed by increasing the dimension to obtain the category $\Disk_{\mathrm{fr}}^{2}$. The objects are finite disjoint unions of $\R^{2}$, and the space of morphisms is defined analogously to the one-dimensional case, namely as a space of framed embeddings.

    A $\Disk_{\mathrm{fr}}^{2}$-algebra is a symmetric monoidal functor $\mathcal{A}$ from $\Disk_{\mathrm{fr}}^{2}$ to $\mathbf{Pr}_{\C}$, and as in remark \ref{rem1}, it gives the data of a category $\mathcal{A}(\mathbb{R})$ together with maps $\mathcal{A}(\overset{n}{\underset{k=1}{\sqcup}} \R^2  \hookrightarrow \R^2): \overset{n}{\underset{k=1}{\otimes}} \mathcal{C} \rightarrow \mathcal{C}$, for every configuration of disjoint discs embedded in $\R^2$. Let us again label the category $\mathcal{A}(\mathbb{R})$ with $\mathcal{C}$; now, we can visually think of $\mathcal{A}$ as in the figure below.

     \begin{figure}[H]
 \centering
    \begin{tikzpicture}
        %rectangle
        \draw[fill=blue!30] (0,0) rectangle (4,3);
        %Labels
        \node at (0.7,2.5) {$\mathcal{A}$};
    \end{tikzpicture}
        \caption{a $\Disk_{\mathrm{fr}}^2$-algebra}
\end{figure}
    
    A monoidal structure can be obtained as the image of any framed embedding of two disjoint open disks; for an example, look at the figure \ref{monoidal structure}. This is well-defined up to an equivalence since all such framed embeddings are isotopic. The isotopy between two different embeddings will induce the equivalence of the functors.
    
\begin{figure}[H]
    \centering
    $\mathcal{A} \, \, \Big(  \quad
    \begin{tikzpicture} [baseline={([yshift=-0.8ex]current bounding box.center)}] 
        %rectangle
        \draw[fill=blue!20] (0,0) rectangle (4,3);
        \draw[dashed,fill=blue!30] (0.5,0.4) rectangle (1.8,1.9);
        \draw[dashed,fill=blue!30] (2.2,1.3) rectangle (3.5,2.4);
    \end{tikzpicture} \quad \Big) : \mathcal{C} \otimes \mathcal{C} \xrightarrow[]{} \mathcal{C}.$
        \caption{monoidal strucutre}
        \label{monoidal structure}
\end{figure}
    
    In the two-dimensional case, we obtain even more data. Using the isotopes between the embedding of two discs and the embedding of the swapped discs, we obtain a braiding on the category $\mathcal{C}$. 

    \begin{figure}[H]
    \centering
    $\mathcal{A} \Big(  \, \,
    \begin{tikzpicture} [baseline={([yshift=-0.8ex]current bounding box.center)}] 
        %rectangle
        \draw[fill=blue!20] (0,0) rectangle (4,3);
        \draw[dashed,fill=blue!30] (0.5,0.4) rectangle (1.8,1.9);
        \draw[dashed,fill=blue!30] (2.2,1.3) rectangle (3.5,2.4);
        \node[blue!70!black] at (1.15,1.15) {1};
        \node[red!90!black] at (2.85,1.85) {2};   
    \end{tikzpicture} \quad
    \begin{tikzpicture} [baseline={([yshift=5ex]current bounding box.center)}, scale=0.6] 
        %rectangle
        \node at (2,3.8) {\text{swap}};
        \node at (2,3.3) {\Large$\Rightarrow$};
        \draw[fill=blue!20] (0,0) rectangle (4,3);
        \draw[dashed,fill=blue!30] (0.5,0.4) rectangle (1.8,1.9);
        \draw[dashed,fill=blue!30] (2.2,1.3) rectangle (3.5,2.4);
        \path [>=triangle 45, <->, blue!70!black, bend right] (2,0.4) edge (3,1.2);
        \path [>=triangle 45, <->, blue!70!black, bend left] (1.2,2.1) edge (2.5,2.6); 
    \end{tikzpicture} 
    \quad
    \begin{tikzpicture} [baseline={([yshift=-0.8ex]current bounding box.center)}] 
        %rectangle
        \draw[fill=blue!20] (0,0) rectangle (4,3);
        \draw[dashed,fill=blue!30] (0.5,0.4) rectangle (1.8,1.9);
        \draw[dashed,fill=blue!30] (2.2,1.3) rectangle (3.5,2.4);
        \node[red!70!black] at (1.15,1.15) {2};
        \node[blue!70!black] at (2.85,1.85) {1};  
    \end{tikzpicture}  \, \, \Big) $
        \caption{braiding}
\end{figure}
 So far, all the examples of $\Disk_{\mathrm{fr}}^{n}$-algebras we looked at have taken place in 2-categories (or rather the underlying $(2,1)$-categories). More generally, we can consider $\Disk_{\mathrm{fr}}^n$-algebras in an $(\infty,1)$-category.

 If we start with a ``good'' $(\infty,k)$-category $\mathbf{S}$, one can consider the higher Morita category, which is an $(\infty, n+k)$-category and whose objects are $\Disk_{\mathrm{fr}}^n$-algebras in $\mathbf{S}$. In the literature, this is often denoted as $\mathrm{Alg}_n (\mathbf{S})$. Roughly, the 1-morphisms are $\Disk_{\mathrm{fr}}^{n-1}$-algebras in the category of bimodules of $\Disk_{\mathrm{fr}}^{n}$-algebras, the 2-morphisms are $\Disk_{\mathrm{fr}}^{n-2}$-algebras in the category of bimodules of  $\Disk_{\mathrm{fr}}^{n-1}$-algebras, and so forth. The $l$-morphisms in $\mathbf{S}$ are used to define the $n+l$-morphisms.

Two notable models for this $(\infty,n+k)$-category exist, both by Johnson-Freyd--Scheimbauer\footnote{T.~Johnson-Freyd and C.~Scheimbauer, \textit{(Op) lax natural transformations, twisted quantum field theories, and “even higher” Morita categories}, \href{https://www.researchgate.net/publication/310787437_Oplax_natural_transformations_twisted_quantum_field_theories_and_even_higher_Morita_categories}{Advances in Mathematics} 307(4):147-223}, extending the constructions of either Haugseng\footnote{R.~Haugseng, \textit{The higher Morita category of $E_n$–algebras}, \href{https://doi.org/10.2140/gt.2017.21.1631}{Geometry and Topology} 21(3)
1631-1730} or Calaque--Scheimbauer\footnote{D.~Calaque and C.~Scheimbauer, \textit{Factorization Homology as a Fully Extended Topological Field Theory}}. The following definitions are based on these constructions. For technical details and the precise definition of ``good'', we refer to these references.
    
\begin{remark} 
    To summarize, in the one-dimensional case, $\Disk_{\mathrm{fr}}^{1}$-algebras in $\mathbf{Pr}_{\mathbb{C}}$ exactly correspond to tensor categories. The higher Morita category of $\Disk_{\mathrm{fr}}^{1}$-algebras in  $\mathbf{Pr}_{\mathbb{C}}$ will produce a 3-category of tensor categories. It is a well-known result that $\Disk_{\mathrm{fr}}^{2}$-algebras in  $\mathbf{Pr}_{\mathbb{C}}$ are identified with braided tensor categories. The higher Morita category of such algebras gives rise to a 4-category of braided tensor categories. 
\end{remark}

    Let's delve into the visual representation of objects and morphisms in the category $\mathrm{Alg}_2 (\mathbf{Pr}_{\mathbb{C}})$. In this context, the 1-morphisms are $\Disk_{\mathrm{fr}}^{1}$-algebras in the category of bimodules, locating them along the line as we have seen in remark \ref{rem1}. Objects, on the other hand, inhabit the plane as they are $\Disk_{\mathrm{fr}}^{2}$-algebras. This makes the situation more intricate. We now consider 1-morphisms in terms of manifolds with defects of codimension one. To elaborate further, let's examine the figure below. The discs embedded on the left side are associated with a category $\mathcal{A}$, while those on the right side are governed by a category $\mathcal{B}$. These correspond to different objects. Now, new objects appear once we embed discs so that they contain the defect, namely the red line on the manifold. Let the image of such a disc map be a category $\mathcal{C}$. The embeddings of disjoint discs containing the defect into a larger disc that also contains the defect will induce the monoidal structure on the category $\mathcal{C}$. This structure mirrors the one-dimensional case, with one-dimensional discs, that was already discussed above.

\begin{figure}[H]
 \centering
    \begin{tikzpicture}
        %rectangle
        \draw[fill=blue!20] (0,0) rectangle (4,3);
        % circles
        \draw[dashed,fill=blue!30] (0.7,1) circle (0.4);
        \draw[dashed,fill=blue!30] (3.3,2) circle (0.4);
        \draw[dashed,fill=blue!30] (2,2.2) circle (0.4);
        \draw[dashed,fill=blue!30] (2,0.6) circle (0.4);
         % line in the middle
        \draw[thick,red] (2,0) -- (2,3);
        %Labels
        \node at (0.7,2.5) {$\mathcal{A}$};
        \node at (3.3,0.5) {$\mathcal{B}$};
        \node at (2.4,2.7) {$\mathcal{C}$};
        \node at (2.4,1.1) {$\mathcal{C}$};
    \end{tikzpicture}
        %\caption{}
        \label{fig:arrow}
\end{figure}

    Let us now consider the additional type of embedding between discs that emerges when considering discs embedded in manifolds with defects. Namely, embedding a disk without the defect into a larger disk that contains a defect. Such embedding corresponds to the functor between the category $\mathcal{A}$ and $\mathcal{C}$. 

\begin{figure}[H] 
\centering
    \begin{tikzpicture}
        %rectangle
        \draw[fill=blue!20] (0,0) rectangle (4,3);
        %circles
        \draw[dashed,fill=blue!30] (1.45,1.5) circle (0.9);
        \draw[dashed,fill=blue!40] (1.2,1.5) circle (0.4);
        % line in the middle
        \draw[thick,red] (2,0) -- (2,3);
        %Labels
        \node at (0.7,2.5) {$\mathcal{A}$};
        \node at (3.3,0.5) {$\mathcal{B}$};
        \node at (2.2,2.4) {$\mathcal{C}$};
    \end{tikzpicture}
    %\caption{Your caption here}
    \label{fig:emb}
\end{figure}

    Moreover, this will be a monoidal functor. The data of the monoidal functor comes from the isotopy between the embeddings on the left and the embeddings on the right in the figure below. 

\begin{figure}[H]  
\centering
    \begin{tikzpicture}
        % Left rectangle
        \draw[fill=blue!20] (0,0) rectangle (5,4);
         % Big circle
        \draw[dashed,fill=blue!30] (2,2) circle (1.7);
        % ellipse
        \draw[dashed,fill=blue!40] (1.7,1.5) ellipse (1.1 and 0.4);
        \draw[dashed,fill=blue!40] (1.7,2.5) ellipse (1.1 and 0.4);
        % Small circles 
        \draw[dashed, fill=blue!50] (1.25,2.5) circle (0.25);
        \draw[dashed, fill=blue!50] (1.25,1.5) circle (0.25);
        % Middle line
        \draw[thick,red] (2.5,0) -- (2.5,4);
         % Labels
        \node at (0.5,3.5) {$\mathcal{A}$};
        \node at (4.5,0.5) {$\mathcal{B}$};
        \node at (3.8,2.8) {$\mathcal{C}$};
        
        % Right rectangle
        \draw[fill=blue!20] (6,0) rectangle (11,4);
       % Big circle
        \draw[dashed,fill=blue!30] (8,2) circle (1.7);
        % ellipse
        \draw[dashed,fill=blue!40] (7.25,2) ellipse (0.5 and 1.2);
        % Small circles and ellipse
        \draw[dashed, fill=blue!50] (7.25,2.5) circle (0.25);
        \draw[dashed, fill=blue!50] (7.25,1.5) circle (0.25);
         % Middle line
        \draw[thick,red] (8.5,0) -- (8.5,4);
         % Labels
        \node at (6.5,3.5) {$\mathcal{A}$};
        \node at (10.5,0.5) {$\mathcal{B}$};
        \node at (9.8,2.8) {$\mathcal{C}$};
        
        % Implication symbol
        \node at (5.5,2) {\Large$\Rightarrow$};
    \end{tikzpicture}
    %\caption{Your caption here}
    \label{fig:monoidal}
\end{figure}

    Analogously, we obtain another monoidal functor from the category $B$ to $\mathcal{C}$. Combining these together, a monoidal functor from $A \otimes B \xrightarrow[]{} \mathcal{C}$ is produced. Finally, using these ideas, one can produce a half braiding. We can conclude that the functor $A \otimes B \xrightarrow[]{} \mathcal{C}$ factors through the Drinfeld center of the category $\mathcal{C}$. The data of this functor is called \textbf{central algebra}.

\[\begin{tikzcd} \label{DC}
	A \otimes B && \mathcal{C} \\
	& \mathcal{Z}_{Dr} (\mathcal{C})
	\arrow["",from=1-1, to=2-2]
	\arrow["",from=2-2, to=1-3]
	\arrow["",shift left, from=1-1, to=1-3]
\end{tikzcd}\]
\begin{definition} 
    The 3-category denoted by $\mathbf{Tens}_{\C}$ is the higher Morita category $\mathrm{Alg}_1 (\mathbf{Pr}_{\C})$. Its objects are tensor categories, its 1-morphisms from $\mathcal{A}$ to $\mathcal{B}$ in $\mathbf{Tens}_{\C}$ are $\mathcal{A}$-$\mathcal{B}$-bimodule categories, and its 2-morphisms are bimodule functors between bimodule categories, and the 3-morphisms are natural transformations.
\end{definition}

\begin{remark}
    The category $\mathbf{Tens}_{\C}$ is home to Turaev-Viro theories. This is where symmetries of 2-dimensional QFTs live.
\end{remark}

\begin{definition}\label{def:Brten}
    The 4-category denoted with $\mathbf{BrTens}_{\C}$ is the higher Morita category $\mathrm{Alg}_2 (\mathbf{Pr}_{\mathbb{C}})$. Its objects are braided tensor categories. The 1-morphisms from $\mathcal{A}$ to $\mathcal{B}$ in $\mathbf{BrTens}_{\C}$ are algebra objects in the 2-category of $\mathcal{A} - \mathcal{B}$-bimodules. Brochier, Jordan, and Snyder showed that these are exactly central tensor categories\footnote{Proposition 3.2 in A.~Brochier, D.~Jordan and N.~Snyder, \textit{On dualizability of braided tensor categories}, \href{https://arxiv.org/pdf/1804.07538}{\tt arXiv:1804.07538}}. The 2-morphisms are centered bimodule categories, and the 3-morphisms are bimodule functors between such. Finally, the 4-morphisms are natural transformations.
\end{definition}

\begin{remark}
    The picture one should have in mind when thinking about $\Disk_{\mathrm{fr}}^3$-algebras is shown in the following figure. 

\begin{figure}[H]
    \centering
    \begin{tikzpicture}
        % Spheres
        \shade[ball color=blue!30] (1,0) circle (0.6);
        \shade[ball color=blue!30] (1,4.5) circle (0.6);
        \shade[ball color=blue!30] (5,0) circle (0.6);
        \shade[ball color=blue!30] (5,4.5) circle (0.6);
        % Labels for spheres
        \node at (0.2,4.7) {$\mathcal{A}$};
        \node at (5.8,4.7) {$\mathcal{A}$};
        \node at (0.2,-0.3) {$\mathcal{B}$};
        \node at (5.8,-0.3) {$\mathcal{B}$};
          % Red rectangle (curved)
        \fill[red!40] (-1,1) -- (7,1) -- (6,3) .. controls (4,3) and (2,3) .. (0,3) -- cycle;
        % Dashed circles
        \draw[dashed, fill=red!50] (1.5,2) circle (0.45);
        \draw[dashed, fill=red!50] (4.5,2) circle (0.45);
        % Labels for dashed circles
        \node at (0.8,2) {$\mathcal{C}$};
        \node at (5.2,2) {$\mathcal{C}$};
        % Dot and circle
        \draw[dashed, fill=red!70] (3,2) circle (0.4);
        \fill[color=green] (3,2) circle (2pt);
        % Green line
        \draw[thick,green!50] (3,3) -- (3,1);
        % Labels for dot and circle
        %\node at (1,1.5) {$\mathcal{I}$};
        %\node at (1,0.5) {$\mathcal{J}$};
    \end{tikzpicture}
    %\caption{}
    \label{fig:3alg}
\end{figure}
\end{remark}

\subsection{Dualizability in Higher Morita Categories}
    We are interested in discussing dualizability in the higher Morita categories mentioned above. Let us collect some of the most important results. Starting with the $\mathbf{Tens}_{\C}$ category, it is well known that every object in $\mathbf{Tens}_{\C}$ is 1-dualizable. 

\begin{theorem}
    Every tensor category is 1-dualizable in $\mathbf{Tens}_{\C}$.
\end{theorem}
\begin{definition}
A monoidal category $\mathcal{A}$ is \textbf{good} if the tensor product functor preserves small sifted colimits, and $\mathcal{A}$ admits small sifted colimits.
\end{definition}

\begin{remark}
     A more general result was proven by Lurie,\footnote{Claim 4.1.14 in J.~Lurie, \emph{On the classification of
topological field theories},
  Current developments in mathematics, 2008, Int. Press, Somerville, MA, 2009,
  pp.~129--280. \href{http://arxiv.org/abs/arXiv:0905.0465}{{\tt
  arXiv:0905.0465}}} showing that for any good category $\mathbf{S}$, every $\Disk_{\mathrm{fr}}^{1}$-algebra in $\Alg_1(\mathbf{S})$ is 1-dualizable. Moreover, Gwilliam and Scheimbauer\footnote{O.~Gwilliam and C.~Scheimbauer, \textit{Duals and adjoints in higher Morita categories}, \href{https://arxiv.org/pdf/1804.10924}{\tt arXiv:1804.10924}} proved that every $\Disk_{\mathrm{fr}}^{n}$-algebra in $\Alg_n(\mathbf{S})$ is $n$-dualizable. The underlying idea is that a certain amount of dualizability comes from topological manipulations.
\end{remark}

\begin{definition}
    A presentable category $\mathcal{A}$ is \textbf{nice} if for every object $x$ in $\mathcal{A}$, $x$ can be written as the following colimit over all the compact projective objects $P$:
    $$x = \underset{P}{\mathrm{colim }} \mathrm{Hom}(P,X) \otimes P$$
\end{definition}

\begin{definition}
    If $\mathcal{A}$ is a nice tensor category, then $\mathcal{A}$ is \textbf{cp-rigid} if all compact projectives have duals.  
\end{definition}

\begin{theorem}\label{thm:cp-rigidis2-dual} 
    Every cp-rigid tensor category is 2-dualizable.\footnote{Theorem 1.8 in A.~Brochier, D.~Jordan, P.~Safronov and N.~Snyder, \textit{Invertible braided tensor categories}, \href{https://msp.org/agt/2021/21-4/p16.xhtml}{Algebraic and Geometric Topology 21.4} (2021): 2107-2140. \href{https://arxiv.org/abs/2003.13812}{\tt arXiv:2003.13812}}
\end{theorem}

\begin{theorem}\label{thm:fusioncategories} 
    Every fusion category over a field of characteristic zero is 3-dualizable. \footnote{Theorem 3.4.3 in C.~Douglas, C.~Schommer-Pries, and N.~Snyder, \textit{Dualizable tensor categories}, \href{https://books.google.de/books?id=gCo0EAAAQBAJ}{American Mathematical Society}}
\end{theorem}
It is widely expected that equipping the framed fully extended 3D TFT implied by Theorem \ref{thm:fusioncategories} with an $\mathrm{SO}(3)$-fixed point structure corresponds to a spherical structure on the fusion category, and that the resulting oriented $3$D TFT coincides with the Turaev-Viro theory.

%%%%%%%%%%%%%%%%%%%%%%%%%%%%%%%%%%%%%%%%%%%%%%%%%%%%%%%%%%%%%%%%%%%%%%%

%Jonte
%\newpage
\section{Lecture 4: TFTs with Anomaly}

We now leave the realm of 3-dimensional TFTs and enter into the study of TFTs in dimension $4$. The target for this class of TFTs is the symmetric monoidal 4-category of braided tensor categories $\textbf{BrTen}_{\mathbb{C}}$ as introduced in Def~\ref{def:Brten}.  To understand this class of TFTs, we need to understand, by the Cobordism Hypothesis, dualizability in $\textbf{BrTen}_{\mathbb{C}}$. Therefore, let us collect some results in this direction.

\begin{theorem}
    Every braided tensor category is $2$-dualizable.\footnote{Theorem 5.1 in O.~Gwilliam and C.~Scheimbauer, \textit{Duals and adjoints in higher morita categories}, \href{https://arxiv.org/pdf/1804.10924}{\tt arXiv:1804.10924} }
\end{theorem}

\begin{theorem}
    Every cp-rigid braided tensor category is 3-dualizable.\footnote{Thm. 5.16 in A.~Brochier, D.~Jordan and N.~Snyder, \emph{On dualizability of braided tensor categories}, \href{https://www.cambridge.org/core/journals/compositio-mathematica/article/on-dualizability-of-braided-tensor-categories/D0401593D715F3416AC263A74BD4ADDD}{Compositio Mathematica 157.3} (2021): 435-483, \href{https://arxiv.org/pdf/1804.07538}{\tt arXiv:1804.07538}}
\end{theorem}

 As in the case of Theorem~\ref{thm:cp-rigidis2-dual} the proofs of this result use the technique of monadic reconstruction. 

\begin{remark}
    The class of TFTs induced by cp-rigid braided tensor categories, which are known in physics as Kapustin-Witten theories, mathematically describe skein modules and character stacks.
\end{remark}

 The analog of Theorem~\ref{thm:fusioncategories} for braided tensor categories is:

\begin{theorem}\label{thm:braidedfusion}
    Every braided fusion category is $4$-dualizable.\footnote{Thm. 5.21. in A.~Brochier, D.~Jordan and N.~Snyder, \emph{On dualizability of braided tensor categories}, \href{https://www.cambridge.org/core/journals/compositio-mathematica/article/on-dualizability-of-braided-tensor-categories/D0401593D715F3416AC263A74BD4ADDD}{Compositio Mathematica 157.3} (2021): 435-483, \href{https://arxiv.org/pdf/1804.07538}{\tt \tt arXiv:1804.07538}}
\end{theorem}

     Note that neither Theorem~\ref{thm:fusioncategories} nor Theorem~\ref{thm:braidedfusion} claim that all $3$- respectively $4$-dualizable objects arise from (braided) fusion categories. In fact, for the symmetric monoidal $3$-category $\mathbf{Tens}_{\mathbb{C}}$ it is still an open question if there exist other non-semisimple examples. On the other hand, for the symmetric monoidal $4$-category $\textbf{BrTen}_{\mathbb{C}}$ it is known that modular tensor categories, which possibly are non-semisimple, form another interesting class of invertible and hence 4-dualizable objects.\footnote{A.~Brochier, D.~Jordan, P.~Safronov and N.~Snyder, \textit{Invertible braided tensor categories}, \href{https://msp.org/agt/2021/21-4/p16.xhtml}{Algebraic and Geometric Topology 21.4} (2021): 2107-2140.\href{https://arxiv.org/abs/2003.13812}{\tt arXiv:2003.13812}}

     \begin{definition}
    Let $\mathcal{A} \in \textbf{BrTen}_{\mathbb{C}}$ be braided tensor category with braiding $\beta$. An object $A\in \mathcal{A}$ is called \textbf{transparent}, if for all objects $X \in \mathcal{A}$ the double braiding with $A$ is trivial, i.e
    \begin{equation}
        \beta_{A,X} \circ \beta_{X,A} = \mathrm{id}_{X \otimes A},.
    \end{equation}
    The \textbf{Müger center or $\mathbb{E}_{2}$-center} of $\mathcal{A}$ is the full subcategory $\mathcal{Z}_{2}(\mathcal{A}) \subset \mathcal{A}$ generated by transparent objects. Note that the unit object $1_{\mathcal{A}}$ of $\mathcal{A}$ is always transparent. A braided tensor category $\mathcal{A}$ is called \textbf{non-degenerate} if the unique $\mathbb{C}$-linear colimit preserving functor $i:\mathrm{Vect}_{\mathbb{C}} \rightarrow \mathcal{Z}_{2}(\mathcal{A})$ that maps $\mathbb{C} \mapsto 1_{\mathcal{A}}$ is an equivalence. We call a non-degenerate braided tensor category $\mathcal{A}$ a \textbf{modular} tensor category.
\end{definition}

The terminology arises from the following example:

\begin{example}
    Let $\mathcal{A}$ be a braided fusion category.  Then $\mathcal{A}$ is non-degenerate if and only if it is a modular fusion category, i.e. its $S$-matrix is invertible.
\end{example}
 
 There connection to fully extended TFTs arises from their interpretation in terms of so-called \textbf{invertible} objects in $\mathbf{BrTen}_{\mathbb{C}}$. Recall that an object in a symmetric monoidal $n$-category is \textbf{invertible} if and only if it is $n$-dualizable and the evaluation and coevaluation, as well as all units and counits are isomorphisms. Let us look at an example to understand what this means:
 
\begin{example}
    Let us again consider the symmetric monoidal category $\textbf{Vect}^{\otimes}_{\mathbb{C}}$ of $\mathbb{C}$-vector spaces from Example~\ref{example:vect}. Recall that for a dualizable vector space $V$ there exists a dual vector space $V^{\vee}$ and a linear map $\mathrm{ev}_{V}:V^{\vee}\otimes V \rightarrow \mathbb{C}$. If $V$ is invertible, the linear map $\mathrm{ev}_{V}$ is an isomorphism. Comparing dimensions of the source and target one observes that $V \simeq \mathbb{C}$. Note that there is no canonical choice of such an isomorphism. Every such isomorphism corresponds to a choice of a non-zero vector $v\in V$ and there is no distinguished choice for $v$. This phenomenon is the origin of anomalies \footnote{D.~S.~Freed, \textit{Anomalies and invertible field theories}, Proc. Symp. Pure Math. Vol. 88. 2014  \href{https://arxiv.org/pdf/1404.7224}{\tt \tt arXiv:1404.7224}}.
\end{example}

\begin{remark}
    Let $\mathcal{A}\in \mathbf{Tens}_{\mathbb{K}}$ be an invertible fusion category over an arbitrary field $\mathbb{K}$. In case that $\mathbb{K}$ is algebraically closed the only invertible $\mathbb{K}$-linear fusion category is given by $\mathrm{Vect}_{\mathbb{K}}$. Indeed, part of the dualizability data of a fusion category $\mathcal{A}$ is the unit inclusion
    \[
   \iota: \mathrm{Vect}_{\mathbb{K}}\simeq \mathcal{Z}_{DR}(\mathrm{Vect}_{\mathbb{K}}) \rightarrow \mathcal{Z}_{DR}(\mathcal{A})
    \]
    into the Drinfeld center of $\mathcal{A}$. If $\mathcal{A}$ is invertible, then $\iota$ is an equivalence. But by Mügers theorem fusion categories with braided equivalent Drinfeld centers are Morita equivalent.
\end{remark}

\begin{theorem}\label{thm:modular}
    Let $\mathcal{A}$ be a cp-rigid finite braided tensor category. Then $\mathcal{A}$ is invertible in $\textbf{BrTen}_{\mathbb{C}}$ if and only if it is non-degenerate.\footnote{Theorem 3.20. in A.~Brochier, D.~Jordan, P.~Safronov and N.~Snyder, \textit{Invertible braided tensor categories}, \href{https://msp.org/agt/2021/21-4/p16.xhtml}{Algebraic and Geometric Topology 21.4} (2021): 2107-2140. \href{https://arxiv.org/abs/2003.13812}{ \tt arXiv:2003.13812}}
\end{theorem}

 We will later see examples of modular tensor categories that are not semisimple. The corresponding fully extended TFTs are called Crane-Yetter theories. These $4$-dimensional topological field theories were historically first constructed without using the Cobordism Hypothesis. Instead, they were constructed downwards starting by first defining the value on 4-manifolds and afterwards extending to manifolds of lower dimension.\footnote{L.~Crane, L.~H.~Kauffman and D.~N.~Yetter, \textit{State-sum invariants of 4-manifolds}, \href{https://www.worldscientific.com/doi/abs/10.1142/S0218216597000145}{Journal of Knot Theory and Its Ramifications 6.02} (1997): 177-234. \href{https://arxiv.org/abs/hep-th/9409167}{\tt arXiv:hep-th/9409167}}  
 
Note that for an invertible braided tensor category $\mathcal{A}$ the associated fully extended TFT
\[
\mathbf{Z}_{\mathcal{A}}: \mathbf{Cob}_{\mathrm{fr}}^{4} \rightarrow \mathbf{BrTens}_{\mathbb{C}}
\]
maps every $k$-morphism in $\mathbf{Cob}_{\mathrm{fr}}^{4}$ to an invertible $k$-morphism in $\mathbf{BrTens}_{\mathbb{C}}$. Indeed, it follows from the definition of being invertible that for any handle the associated morphism from Example~\ref{example:handles} is invertible. Since every morphism in $\mathbf{Cob}_{\mathrm{fr}}^{4}$ can be constructed from such a handle, it will itself be invertible. In particular, the value of $\mathbf{Z}_{\mathcal{A}}$ on an arbitrary $3$-manifold is a $1$-dimensional vector space and on an arbitrary $4$-manifold a non-zero number. 

This observation imposes strong restriction on the invariants described by invertible TFTs. In fact, Galatius, Madsen, Tillmann and Weiss observed in their study of invertible 4d-TFTs, that the invariant associated to a $4$-manifold $M^{4}$ only depends on the homotopy type of $M^{4}$. The associated invariant can explicitly be described via the signature $\sigma(M^{4})$ and  Euler-characteristic $E(M^{4})$ of $M^{4}$.\footnote{S.~Galatius, I.~Madsen, U.~Tillmann, and M.~Weiss, \textit{The homotopy type of the cobordism category}, \href{https://arxiv.org/abs/math/0605249}{ \tt arXiv:math/0605249} } For a $4$-manifold that admits a framing, these invariants are trivial. However, we know that for every non-degenerate braided tensor category there exists a non-trivial fully extended invertible 4d framed TFT. This observation is called the \textbf{Crane-Yetter paradox}. 

The solution to this problem is, as we will see in the following lectures, that Crane-Yetter shows its full potential when studied relatively, that is, as an anomaly of lower dimensional TFTs.

\subsection{Anomalies as relative TFTs}\label{subsect:Anomalies}

One of the first TFTs mathematically constructed is called Witten-Reshetikhin-Turaev theory (WRT).\footnote{N.~Reshetikhin and V.~G.~Turaev, \textit{Invariants of 3-manifolds via link polynomials and
quantum groups}, \href{https://link.springer.com/article/10.1007/BF01239527}{Inventiones mathematicae 103.1} (1991): 547-597.} This is a $3$d-TFT that uses a modular fusion category $\mathcal{A}$ as an input. In the case where the modular fusion category is the representation category of a quantum group\footnote{Technically, WRT uses a semisimplification of the full representation category.}, this theory is believed to be a mathematical model for Chern-Simons theory. As Crane-Yetter theory, WRT-theory was historically constructed first by defining the value of the TFT on $3$-manifolds using surgery. Let us see what we can learn about WRT-theory using fully extended TFTs.

In contrast to all other TFTs discussed in this lecture, WRT-theory is not always fully extended. From the physics perspective, this is not surprising, as Chern-Simons theory has a so-called anomaly.\footnote{See also Section \ref{sec:3} of Freed's lecture for more on anomalies.} Indeed, its Lagrangian is only determined as a function 

\begin{equation}
    \mathcal{L}:\{Fields\} \rightarrow \mathbb{R}/(2\pi \mathbb{Z})
\end{equation}
and therefore the partition function $\mathbf{Z}_{CS}$ is only determined up to a phase. 

Mathematically, we describe this by saying that the partition function $\mathbf{Z}_{CS}$ of Chern-Simons theory associates to every closed $3$-manifold $M^{3}$ an element $\mathbf{Z}_{CS}(M^{3})$ of an invertible vector space $\mathbf{Z}_{inv}(M^{3})$ instead of a number. This collection of invertible vector spaces $\mathbf{Z}_{inv}(M^{3})$ arises itself from an invertible TFT $\mathbf{Z}_{inv}$ that describes the anomaly.   

At least conjecturally, we have control over the anomaly theory of WRT-theory:

\begin{conjecture}[Walker, Freed, Telemann]\label{Conj:Chern-Simons}
    The 4D Crane-Yetter theory from Theorem~\ref{thm:modular} attached to the modular fusion category $\mathcal{A}$ is the anomaly of the 3d WRT-theory associated to $\mathcal{A}$.
\end{conjecture}

 To understand the consequences of this conjecture, let us delve a bit further into the mathematical description of anomalous TFTs. 

\begin{definition}
    Let $\mathbf{C}$ be a symmetric monoidal $(\infty,n+1)$-category and $\alpha:\mathbf{Cob}^{n+1}_{\mathrm{fr}} \rightarrow \mathbf{C}$ an invertible $(n+1)$-dimensional TFT. An $n$-dimensional \textbf{fully extended TFT with anomaly} $\alpha$ is a symmetric monoidal oplax natural transformation 
    \[
    \mathbf{Z}:\mathbf{1}_{n} \rightarrow \tau_{\leq n}\alpha
    \]
    from the trivial theory $n$-dimensional theory to the restriction of $\alpha$ to $\mathbf{Cob}_{\mathrm{fr}}^{n}$.
\end{definition}

\begin{example}
Using the above definition of an anomaly, Conjecture~\ref{Conj:Chern-Simons} states that for every modular fusion category $\mathcal{A}$, the associated WRT-theory defines an oplax natural transformation 
\[
   \mathbf{1}_{3} \xrightarrow{\mathbf{Z}_{WRT}} \tau_{\leq 3} \mathbf{Z}_{CY} 
\]
from the trivial $3$-dimensional theory to Crane-Yetter theory. In particular, the partition function of a framed $3$-manifold $M^{3}$ would be given by a linear map
\[
 \mathbf{Z}_{WRT}(M^{3}):\mathbb{C} \rightarrow \mathbf{Z}_{CY}(M^{3})
\]
of vector spaces, i.e by an element $\mathbf{Z}_{WRT}(M^{3}) $ in the invertible vector space $\mathbf{Z}_{CY}(M^{3})$.
\end{example}
   The by-hand construction of an oplax transformation between two fully extended TFTs is in general at least as hard as the by-hand construction of a fully extended TFT itself. Remarkably, we can also reduce the question about the existence of an oplax transformation to the Cobordism Hypothesis by introducing a new target\footnote{Definition 5.14 in T.~Johnson-Freyd and C.~Scheimbauer, \textit{(Op) lax natural transformations, twisted quantum field theories, and “even higher” Morita categories}. To every symmetric monoidal $(\infty,n)$-category $\mathcal{C}$ we can associate a new symmetric monoidal $(\infty,n)$-category $\mathcal{C}^{\rightarrow}$, \href{https://www.researchgate.net/publication/310787437_Oplax_natural_transformations_twisted_quantum_field_theories_and_even_higher_Morita_categories}{Advances in Mathematics} 307(4):147-223}  with
\begin{itemize}
    \item objects given by morphisms $F:C_{0} \rightarrow C_{1}$ in $\mathcal{C}$
    \item 1-morphisms $\Theta:= (\Theta_{0},\Theta_{1},\Theta_{2}):F \rightarrow G$ given by oplax commutative squares
    \[
    \begin{tikzcd}
        C_{0} \arrow[r, " \Theta_{0}"] \arrow[d, "F"] & D_{0} \arrow[d, "G"] \\
        C_{1} \arrow[r, "\Theta_{1}"] \arrow[ur, Rightarrow] & D_{1} 
    \end{tikzcd}
    \]
    where $\Theta_{2}:\Theta_{1}\circ F \Rightarrow G\circ \Theta_{0}$ is a $2$-morphism in $\mathcal{C}$.
    \item higher morphisms by higher oplax commutative squares
\end{itemize}
The symmetric monoidal structure is given on objects by mapping two arrows $F:C_{0} \rightarrow C_{1}$ and $G:D_{0} \rightarrow D_{1}$ to the monoidal product of the arrows $F\otimes G: C_{0} \otimes D_{0} \rightarrow C_{1} \otimes D_{1}$ in $\mathcal{C}$. The symmetric monoidal $(\infty,n)$-category $\mathcal{C}^{\rightarrow}$ is called the $(\infty,n)$-category of \textbf{oplax arrows}.  

Using this modified target category, we can now understand oplax transformation by applying the Cobordism Hypothesis. For the examples of WRT-theory, we consider $\textbf{BrTen}_{\mathbb{C}}^{\rightarrow}$ the symmetric monoidal $3$-category of oplax arrows in the $4$-category $\textbf{BrTen}_{\mathbb{C}}$. For every modular tensor category $\mathcal{A}$ the braided monoidal functor $\eta: \textbf{Vect}_{\mathbb{C}} \rightarrow \mathcal{A}$ given by the unit inclusion equips $\mathcal{A}$ with the structure of a $\textbf{Vect}_{\mathbb{C}}-\mathcal{A}$ central algebra denoted $\mathcal{A}_{\eta}$. This defines an object in $\textbf{BrTen}_{\mathcal{C}}^{\rightarrow}$ that satisfies the following properties:

\begin{theorem}\footnote{Cor. 1.2 in B.~Haïoun, \textit{Unit inclusion in a (non-semisimple) braided tensor category and (non-compact)
relative TQFTs}, \href{https://arxiv.org/abs/2304.12167}{ 	\tt arXiv:2304.12167}}
\label{thm:unit}
    Let $\textbf{BrTen}_{\mathbb{C}}^{\rightarrow}$ be the $3$-category of oplax arrows in the $4$-category $\textbf{BrTen}_{\mathbb{C}}$ and let $\mathcal{A}$ be a modular tensor category. For the central algebra $\mathcal{A}_{\eta}$ the following holds: 
    \begin{itemize}
        \item[(1)] If $\mathcal{A}$ is semisimple, then $\mathcal{A}_{\eta}$ is $3$-dualizable and $\mathcal{A}_{\eta}$ induces a fully extended $3$d-TFT with target $\textbf{BrTen}_{\mathbb{C}}^{\rightarrow}$
        
        \item[(2)] If $\mathcal{A}$ is not semisimple, then $\mathcal{A}_{\eta}$ is not 3-dualizable, but it is non-compact 3-dualizable and induces a non-compact $3$-dimensional TFT with target $\textbf{BrTen}_{\mathbb{C}}$.
    
    \end{itemize}
 \end{theorem}
\begin{remark}
A non-compact $n$-dimensional fully extended TFT is a fully extended TFT that can only be defined on top dimensional manifolds with boundary.\footnote{Definition 4.2.10 in J.~Lurie, \textit{On the classification of topological field theories}, Current developments in mathematics 2008.1 (2008): 129-280 \href{https://arxiv.org/abs/0905.0465}{ \tt arXiv:0905.0465}} In general non-compact theories occur when certain state spaces are not finite-dimensional. As a consequence, it is impossible to define the TFT on $n$-handles.
\end{remark}
 The theorem implies that every modular tensor category $\mathcal{A}$ induces an oplax transformation of non-compact $3$-dimensional TFTs
\[
\mathbf{Z}_{WRT}:\mathbf{1}_{3} \rightarrow \tau_{\leq 3}\mathbf{Z}_{CY} 
\]
that extends to a transformation of fully extended $3d$-TFTs if $\mathcal{A}$ is semisimple. As a generalization of Conjecture~\ref{Conj:Chern-Simons}, we call the non-compact $3$-dimensional TFT with anomaly $\tau_{\leq 3}\mathbf{Z}_{CY}$ the \textbf{framed WRT-theory} associated to the modular tensor category $\mathcal{A}$.
\begin{remark}
    Reutter and Walker announced in University Quantum Symmetries Lectures (UQSL) that the reason for the non-existence of $3$-handles for a non-semisimple modular category $\mathcal{A}$ is that the datum associated by the TFT to the $4$-ball, viewed as a cobordism $B^{4}:S^{3} \rightarrow \varnothing$, has to be the zero map (degeneracy of traces). A partial solution to this problem is expected to be given by the theory of modified traces, which give rise to non-semisimple WRT invariants.\footnote{C.~Blanchet, F.~Costantino, N.~Geer and B.~Patureau-Mirand, \textit{Non-semisimple tqfts, Reidemeister torsion and Kashaev’s invariants}, \href{https://www.sciencedirect.com/science/article/pii/S0001870816307174}{Advances in Mathematics 301} (2016): 1-78. \href{https://arxiv.org/abs/1404.7289}{\tt arXiv:1404.7289}}
\end{remark}

\section{Lecture 5: A new perspective on WRT-theory}

In the last lecture, given any modular tensor category $\mathcal{A}$, using the Cobordism Hypothesis we constructed a framed version of WRT-theory. On the other hand, classically WRT-theory is known to be an oriented TFT. To close this gap, we have to study, as we have learned from Lemma~\ref{lemma:orientedcob}, homotopy fixed points in the space of fully dualizable objects of the symmetric monoidal 4-category $\mathbf{BrTens}_{\mathbb{C}}$. As a warm-up, let us first consider homotopy fixed points on the symmetric monoidal $3$-category $\mathbf{Tens}_{\mathbb{C}}$

\begin{example}
    Consider the $3$-category $\mathbf{Tens}_{\mathbb{C}}$. We denote by $\mathbf{Fus}_{\mathbb{C}}$ the full subspace of the space of fully dualizable objects $\mathbf{Tens}_{\mathbb{C}}^{fd}$ generated by fusion categories. 
     This space admits an $\mathrm{SO}(2)$-action analogous to the one in Example~\ref{example:so2}. The Serre automorphism of a fusion category $\mathcal{A}$ is given by the bimodule $_{\mathcal{A}^{\vee\vee}}\mathcal{A}_{\mathcal{A}}$, where the left $\mathcal{A}$-action is twisted by the monoidal double dual functor. More precisely the action of $a \in \mathcal{A}^{\vee\vee}$ on $a'\in \mathcal{A}$ is given by $a\triangleright a':= a^{\vee}\otimes a$.
     
    As we have learned in Example~\ref{example:so2} an $\mathrm{SO}(2)$-homotopy fixed point is given by a trivialization of the Serre automorphism. Unraveling definitions, such a trivialization is equivalent to the datum of an invertible object $E \in \mathcal{A}$ together with a monoidal natural isomorphism $\eta:id_{\mathcal{C}}\simeq E^{-1}\otimes (-)^{\vee\vee} \otimes E$. Such a datum is called a \textit{weak pivotal structure}. 
\end{example}

\begin{example}\label{example:fusionso3}
    Since all $\mathbb{C}$-linear fusion categories are $3$-dualizable, the space $\mathbf{Fus}_{\mathbb{C}}$ also admits a $\mathrm{SO}(3)$-action.  Similarly to Example~\ref{example:so2}, we can describe this action by considering an explicit CW-decomposition of the classifying space $B\mathrm{SO}(3)$. This decomposition admits a unique $2$-cell  $e_{2}^{\mathrm{SO}(3)}$, that maps under the continuous map $\iota: B\mathrm{SO}(2) \rightarrow B\mathrm{SO}(3)$ to unique 2-cell $e_{2}^{\mathrm{SO}(2)}$ of $\mathrm{SO}(2)$. Further, the decomposition has a unique 3-cell $e_{3}^{\mathrm{SO}(3)}$, which attaches to $e_{2}^{\mathrm{SO}(3)} \cup e_{0}^{\mathrm{SO}(3)}$ in such a way that $\pi_{2}(B\mathrm{SO}(3)) \simeq \mathbb{Z}/2\mathbb{Z}$ and the induced map
    \[
        \pi_{2}(\iota):\pi_{2}(B\mathrm{SO}(2))= \mathbb{Z} \rightarrow \mathbb{Z}/2\mathbb{Z} = \pi_{2}(B\mathrm{SO}(3))
   \]
    becomes the canonical projection for the choice of generators $e_{2}^{\mathrm{SO}(2)}$ and $e_{2}^{\mathrm{SO}(3)}$. As a consequence, it follows that for every fusion category $\mathcal{A}$ the square of the Serre automorphism $_{\mathcal{A}^{\vee\vee\vee\vee}}\mathcal{A}_{\mathcal{A}}$ admits a canonical trivialization. This observation is known as \textbf{Radford's theorem}:
    
    \begin{theorem}\footnote{Proposition 4.5 in P.~Etingof, D.~Nikshych and V.~Ostrik, \textit{An analogue of Radford’s $\mathrm{S}^4$-formula for finite tensor
categories}, \href{https://academic.oup.com/imrn/article/2004/54/2915/719357}{International Mathematics Research Notices 2004.54} (2004): 2915-2933. \href{https://arxiv.org/abs/math/0404504}{ \tt arXiv:math/0404504}}
        In any fusion category $\mathcal{C}$ there exists a canonical monoidal natural isomorphism $(-)^{\vee\vee\vee\vee}\simeq (-)$ called the Radford isomorphism. In particular, $E^{2}\simeq 1_{\mathcal{C}}$.
    \end{theorem}
 Experts widely believe that even the following stronger statement is true:
    \begin{conjecture}
        Every fusion category admits a monoidal natural isomorphism $(-)^{\vee\vee}\simeq (-)$, called a \textbf{pivotal structure}.
    \end{conjecture}
 To describe a full CW-decomposition of $B\mathrm{SO}(3)$ and consequently a full $\mathrm{SO}(3)$-homotopy fixed point, we also need to include cells of dimension higher than $3$. It is expected, that for a fusion category $\mathcal{A}$ a homotopy fixed point for the full $\mathrm{SO}(3)$-action corresponds to a so-called \textbf{spherical structure}.\footnote{Definition 3.5.2 in C.~Douglas, C.~Schommer-Pries and N.~Snyder, \emph{Dualizable tensor categories}, Vol. 268. No. 1308. \href{https://www.ams.org/books/memo/1308/}{American Mathematical Society}, 2020, \href{https://arxiv.org/abs/1312.7188}{ \tt arXiv:1312.7188}}.
\end{example}

 For the symmetric monoidal $4$-category the homotopy fixed point data has not been worked out so far, but let us nevertheless collect some expectations.
\begin{expectation}
    Let $\mathbf{BrTens}_{\mathbb{C}}^{cpr}$ denote the full subspace of the space of $3$-dualizable objects $\mathbf{BrTens}_{\mathbb{C}}^{3-dual}$ generated by cp-rigid braided tensor categories. This subspace admits an $\mathrm{SO}(3)$-action induced from $\mathbf{BrTens}_{\mathbb{C}}^{3-dual}$. We can embed $\mathrm{SO}(2)$ into $\mathrm{SO}(3)$ by either associating to a matrix $M\in \mathrm{SO}(2)$ the matrix in $\mathrm{SO}(3)$ with $M$ embedded in the upper left or bottom right corner. This induces two different $\mathrm{SO}(2)$-actions. If we consider the first action, we are in the situation of Example~\ref{example:fusionso3}. Consequently, a trivialization of this action induces a weak pivotal structure and a Radford isomorphism on the cp-rigid braided tensor category $\mathcal{A}$ as before. Note that in this case, the Radford isomorphism exists for any braided tensor category, even if it is not fusion.  
    A trivialization of the $\mathrm{SO}(2)$-action induced by the second inclusion can be understood in terms of oriented disk algebras.
    The topological category $\mathrm{Disk}_{\mathrm{fr}}^{2}$ of framed disks has an analog for oriented disks called the \textbf{category of oriented disks} $\mathrm{Disk}_{\mathrm{or}}^{2}$. Its set of objects consists of finite disjoint unions of oriented discs and its space of morphisms is the space of \textit{orientation preserving} embeddings. We depict the unique orientation on a disc as:
    \[
    \begin{tikzpicture}[auto]
         \draw[fill=blue!20] (0,0) circle [radius=1cm];
         \node[rectangle, fill, scale=0.75] at (0,1) {} ;
         \draw[thick,->] (0,0) to (0,0.5);
         \node[scale=0.6] at (-0.25,0.4) {$2$};
         \draw[thick,->] (0,0) to (0.5,0);
         \node[scale=0.6] at (0.4,-0.25) {$1$};
    \end{tikzpicture} 
    \]
    where the black rectangle marks the direction of the second basis vector. The first basis vector is always assumed to be rotated by $90^{\circ}$ in the clockwise direction. 
    
    This convention is specifically useful for drawing oriented embeddings. A general oriented embedding consists of a framed embedding and a rotation. We can depict the rotation performed on an embedded disc by looking at the relative angle between the markings of the ambient and embedded disc. For example, the following depicts an oriented embedding  of $3$ discs:

\begin{figure}[H]  
\centering
    \begin{tikzpicture}
        % Left rectangle
        \draw[fill=blue!20] (0,0) circle [radius=1.75cm];
        \node[rectangle, fill] at (0,1.75) {} ;
        \draw[dashed,fill=blue!30] (-1,0) circle [radius=0.5cm];
        \draw[dashed, fill=blue!30] (0.5,0.5) circle [radius=0.75cm];
        \draw[dashed, fill=blue!30] (0.25,-1) circle [radius=0.6cm];
        \node[rectangle, fill,scale=0.5] at (-1,0.5) {} ;
        \node[rectangle, fill, scale=0.5] at (1.25,0.5) {} ;
        \node[rectangle, fill, scale=0.5] at (0.25,-1.6) {} ;
    \end{tikzpicture}
    %\caption{Your caption here}
    \label{fig:monoidal}
\end{figure}

 Let us now try to understand the data encoded in a $\mathrm{Disk^{2}_{\mathrm{or}}}$-algebra $\mathcal{A}:\mathrm{Disk^{2}_{\mathrm{or}}}\rightarrow \mathbf{Pr}$. First, $\mathcal{A}$ has the structure of a $\mathrm{Disk^{2}_{\mathrm{fr}}}$-algebra by only considering those embeddings where the marking of the embedded discs align with the marking of the ambient disc. But it also has new operations and relations induced by rotations. For example, $360^{\circ}$-degree rotation of the ambient disc induces an orientation preserving isotopy:
\[
\begin{tikzpicture}
    \draw[fill=blue!20] (0,0) circle [radius=1.5cm];
        \node[rectangle, fill] at (0,1.5) {} ;
        \draw[dashed,fill=blue!30] (0,0) circle [radius=1cm];
        \node[rectangle, fill,scale=0.5] at (0,1) {} ;
         \draw[fill=blue!20] (7,0) circle [radius=1.5cm];
        \node[rectangle, fill] at (7,1.5) {} ;
        \draw[dashed,fill=blue!30] (7,0) circle [radius=1cm];
        \node[rectangle, fill,scale=0.5] at (7,1) {} ;
        \node at (3.5,0) {\Large $\Rightarrow$};
        \node at (3.5,0.5) {$360^{\circ}$-rotation};
        \draw[fill=blue!20] (3.5,-1.25) circle [radius=0.75cm];
        \node[rectangle, fill,scale=0.5] at (2.75,-1.25) {} ;
        \draw[dashed,fill=blue!30] (3.5,-1.25) circle [radius=0.4cm];
        \node[rectangle, fill,scale=0.3] at (3.5,-0.85) {} ;
        \draw[->] (3.5,-0.3) to [bend right=45] (2.55,-1.25);
        
\end{tikzpicture}
\]
whose image under the functor $\mathcal{A}$ induces an invertible natural transformation
\[
\begin{tikzcd}
    \mathcal{A}(\mathbb{R}^{2}) \arrow[r, swap, bend right=50, "id_{\mathcal{A}(\mathbb{R}^{2})}", ""{name=D, above}] \arrow[r, bend left= 50, "id_{\mathcal{A}(\mathbb{R}^{2})}", ""{name=U,below}] & \mathcal{A}(\mathbb{R}^{2}) \arrow[Rightarrow, from=U, to=D, "\Theta"]
\end{tikzcd}
\]
that we denote by $\Theta$. This automorphism satisfies a non-trivial relation that can be encoded in terms of oriented embeddings of two discs. On the one hand, rotating the ambient disc by $360^{\circ}$ induces an orientation preserving isotopy:
\[
\begin{tikzpicture}
    \draw[fill=blue!20] (0,0) circle [radius=1.5cm];
        \node[rectangle, fill] at (0,1.5) {} ;
        \draw[dashed,fill=blue!30] (-0.75,0) circle [radius=0.5cm];
        \node[rectangle, fill,scale=0.5] at (-0.75,0.5) {} ;
         \draw[dashed,fill=blue!30] (0.75,0) circle [radius=0.5cm];
        \node[rectangle, fill,scale=0.5] at (0.75,0.5) {} ;
        
           \draw[fill=blue!20] (7,0) circle [radius=1.5cm];
           \node[rectangle, fill] at (7,1.5) {} ;
        \draw[dashed,fill=blue!30] (6.25,0) circle [radius=0.5cm];
        \node[rectangle, fill,scale=0.5] at (6.25,0.5) {} ;
         \draw[dashed,fill=blue!30] (7.75,0) circle [radius=0.5cm];
        \node[rectangle, fill,scale=0.5] at (7.75,0.5) {} ;
        \node at (3.5,0) {\Large $\Rightarrow$};
        \node at (3.5,0.5) {$360^{\circ}$-rotation};
        \draw[fill=blue!20] (3.5,-1.25) circle [radius=0.75cm];
        \node[rectangle, fill,scale=0.5] at (2.75,-1.25) {} ;
        \draw[dashed,fill=blue!30] (3.2,-1.25) circle [radius=0.2cm];
        \node[rectangle, fill,scale=0.3] at (3.2,-1.05) {} ;
        \draw[dashed,fill=blue!30] (3.8,-1.25) circle [radius=0.2cm];
        \node[rectangle, fill,scale=0.3] at (3.8,-1.05) {} ;
        \draw[->] (3.5,-0.3) to [bend right=45] (2.55,-1.25);
\end{tikzpicture}
\]
whose image under $\mathcal{A}$ is given by the horizontal composite natural transformation:
\[
\begin{tikzcd}
    \mathcal{A}(\mathbb{R}^{2})\otimes \mathcal{A}(\mathbb{R}^{2}) \arrow[r, "\otimes"] & \mathcal{A}(\mathbb{R}^{2}) \arrow[r, swap, bend right=50, "id_{\mathcal{A}(\mathbb{R}^{2})}", ""{name=D, above}] \arrow[r, bend left= 50, "id_{\mathcal{A}(\mathbb{R}^{2})}", ""{name=U,below}] & \mathcal{A}(\mathbb{R}^{2}) \arrow[Rightarrow, from=U, to=D, "\Theta"]
\end{tikzcd}
\]
On the other hand, up to an orientation preserving isotopy of these isotopies, we can decompose the full rotation of the ambient disc into a composition of braidings and rotations of the embedded discs:
\[
\begin{tikzpicture}
    \node at (0,0) { \begin{tikzpicture}
        \draw[fill=blue!20] (0,0) circle [radius=1cm];
        \node[rectangle, fill, scale=0.6] at (0,1) {} ;
        \draw[dashed,fill=blue!30] (-0.5,0) circle [radius=0.4cm];
        \node[rectangle, fill,scale=0.4] at (-0.5,0.4) {} ;
         \draw[dashed,fill=blue!30] (0.5,0) circle [radius=0.4cm];
        \node[rectangle, fill,scale=0.4] at (0.5,0.4) {} ;
        \end{tikzpicture}};

        \node at (2.5,-0.6) {
        \begin{tikzpicture}
        \node at (0,0) {\Large $\Rightarrow$};
        \node at (0,0.5) {$2\times$-swap};
        \draw[fill=blue!20] (0,-1.25) circle [radius=0.75cm];
        \node[rectangle, fill,scale=0.5] at (0,-0.5) {} ;
        \draw[dashed,fill=blue!30] (-0.3,-1.5) circle [radius=0.2cm];
        \node[rectangle, fill,scale=0.3] at (-0.3,-1.3) {} ;
        \draw[dashed,fill=blue!30] (0.3,-1) circle [radius=0.2cm];
        \node[rectangle, fill,scale=0.3] at (0.3,-0.8) {} ;
        \draw[<->, bend left=50] (-0.4,-1.2) to (0.1,-0.8);
        \draw[<->, bend right=50] (0,-1.7) to (0.4,-1.3);
        
          \end{tikzpicture}
        };

        \node at (5,0) {
        \begin{tikzpicture}
            \draw[fill=blue!20] (0,0) circle [radius=1cm];
        \node[rectangle, fill, scale=0.6] at (0,1) {} ;
        \draw[dashed,fill=blue!30] (-0.5,0) circle [radius=0.4cm];
        \node[rectangle, fill,scale=0.4] at (-0.5,0.4) {} ;
         \draw[dashed,fill=blue!30] (0.5,0) circle [radius=0.4cm];
        \node[rectangle, fill,scale=0.4] at (0.5,0.4) {} ;
        \end{tikzpicture}
        };

     \node at (7.5,-0.6) {
        \begin{tikzpicture}
        \node at (0,0) {\Large $\Rightarrow$};
        \node at (0,0.5) {$360^{\circ}$-rotation};
        \draw[fill=blue!20] (0,-1.25) circle [radius=0.75cm];
        \node[rectangle, fill,scale=0.5] at (0,-0.5) {} ;
        \draw[dashed,fill=blue!30] (-0.35,-1.25) circle [radius=0.2cm];
        \node[rectangle, fill,scale=0.3] at (-0.55,-1.25) {} ;
        \draw[dashed,fill=blue!30] (0.35,-1.25) circle [radius=0.2cm];
        \node[rectangle, fill,scale=0.3] at (0.15,-1.25) {} ;
        \draw[->] (-0.35,-0.95) to [bend right=45] (-0.65,-1.25);
        \draw[->] (0.35,-0.95) to [bend right=45] (0.05,-1.25);
          \end{tikzpicture}
        };

         \node at (10,0) {
        \begin{tikzpicture}
            \draw[fill=blue!20] (0,0) circle [radius=1cm];
        \node[rectangle, fill, scale=0.6] at (0,1) {} ;
        \draw[dashed,fill=blue!30] (-0.5,0) circle [radius=0.4cm];
        \node[rectangle, fill,scale=0.4] at (-0.5,0.4) {} ;
         \draw[dashed,fill=blue!30] (0.5,0) circle [radius=0.4cm];
        \node[rectangle, fill,scale=0.4] at (0.5,0.4) {} ;
        \end{tikzpicture}
        };

\end{tikzpicture}
\]
Its image under $\mathcal{A}$ is given by the composite natural transformation:
\[
\begin{tikzcd}
    \mathcal{A}(\mathbb{R}^{2}) \otimes\mathcal{A}(\mathbb{R}^{2}) \arrow[r, swap, bend right=70, "\otimes", ""{name=C, above}] \arrow[r, bend left= 70, "\otimes", ""{name=A, below}] \arrow[r, "\otimes^{op}"{name=B1, above}, ""{name=B2,below}, ] &  \mathcal{A}(\mathbb{R}^{2}) \arrow[Rightarrow,from=A, to=B1, "\sigma"] \arrow[Rightarrow, "\sigma", from=B2, to=C] \arrow[r, swap, bend right=50, "id_{\mathcal{A}(\mathbb{R}^{2})}", ""{name=D, above}] \arrow[r, bend left= 50, "id_{\mathcal{A}(\mathbb{R}^{2})}", ""{name=U,below}] & \mathcal{A}(\mathbb{R}^{2}) \arrow[Rightarrow, from=U, to=D, "\Theta"]
\end{tikzcd}
\]
     Since the above isotopies are itself related by an isotopy of isotopies, the natural transformations associated to these isotopies by $\mathcal{A}$ are equivalent. Evaluated on objects $V,W\in \mathcal{A}$ the relation reads as follows:
    \begin{equation}\label{eq:balancing}
    \Theta_{V\otimes W}= \sigma_{W,V} \circ \sigma_{V,W} \circ (\Theta_{V}\otimes\Theta_{W}) 
     \end{equation}
    A natural automorphism $\Theta$ of the identity functor $id_{\mathcal{A}}$ that satisfies Equation~\ref{eq:balancing} is called a \textbf{balancing}. One can check that lifting a $\mathrm{Disk}_{\mathrm{fr}}^{2}$-algebra $\mathcal{A}$ to a $\mathrm{Disk}_{\mathrm{or}}^{2}$-algebra in $\mathbf{Pr}$ is equivalent to choosing a balancing for $\mathcal{A}$. Therefore, the expectation is that a homotopy fixed point structure with respect to the second $\mathrm{SO}(2)$-action on a cp-rigid braided tensor category $\mathcal{A}$ is equivalent to a choice of balancing. 
    
    Let us finally also comment on $\mathrm{SO}(3)$-homotopy fixed points. A balancing on a cp-rigid braided tensor category $(\mathcal{A},\Theta)$ is called a \textbf{ribbon structure} if it is compatible with the duality, i.e for all compact-projective objects $V\in \mathcal{A}$
    \[
    \Theta_{V^{\vee}} = \Theta_{V}^{\vee}
    \]
    where $\Theta_{V}^{\vee}$ denotes the dual morphism. It is expected that the structure full $\mathrm{SO}(3)$ homotopy fixed point corresponds to a ribbon structure.
\end{expectation}

 After this short interlude on homotopy fixed points, let us now get back to framed WRT-theories. We leave it to the reader to apply the above discussion on homotopy fixed points to WRT-theories and instead focus on clarifying a common misconception.

\subsection{WRT Theory as a sheaf on the character stack}

As we have remarked in the beginning of Lecture~\ref{subsect:Anomalies}, the WRT-theory associated to the representation category of a quantum group is widely believed to be a mathematical model for physical Chern-Simons theory. \textbf{But this is a common misconception!} In fact, it is more accurate to say that the corresponding WRT-theory only is a mathematical model of Chern-Simons theory in a \textit{formal neighborhood} of the trivial local system. In other words, we have described Chern-Simons theory \textit{perturbatively} around a specific critical point of the Chern-Simons functional. To describe non-perturbative Chern-Simons theory, we need to consider other critical points as well, or in other words we need to vary WRT-theory over the critical locus of the Chern-Simons functional. 

The critical locus of the Chern-Simons functional, with structure group $G$, on a $3$-manifold $M$ is given by the \textbf{character stack} $\mathrm{Ch}_{G}(M)$ of $M$. This can be defined mathematically as the mapping stack $\mathrm{Map}(M,BG)$ from the manifold $M$ into the classifying stack $BG$ of the group $G$. Alternatively it can be described as the quotient stack associated to the action of $G$ via conjugation on the representation variety $\mathrm{Hom}_\mathrm{Grp}(\pi_{1}(M),G)$. 

Before we can start varying the WRT-theory associated to a quantum group over the character stack, we need to recall some facts about quantum groups.\footnote{G.~Lusztig, \textit{Introduction to quantum groups}, Springer Science and Business Media, 2010, \href{https://link.springer.com/book/10.1007/978-0-8176-4717-9}{https://doi.org/10.1007/978-0-8176-4717-9}} The following discussion works analogously for any semisimple complex algebraic group $G$. For simplicity we only discuss the case $G= SL_{2}(\mathbb{C})$.

\textbf{Lusztig's divided power quantum group} for $\mathfrak{sl}_{2}$ denoted $\mathcal{U}^{L}_{q}(\mathfrak{sl}_{2})$ is the $\mathbb{Z}[q,q^{-1}]$-algebra generated by symbols $K^{(r)}$ with $r \in \mathbb{Z}$ and $E^{(r)}$, $F^{(r)}$ with $r \in \mathbb{Z}_{\geq 0}$. These generators are subject to relations of the form 
\[
    E^{(m)} E^{(n)} = \begin{bmatrix}
m+n \\
m,n 
\end{bmatrix}_{q} E^{(m+n)}
\]
 where 
\[
    \begin{bmatrix}
m+n \\
m,n 
\end{bmatrix}_{q} = \frac{[m+n]_{q}}{[m]_{q}[n]_{q}}
\]
is the q-trinomial coefficient. We can recover the usual quantum group of $\mathfrak{sl}_{2}$ at generic $q$ denoted $\mathcal{U}_{q}(\mathfrak{sl}_{2})$ from $\mathcal{U}^{L}_{q}(\mathfrak{sl}_{2})$. More precisely there exists an isomorphism of $\mathbb{Q}(q)$-Hopf algebras.
\begin{equation}\label{eq:integralform}
   \mathcal{U}^{L}_{q}(\mathfrak{sl}_{2}) \otimes_{\mathbb{Z}[q,q^{-1}]} \mathbb{Q}(q) \simeq \mathcal{U}_{q}(\mathfrak{sl}_{2})
\end{equation}
A $\mathbb{Z}[q,q^{-1}]$-Hopf algebra that satisfies the above property is called an integral form of $\mathcal{U}_{q}(\mathfrak{sl}_{2})$ . In the following, we denote the generators of $\mathcal{U}^{L}_{q}(\mathfrak{sl}_{2})$ by $E,F,K,K^{-1}$. Under the equivalence of Equation~\ref{eq:integralform} the generators $E^{(r)}$ of Lusztig's quantum group $\mathcal{U}^{L}_{q}(\mathfrak{sl}_{2})$ are identified with
\[
    E^{(r)}= \frac{E^{r}}{[r]_{q}}
\]
in the usual quantum group $\mathcal{U}^{L}_{q}(\mathfrak{sl}_{2})$. Moreover, we can also construct quantum groups at roots of unity from Lusztig's integral form. Therefore, fix some $l\in \mathbb{N}_{>1}$ and denote by $\mathbb{Q}(\epsilon)$ the algebra of rational functions in $\epsilon$, subject to the relation $\epsilon^{l} = 1$. The \textbf{quantum group at an $l$-th root of unity} is the $\mathbb{Q}(\epsilon)$-algebra

\begin{equation}
\mathcal{U}^{L}_{q^{l}=1}(\mathfrak{sl}_{2}) :=\mathcal{U}^{L}_{q}(\mathfrak{sl}_{2}) \otimes_{\mathbb{Z}[q,q^{-1}]} \mathbb{Q}(\epsilon)
\end{equation}
This construction imposes the relation that $q$ is a $l$-th-root of unity $q^{l}=1$ and induces, since $[l]_{q}$ vanishes, the following relation
\begin{equation}\label{eq:vanish}
    E^{l} = [l]_{q} E^{(l)} = 0
\end{equation}
on the generators. 

Quantum groups at roots of unity admit an interesting Hopf subalgebra called the small quantum group $u_{q}(\mathfrak{sl}_{2})$. This is defined as the Hopf subalgebra of $\mathcal{U}^{L}_{q^{l}=1}(\mathfrak{sl}_{2}) $ generated by  $E,F,K^{\pm}$. 
If we denote by $\eta$ the counit of $u_{q}(\mathfrak{sl}_{2})$, then the following holds:
\begin{theorem}
    There exists a Hopf $\mathbb{C}$-algebra morphism $\pi:\mathcal{U}_{q^{l}=1}^{L}(\mathfrak{sl}_{2}) \rightarrow U(\mathfrak{sl}_{2})$ that induce an isomorphism of Hopf $\mathbb{C}$-algebras 
    \begin{equation}
    \mathcal{U}_{q}^{L}(\mathfrak{sl}_{2})/\mathrm{ker}(\eta) \simeq U(\mathfrak{sl}_{2})    
    \end{equation}
    where $U(\mathfrak{sl}_{2})$ denotes the usual universal enveloping algebra of $\mathfrak{sl}_{2}$
\end{theorem}
\begin{remark}
    It is possible to do a similar construction for a general semisimple Lie algebra $\mathfrak{g}$. But in this case, the quotient $ \mathcal{U}_{q}^{L}(\mathfrak{g})/\mathrm{ker}(\eta_{\mathfrak{g}})$ is only a finite extension of the universal enveloping algebra $U(\mathfrak{g})$. Regardlessly, an analogue of the Hopf algebra morphism $\pi$ exists for a general class of Lie algebras and roots of unities.\footnote{C.Negron, \emph{Quantum Frobenius and modularity for quantum groups at arbitrary roots of 1},
   \href{https://arxiv.org/abs/2311.13797}{{\tt
   	arXiv:2311.13797}}}  
\end{remark}
 For our application to Chern-Simons theory, we are interested in the braided tensor categories that arise as the representation categories of these quantum groups. Therefore, we denote by $\mathrm{Rep}_{q}(SL_{2}(\mathbb{C}))$ the \textbf{representation category of} $\mathcal{U}^{L}_{q^{l}=1}(\mathfrak{sl}_{2})$. Since $\mathcal{U}^{L}_{q^{l}=1}(\mathfrak{sl}_{2})$ is a Hopf algebra, it follows that this category is monoidal and cp-rigid. 
 
Further, the Hopf algebra $\mathcal{U}^{L}_{q^{l}=1}(\mathfrak{sl}_{2})$ can be equipped with an $R$-matrix given by
\[
    R= q^{\frac{H\otimes H}{2}}\sum_{k=0}^{\infty} \frac{(E\otimes F)^{k}}{[k]_{q}} = q^{\frac{H\otimes H}{2}}\sum_{k=0}^{\infty} [k]_{q} E^{(k)}\otimes F^{(k)} \in \mathcal{U}^{L}_{q^{l}=1}(\mathfrak{sl}_{2})\otimes \mathcal{U}^{L}_{q^{l}=1}(\mathfrak{sl}_{2})
\]
Note that it follows from Equation~\ref{eq:vanish} that only finitely many summands are non-zero. This $R$-matrix equips the monoidal category $\mathrm{Rep}_{q}(SL_{2}(\mathbb{C}))$ with a braiding. 

We further denote by $\mathrm{Rep}(SL_{2}(\mathbb{C}))$ the representation category of $U(\mathfrak{sl}_{2})$. The cocommutative Hopf algebra structure on $U(\mathfrak{sl}_{2})$ equips this category with the structure of a symmetric monoidal cp-rigid category. 
 This monoidal category is related to $\mathrm{Rep}_{q}(SL_{2}(\mathbb{C}))$ by a monoidal functor
 \[
 \pi^{\ast}: \mathrm{Rep}(SL_{2}(\mathbb{C}))\rightarrow \mathrm{Rep}_{q}(SL_{2}(\mathbb{C}))
 \]
 induced by the Hopf algebra morphism $\pi$. This functor pulls back $U(\mathfrak{sl}_{2})$-representation along $\pi$. Furthermore, it follows from the explicit form of the $R$-matrix that all objects $\pi^{\ast}(V)$ in the essential image of $\pi^{\ast}$ are transparent and hence the functor $\pi^{\ast}:\mathrm{Rep}(SL_{2}(\mathbb{C}))\rightarrow \mathcal{Z}_{2}(\mathrm{Rep}_{q}(SL_{2}(\mathbb{C})))$ factors through the Müger center. These structures fit together and induce the following equivalence of braided tensor categories
\begin{equation}\label{eq:quantum}
    \mathrm{Rep}_{q}(G)\boxtimes_{\mathrm{Rep}(G)} \mathrm{Vect}_{\mathbb{C}} \simeq \mathrm{Rep}(u_{q}(\mathfrak{g}))
\end{equation}
 
 We now apply these results on quantum groups to describe how the anomaly of Chern-Simons theory varies over the character stack $\mathrm{Ch}_G(M)$ of a closed manifold $M$. In a formal neighborhood of a critical point of the Chern-Simons functional the anomaly is described by the CY-theory associated to the modular tensor category $\mathrm{Rep}(u_{q}(\mathfrak{g}))$. Similarly in a disk-like neighborhood $U\subset M$, the anomaly should be described by a sheaf of braided tensor categories on $\mathrm{Ch}_{G}(U)\simeq BG$, whose stalk at the unique point is given by $\mathrm{Rep}(u_{q}(\mathfrak{g}))$. Interpreting Equation~\ref{eq:quantum} in terms of algebraic geometry means that the braided monoidal category $\mathrm{Rep}_{q}(G)$ together with its $\mathrm{Rep}(G)$-action precisely defines such a sheaf. The relation to anomalies comes from the following:

\begin{conjecture}[Ben-Zvi, Jordan, Safronov] 
The sheaf of $\mathbb{E}_{2}$-algebras $\mathrm{Rep}_{q}(G)$ is an \textbf{invertible} sheaf and hence induces an \textbf{invertible} TFT.
\end{conjecture}

 We spend the rest of this lecture unraveling this idea in more detail and explaining its consequences. Analogously to the symmetric monoidal $4$-category $\mathbf{BrTens}_{\mathbb{C}}$, there exists a symmetric monoidal $5$-category $\mathbf{SymTens}_{\mathbb{{C}}} = \Alg_3(\Pr)$, whose objects are symmetric monoidal categories and whose morphisms are bimodules of braided tensor categories. In particular, since $\mathrm{Rep}_{q}(G)$ defines a sheaf of $\mathbb{E}_{2}$-algebras on $BG$ it induces an endomorphism of $\mathrm{Rep}(G)$ in $\mathbf{SymTens}_{\mathbb{C}}$. The mathematical incarnation of the above idea is then given by the following theorem:

\begin{theorem}\label{thm:relCY}
    Regard $\mathrm{Rep}_{q}(G)\in \mathrm{End}_{\mathbf{SymTens}_{\mathbb{C}}}(\mathrm{Rep}(G))$ as an endomorphism in the 5-category $\mathbf{SymTens}_{\mathbb{C}}$. Then $\mathrm{Rep}_{q}(G)$ is an automorphism.\footnote{Theorem 1.1 in P.~Kinnear, \emph{Non-semisimple Crane-Yetter theory varying over the character stack}, \href{https://arxiv.org/abs/2404.19667}{\tt arXiv:2404.19667}}
\end{theorem}
 This theorem forms an upgrade of the fact that $\mathrm{Rep}(u_{q})$ is invertible in $\mathbf{BrTen}_{\mathbb{{C}}}$ to the fact that $\mathrm{Rep}_{q}(G)$ is relative invertible in $\mathbf{SymTen}_{\mathbb{C}}$. 
Its consequence for fully extended TFTs is that there exists a relative invertible 4d-Crane-Yetter theory 
\begin{equation}
    \mathbf{Z}_{RCY}:\mathbf{Cob}^{n}_{\mathrm{fr}} \rightarrow \mathbf{SymTen}_{\mathbb{C}}^{\rightarrow}
\end{equation}
that associates to the point the braided tensor category $\mathrm{Rep}_{q}(G)$ viewed as a $\mathrm{Rep}(G)$-bimodule. This TFT is expected to describe how the anomaly varies over the character stack. The difference between this TFT and classical CY-theory is, that this TFT associates to manifolds sheaves of vector spaces, categories etc.\ on the character stack $\mathrm{Ch}_{G}$ instead of individual vector spaces, categories, and so on. We can then compute classical CY-theory from relative CY-theory by taking global sections of this sheaves. Let us look at some consequences of this:
\begin{example}
Relative $CY$-theory associates to every $3$-manifold $M^{3}$ a vector bundle $Z_{RCY}(M^{3}) \in \mathbf{Vec}(\mathrm{Ch}_{G}(M^{3}))$ on the character stack. It follows from Theorem~\ref{thm:relCY} that this is in fact a line bundle. So it is $1$-dimensional, in particular finite dimensional, locally in a neighborhood of every point. In contrast, the vector space of global sections of this line bundle, i.e. the value of classical $CY$-theory, can in fact be $\infty$-dimensional.
\end{example}
\begin{example}
    Relative CY-theory associates to every $2$-manifold $\Sigma^{2}$ a sheaf of categories $Z_{RCY}(\Sigma^{2})$ on the character stack $\mathrm{Ch}_{G}(\Sigma)$. This sheaf of categories is expected to be associated to the unicity conjecture of Bonahon and Wong \footnote{F.~Bonahon and H.~Wong, \textit{Representations of the Kauffman bracket skein algebra III: closed
surfaces and naturality}, \href{https://ems.press/journals/qt/articles/16200}{Quantum Topology 10.2} (2019): 325-398. \href{https://arxiv.org/abs/1505.01522}{ 	\tt arXiv:1505.01522}}:
    \begin{conjecture}
        There is a 1-to-1 correspondence between irreducible representations of the Skein algebra at $q^{l}=1$ and points of $\mathrm{Ch}_{G}(\Sigma)$
    \end{conjecture}
This conjecture was solved step by step by different authors. First the generic case was  solved.\footnote{C.~Frohman, J.~Kania-Bartoszynska and T.~Lê, \textit{Unicity for representations of the Kauffman bracket skein algebra}, \href{https://link.springer.com/article/10.1007/s00222-018-0833-x}{Inventiones mathematicae 215} (2019): 609-650. \href{https://arxiv.org/abs/1707.09234}{\tt arXiv:1707.09234}} Afterwards their result was extended to the smooth locus\footnote{I.~Ganev, D.~Jordan and P.~Safronov, \textit{The quantum frobenius for character varieties and
multiplicative quiver varieties}, \href{https://ems.press/journals/jems/articles/14263112}{Journal of the European Mathematical Society} (2024). \href{https://arxiv.org/abs/1901.11450}{\tt arXiv:1707.09234}} and finally proven in complete generality \footnote{H.~Karuo and J.~Korinman, \textit{Azumaya loci of skein algebras}, \href{https://arxiv.org/abs/2211.13700}{\tt arXiv:2211.13700}}.
\end{example}
Let us finish with a general expectation on how this extends to Chern-Simons theory:
\begin{expectation}
    The expectation is that there exists a fully extended Chern-Simons theory with values in $(\mathbf{SymTen}_{\mathbb{C}}^{\rightarrow})^{\rightarrow}$. Thinking of objects in $\mathrm{End}_{\mathbf{SymTen}_{\mathbb{C}}}(\mathrm{Rep}(G))$
    as sheaves of TFTs on the character stack $\mathrm{Ch}_{G}$, this fully extended Chern-Simons theory should be a morphism $\mathcal{O}_{\mathrm{Ch}_{G}}\rightarrow \mathbf{Z}_{RCY}$, from the "structure sheaf",which is associated to the trivial TFT, to the relative CY-theory. As a consequence, the value of this theory on a $3$-manifold $M^{3}$ would be a function $f_{CS}^{M^{3}}$ on the character variety instead of a number. The hope is that the Chern-Simons invariant $\mathbf{Z}_{CS}(M^{3})$ of $M^{3}$ can finally be extracted from $f_{CS}^{M^{3}}$ as the mathematically not well defined integral over the character variety
    \begin{equation}
        \mathbf{Z}_{CS}(M^{3}) = \int_{\mathrm{Ch}_{G}(M^{3})} f_{CS}^{M^{3}}
    \end{equation}

\end{expectation}

}

\newpage

{\clearpage\newcourse{Finite symmetry in QFT}{Daniel S.~Freed}

\renewcommand{\:}{\colon}
% %%%%%%%%%%%%%%%%%%%%%%%%%%%%%%%%%%%%%
% \renewcommand{\AA}{{\mathbb A}}
\newcommand{\Ahat}{{\hat A}}
\newcommand{\CC}{{\mathbb C}}
\newcommand{\CP}{{\mathbb C\mathbb P}}
\newcommand{\EE}{\mathbb E}
\newcommand{\FF}{\mathbb F}
\newcommand{\HH}{{\mathbb H}}
\newcommand{\NN}{{\mathbb N}}
\newcommand{\PP}{{\mathbb P}}
\newcommand{\QQ}{{\mathbb Q}}
\newcommand{\RP}{{\mathbb R\mathbb P}}
\newcommand{\RR}{{\mathbb R}}
\newcommand{\TT}{\mathbb T}
\newcommand{\ZZ}{{\mathbb Z}}
\renewcommand{\chiup}{\raise.5ex\hbox{$\chi$}}
\renewcommand{\cir}{S^1}
\newcommand{\dbar}{{\overline\partial}}
\newcommand{\inv}{^{-1}}
\newcommand{\mlstrut}{_{\vphantom{1*\prime y}}}
\newcommand{\res}[1]{\negmedspace\bigm|\mstrut_{#1}}
\renewcommand{\Res}[1]{\negmedspace\biggm|\mstrut_{#1}}
\newcommand{\temsquare}{\raise3.5pt\hbox{\boxed{ }}}
\newcommand{\zmod}[1]{\ZZ/#1\ZZ}
\newcommand{\zn}{\zmod{n}}
\newcommand{\zt}{\zmod2}
% %\let\germ\frak

\newcommand{\theprotag}[2]{#2~#1}
\newcommand{\longhookrightarrow}{\lhook\joinrel\longrightarrow}
\newcommand{\hneg}{\mkern-.5\thinmuskip}
 
% \DeclareFontFamily{U}{mathx}{}
% \DeclareFontShape{U}{mathx}{m}{n}{<-> mathx10}{}
% \DeclareSymbolFont{mathx}{U}{mathx}{m}{n}
% \DeclareMathAccent{\widehat}{0}{mathx}{"70}
% \DeclareMathAccent{\widecheck}{0}{mathx}{"71}
 
% \DeclareMathSymbol{\bigtimes}{1}{mathx}{"91}

\newcommand{\upplus}{^{>0}}
\newcommand{\upnn}{^{\ge0}} 
\newcommand{\Zp}{\ZZ\upplus}
\newcommand{\Znn}{\ZZ\upnn}
\newcommand{\Rp}{\RR\upplus}
\newcommand{\Rnn}{\RR\upnn}

\newcommand{\Ar}{A\mstrut _{\! A}}
\newcommand{\BNG}{B^{}_{\nabla }G}
\newcommand{\BnA}[1]{B^{#1}\hneg A}
\newcommand{\BnF}{\Bord_n(\sF)}
\newcommand{\BtA}{B^2\!A}
\newcommand{\Cx}{\CC^{\times}}
\newcommand{\FIR}{F_{\textnormal{IR}}}
\newcommand{\FUV}{F_{\textnormal{UV}}}
\newcommand{\GA}{\CC[G]}
\newcommand{\LlX}{\sX^{S^{\ell -1}}}
\newcommand{\MTSO}{MT\!\SO}
\newcommand{\OC}{\Omega\sC}
\newcommand{\OX}{\Omega \sX}
\newcommand{\OlX}{\Omega ^\ell \sX}
\renewcommand{\PV}{\PP V}
\newcommand{\RAq}{R_{A',q}}
\newcommand{\RepcA}{\Rep_c(A)}
\newcommand{\STd}{\Sigma ^n\sTd}
\newcommand{\VV}{\mathbb{V}}
\newcommand{\XTd}{\sX\mstrut _{\Sigma ^n\sTd}}
\newcommand{\XT}{\sX\mstrut _{\sT}}
\newcommand{\Xln}[1]{(\sX_{#1},\lambda _{#1})}
\newcommand{\Xl}{(\sX,\lambda )}
\newcommand{\Xm}[1]{\sX^{#1}}
\newcommand{\bG}{\overline{G}}
\newcommand{\bH}{\overline{H}}
\newcommand{\bNl}{\overline{N}^\ell }
\newcommand{\bT}{\overline{T}}
\newcommand{\ba}{\bar{\alpha }}
\newcommand{\bmul}{\bmu{\!\ell}}
\newcommand{\bn}{\bar{\nu }}
\newcommand{\bone}{\mathbbm{1}}
\newcommand{\bsF}{\overline{\sF}}
\newcommand{\bs}{\!\!\!\bigm/\!\!\sigma }
\newcommand{\dual}{^\vee}
\newcommand{\eA}{\epsilon \mstrut _{\hneg A}}
\newcommand{\eP}{\epsilon \mstrut _{\Phi }}
\newcommand{\ed}{\epsilon \dual}
\newcommand{\hF}{\widehat{F}}
\newcommand{\hH}{\widehat{H}}
\newcommand{\hdD}{\hat{\delta }_D}
\newcommand{\hth}{\widehat{\theta }}
\renewcommand{\op}{^{\textnormal{op}}}
\newcommand{\rL}{\textnormal{L}}
\newcommand{\rM}{\textnormal{M}}
\newcommand{\rT}{\textnormal{T}}
\newcommand{\rd}{\rho ^{\vee}}
\newcommand{\sA}{\mathscr{A}}
\newcommand{\sB}{\mathscr{B}}
\newcommand{\sC}{\mathscr{C}}
\newcommand{\sF}{\mathcal{F}}
\newcommand{\sG}{\mathscr{G}}
\newcommand{\sHl}{{}\mstrut _{\GA}\sH}
\newcommand{\sH}{\mathscr{H}}
\newcommand{\sK}{\mathcal{K}}
\newcommand{\sL}{\mathcal{L}}
\newcommand{\sM}{\mathscr{M}}
\newcommand{\sP}{\sigma\mstrut _{\!\Phi }}
\newcommand{\sTd}{\sT\dual}
\newcommand{\sT}{\boldsymbol{T}}
\newcommand{\sXd}[2]{\sigma _{#2}^{(#1)}}
\newcommand{\sXn}{\sigma _{\sX}^{(n+1)}}
\newcommand{\sX}{\mathscr{X}}
\newcommand{\sY}{\mathscr{Y}}
\newcommand{\sZ}{\mathscr{Z}}
\newcommand{\scP}{\mathcal{P}}
\newcommand{\scaP}{\mathscr{P}}
\newcommand{\scrT}{\mathscr{T}}
\newcommand{\sdrd}{(\sd ,\rd )}
\newcommand{\sd}{\sigma ^{\vee}}
\newcommand{\srtF}{(\sigma ,\rho ,\tF)}
\newcommand{\sr}{(\sigma ,\rho )}
\newcommand{\tFt}{(\tF,\theta )}
\newcommand{\tF}{\widetilde{F}}
\newcommand{\tZ}{\widetilde{Z}}
\newcommand{\tll}{\tau ^{\ell -1}\lambda }
\newcommand{\tl}{\tilde{\lambda }}
\newcommand{\tsF}{\widetilde{\sF}}
\newcommand{\xreg}{x_{\text{reg}}}
\newcommand{\zoh}{[0,1)}
\newcommand{\zo}{[0,1]}

\newcommand{\emod}{\hspace{-.25em}\raisebox{-0.88em}{\scriptsize$\epsilon $}\!\!\bigm/\hneg\!}

\newsavebox{\bmuubox}
\savebox{\bmuubox}{$\raisebox{-.09em}{\scalebox{.7}{\rotatebox{7.9}{\bf/}}}\hspace{-0.25em}\mu$}
\newcommand{\bmuu}{\scalerel*{\usebox{\bmuubox}}{\mu}}
\newcommand{\bmu}[1]{\bmuu_{#1}}
\newcommand{\bmut}{\bmu{2}}

\newcommand{\lact}{\;\rotatebox[origin=c]{-90}{$\circlearrowleft$}\;} 
\newcommand{\ract}{\;\rotatebox[origin=c]{+90}{$\circlearrowleft$}\;} 
\newcommand{\squig}{{\scriptstyle\sim\mkern-3.9mu}}
\newcommand{\lsquigend}{{\scriptstyle\lhd\mkern-3mu}}
\newcommand{\rsquigend}{{\scriptstyle\rule{.1ex}{0ex}\rhd}}
\newcounter{sqindex}
\newcommand\squigs[1]{%
  \setcounter{sqindex}{0}%
  \whiledo {\value{sqindex}< #1}{\addtocounter{sqindex}{1}\squig}%
}
\newcommand\rsquigarrow[2]{%
  \mathbin{\stackon[2pt]{\squigs{#2}\rsquigend}{\scriptscriptstyle\text{#1\,}}}%
}
\newcommand\lsquigarrow[2]{%
  \mathbin{\stackon[2pt]{\lsquigend\squigs{#2}}{\scriptscriptstyle\text{\,#1}}}%
}
\makeatother

\setbox0\hbox{(k)\ } \newdimen\secwidth \secwidth=\wd0
\newcount\probno \probno=0 \newcount\probsecno
\def\probsec{\futurelet\next\probsecc}
\def\probsecc{\bigbreak \global\advance\probno by 1
  \noindent \begingroup  \parskip2pt plus1pt minus1pt \parindent0pt
  \probsecno=0 
  \ifx\next\sec\firstsec\else\firsttext\fi}
\long\def\firsttext#1\sec{#1\sec}
\def\firstsec\else\firsttext\fi\sec#1\sec{\fi\advance\probsecno by 1 
\leftskip\secwidth\leavevmode\llap{\number\probno.\enspace 
  (\secno\unskip)\ }#1\sec}
\long\def\sec{\par\leftskip\secwidth\smallbreak\advance\probsecno by 1 
  \leavevmode\llap{\hbox to \secwidth{(\secno\unskip)\hfil}}}
\def\secno{\ifcase\probsecno\or a \or b \or c \or d \or e \or f \or g \or h
  \or i \or j \or k \or l \or m \or n \or o \or p \or q \or r \or s \or t \or
    u \or v \or w \or x \or y \or z\else\fi}
\def\endprobsec{\par\endgroup\bigbreak}

% TODO: Include
%  \title[Topological symmetry in QFT]{Four Lectures on Topological symmetry in QFT} %% first is short title
%  \author[D. S. Freed]{Daniel S.~Freed}
%  \address{Harvard University \\ Department of Mathematics \\ Science Center
% Room 325 \\ 1 Oxford Street \\ Cambridge, MA 02138}
%  \email{dafr@math.harvard.edu}
%  \thanks{This material is based upon work supported by the National Science
% Foundation under Grant Number DMS-2005286 and by the Simons Foundation Award
% 888988 as part of the Simons Collaboration on Global Categorical Symmetries.
% Research at Perimeter Institute is supported in part by the
% Government of Canada through the Department of Innovation, Science and
% Economic Development Canada and by the Province of Ontario through the
% Ministry of Colleges and Universities.}
%  \date{June 1, 2024}
% \maketitle

These notes are based on lectures given at the \emph{Global Categorical
Symmetries Summer School} at the Perimeter Institute, June 13--17, 2022.  The
ideas discussed here about symmetry in field theory represent joint work with
Constantin Teleman and Greg Moore.  Our joint
paper\footnote{\label{FMT}Daniel~S. Freed, Gregory~W. Moore, and Constantin
Teleman, \emph{Topological
  symmetry in quantum field theory},
  \href{http://arxiv.org/abs/arXiv:2209.07471}{{\tt arXiv:2209.07471}}}
appeared shortly after these lectures were given.  That paper includes many
references; these notes only give minimal direction for further exploration,
and we encourage the reader to seek out the references in~\footref{FMT}.  The
expository paper~\footnote{\label{F1}Daniel~S. Freed, \emph{Introduction to
topological symmetry in QFT}, \href{http://arxiv.org/abs/arXiv:2212.00195}{{\tt
arXiv:2212.00195}}} is another introduction to this material.  The current text
has much overlap with these references, including long verbatim passages, but
it also contains additional material and motivation, and, as befits summer
school notes, there are also problems for the reader. 
 
Motivated by concepts for algebras~\ref{subsec:1.6}, in~\ref{subsec:c3.1} we
introduce an abstract notion of symmetry in a field theory, which we call a
\emph{quiche}.  This separation of abstract symmetry from concrete
realizations, which we discuss in~\ref{subsec:c3.2}, is one of our main themes.
Another is the calculus of defects, which can be used to make universal
computations.  Illustrations may be found in \autoref{thm:c14},
\ref{subsec:3.9}, \autoref{thm:c49}, and~\ref{subsec:4.17}, among others.
Quotients and quotient defects are natural in our approach.  The discussion
that begins in~\ref{subsec:3.15} is another demonstration of computations with
abstract symmetry applied to the realization in quantum field theories, as is
the discussion of electromagnetic duality in~\autoref{subsec:4.10}
and~\ref{subsec:4.11}.
 
I thank my collaborators Constantin and Greg, and we all thank many people for
comments on our paper, including Ibou Bah, David Ben-Zvi, Mike Freedman, Dan
Friedan, Mike Hopkins, Theo Johnson-Freyd, Alexei Kitaev, Justin Kulp, Kiran
Luecke, Ingo Runkel, Will Stewart, and Jingxiang Wu.

This material is based upon work supported by the National Science
Foundation under Grant Number DMS-2005286 and by the Simons Foundation Award
888988 as part of the Simons Collaboration on Global Categorical Symmetries.
Research at Perimeter Institute is supported in part by the
Government of Canada through the Department of Innovation, Science and
Economic Development Canada and by the Province of Ontario through the
Ministry of Colleges and Universities.

\newpage
   \section{Lecture 1: Variations on the Theme of Symmetry}\label{sec:1}
% lastsubsec@ 16

  \subsection*{Introduction}\label{subsec:1.1}

We give a conceptual framework for some of the developments of the past few
years around ``global categorical symmetries''.  A few immediate comments to
give some perspective.

  \begin{remark}[]\label{thm:1}
 \ 
 \begin{enumerate}[label=\textnormal{(\arabic*)}]

 \item The word `global' can be dropped: we are discussing symmetries of a
field theory that are analogous to symmetries of any other mathematical
structure.  The word `global' is often used in contradistinction to `gauge'
symmetries, but gauge symmetries are not symmetries of theories: they are
encoded in the (higher) groupoid structure of fields.

 \item The word `categorical' can also be dropped.  Mathematics is
traditionally expressed in the language of sets and functions, and when
mathematical objects have internal symmetries they organize into (higher)
categories rather than sets.  The symmetries we discuss often have this
higher structure, so naturally involve categories.  

 \item There is a large amount of work over the past few years on the topic
of symmetry in quantum field theory, and in particular the application of
``global categorical symmetries'' to dynamics and other questions.  We give
some excerpts from this literature to illustrate how the framework we develop
here applies.  However, we emphasize that the topic of symmetry in quantum
field theory is a large one with many facets, and the framework here does not
apply to all of it. 
 
 \item We restrict to the analog of \emph{finite} group symmetry, including
homotopical versions;\footnote{which include ``higher form symmetries''
(though we do not use this term: there are no differential forms in the
finite case) and ``2-group symmetries'' } it will be interesting to
generalize to the analog of Lie group symmetry.  It is more natural in
quantum theory to have \emph{algebras} of symmetries, rather than
\emph{groups} of symmetries, and so in particular we encounter non-invertible
symmetries.

 \end{enumerate}
  \end{remark}

  \subsection{Main idea}\label{subsec:1.2}

The motivating thought is simple:  
  \begin{equation}\label{eq:41}
  \begin{gathered}
       \textnormal{Separate out the abstract structure of symmetry}\\[-.5em]
       \textnormal{from its 
     concrete manifestations as actions or representations} 
  \end{gathered}
  \end{equation}
Historically, the concept of an abstract group was introduced to synthesize and
further develop diverse instances of group symmetry in geometry (Klein), in
algebra (Galois), in number theory (Gauss), etc.  Perhaps it is Arthur Cayley
in 1854 who first articulated the definition of an abstract group---I'm no
historian---and now every student of mathematics learns this concept early on.
The structure of groups is then used to study representations---linear and
nonlinear.  Similar comments apply to algebras.  The elements of an algebra act
as linear operators on any module.  In the context of field theory, the analog
of an algebra of symmetries is, as we will argue, a \emph{topological} field
theory together with a boundary theory.  The analog of elements of an algebra
are \emph{defects} in the topological field theory, which act on any quantum
field theory.  In this sense, the quantum field theory is a ``module'' over the
topological field theory.
 
In this lecture we begin with a brief discussion of groups and algebras
before turning to symmetry in field theory.  We return at the end to further
aspects of groups and algebras whose analogs in field theory play an
important role in later lectures.

  \subsection*{Groups}

  \subsection{Taxonomy}\label{subsec:1.3}

 The simplest dichotomy is between (1)~discrete groups, and (2)~groups which
have a nontrivial topology.  For the former we distinguish according to the
trichotomy of cardinalities: finite, countable, uncountable.  The nature of
finite groups is quite different from that of infinite discrete groups; the
phrase `discrete group' evokes very different images from the phrase `finite
group', even though finite groups form a subset of discrete groups.  For
example, linear representations of a finite group are rigid---they do not
deform---whereas, for instance, the infinite cyclic group~$\ZZ$ has a
continuous family of 1-dimensional complex unitary representations $n\mapsto
e^{ixn}$ parametrized by~$x\in \RR$.  Among topological groups, the nicest are
Lie groups.  Here too there is an important dichotomy: compact vs.\ noncompact.
Compact Lie groups, which include finite groups, have a well-established
structure theory and representation theory, especially for connected compact
groups; again, representations are rigid.  Noncompact Lie groups, which include
countable discrete groups, also enjoy a robust structure and representation
theory, but of a very different nature.  Moving on, there are infinite
dimensional Lie groups as well as topological groups that do not admit a
manifold structure.  In a different direction, there are homotopical
groups---\emph{grouplike $A_{\infty }$-spaces}, that are higher generalizations
of more classical \emph{$H$-groups}---which may be (homotopy) finite or
infinite.  To wit, if $\sX$~is any pointed topological space, then the
space~$\OX$ of loops at the basepoint has the composition law of concatenation
of loops, and this makes $\OX$~ a group up to homotopy.
 
The field theory symmetry structure we study is analogous to that of a
\emph{finite} group, or, more generally, to a homotopical group that is
\emph{$\pi $-finite} in the sense that it has only finitely many nonzero
homotopy groups, each of which is finite.

  \subsection{Fibering over~$BG$}\label{subsec:1.4}

Let $G$~be a finite group.  A \emph{classifying space}~$BG$ is derived from a
contractible topological space~$EG$ equipped with a free $G$-action by taking
the quotient; the homotopy type of~$BG$ is independent of choices.  If $X$~is
a topological space equipped with a $G$-action, then the \emph{Borel
construction} is the total space of a fiber bundle
  \begin{equation}\label{eq:c1}
     \begin{gathered} \xymatrix@C-28pt{X_G\ar[d]^{\pi }&=EG\times \mstrut _GX\\BG}
     \end{gathered} 
  \end{equation}
with fiber~$X$.  If $*\in BG$ is a chosen point, and we choose a basepoint in
the $G$-orbit in~$EG$ labeled by~$*$, then the fiber~$\pi \inv (*)$ is
canonically identified with~$X$.  We say the \emph{abstract symmetry data}
(in the sense of groups) is the pair~$(BG,*)$, and a \emph{realization} of
the symmetry~$(BG,*)$ on~$X$ is a pair consisting of a fiber
bundle~\eqref{eq:c1} over~$BG$ together with an identification of the fiber
over~$*\in BG$ with~$X$.

  \begin{remark}[]\label{thm:2}
 We are already moving to homotopy theory, and it is more natural to take the
\emph{homotopy fiber}, which is a special case of a \emph{homotopy fiber
product}.  For continuous maps~$f,g$ we can realize the homotopy fiber
product as the space~$F$ and dotted maps indicated in the diagram
  \begin{equation}\label{eq:6}
     \begin{gathered} \xymatrix{&F\ar@{-->}[dr]\ar@{-->}[dl]\\
     X\ar[dr]_{f}&&Z\ar[dl]^{g}\\ &Y} \end{gathered} 
  \end{equation}
A point of~$F$ is a triple~$(x,z,\gamma )$ in which $x\in X$, $z\in z$, and
$\gamma $~is a path in~$Y$ from $f(x)$ to~$g(z)$.  The homotopy fiber~$Z$ over
the basepoint in~\eqref{eq:c1} is the homotopy fiber product
  \begin{equation}\label{eq:1}
     \begin{gathered} \xymatrix{&Z\ar@{-->}[dr]\ar@{-->}[dl]\\
     \ast\ar[dr]&&X_G\ar[dl]\\ &BG} \end{gathered} 
  \end{equation}
  \end{remark} 

  \begin{exercise}[]\label{thm:12}
 Construct a homotopy equivalence $X\xrightarrow{\;\simeq \;} Z$.  (You may
want to know the \emph{homotopy lifting property} for the fiber bundle
$X_G\to BG$.) 
  \end{exercise}

  \subsection[Homotopical groups]{Homotopical groups\footnote{By `homotopical group' we mean a a
\emph{grouplike $A_{\infty }$-space}, which is the higher generalization of
the more classical \emph{$H$-group}; see Edwin~H. Spanier, \emph{Algebraic
topology}, Springer-Verlag, New York, 1981 for the classical notion; the
higher coherent generalization was introduced by James Stasheff in his PhD
thesis.  This nomenclature pertains if $\sX$~is path connected; see
Remark~\ref{thm:c55} below.  But we allow disconnected spaces
too.}}\label{subsec:1.5}

  A pair~$(\sX,*)$ consisting of a $\pi $-finite topological space~$\sX$ and
a basepoint~$*\in \sX$ is a generalization of~$(BG,*)$.  Here is the formal
definition of the finiteness we assume.
 
  \begin{definition}[]\label{thm:c53}
 \ 
 \begin{enumerate}[label=\textnormal{(\arabic*)}]

 \item A topological space~$\sX$ is \emph{$\pi $-finite} if (i)~$\pi
_0\sX$~is a finite set, (ii)~for all~$x\in \sX$, the homotopy group $\pi
_q(\sX,x)$, $q\ge1$, is finite, and (iii)~there exists $Q\in \ZZ^{>0}$ such
that $\pi _q(\sX,x)=0$ for all $q> Q$, $x\in \sX$.  (For a fixed bound~$Q$
we say that $\sX$~is $Q$-finite.)

 \item A continuous map $f\:\sY\to \sZ$ of topological spaces is \emph{$\pi
$-finite} if for all~$z\in \sZ$ the homotopy fiber\footnote{\label{hq}As
in~\eqref{eq:1}, the homotopy fiber over~$z\in \sZ$ consists of
pairs~$(y,\gamma )$ of a point~$y\in \sY$ and a path~$\gamma $ in~$\sZ$
from~$z$ to~$f(y)$.} over~$z$ is a $\pi $-finite space.

 \item A spectrum\footnote{A spectrum is a sequence of pointed topological
spaces $\{E_q\}_{q\in \ZZ}$ and maps $\Sigma E_q\to E_{q+1}$.}~$E$ is
\emph{$\pi $-finite} if each space in the spectrum is a $\pi $-finite space.

 \end{enumerate} 
  \end{definition}

  \begin{example}[]\label{thm:c54}
 An Eilenberg-MacLane space~$K(\pi ,q)$ is $\pi $-finite if $\pi $~is a finite
group.  We use notation which emphasizes the role of~$\sX$ as a classifying
space: if~$q=1$ we denote $K(\pi ,1)$ by~$B\pi $, and if~$q\ge 1$ and $A$~is a
finite abelian group, we denote~$K(A,q)$ by~$\BnA q$.  Just as there are group
extensions of ordinary groups, so too there are extensions of homotopical
groups.  These are Postnikov towers.  For example, let $G$~be a finite group
and let $A$~be a finite abelian group.  Then extensions\footnote{The
sequence~\eqref{eq:2} is the Postnikov tower for a space~$\sX$ with $\pi
_q\sX=0$, $q\neq 1,2$.}  of the form
  \begin{equation}\label{eq:2}
     1\longrightarrow \BnA2\longrightarrow \sX\longrightarrow
     BG\longrightarrow 1 
  \end{equation}
are classified by group actions of~$G$ on~$A$ together with a cohomology
class in~$H^3(BG;A)$, where the coefficients~$A$ are twisted by the group
action.  Thus $\sX$~is a topological space with only two nonzero homotopy
groups: $\pi _1\sX=G$, $\pi _2\sX=A$.  This class of spaces was studied long
ago by George Whitehead.  Nowadays one might say that $\sX$~is the classifying
space of a \emph{2-group}.
  \end{example}

  \begin{remark}[]\label{thm:c55}
 If $\sX$~is a path connected topological space with basepoint~$*\in \sX$, then
$\sX$~is the classifying space of its based loop space~$\OX$, where the latter
is a grouplike $A_\infty $-space by composition of based loops.
  \end{remark}

  \begin{remark}[]\label{thm:c56}
 \ 

 \begin{enumerate}[label=\textnormal{(\arabic*)}]

 \item A topological space~$\sX$ gives rise to a sequence of higher groupoids
$\pi _0\sX$, $\pi _{\le1}\sX$, $\pi _{\le2}\sX$, \dots, or indeed to an $\infty
$-groupoid.  There is a classifying space construction which passes in the
opposite direction from higher groupoids to topological spaces.  An $\infty
$-groupoid is $\pi $-finite if it satisfies the conditions in
Definition~\ref{thm:c53}(1), which hold iff the corresponding topological space
is $\pi $-finite.
 \item In a similar way, one can define $\pi $-finiteness for a simplicial
set. 

 \end{enumerate} 
  \end{remark}

  \subsection*{Algebras}

  \subsection{The sandwich}\label{subsec:1.6}

Let $A$~be an algebra, and for definiteness suppose that the ground field
is~$\CC$.  Partly for simplicity, and partly by the analogy with
\emph{finite} groups, assume that $A$~and the modules that follow are finite
dimensional.  Let $R$~be the \emph{regular} right $A$-module, i.e., the
vector space~$A$ furnished with the right action of~$A$ by multiplication.
The pair~$(A,R)$ is \emph{abstract symmetry data} (in the sense of algebras):
the \emph{realization} of~$(A,R)$ on a vector space~$V$ is a pair~$(L,\theta
)$ consisting of a left $A$-module~$L$ together with an isomorphism of vector
spaces
  \begin{equation}\label{eq:c2}
     \theta \:R\,\otimes  \mstrut _AL\xrightarrow{\;\;\cong
     \;\;}V.  
  \end{equation}
The tensor product in~\eqref{eq:c2}---an algebra sandwiched between a right
and left module---is a general structure that recurs in these lectures. 

  \begin{remark}[]\label{thm:8}
 It may seem pedantic to introduce the module~$R$ here; one usually simply
talks about a left module over~$A$.  But I want to emphasize the distinction
between the abstract symmetry structure and its concrete action on a vector
space, and for this we need to be able to recover the underlying vector space
from the left module.
  \end{remark}

  \begin{remark}[]\label{thm:57}
 More generally, in abstract symmetry data $R$~can be \emph{any} right module;
it needn't be the regular module. 
  \end{remark}

Observe that the right regular module satisfies the algebra isomorphism 
  \begin{equation}\label{eq:c3}
     \End_A(R)\cong A, 
  \end{equation}
where the left hand side is the algebra of linear maps $R\to R$ that commute
with the right $A$-action, i.e., the left action of~$A$ on itself.

  \subsection{Left vs.~right}\label{subsec:1.16}

  The choice of left vs.~right in a given situation is made by choice or
convention.  My convention always puts \emph{structural} actions on the right
and \emph{geometric} actions on the left.  Here the right module~$R$ is part of
the symmetry structure, and the left module is the geometric action on the
vector space.  As another example, if $V$~is a finite dimensional real vector
space of dimension~$n$, then the space of bases~$\sB(V)$---the set of
isomorphisms $\RR^n\to V$---carries a right structural action of~$\GL_n\!\RR$
and a left geometric action of~$\Aut(V)$.

  \subsection{The group algebra}\label{subsec:1.7}

Let $G$~be a finite group.  The \emph{group algebra} $A=\CC[G]$ is the free
vector space on the set~$G$, which is then a linear basis of~$A$; multiply
basis elements according to the group law in~$G$.  A left $A$-module~$L$
restricts to a linear representation of~$G$.  The tensor product
in~\eqref{eq:c2} recovers the vector space that underlies the representation.
In the setup of~\ref{subsec:1.4}, take~$X=L$ to construct a vector bundle
$L_G\to BG$ whose fiber over~$*\in BG$ is~$L$.
 
There is an inclusion $G\subset \GA$ whose image consists of \emph{units},
i.e., of invertible elements in the algebra.  But the typical element of~$\GA$
is not invertible.  For example, the sum $g_1+\dots +g_k$ over a conjugacy
class in~$G$ is not invertible unless~$k=1$.  In general, this sum is a central
element.  In fact, the center of~$\GA$ is generated by these elements.
 
Noninvertible elements in~$\GA$ play an important role in the study of
$G$-symmetry.  For example, when $G$~is the symmetric group on $n$~letters,
then the theory of irreducible representations and their associated Young
tableaux is developed in terms of certain projectors in~$\GA$. 

  \begin{example}[]\label{thm:4}
 Consider the Lie algebra~ $\mathfrak{s}\mathfrak{u}_3$, and let
$A=U(\mathfrak{s}\mathfrak{u}_3)$ be its universal enveloping algebra
(over~$\CC$).  The center of~$A$ is isomorphic to a polynomial algebra in
2~variables; it is generated by the Casimir elements~$x_2,x_3\in A$.  These
Casimirs act as linear operators on any $A$-module---i.e., on a
representation of~$\SU_3$---and by Schur's lemma they act by a scalar if the
module is irreducible.  So these operators can be used to decompose an
arbitrary $A$-module into a direct sum of isotypical submodules.  This is
simply another illustration of: (1)~the importance of noninvertible elements
in an algebra, and (2)~the use of particular elements in an abstract algebras
(here central elements) in concrete realizations.
  \end{example}

  \subsection{Higher algebra}\label{subsec:c1.5}

The higher versions of finite groups in~\ref{subsec:1.5} have an analog in
algebras as well.  For example, a \emph{fusion category}~$\sA$ is a ``once
higher'' version of a finite dimensional semisimple algebra, and there is a
well-developed theory of modules over a fusion category.\footnote{Pavel
Etingof, Shlomo Gelaki, Dmitri Nikshych, and Victor Ostrik,
  \href{http://dx.doi.org/10.1090/surv/205}{\emph{Tensor categories}},
  Mathematical Surveys and Monographs, vol. 205, American Mathematical Society,
  Providence, RI, 2015.}  In particular,
$\sA$~is a right module over itself, the right regular module.  A finite
group~$G$ gives rise to the fusion category $\sA=\Vect[G]$ of finite rank
vector bundles over~$G$ with convolution product.

  \subsection*{Main Definitions}

These are Definition~\ref{thm:c13} and Definition~\ref{thm:c11} below.

  \subsection{Remarks about field theory}\label{subsec:1.13}

We begin with a few general remarks, and we defer to Lecture~\ref{sec:2} for
a more in-depth discussion.
 
Perhaps the first point to make is the metaphor of a field theory as a
representation of a Lie group, or better
  \begin{equation}\label{eq:5}
     \textnormal{field theory $\sim$ module over an algebra} 
  \end{equation}
This is of course only a very rough analogy, but nonetheless it provides
useful guidance and language.  (Our language sometimes seems to assume the
module has an algebra structure, but that is not an assumption we make.)  We
do always work in the Wick-rotated context, so for a quantum field theory we
work on Riemannian manifolds rather than Lorentz manifolds.  As pioneered by
Graeme Segal, a Wick-rotated theory is a linear representation of a bordism
category; it is the bordism category which plays the role of the algebra
in~\eqref{eq:5}.    

  \begin{remark}[]\label{thm:58}
 We do not persist with the metaphor~\eqref{eq:5} in this text since we use the
module metaphor for domain walls and boundary theories;
see~\ref{subsec:1.14}. 
  \end{remark}

The field theories that encode finite symmetries are \emph{topological}, and
we bring to bear the mathematical development of topological field theory.
In particular, we work with \emph{fully local} (also known as \emph{fully
extended}) topological field theories.  In the axioms this is realized by
having the theory defined on a higher bordism category of manifolds with
corners of all codimension.  The field theories on which the symmetry acts
are typically not topological, and for general quantum field theories the
fully local aspect has yet to be fully developed.  Nonetheless, our
exposition often implicitly assumes full locality.

  \begin{remark}[]\label{thm:6}
 Just as one specifies a Lie group to talk about its representations, one
must specify a bordism category to talk about its representations (field
theories).  There are two sorts of ``discrete parameters''.  First, there is a
dimension~$n$, which in the physical anti-Wick-rotated theory is the
dimension of spacetime.  Second there is a collection~$\sF$ of background
fields.  We use the terminology `$n$-dimensional field theory on~$\sF$' or
`$n$-dimensional field theory over~$\sF$'.  We define background fields in
the next lecture (Definition~\ref{thm:20}); for today's lecture they remain
in the deep background.  We often work in shorthand, illustrated by the
following for a gauged nonlinear $\sigma $-model with target~$S^2$:
  \begin{equation}\label{eq:7}
     \sF=\{\textnormal{orientation, Riemannian metric, $\SO_3$-connection,
     section of twisted $S^2$-bundle}\} 
  \end{equation}
  \end{remark}

  \begin{figure}[ht]
  \centering
  \includegraphics[scale=.4]{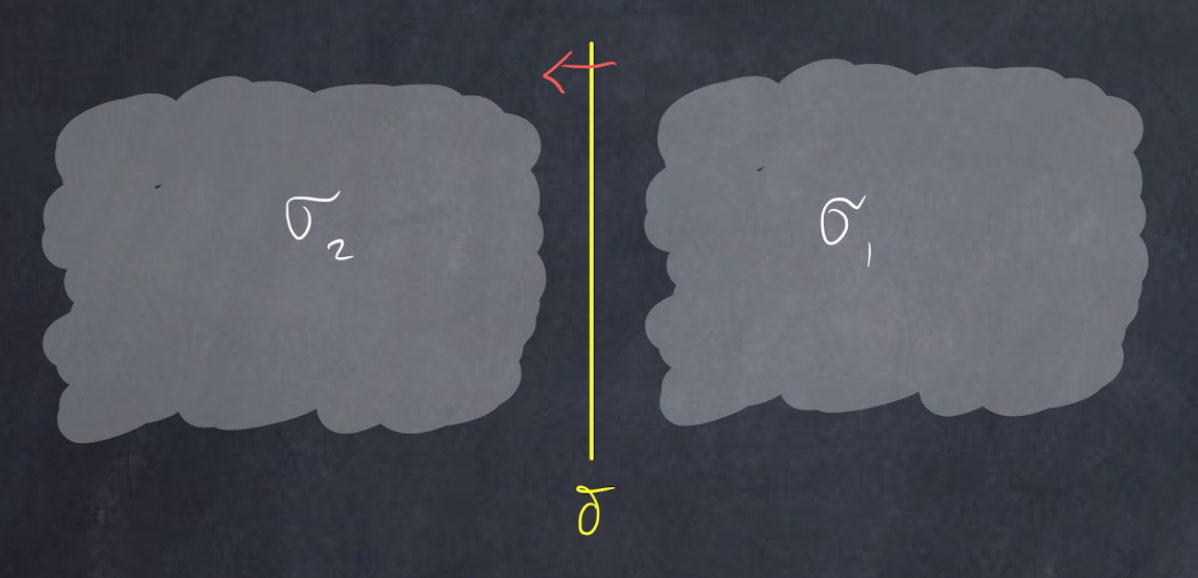}
  \vskip -.5pc
  \caption{A domain wall $\delta \:\sigma _1\to \sigma _2$}\label{fig:c1}
  \end{figure}

  \subsection{Domain walls and boundary theories}\label{subsec:1.14}

Let $\sigma _1,\sigma _2$ be $(n+1)$-dimensional theories on background
fields~$\sF_1,\sF_2$.  A \emph{domain wall} $\delta \:\sigma _1\to \sigma _2$
is the analog\footnote{We emphasize that $\sigma _1$~and $\sigma _2$~are not
assumed to be algebra objects in the symmetric monoidal category of field
theories.} of a ``$(\sigma _2,\sigma _1)$-bimodule''; see Figure~\ref{fig:c1}
for a depiction.  We remove the scare quotes and use the convenient
terminology `$(\sigma _2,\sigma _1)$-bimodule' for a domain wall.  The triple
$(\sigma _1,\sigma _2,\delta )$ is formally a functor with domain a bordism
category of smooth $(n+1)$-dimensional manifolds with corners that are
equipped with a partition into regions labeled~`1' and~`2' separated by a
cooriented codimension one submanifold (with corners) that is ``$\delta
$-colored''.  This is illustrated in Figure~\ref{fig:c2}.  As a special case,
a domain wall from the tensor unit theory~$\bone$ to itself is an
$n$-dimensional (absolute, standalone) theory.  More generally, we can tensor
any domain wall $\delta \:\sigma _1\to \sigma _2$ with an $n$-dimensional
theory to obtain a new domain wall.  There is a composition law on
topological domain walls that are ``parallel'' (in the sense that they have
trivial normal bundles and one is the image of a nonzero section of the
normal bundle of the other, using a tubular neighborhood):
  \begin{equation}\label{eq:c13}
     \begin{gathered} \xymatrix{\sigma _1\ar[r]^{\delta '}\ar@/_1pc/[rr]_{\delta
     ''\circ \,\delta '} & \sigma _2\ar[r]^{\delta ''}& \sigma _3}
     \end{gathered} 
  \end{equation}

  \begin{figure}[ht]
  \centering
  \includegraphics[scale=1.7]{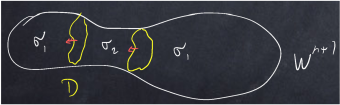}
  \vskip -.5pc
  \caption{Domain walls in the manifold~$W$}\label{fig:c2}
  \end{figure}

  \subsection{Boundary theories}\label{subsec:1.15}

Following the metaphor of domain wall as bimodule, there are special cases of
right or left modules.  For field theory these are called \emph{right
boundary theories} or \emph{left boundary theories}, as depicted in
Figure~\ref{fig:c3}.  (Normally, we omit the region labeled~`$\bone$' in the
drawings: it is transparent.)  A right boundary theory of~$\sigma $ is a
domain wall $\sigma \to \bone$; a left boundary theory is a domain wall
$\bone\to \sigma $. 

  \begin{remark}[]\label{thm:9}
 The nomenclature of right vs.~left may at first be confusing; it does follow
standard usage for modules over an algebra---the direction (right or left) is
that of the action of the algebra on the module.  In fact, following our
general usage for domain walls, we use the terms `right $\sigma $-module' and
`left $\sigma $-module' for right and left boundary theories.  But a right
boundary theory appears on the left in drawings, just as a right module~$R$
over an algebra~$A$ appears to the left of the algebra in the expression
`$R_A$'.
  \end{remark}

  \begin{figure}[ht]
  \centering
  \includegraphics[scale=.35]{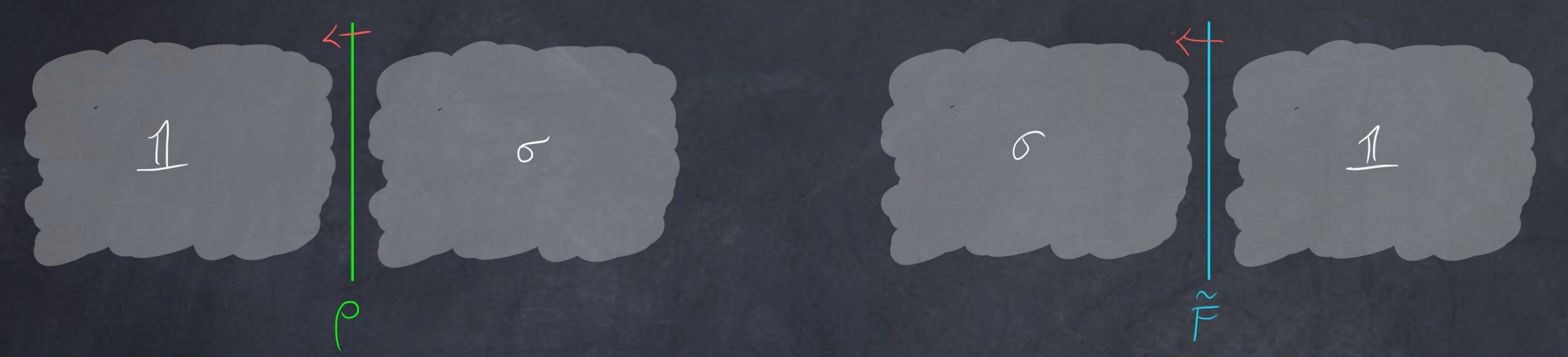}
  \vskip -.5pc
  \caption{Left: a right boundary theory.  Right: a left boundary
  theory}\label{fig:c3} 
  \end{figure}

  \subsection{Abstract finite symmetry in field theory}\label{subsec:c3.1}

We turn now to the central concept in these lectures.

  \begin{definition}[]\label{thm:c9}
 Fix~$n\in \ZZ^{\ge0}$.  An \emph{$n$-dimensional quiche} is a pair~$(\sigma
,\rho )$, where
$$\sigma \:\Bord_{n+1}(\sF)\to \sC$$
is an $(n+1)$-dimensional
topological field theory and $\rho $~is a right topological $\sigma $-module.
  \end{definition}

\noindent
 The dimension~$n$ pertains to the theories on which~$(\sigma ,\rho )$ acts,
not to the dimension of the field theory~$\sigma $.  The terminology `quiche' is
meant to evoke~ $\sigma $ as the filling and~$\rho $ as the crust.  One might
want to assume that $\rho $~is nonzero, which is true for the particular
boundaries in Definition~\ref{thm:c13} below.  Our statement of
Definition~\ref{thm:c9} has not made explicit the background fields, but they
are there in the background (and there are issues to address concerning them).
 
This definition is extremely general.  The following singles out a class of
boundary theories which more closely models the discussions
in~\ref{subsec:1.4} and~\ref{subsec:1.6}.  We freely use the language and
setting of fully local topological field theory.  Recall that if $\sC'$~is a
symmetric monoidal $n$-category, then there is a symmetric monoidal Morita
$(n+1)$-category $\Alg(\sC')$ whose objects are objects in~$\sC'$ equipped
with an algebra structure and whose 1-morphisms $A_0\to A_1$ are
$(A_1,A_0)$-bimodules.

  \begin{definition}[]\label{thm:c13}
 Suppose $\sC'$~is a symmetric monoidal $n$-category and $\sigma $~is an
$(n+1)$-dimensional topological field theory with codomain $\sC=\Alg(\sC')$.
Let $A=\sigma (\pt)$.  Then $A$~is an algebra in~$\sC'$ which, as an object
in~$\sC$, is $(n+1)$-dualizable.  Assume that the right regular module~$\Ar$
is $n$-dualizable as a 1-morphism in~$\sC$.  Then the boundary theory~$\rho $
determined by~$\Ar$ is the \emph{right regular boundary theory} of~$\sigma $,
or the \emph{right regular $\sigma $-module}.  See \footnote{Jacob Lurie, \emph{On the classification of
topological field theories},
  Current developments in mathematics, 2008, Int. Press, Somerville, MA, 2009,
  pp.~129--280. \href{http://arxiv.org/abs/arXiv:0905.0465}{{\tt
  arXiv:0905.0465}}} for discussions of dualizability (\S2.3) and boundary
  theories~(\S4.3). 
  \end{definition}

\noindent
 We use an extension of the cobordism hypothesis to generate the boundary
theory~$\rho $ from the right regular module~$\Ar$.  Observe that $\Ar$~is
the value of the pair~$(\sigma ,\rho )$ on the bordism depicted in
Figure~\ref{fig:c5}; the white point is incoming, so the depicted bordism
maps $\pt\to \emptyset $.

  \begin{figure}[ht]
  \centering
  \includegraphics[scale=1.6]{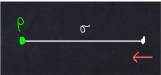}
  \vskip -.5pc
  \caption{The bordism which computes~$A_A$}\label{fig:c5}
  \end{figure}

  \begin{remark}[]\label{thm:c10}
 \ 

 \begin{enumerate}[label=\textnormal{(\arabic*)}]

 \item The right regular $\sigma $-module~$\rho $ satisfies $\End_\sigma (\rho
)\cong \rho $; compare~\eqref{eq:c3}.

 \item The regular boundary theory is also called a \emph{Dirichlet boundary
theory}.

 \item Not every topological field theory~$\sigma $ can appear in
Definition~\ref{thm:c9}.  For example, the main theorem in
\footnote{\label{FT}Daniel~S. Freed and Constantin Teleman, \emph{Gapped
boundary theories in three dimensions},
  \href{http://dx.doi.org/10.1007/s00220-021-04192-x}{Comm. Math. Phys.
  \textbf{388} (2021)}, no.~2, 845--892,
  \href{http://arxiv.org/abs/arXiv:2006.10200}{{\tt arXiv:2006.10200}}.

} asserts that ``most'' 3-dimensional
Reshetikhin-Turaev theories do not admit any nonzero topological boundary
theory, hence they cannot act as symmetries of a 2-dimensional field theory.
On the other hand, the Turaev-Viro theory~$\sP $ formed from a (spherical)
fusion category~$\Phi $ takes values in the 3-category~$\Alg(\Cat)$ for a
suitable 2-category~$\Cat$ of linear categories.  Thus $\sP$~admits the right
regular $\sigma $-module defined by the right regular module~$\Phi \mstrut
_{\hneg\Phi }$.

 \item Let $G$~be a finite group.  Then $G$-symmetry in an $n$-dimensional
quantum field theory is realized via $(n+1)$-dimensional finite gauge theory.
The partition function counts principal $G$-bundles, weighted by the
reciprocal of the order of the automorphism group.  The regular boundary
theory has an additional fluctuating field: a section of the principal
$G$-bundle, i.e., a trivialization of the restriction of the principal
$G$-bundle to the boundary.

 \item The topological field theory~$\sigma $ need only be a once-categorified
$n$-dimensional theory, not a full $(n+1)$-dimensional theory.  This relaxation
permits more examples, but we do not pursue them in these lectures.

 \end{enumerate}
  \end{remark}

  \subsection{Concrete realization of finite symmetry in field theory}\label{subsec:c3.2}

We now define a concrete realization of~$\sr$ as symmetries of a quantum
field theory.

  \begin{definition}[]\label{thm:c11}
 Let $(\sigma ,\rho )$ be an $n$-dimensional quiche.  Let $F$~be an
$n$-dimensional field theory.  A \emph{$\sr$-module structure} on~$F$ is a
pair~$\tFt$ in which $\tF$~is a left $\sigma $-module and $\theta $~is an
isomorphism
  \begin{equation}\label{eq:c21}
     \theta \:\rho \otimes \mstrut _{\sigma }\tF\xrightarrow{\;\;\cong \;\;}F 
  \end{equation}
of absolute $n$-dimensional theories. 
  \end{definition}

  \begin{figure}[ht]
  \centering
  \includegraphics[scale=1.6]{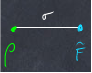}
  \vskip -.5pc
  \caption{The ``sandwich''}\label{fig:c6}
  \end{figure}

\noindent
 Here `$\rho \otimes \mstrut _{\sigma }\tF$' notates the dimensional
reduction of~$\sigma $ along the closed interval with boundaries colored
with~$\rho $ and~$\tF$; see Figure~\ref{fig:c6}.  The bulk theory~$\sigma $ with
its right and left boundary theories~$\rho $ and~$\tF$ is sometimes called a
``sandwich''.   

  \begin{remark}[]\label{thm:c12}
  \  
 \begin{enumerate}[label=\textnormal{(\arabic*)}]

 \item The theory~$F$ and so the boundary theory~$\tF$ may be topological or
nontopological, and we allow it to be not fully extended (in which case we
use truncations of~$\sigma $ and~$\rho $).

 \item The sandwich picture of~$F$ as $\rho \otimes _\sigma \tF$ separates
out the topological part~$\sr$ of the theory from the potentially
nontopological part~$\tF$ of the theory.  This is advantageous, for example
in the study of defects.  It allows general computations in the abstract
symmetry data which apply to every realization as a symmetry of a field
theory.

 \item Typically, symmetry persists under renormalization group flow, hence a
low energy approximation to~$F$ should also be a $\sr$-module.  If $F$~is
gapped, then at low energies we expect a topological theory (up to an
invertible theory), so we can bring to bear powerful methods and theorems in
topological field theory to investigate \emph{topological} left $\sigma
$-modules.  This leads to dynamical predictions, as we illustrate in
Lecture~\ref{sec:4}.

 \end{enumerate}
  \end{remark}

  \subsection*{Examples}

  \begin{example}[quantum mechanics $n=1$]\label{thm:c14}
 Consider a quantum mechanical system defined by a Hilbert space~$\sH$ and a
time-independent Hamiltonian~$H$.  The Wick-rotated theory~$F$ is regarded as
a map with domain $\Bord_{\langle 0,1 \rangle}(\sF)$ for
  \begin{equation}\label{eq:c23}
      \sF=\{\textnormal{orientation, Riemannian metric}\}. 
  \end{equation}
Roughly speaking, $F(\pt_+)=\sH$ and $F(X)= e^{-\tau H/\hbar}$ for~$\tau \in
\RR^{>0}$ and $X=[0,\tau ]$ with the standard orientation and Riemannian
metric.  We refer to a recent preprint of Kontsevich-Segal
\footnote{\label{ks}Maxim Kontsevich and Graeme Segal, \emph{Wick rotation
and the positivity of energy in quantum field theory},
\href{http://arxiv.org/abs/arXiv:2105.10161}{{\tt arXiv:2105.10161}}.  } for
more precise statements, in particular that the 0-manifold~$\pt_+$ is
embedded in the germ of an oriented Riemannian 1-manifold.
 
Now suppose $G$~is a finite group equipped with a unitary representation
$S\:G\to U(\sH)$, and assume that the $G$-action commutes with the
Hamiltonian~$H$.  To express this symmetry in terms of
Definition~\ref{thm:c9} and Definition~\ref{thm:c11}, let $\sigma $~be the
2-dimensional finite gauge theory with gauge group~$G$.  If we were only
concerned with~$\sigma $ we might set the codomain of~$\sigma $ to be
$\sC=\Alg(\sC')$ for $\sC'$~the category of finite dimensional complex vector
spaces and linear maps.  But to accommodate the boundary theory~$\tF$ for
quantum mechanics, we let~$\sC'$ be a suitable category of topological vector
spaces.  The pair~$\sr$ is defined on $\Bord_2=\Bord_{\langle 0,1,2 \rangle}$
with no background fields.  Then $\sigma (\pt)=\GA$ is the complex group
algebra of~$G$, and $\rho (\pt)$~is its right regular module.

  \begin{figure}[ht]
  \centering
  \includegraphics[scale=1.5]{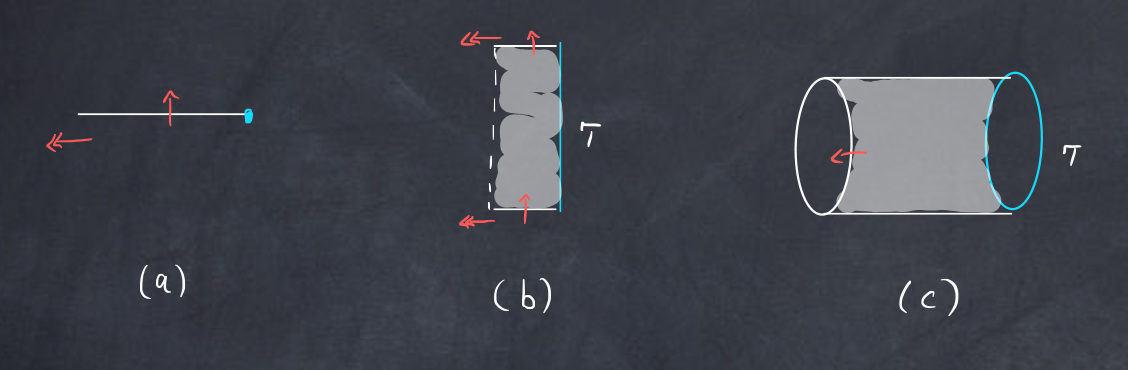}
  \vskip -.5pc
  \caption{Three bordisms evaluated in~\eqref{eq:c22} in the theory~$(\sigma ,\tF)$}\label{fig:c8}
  \end{figure}

Now we describe the left boundary theory~$\tF$, which has as background
fields~\eqref{eq:c23}, as does the (absolute) quantum mechanical theory~$F$.
Observe that by cutting out a collar neighborhood it suffices to define~$\tF$
on cylinders (products with~$[0,1]$) over $\tF$-colored boundaries.  The
bordisms in Figure~\ref{fig:c8} do not have a well-defined width since there
is a Riemannian metric only on the colored boundary.  That boundary has a
well-defined length~$\tau $ in~(b) and~(c).  We refer to \S2.1.1 of footnote~
\footref{FT} for the conventions about arrows of time and dashed lines.
Evaluation of these bordisms under~$(\sigma ,\tF)$ gives:
  \begin{equation}\label{eq:c22}
     \begin{aligned} &\textnormal{(a)~the left module }\sHl \\
      &\textnormal{(b) }e^{-\tau H/\hbar}\:\sHl\longrightarrow \sHl \\
      &\textnormal{(c)~the central function
      $g\longmapsto\Tr\mstrut _{\sH}\bigl(S(g)e^{-\tau H/\hbar}\bigr)$ on~$G$}
      \end{aligned} 
  \end{equation}
Of course, this is not a complete construction of the nontopological $\sigma
$-module~$\tF$, but it gives some intuition for that theory.

  \begin{figure}[ht]
  \centering
  \includegraphics[scale=.55]{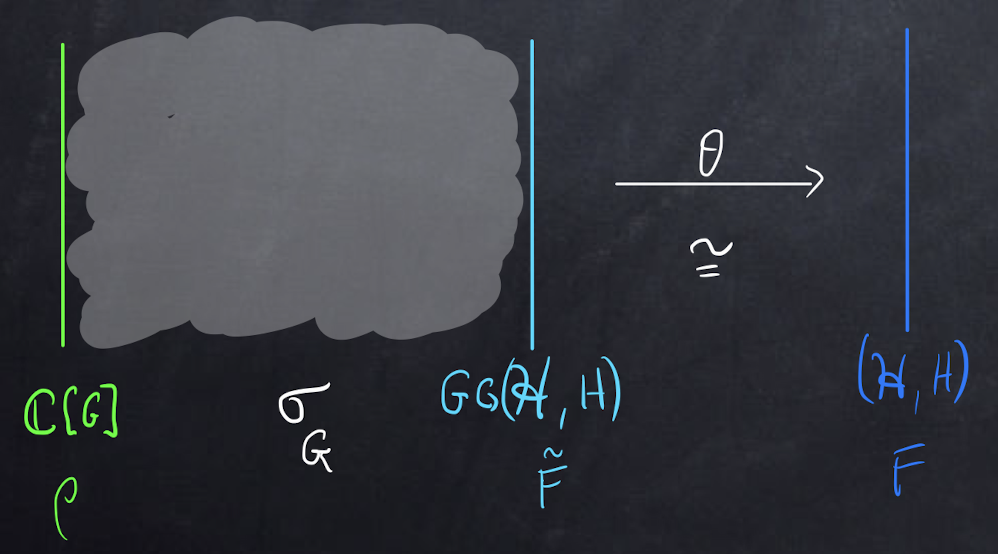}
  \vskip -.5pc
  \caption{Quantum mechanics with $G$-symmetry}\label{fig:1}
  \end{figure}

Figure~\ref{fig:1} illustrates the $G$-action on the quantum mechanics
theory~$F$.  Although we have not discussed defects yet, we illustrate how
they work in this basic example.  A point defect in~$F$ is what is usually
called an \emph{observable}.  Think of time running up vertically, and then a
point defect~$\delta $ is the insertion of an observable, or operator, at a
given time, as depicted in Figure~\ref{fig:2}.  Also shown is the \emph{link}
of the point, which is a 0-sphere.  The possible defects on the point are the
elements of the topological vector space 
  \begin{equation}\label{eq:8}
     \varprojlim_{\epsilon \to 0}\Hom\bigl(1,F(S^0_\epsilon ) \bigr). 
  \end{equation}
Here $\epsilon $~measures the size of the linking 0-sphere, and one takes an
inverse limit as this size shrinks to zero.  That inverse limit is a space of
singular operators on~$\sH$; see footnote~\footref{ks}.  In these lectures we mostly
compute with defects in topological theories, and for these there is no need
to take a limit.  Here, for ease of notation and because we are after more
formal points, we denote this space of operators as `$\End(\sH)$', even
though the notation suggests bounded operators.

  \begin{figure}[ht]
  \centering
  \includegraphics[scale=.6]{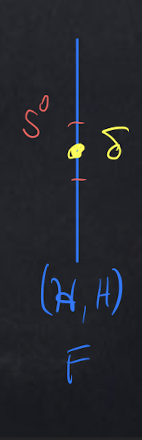}
  \vskip -.5pc
  \caption{A point observable in quantum mechanics}\label{fig:2}
  \end{figure}

Now we look at defects in the sandwich picture in Figure~\ref{fig:1} and
transport to point defects in the theory~$F$; see Figure~\ref{fig:2}.  First
consider a point defect on the $\tF$-colored boundary, as in
Figure~\ref{fig:3}.  The link of the point is depicted, and its value under
the pair $(\sigma ,\sF)$ is computed at the bottom of the figure.  (There
should be an inverse limit, which is omitted.)  The result is that such a
defect is an operator on~$\sH$ which commutes with the $G$-action.  Of
course, such an operator need not be invertible.  Also, since $\tF$~is not
topological, this is not a topological defect.

  \begin{figure}[ht]
  \centering
  \includegraphics[scale=.6]{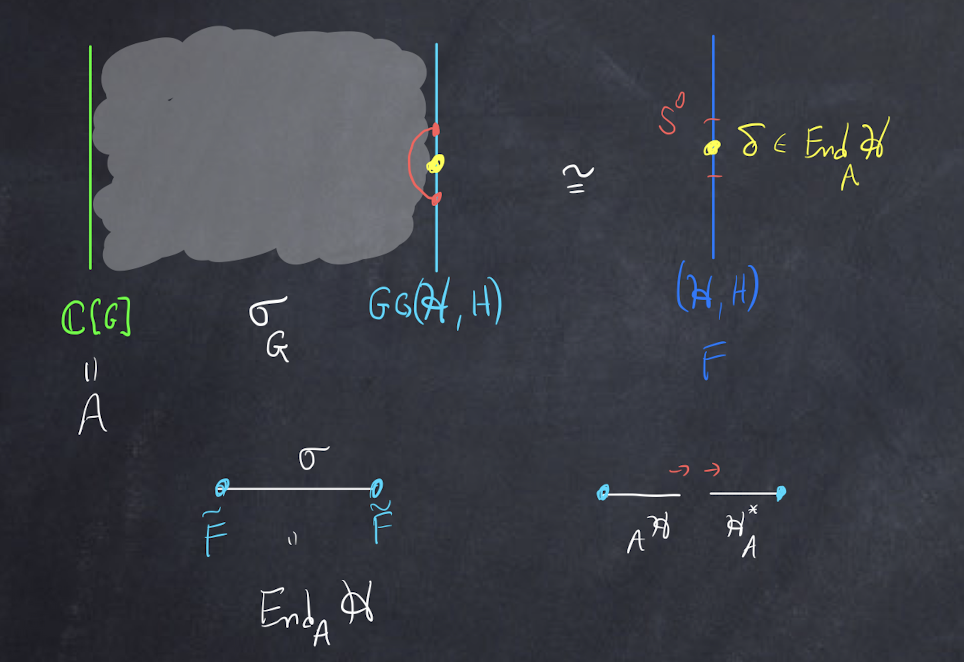}
  \vskip -.5pc
  \caption{A point defect on the $\tF$-boundary}\label{fig:3}
  \end{figure}

  \begin{figure}[ht]
  \centering
  \includegraphics[scale=.65]{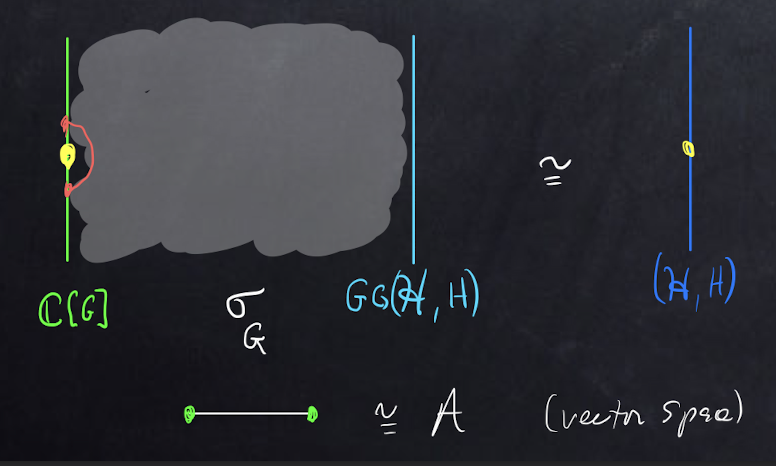}
  \vskip -.5pc
  \caption{A point $\rho $-defect}\label{fig:4}
  \end{figure}

At the other extreme is a point defect on the $\rho $-colored boundary, which
we call a \emph{point $\rho $-defect}; see Figure~\ref{fig:4}.  Now the link
is in the topological field theory $(\sigma ,\rho )$; that is, the bulk
topological theory~$\sigma $ with topological boundary theory~$\rho $.  The
value of the link is the vector space which underlies the group algebra
$A=\GA$.  Hence the point defect is labeled by an element of~$A$.  This may
be an element of the group, which is a unit in~$A$, or it may be a nonunit
(noninvertible element) in~$A$.  This is a topological defect, as it is a
defect in the topological field theory.  Now imagine having both of these
point defects.  Since the point $\rho $-defect is topological, it can be
moved in time without changing any correlation functions.  Visibly it
commutes with the point defect on the $\tF$-boundary, which recall is an
operator that commutes with~$A$.

  \begin{figure}[ht]
  \centering
  \includegraphics[scale=.7]{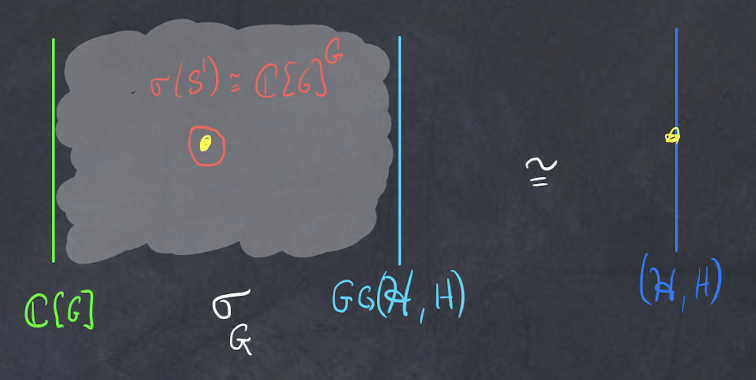}
  \vskip -.5pc
  \caption{A central point defect}\label{fig:5}
  \end{figure}

One can have a point defect in the bulk theory, as in Figure~\ref{fig:5}.
The link of this point is a circle, and the value~$\sigma (\cir)$ of the
finite gauge theory on the circle is a vector space which may be identified
with the center of the group algebra~$\GA$.  As stated earlier, it has a
basis labeled by conjugacy classes: the central element of~$\GA$ is the sum
of elements in the conjugacy class.  These topological defects commute with
the point defects in Figures~\ref{fig:3} and ~\ref{fig:4}. 

  \begin{figure}[ht]
  \centering
  \includegraphics[scale=.65]{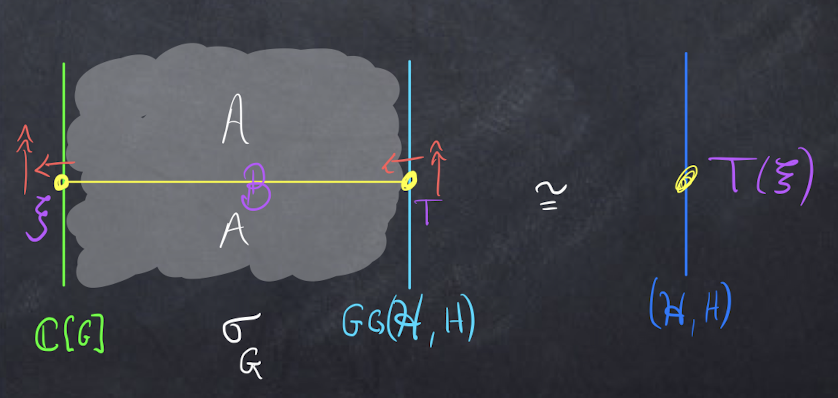}
  \vskip -.5pc
  \caption{A general point defect}\label{fig:6}
  \end{figure}

The general point defect in~$F$ can be realized by a defect on the closed
interval depicted in Figure~\ref{fig:6}.  When working on a stratified
manifold, we evaluate on links working from higher dimensional strata to
lower dimensional strata.  The space of labels on a given stratum may depend
on the labels chosen on higher dimensional strata, as they do at the
endpoints in this case.  We do not explain the evaluation of the links in
detail here, but simply report that: (1)~the label in the interior of the
defect is an $(A,A)$-bimodule~$B$; (2)~the label at the endpoint on the $\rho
$-colored boundary is a vector~$\xi \in B$; and (3)~the label at the endpoint
on the $\tF $-colored boundary is an $(A,A)$-bimodule map $B\to \End(\sH)$.
Under the isomorphism~$\theta $, this maps to the point defect labeled
by~$T(\xi )$ in the theory~$F$.

  \begin{exercise}[]\label{thm:11}
 What happens if~$B=A$?  (This is the transparent defect in the interior.) 
  \end{exercise}

  \begin{figure}[ht]
  \centering
  \includegraphics[scale=.45]{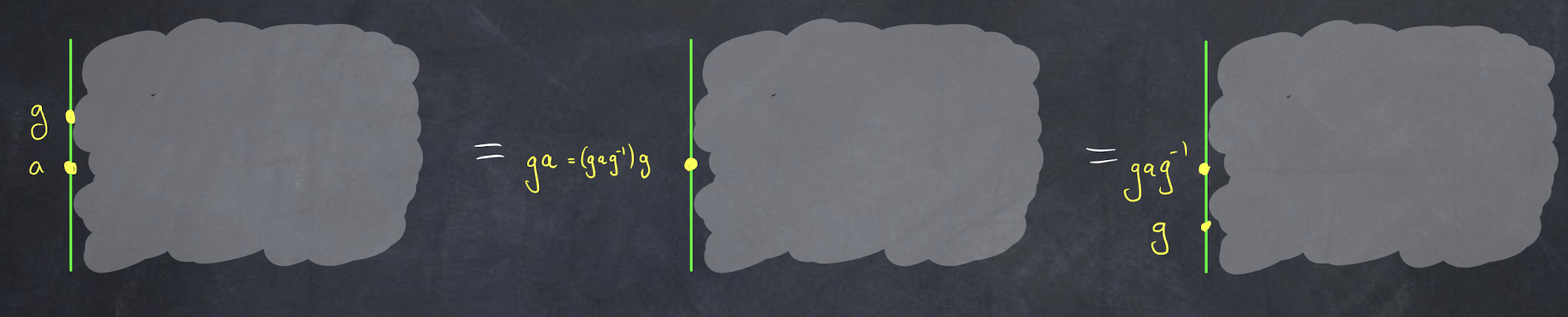}
  \vskip -.5pc
  \caption{Commuting point $\rho $-defects}\label{fig:7}
  \end{figure}

We compute the action of the point $\rho $-defect labeled by~$g\in G\subset
\GA$ on a general point defect in~$F$.  This is a computation in the
topological field theory~$\sr$; it applies universally to any left
$\sr$-module.  First, consider the special case of a topological point defect
labeled by~$a\in A=\GA$, as illustrated in Figure~\ref{fig:7}.  Passing from
the first picture to the second is the fusion of point defects, which we will
compute later is multiplication in~$A$.  The same fusion applies when passing
from the third picture to the second.  So the effect of moving the $g$-defect
past the $a$-defect is conjugation of~$a$.  In Figure~\ref{fig:8} we
illustrate how the $g$-defect moves past a general (nontopological) point
defect.  Now the label at the $\rho $-colored boundary is a vector~$\xi \in
B$, where $B$~is an $(A,A)$-bimodule, and again the effect is to
conjugate~$\xi $ by~$g$.  Applying the isomorphism~$\theta $ from the
sandwich theory to the theory~$F$, we pass from~$T(\xi )$ to $T(g\xi g\inv)
=gT(\xi )g\inv $, since $T$~is an $(A,A)$-bimodule map.  This is the expected
action on point defects.

  \begin{figure}[ht]
  \centering
  \includegraphics[scale=1.04]{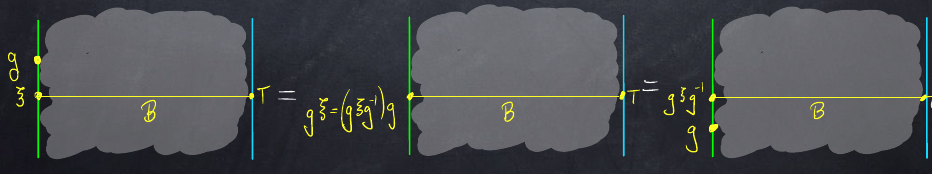}
  \vskip -.5pc
  \caption{Action of a topological point defect on a general point defect}\label{fig:8}
  \end{figure}

  \end{example}

  \begin{remark}[]\label{thm:c15}
 The finite gauge theory~$\sigma $ can be constructed via a finite path
integral from the $\pi $-finite space~$BG$; we discuss finite path integrals in
the next lecture.  Similarly, the boundary theory~$\rho $ can be constructed
from a basepoint $*\to BG$: the principal $G$-bundles are equipped with a
trivialization on $\rho $-colored boundaries.  A traditional picture of the
$G$-symmetry of the theory~$F$ uses this \emph{classical} picture: the sheaf of
background fields~$\sF$ is augmented to the sheaf
$\tsF=\{\textnormal{orientation, Riemannian metric, $G$-bundle}\}$, which
fibers over the sheaf $B_{\bullet }G=\{\textnormal{$G$-bundle}\}$, so in that
sense it ``fibers over~$BG$'' as in~\ref{subsec:1.4}.  There is an absolute
field theory on~$\tsF$ which is the ``coupling of~$F$ to a background gauge
field'' for the symmetry group~$G$.  The framework we are advocating here of
$F$~as a $\sr$-module uses the \emph{quantum} finite gauge theory~$\sigma $
that sums over principal $G$-bundles.
  \end{remark}

  \begin{remark}[]\label{thm:c16}
 The finite path integral construction of the regular (Dirichlet) boundary
theory makes the isomorphism~$\theta $ in~\eqref{eq:c21} apparent.  Namely, to
evaluate~$(\sigma ,\rho )$ we sum over $G$-bundles equipped with a
trivialization on $\rho $-colored boundaries.  Since the trivialization
propagates across an interval, the sandwich theory (Figure~\ref{fig:c6}) is
the original theory~$F$ without the explicit $G$-symmetry.
  \end{remark}

  \begin{example}[a homotopical symmetry]\label{thm:c18}
 Let $H$~be a connected compact Lie group, and suppose $A\subset H$ is a finite
subgroup of the center of~$H$.  Let $\bH=H/A$.  Then a $H$-gauge theory in,
say, 4~dimensions---for example, pure Yang-Mills theory---has a $BA$~symmetry.
In this case we take $\sigma =\sXd3{\BtA}$ to be the 3-dimensional $A$-gerbe
theory based on the $\pi $-finite space~$\BtA$, and we take~ $\rho $ to be the
regular boundary theory constructed from a basepoint $* \to \BtA$.  The left
$\sigma $-module~$\tF$ is a $\bH$-gauge theory.  (Aspects of this example are
discussed in more detail in \footnote{Daniel~S. Freed and Constantin Teleman,
\emph{Relative quantum field theory},
\href{http://dx.doi.org/10.1007/s00220-013-1880-1}{Comm. Math. Phys.
\textbf{326} (2014)}, no.~2, 459--476,
\href{http://arxiv.org/abs/arXiv:1212.1692}{{\tt arXiv:1212.1692}}.} and in
\footref{F1}.)
  \end{example}

  \subsection*{Quotients}

For the remainder of this lecture we turn back to a general discussion of
symmetry to remind about two aspects that we will take up in field theory in
subsequent lectures, particularly Lecture~\ref{sec:3}: quotients and
projective symmetries.

If $X$~is a set equipped with the action of a group~$G$, then there is a
quotient set~$X/G$: a point of~$X/G$ is a $G$-orbit in~$X$.  We now give
analogs in the homotopy and algebra settings.

  \subsection{The homotopy quotient}\label{subsec:1.8}

In the topological setting of~\ref{subsec:1.4}, the total space~$X_G$ of
the Borel construction plays the role of the quotient space~$X/G$.  Indeed,
if $G$~acts freely on~$X$, then there is a homotopy equivalence $X_G\simeq
X/G$; in general, $X_G$~is the \emph{homotopy quotient}.   
 
For any map $f\:Y\to BG$ of topological spaces we form the \emph{homotopy
pullback} (see~\eqref{eq:6})
  \begin{equation}\label{eq:c4}
     \begin{gathered}
     \xymatrix{&Z\ar@{-->}[dl]\ar@{-->}[dr]\\Y\ar[dr]_{f}&&X_G\ar[dl]^{\pi
     }\\&BG} 
     \end{gathered} 
  \end{equation}
If $Y$~is path connected and pointed, then there is a homotopy equivalence
$Y\simeq B(\Omega Y)$.  If $BG$~also has a basepoint, and if the map $f\:Y\to
BG$ is basepoint-preserving, then $f$~ is the classifying map of a
homomorphism $\Omega Y\to G=\Omega BG$, at least in the homotopical sense.
In this case $Z$~is the homotopy quotient of~$X$ by the action of~$\Omega Y$.
As a special case, if $G'\subset G$~is a subgroup, and $Y=BG'\to BG$ is the
classifying map of the inclusion, then $Z$~is homotopy equivalent to the
total space of the Borel construction~$X_{G'}$.  Hence \eqref{eq:c4}~is a
generalized quotient construction.  For~$G'=\{e\}$ we have~$Y=*$ and we
recover~$Z=X$, as in~\eqref{eq:1}: no quotient at all.

  \subsection{Augmentations of algebras}\label{subsec:1.9}

There is an analogous story in the setting~\ref{subsec:1.6} of algebras.
An \emph{augmentation} of an algebra~$A$ is an algebra homomorphism $\epsilon
\:A\to \CC$.  Use~$\epsilon $ to endow the scalars~$\CC$ with a right
$A$-module structure: set $\lambda \cdot a=\lambda \epsilon (a)$ for $\lambda
\in \CC$,\,$a\in A$.  If $L$~is a left $A$-module, the vector space 
  \begin{equation}\label{eq:c5}
     Q=\CC\otimes  \mstrut _AL=\CC\otimes \mstrut _\epsilon
     L 
  \end{equation}
plays the role of the ``quotient'' of~$L$ by~$A$. 

  \begin{example}[]\label{thm:c3}
 For the group algebra~$\GA$ of a finite group~$G$, there is a natural
augmentation
  \begin{equation}\label{eq:c6}
     \begin{aligned} \epsilon \:\CC[G]&\longrightarrow \CC \\
      \sum\limits_{g\in G}\lambda _gg&\longmapsto \sum\limits_{g\in G}\lambda
      _g\end{aligned} 
  \end{equation}
where $\lambda _g\in \CC$.  The augmentation is the pushforward on functions
under the map $G\to *$.  If $L$~is a representation of~$G$, extended to a
left $\CC[G]$-module, then the tensor product~\eqref{eq:c5} is the vector
space of \emph{coinvariants}; the relation
  \begin{equation}\label{eq:c7}
     1\otimes \ell =1\otimes g\cdot \ell =1\otimes g'\cdot \ell ,\qquad \ell
     \in L, \quad g,g'\in G,  
  \end{equation}
holds in the tensor product with the augmentation.  As a particular case, let
$S$~be a finite set equipped with a left $G$-action, and let $L=\CC\langle S
\rangle$ be the free vector space generated by~$S$.  Then $\CC\otimes \mstrut
_\epsilon L$ can be identified with~$\CC\langle S/G \rangle$, the free vector
space on the quotient set.
  \end{example}

  \begin{exercise}[]\label{thm:13}
 Prove this last assertion. 
  \end{exercise}

  \begin{exercise}[]\label{thm:14}
 For the finite group~$G$ acting on the finite set~$S$, consider the Borel
construction $S_G\to BG$.  Construct an isomorphism $S/G\to \pi _0(S_G)$.
Compare the information content of~$S_G$ and $\CC\otimes _\epsilon \CC\langle
S  \rangle$.  Which has more information?  How can you alter the other to
recover more information?
  \end{exercise}

  \begin{example}[]\label{thm:59}
 Any character of~$G$ (homomorphism $G\to \Cx$) induces an augmentation of the
group algebra.  Investigate the quotient~\eqref{eq:c5} using this augmentation.
Can you give a description analogous to `coinvariants'? 
  \end{example}

  \begin{remark}[]\label{thm:31}
 Recall the fusion category~$\sA$ in~\ref{subsec:c1.5}.  The analog of an
augmentation is a \emph{fiber functor} on~$\sA$: a homomorphism $\sA\to
\Vect$.  For $\sA=\Vect[G]$ the natural choice is pushforward under the map
$G\to *$ to a point.  
  \end{remark}

  \subsection{Quotient by a subgroup}\label{subsec:1.10}

We can form the ``sandwich''~\eqref{eq:c5} with any right $A$-module in place
of the augmentation.  For $A=\CC[G]$, if $G'\subset G$~is a subgroup, then
$\CC\langle G'\backslash G  \rangle$ is a right $G$-module; for~$G'=G$ it
reduces to the augmentation module~\eqref{eq:c6}.  If $L$~is a
$G$-representation, then the tensor product
  \begin{equation}\label{eq:c8}
     \CC\langle G'\backslash G \rangle\otimes \mstrut _{\CC[G]}L\cong \CC\otimes
     \mstrut _{\CC[G']}L 
  \end{equation}
is the vector space of coinvariants of the restricted $G'$-representation.
This represents the quotient by the subgroup~$G'$.

  \subsection*{Central extensions and projective actions}

  \subsection{Projective representations}\label{subsec:1.11}

There are many situations in which one encounters projective representations
of groups.  For example, suppose $A$~is an algebra and $L$~is an
\emph{irreducible} left module.  Let $G$~be a finite group that acts on~$A$
by algebra automorphisms, i.e., via a group homomorphism $\alpha \:G\to \Aut
A$.  They, typically, we can implement these symmetries on the module~$L$: if
$g\in G$ then we can find a linear automorphism $t\:L\to L$ such that 
  \begin{equation}\label{eq:3}
     t(a\xi ) = \bigl(\alpha (g)a\bigr)t(\xi ),\qquad a\in A,\quad \xi \in
     L. 
  \end{equation}
The map~ $t$ exists if the twisted $A$-module~$L^{\alpha }$ is isomorphic
to~$L$, and by Schur's lemma $t$~is determined up to a scalar.  In other
words, each~$g\in G$ determines a $\Cx$-torsor $T_g$, and the torsors depend
multiplicatively on~$G$.  They fit together into a group~$G^\tau $ which is a
central extension of~$G$ by~$\Cx$:
  \begin{equation}\label{eq:4}
     1\longrightarrow \Cx\longrightarrow G^\tau \longrightarrow
     G\longrightarrow 1 
  \end{equation}
A familiar example\footnote{In this example one uses $\zt$-gradings
everywhere.} has $A$~a Clifford algebra, $L$~an irreducible module, and
$G$~is the orthogonal group.  Then $G$~acts projectively on the Clifford
module, and one obtains the (s)pin central extension of the orthogonal
group. 

  \begin{remark}[]\label{thm:10}
 For a group extension~\eqref{eq:4} one considers representations $\rho
\:G^\tau \to \Aut(V)$ for which $\rho \res{\Cx}$~is scalar multiplication. 
  \end{remark}

  \subsection{The twisted group algebra}\label{subsec:1.12}

Suppose $G$~in~\eqref{eq:4} is a finite group.  Let $L^\tau \to G$ be the
complex line bundle associated to the principal $\Cx$-bundle~\eqref{eq:4}.
Define the \emph{twisted group algebra}
  \begin{equation}\label{eq:c10}
     A^\tau =\bigoplus\limits_{g\in G}L^\tau _g. 
  \end{equation}
Then $A^\tau $~inherits an algebra structure from the group structure of~$G$.
Furthermore, $G^\tau \subset A^\tau $ lies in the group of units.  An $A^\tau
$-module restricts to a linear representation of~$G^\tau $ on which the
center~$\Cx$ acts by scalar multiplication, and vice versa.  Observe that
there is no analog of the augmentation~\eqref{eq:c6} unless we are given a
splitting of the central extension~\eqref{eq:4}.  More generally, if
$H\subset G$ is a subgroup, then a splitting of the restriction
of~\eqref{eq:4} over~$H$ induces an $A^\tau $-module structure on~$\CC\langle
H\backslash G \rangle$, and we can use this to define the quotient by~$H$, as
in~\eqref{eq:c8}.  Absent the splitting, the projectivity obstructs the
quotient construction.

  \begin{exercise}[]\label{thm:15}
 Given an algebra homomorphism $\epsilon \:A^\tau \to \CC$, construct a
splitting of the central extension~\eqref{eq:4}.
  \end{exercise}

  \begin{remark}[]\label{thm:5}
 In field theory, the analog of an action by the central extension of a group
is called an ('t~Hooft) \emph{anomalous symmetry}, and the central
extension~\eqref{eq:4} is called the \emph{anomaly}.  In that context too,
the central extension obstructs the quotient construction (often called
\emph{gauging}).  See the discussion in the second half of
Lecture~\ref{sec:3}. 
  \end{remark}

  \subsection*{Some additional problems}

  \begin{figure}[ht]
  \centering
  \includegraphics[scale=.5]{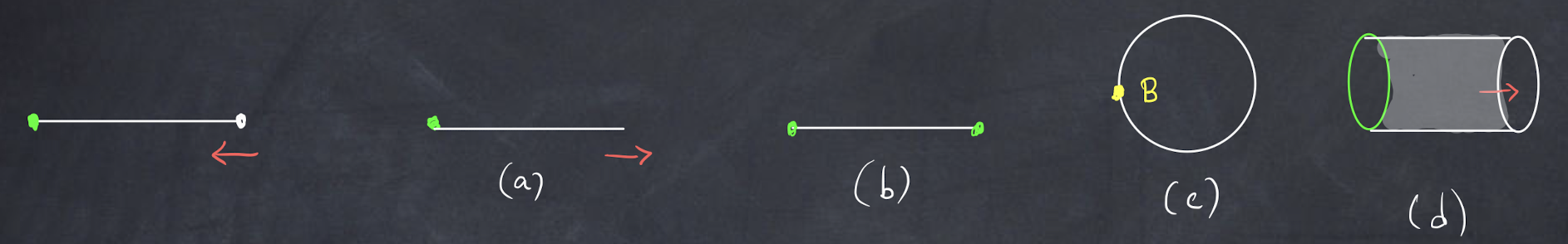}
  \vskip -.5pc
  \caption{Some bordisms in the topological field theory $\sr$}\label{fig:h1}
  \end{figure}

 \problem
 Let $G$~be a finite group, and let $\sigma \:\Bord_2\to \sC$ be the
2-dimensional finite gauge theory.  You can take $\sC=\Alg_1(\Vect)$ the
Morita 2-category of complex algebras, or $\sC=\Cat$ a 2-category of complex
linear categories.  In the former case $\sigma (\pt)=\GA$ is the group
algebra of~$G$; in the latter case $\sigma (\pt)=\Rep(G)$ is the category of
linear representations of~$G$.  Let $\rho $~be the right regular boundary
theory; then the first bordism in Figure~\ref{fig:h1} evaluates to the right
regular module~$A_A$ (or to the functor $\Rep(G)\to \Vect$ which maps a
$G$-module to its underlying vector space.)  Let $B$~be an $(A,A)$-bimodule.
The red arrow indicates incoming vs.~outgoing boundary components.  Compute
the value of~$\sr$ on the bordisms (a), (b), (c), and~(d) in
Figure~\ref{fig:h1}.  You may need to specify more data to achieve an
unambiguous answer.
 \endproblem

  \begin{figure}[ht]
  \centering
  \includegraphics[scale=.5]{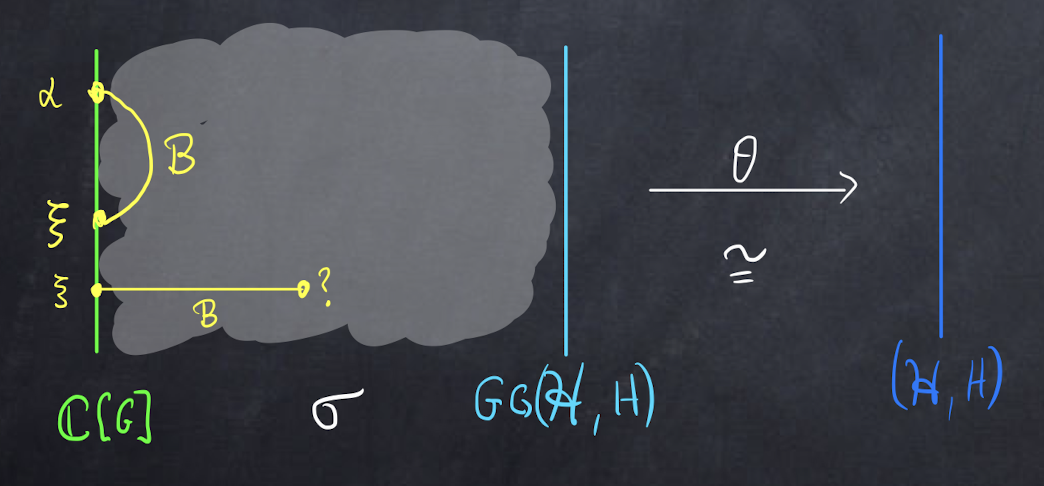}
  \vskip -.5pc
  \caption{Two defects in quantum mechanics}\label{fig:h2}
  \end{figure}

 \problem 
 Figure~\ref{fig:h2} uses the pictorial notation from the lecture.  Here we
are working with the 2-dimensional finite $G$-gauge theory of the previous
problem, which acts on a quantum mechanical system given by a Hilbert
space~$\sH$ and a Hamiltonian~$H$.  In the figure, $B$~is a (dualizable)
$(A,A)$-bimodule, $\xi \in B$ is a vector, and $\alpha \in B^*$ is a
functional.  Take the vertical line to be imaginary time.  The bottom defect
is at a fixed time, but data is missing at the right endpoint.  What data
goes there?  (It is an element of $\Hom\bigl(1,\sigma (L) \bigr)$, where
$L$~is the link of the point.)  What is the image of that defect
under~$\theta $?  (Note that it is a $\sr$-defect.)  For the top defect,
suppose that $\xi ,\alpha $ are at some times~$t_1,t_2$.  Compute the image
of this defect under~$\theta $.
 \endproblem

\newpage
\setcounter{section}{1}
  \section{Lecture 2: Formal structures in field theory; finite homotopy theories}\label{sec:2}
%\setcounter{figure}{16}
%\setcounter{footnote}{9}
% lastsubsec@ 23

In the first part of this lecture we recall some general structures in field
theory.  The framework we use is that introduced by Segal, adapted by Atiyah
for topological theories, and more recently exposed for general quantum field
theories by Kontsevich-Segal; see~\footref{ks}.  This framework is sometimes
referred to as ``functorial field theory'' (but see Remark~\ref{thm:1}(2),
which strongly suggests dropping `functorial').  What we say is developed most
in the literature for \emph{topological} field theories, though the structures
should exist in some form for general field theories.  In particular, the
theory of fully local---or fully extended---field theories is only at its
incipient stages in the nontopological case.  We introduce a class of
topological field theories called \emph{finite homotopy theories}.  These offer
a kind of semiclassical calculus of defects which is quite convenient and
powerful when it applies.  These theories are also a nice playground for
general ideas in field theory.  In the last part of the lecture we return to
finite symmetry in field theory and the central Definitions~\ref{thm:c9}
and~\ref{thm:c11}.
 
In the next lecture we take up one more general piece of
structure---quotients---and then illustrate these ideas in many examples.

  \subsection*{Basics}

  \subsection{A metaphor}\label{subsec:2.1}

This is a reprise of~\ref{subsec:1.13}, now in group form rather than algebra
form.  As with all analogies, not a perfect one:
  \begin{equation}\label{eq:c95}
     \textnormal{field theory $\sim$  representation of a Lie
     group} 
  \end{equation}
One imperfection can be improved if we take a module over an algebra instead of
a representation of a group, as in~\eqref{eq:5}, but we use the word `module'
below in a different sense.  Despite its drawbacks, this metaphor can guide us
in some limited way.

  \subsection{Discrete parameters}\label{subsec:2.2}

The discrete parameters on the right of~\eqref{eq:c95} might be considered to
be a dimension~$n$ and a Lie group~$G$ of dimension~$n$.  One would not just
fix a dimension and, say, study representations of a Lie group of
dimension~8!  Surely, one would distinguish between different Lie groups of
that dimension to fix the ``type'' of representation.  In field theory, too,
one can consider there to be two discrete parameters to fix the ``type'' of
field theory: (1)~the dimension~$n$ of \emph{spacetime}, and (2)~the
collection~$\sF$ of \emph{background fields}.  We can then speak of an
$n$-dimensional field theory on/over $\sF$.

  \begin{remark}[]\label{thm:25}
 The word `discrete parameter' is potentially confusing, since the choice of
background fields may include non-discrete fields, such as a Riemannian
metric.  It is the \emph{choice} of which background fields to include which
is discrete.  It may help to observe that discrete parameters are what we fix
in geometry to construct moduli spaces, for example the moduli space of
curves (of a fixed genus=discrete parameter).
  \end{remark}

  \subsection{Background fields}\label{subsec:2.3}

Informally, background fields of dimension~$n$ are sets assigned to
$n$-dimensional manifolds~$X$ together with a pullback under local
diffeomorphisms $f\:X'\to X$; they are required to depend locally on~$X$.
However, we need more than set-valued fields, since fields---such as
connections---may have internal symmetries.  In the case of ``$B$-fields''
there are higher internal symmetries: automorphisms of automorphisms.  Thus
we could take higher groupoids as the codomain for fields.  Instead we choose
simplicial sets.  Formally, let $\Man_n$ be the category whose objects are
smooth $n$-manifolds and whose morphisms are local diffeomorphisms of
$n$-manifolds.  There is a Grothendieck topology of open covers.

  \begin{definition}[]\label{thm:20}
 A \emph{field} in dimension~$n$ is a sheaf $\sF\:\Man_n\op\to \Set_\Delta $
with values in the category of simplicial sets. 
  \end{definition}

\noindent
 We have not spelled out the sheaf condition, which is the encoding of
locality.  Here $\sF$~could be a single field or a collection of fields; we
do not attempt to define irreducibility here.  A field on an $n$-manifold~$X$
is a 0-simplex in~$\sF(X)$.  An example of a collection of fields
is~\eqref{eq:7}, which we repeat here for convenience:
  \begin{equation}\label{eq:9}
     \sF=\{\textnormal{orientation, Riemannian metric, $\SO_3$-connection,
     section of twisted $S^2$-bundle}\} 
  \end{equation}
Observe that the last two components are sheaves that extend to~$\Man$, the
category of smooth manifolds and smooth maps, but the first two do not
extend.  See~\footnote{\label{FHop}Daniel~S. Freed and Michael~J. Hopkins,
\emph{Chern--{W}eil forms and abstract
  homotopy theory},
  \href{http://dx.doi.org/10.1090/S0273-0979-2013-01415-0}{Bull. Amer. Math.
  Soc. (N.S.) \textbf{50} (2013)}, no.~3, 431--468,
  \href{http://arxiv.org/abs/arXiv:1301.5959}{{\tt arXiv:1301.5959}}.
} for an exposition of simplicial sheaves on $\Man$, and in particular a
discussion of the sheaf condition.

  \begin{remark}[]\label{thm:16}
 One should think of~$\sF$ as a specification of type, not a choice of
specific fields on a specific manifold.  Rather, $\sF(X)$~is the simplicial
set of fields on a manifold~$X$, and---as said above---a choice of specific
fields on~$X$ is a 0-simplex in~$\sF(X)$.
  \end{remark}

  \begin{remark}[]\label{thm:18}
 The sheaf on~$\Man_n$ tells what a field on a \emph{single} manifold is; one
also needs to know what a smooth family of fields parametrized by an
arbitrary smooth manifold is, as well as how to base change such families.
In other words, we must sheafify over~$\Man$.  We do not dwell on this point
in these lectures.
  \end{remark}

  \begin{remark}[]\label{thm:17}
 The Axiom System we use for field theory does not include fluctuating
fields, nor does it include lagrangians; it supposes that all quantization
has already been executed.  In that sense it is a purely quantum axiom
system, as are the earlier axiom systems of Wightman and Haag-Kastler in
Minkowski spacetime, or for that matter those of Dirac, von Neumann, and
Irving Segal, and Mackey for quantum mechanics.  We can, however, contemplate
fiber bundles $\sF\to \overline{\sF}$ of fields and a quantization process
that passes from a field theory over~$\sF$ to a field theory over~$\bsF$.
  \end{remark}

  \begin{exercise}[]\label{thm:23}
 Identify the background fields (and dimension) in familiar quantum field
theories, such as:

 \begin{enumerate}[label=\textnormal{(\arabic*)}]

 \item A 2-dimensional $\sigma $-model with target~$S^2$

 \item Quantum mechanics of a particle on a ring

 \item 4-dimensional QCD and its low energy approximation, the pion theory

 \item 4-dimensional pure Yang-Mills theory

 \item Supersymmetric Yang-Mills theory 
 
 \item Dijkgraaf-Witten theory 
 
 \end{enumerate}
You may want to also identify the fields in the classical theory and the
fiber bundle that maps background fields in the classical theory to those in
the quantum theory.

  \end{exercise}

  \subsection{Field theory}\label{subsec:2.4}

These axioms capture a \emph{Wick-rotated} field theory, formulated on
compact Riemannian manifolds (if there is a Riemannian metric among the
background fields).  Fix $n\in \ZZ^{\ge1}$ and a collection~$\sF$ of
$n$-dimensional fields.  

A field theory is expressed in the language of sets and functions, but because
there are multiple layers of structure it is rather in the language of
categories and functors.  The domain is a bordism category~$\Bord_n(\sF)$ of
$n$-dimensional smooth manifolds with corners equipped with a choice of fields
from~$\sF$.  In the literature one finds detailed constructions for the fully
extended topological case, say in works of Lurie and
Calaque-Scheimbauer;\footnote{Jacob Lurie, \emph{On the classification of
topological field theories},
  Current developments in mathematics, 2008, Int. Press, Somerville, MA, 2009,
  pp.~129--280. \href{http://arxiv.org/abs/arXiv:0905.0465}{{\tt
  arXiv:0905.0465}}.
\newline
Damien Calaque and Claudia Scheimbauer, \emph{A note on the
  {$(\infty,n)$}-category of cobordisms},
  \href{http://dx.doi.org/10.2140/agt.2019.19.533}{Algebr. Geom. Topol.
  \textbf{19} (2019)}, no.~2, 533--655,
  \href{http://arxiv.org/abs/arXiv:1509.08906}{{\tt arXiv:1509.08906}} }
see~\footref{ks} for the nonextended general case.  We assume that all
\emph{topological} theories are fully extended downward in dimension, in which
case $\Bord_n(\sF)$~is a symmetric monoidal $n$-category.  One hopes a similar
strong locality is possible for \emph{nontopological} theories, but that awaits
further development.  In the nonextended case, we interpret `$\Bord_n(\sF)$'
as a 1-category $\Bord_{\langle n-1,n \rangle}(\sF)$ whose objects are closed
$(n-1)$-manifolds and whose morphisms are bordisms between them.

The codomain~$\sC$ of a field theory is a symmetric monoidal
$n$-category.\footnote{One can replace `$n$-category' with `$(\infty
,n)$-category' in our exposition.  Most of our examples are quite finite and
semisimple, but one should also use examples built on derived geometry.}  For
physical applications one takes $\sC$~to be \emph{complex linear}: the
linearity from superposition in quantum mechanics, and the complex ground
field from interference.  Thus we usually have $\Omega ^n\sC=\CC$ and $\Omega
^{n-1}\sC$ equivalent to the category~$\Vect$ of complex vector spaces or to
the category of $\zt$-graded complex vector spaces, discrete vector spaces
for topological theories and topological vector spaces for nontopological
theories.\footnote{The \emph{looping}~$\Omega \sC$ of an $n$-category is the
$(n-1)$-category $\Hom_{\sC}(1,1)$.  If $\sC$~is symmetric monoidal, as it is
in our case, then so too is $\Hom_{\sC}(1,1)$.  Hence one can iterate.}
These assumptions can be relaxed for applications outside of physics.

In the topological case (so $\sF$~consists of ``topological fields'', i.e., the
sheaf~$\sF$ is locally constant) we have the following.
  \begin{definition}[]\label{thm:19}
 A \emph{topological field theory} of dimension~$n$ over~$\sF$ is a symmetric
monoidal functor
  \begin{equation}\label{eq:c11}
     F\:\Bord_n(\sF)\longrightarrow \sC. 
  \end{equation}
  \end{definition}
We would like to think that a suitable variation of this definition applies
to all field theories.  This is clearest in the nonextended nontopological
case, in which we replace~$\sC$ by the 1-category~$t\!\Vect$ of suitable
complex topological vector spaces under tensor product, as in footnote~\footref{ks}.
 
  \begin{remark}[]\label{thm:c4}
 \ 
 \begin{enumerate}[label=\textnormal{(\arabic*)}]

 \item A field theory may be evaluated in a smooth family of manifolds and
background fields parametrized by a smooth manifold~$S$; the result is smooth
and should behave well under base change.  For example, typically correlation
functions are written as smooth functions of parameters---it is these spaces
of parameters that are the missing piece of structure.  Therefore,
\eqref{eq:c11}~should be sheafified over~$\Man$, the site of smooth manifolds
and smooth maps.  (This point has been emphasized by Stolz and
Teichner.\footnote{Stephan Stolz and Peter Teichner,
  \href{http://dx.doi.org/10.1090/pspum/083/2742432}{\emph{Supersymmetric field
  theories and generalized cohomology}}, Mathematical foundations of quantum
  field theory and perturbative string theory, Proc. Sympos. Pure Math.,
  vol.~83, Amer. Math. Soc., Providence, RI, 2011, pp.~279--340.
  \href{http://arxiv.org/abs/arXiv:1108.0189}{{\tt arXiv:1108.0189}}.  })
We already remarked on the need to sheafify~$\sF$ over~$\Man$ in
Remark~\ref{thm:18}.  We also need to sheafify the domain~$\Bord_n(\sF)$ and
the codomain~$\sC$ over~$\Man$.  

 \item A topological field theory is constrained by strong finiteness
properties that follow from Definition~\ref{thm:19}.  In the
metaphor~\eqref{eq:c95}, a \emph{topological} field theory is analogous to a
representation of a \emph{finite} group, though generalizations are possible.

 \item Definition~\ref{thm:19} does not incorporate unitarity, or rather its
Wick-rotated form: reflection positivity.  One needs extra structure to do so,
and it is an open problem to formulate \emph{extended} reflection
positivity.  Observe that not every representation of a semisimple Lie group
is unitarizable, and similarly for field theories, so one does not want
unitarity incorporated into the general definition.
 
 \item The notion of a \emph{free} theory is not evident from
Definition~\ref{thm:19}.  

 \item The collection of field theories of a fixed dimension~$n$ on a fixed
collection~$\sF$ of background fields has an associative composition law:
juxtaposition of quantum systems with no interaction, sometimes called
`stacking'.  There is a unit theory~$\bone$ for this operation.  For example,
if $F_1,F_2$~are theories, and $Y$~is a closed $(n-1)$-manifold with
background fields, then $(F_1\otimes F_2)(Y)=F_1(Y)\otimes F_2(Y)$.  The unit
theory has $\bone(Y)=\CC$; there is a single state on every space.  A unit
for the composition law is called an \emph{invertible field theory}.

 \end{enumerate}
  \end{remark}

  \begin{figure}[ht]
  \centering
  \includegraphics[scale=.5]{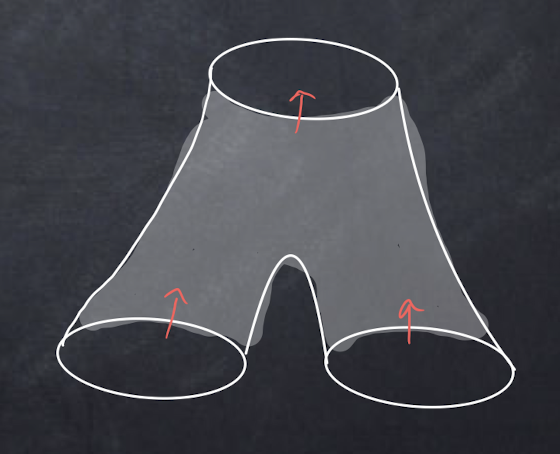}
  \vskip -.5pc
  \caption{The pair of pants bordism}\label{fig:9}
  \end{figure}

  \subsection{Product structures}\label{subsec:2.5}

Let $F$~be a field theory, as in~\eqref{eq:c11}.  As with any homomorphism in
algebra, structures and equations in the domain transport to the codomain.
One example is the product structure on~$F(S^k)$, $0\le k\le n-1$.  Assume
first that $F$~is a \emph{topological} theory.  To illustrate, let $n=2$ and
$k=1$.  Then the ``pair of pants'' bordism~$P$ in Figure~\ref{fig:9} induces
an algebra structure on the vector space~$F(S^1)$.  There is a diffeomorphism
of~$P$ that exchanges the two incoming circles, and this implies that the
multiplication is commutative.  For a 3-dimensional theory~$F'$, the
value~$F'(S^1)$ of the theory on a circle is a linear category, say, and
$P$~induces a monoidal structure.  The aforementioned diffeomorphism induces
\emph{data}, not a \emph{condition}, on~$F'(S^1)$, namely the data of a
\emph{braiding} for the monoidal structure.  As we go higher in category
number, the data and conditions proliferate. 
 
From a more sophisticated viewpoint, the sphere~$S^k$ is an
\emph{$E_{k+1}$-object} in the bordism category, and the
functor~\eqref{eq:c11} induces an $E_{k+1}$ structure on its image.

In a nontopological theory one also has these product structures, but now
they depend continuously on background fields, such as a conformal structure
or Riemannian metric.  So on the topological vector space attached
to~$S^{n-1}$ in a limit of zero radius we obtain some version of an operator
product expansion.  In 2-dimensional conformal field theories this leads to
vertex operator algebras. 

  \begin{exercise}[]\label{thm:24}
 Draw pictures that illustrate the algebra structure on $S^0$ in the bordism
category.  Be sure to verify associativity and also to identify the unit.
Include topological background fields, such as an orientation or spin
structure.
  \end{exercise}

  \subsection*{Finite homotopy theories I}

As mentioned in the introduction, this is a useful class of theories, both in
applications and for general theory.  One constructs a finite homotopy theory
via a version of the Feynman path integral.  Such a theory is simultaneously
more special than a general quantum field theory---because one has finite sums
rather than integrals over infinite dimensional spaces---and also more
general---because one integrates in all codimensions, and in positive
codimension one sums in a higher category rather than simply summing complex
numbers.
 
This class of topological field theories was introduced by Kontsevich in~1988
and was picked up by Quinn a few years later.  They are also the subject of a
series of a series of papers by Turaev in the early 2000's.  These
\emph{finite homotopy theories} lend themselves to explicit computation using
topological techniques.

The ``finite path integral'' quantization in \emph{extended} field theory was
introduced in~1992 in~\footnote{Daniel~S. Freed, \emph{Higher algebraic
structures and quantization}, Comm.  Math. Phys. \textbf{159} (1994), no.~2,
343--398, \href{http://arxiv.org/abs/arXiv:hep-th/9212115}{{\tt
arXiv:hep-th/9212115}}.  }; see also \S3 and \S8
of~\footnote{Daniel~S. Freed, Michael~J. Hopkins, Jacob Lurie, and Constantin
Teleman,
  \emph{Topological quantum field theories from compact {L}ie groups}, A
  celebration of the mathematical legacy of {R}aoul {B}ott, CRM Proc. Lecture
  Notes, vol.~50, Amer. Math. Soc., Providence, RI, 2010, pp.~367--403.
  \href{http://arxiv.org/abs/arXiv:0905.0731}{{\tt arXiv:0905.0731}}.  }.
The modern approach uses \emph{ambidexterity} or \emph{higher
semiadditivity}, as introduced by Hopkins-Lurie\footnote{Michael~J. Hopkins
and Jacob Lurie, \emph{Ambidexterity in K(n)-local stable
  homotopy theory}.
  \url{https://www.math.ias.edu/~lurie/papers/Ambidexterity.pdf}. preprint. 
See also Gijs Heuts and Jacob Lurie,
  \href{http://dx.doi.org/10.1090/conm/613/12236}{\emph{Ambidexterity}},
  Topology and Field Theories, Contemp. Math., vol. 613, Amer. Math. Soc.,
  Providence, RI, 2014, pp.~79--110 as well as Yonatan Harpaz, \emph{Ambidexterity and the universality of finite spans},
  \href{http://dx.doi.org/10.1112/plms.12367}{Proc. Lond. Math. Soc. (3)
  \textbf{121} (2020)}, no.~5, 1121--1170,
  \href{http://arxiv.org/abs/arXiv:1703.09764}{{\tt arXiv:1703.09764}} and Shachar Carmeli, Tomer~M. Schlank, and Lior Yanovski, \emph{Ambidexterity in
  chromatic homotopy theory},
  \href{http://dx.doi.org/10.1007/s00222-022-01099-9}{Invent. Math.
  \textbf{228} (2022)}, no.~3, 1145--1254,
  \href{http://arxiv.org/abs/arXiv:1811.02057}{{\tt arXiv:1811.02057}}.  }.
We do not say much about quantization in these lectures.

In these lectures we use finite homotopy theories as examples of
$(n+1)$-dimensional topological field theories~$\sigma $ which act on
$n$-dimensional quantum field theories.  For that reason, in this section we
denote the dimension by~`$m$' rather than~`$n$'; in the application to
symmetry, $m=n+1$.

  \subsection{$\pi $-finite spaces, maps, simplicial sheaves, \dots
}\label{subsec:2.6}

We defined $\pi $-finite spaces and $\pi $-finite maps in
Definition~\ref{thm:c53}.  I invite you to review it now.  We will also use
$\pi $-finite infinite loop spaces and $\pi $-finite infinite loop maps.

  \begin{remark}[]\label{thm:c56p}
 \ 

 \begin{enumerate}[label=\textnormal{(\arabic*)}]

 \item A simplicial sheaf is $\pi $-finite if its values are $\pi $-finite
simplicial sets.

 \item These variations pertain to the relative cases of maps, as in
Definition~\ref{thm:c53}(2).  

 \end{enumerate} 
  \end{remark}

  \begin{example}[]\label{thm:c57}
 Fix $m\in \ZZ^{\ge1}$ and consider the simplicial sheaves of fields which
assign to an $m$-manifold~$W$:
  \begin{equation}\label{eq:c64}
     \begin{aligned} \tsF(W) &= \{\textnormal{Riemannian metric,
      $\SU_2$-connection}\} \\ \sF(W) &= \{\textnormal{Riemannian metric,
      $\SO_3$-connection}\} \\ \end{aligned} 
  \end{equation}
There is a map $p\:\tsF\to \sF$ which takes an $\SU_2$-connection to the
associated $\SO_3$-connection.  The map~$p$ is a fiber bundle of simplicial
sheaves.  Neither $\tsF$~nor $\sF$~is $\pi $-finite, but the map~$p$ is $\pi
$-finite.  The fiber over a principal $\SO_3$-bundle $\overline{P}\to W$ is the
groupoid of lifts to a principal $\SU_2$-bundle $P\to W$ (which may be empty).
If nonempty, the fiber is a torsor over the groupoid of double covers of~$W$,
i.e., a gerbe.  The groupoid of double covers is the fundamental groupoid of
the mapping space $\Map(W,B\!\bmut)$, where $\bmut=\{\pm1\}$~is the center
of~$\SU_2$.  Observe that $B\!\bmut\simeq \RP^{\infty}$ is a $\pi $-finite
space, in fact a $\pi $-finite infinite loop space.  As a matter of notation,
we write `$\BNG$' for the sheaf of $G$-connections---here $G$~is a Lie
group---and the group extension $\bmut\to \SU_2\to \SO_3$ leads to the fiber
bundle
  \begin{equation}\label{eq:10}
     \begin{gathered} \xymatrix{B^{}_{\nabla }\hneg\SU_2\ar[r] & B^{}_{\nabla
     }\hneg\SO_3\ar[d]^{w_2}\\&B^2\!\bmut} \end{gathered} 
  \end{equation}
of simplicial sheaves on~$\Man$ (which we restrict to~$\Man_m$).  The
vertical map is a geometric form of the second Stiefel-Whitney class.
  \end{example}

\subsection{Cocycles}\label{subsec:2.7}

We abuse the term `cocycle', which should be reserved for cohomology theories
described in terms of cochain complexes.  We use it for any `geometric
representative' of a cohomology class.  Thus, if $\{E_q\}$~is a spectrum---a
sequence of pointed infinite loop spaces that represent a cohomology
theory---then a ``cocycle'' of degree~$q$ on a space~$X$ can be taken to be a
continuous map $X\to E_q$.  In any model there is a zero cocycle; in this model
it is the constant map to the basepoint of~$E_q$.
 
The data that determines a finite homotopy theory is a triple
$(m,\sX,\lambda )$ in which $m\in \ZZ^{\ge1}$, the space~$\sX$ is $\pi
$-finite, and $\lambda $~is a cocycle on~$\sX$ of degree~$m$.  We denote the
theory as `$\sXd m{(\sX,\lambda )}$' or `$\sXd m{\sX}$', the latter in case
$\lambda =0$.  A nonzero cocycle~$\lambda $ encodes an \emph{'t~Hooft
anomaly} in case $\sXd m{(\sX,\lambda )}$ is part of symmetry data.

  \subsection{Examples}\label{subsec:2.15}

 I hope the following list helps to relate this account to familiar terrain.

 \begin{enumerate}[label=\textnormal{(\arabic*)}]

 \item Let $G$~be a finite group.  Its classifying space~$\sX=BG$ is an
Eilenberg-MacLane space, so is $\pi $-finite.  Set~$\lambda =0$.  Then for
any~$m$ the triple $(m,BG,0)$ gives rise to finite $G$-gauge theory in
dimension~$m$. 

 \item Now suppose $\lambda $~represents a cohomology class in $H^m(BG;\Cx)$.
Then $\sXd m{(BG,\lambda )}$ is a twisted finite $G$-gauge theory, a
Dijkgraaf-Witten theory.

 \item Let $A$~be a finite abelian group and $p\in \ZZ^{\ge0}$.  Set
$\sX=\BnA{p+1}$, an Eilenberg-MacLane space $K(A,p+1)$.  A map into
$\BnA{p+1}$ represents a higher $A$-gerbe, which is a background field for a
``$p$-form'' symmetry.  More standardly, this is the symmetry group~$\BnA p$,
which is a homotopical form of a group.  For any $m\in \ZZ^{\ge1}$, the
theory $\sXd m{\BnA{p+1}}$ counts these higher $A$-gerbes.  In the context of
Definitions~\ref{thm:c9} and ~\ref{thm:c11} it encodes $\BnA p$-symmetry.

 \item Example~\ref{thm:c54} describes the classifying space of a
\emph{2-group}, which is a (general) path-connected 2-finite space. 
 
 \end{enumerate} 

  \begin{exercise}[]\label{thm:26}
 Identify the triple $(m,\sX,\lambda )$ for a spin Chern-Simons theory with
finite gauge group. 
  \end{exercise}

  \subsection*{Finite homotopy theories II}

Fix~$m\in \ZZ^{\ge1}$ and suppose $p\:\tsF\to \sF$ is a $\pi $-finite fiber
bundle of simplicial sheaves $\Man_m\op\to \Set_\Delta $; see~\eqref{eq:10} for
an example.  The basic idea is that there is a finite process that takes an
$m$-dimensional field theory~$\tF$ over~$\tsF$ as input and produces an
$m$-dimensional field theory~$F$ over~$\sF$ as output.  One obtains~$F$
from~$\tF$ by summing over the (fluctuating) fields in the fibers of~$p$.
Since $p$~is $\pi $-finite, this is a finite sum---a finite version of the
Feynman path integral.

  \begin{remark}[]\label{thm:c58}
 \ 

 \begin{enumerate}[label=\textnormal{(\arabic*)}]

 \item It often happens that the theory~$\tF$ is\footnote{One point of view,
advocated by Nathan Seiberg, is:  `classical' field theory $=$ invertible
field theory.} ``classical'', in which case it is an invertible field theory.
Then $F$~is its quantization.

 \item The framework is most developed for \emph{topological} field theories,
in which case we can work in \emph{extended} field theory.

 \end{enumerate} 
  \end{remark}

We do not give a systematic treatment of this quantization.  Rather, we
illustrate through an example that brings in the semiclassical mapping
spaces that are our focus.  This example is relevant for many 4-dimensional
gauge theories, in which case the abelian group~$A$ in the example is a
subgroup of the center of the gauge group.  (One can change the numbers to
apply this example in any dimension.)

  \begin{example}[]\label{thm:c59}
 Let $A$~be a finite abelian group and set~$\sX=\BnA2$.  For definiteness fix
dimension~$m=5$.  Our aim is to construct a 5-dimensional topological field
theory~$F=\sXd 5{\BnA2}$.  In the terms above: $\tsF$~is the simplicial sheaf
on~$\Man_5$ that assigns to a 5-manifold~$W$ the 2-groupoid $\pi
_{\le2}\Map(W,\BnA2)$ (made into a simplicial set), $\tF$~is the tensor unit
theory, and $\sF$~is the trivial simplicial sheaf that assigns a point to
each 5-manifold~$W$.  (The theory~$F$ is unoriented: there are no background
fields.)  We have not specified the codomain~$\sC$ of the theory, and one has
latitude in this choice.  For our purposes we assume standard choices at the
top three levels: $\Omega ^3\sC=\Cat$ is a linear 2-category of complex
linear categories, from which it follows that $\Omega ^4\sC=\Vect$ is a
linear 1-category of complex vector spaces and $\Omega ^5\sC=\CC$.
 
Let $M$~be a closed manifold.  Then $F(M)$~is the quantization of the mapping
space 
  \begin{equation}\label{eq:c65}
     \Xm M=\Map(M,\sX) 
  \end{equation}
The nature of that quantization depends on~$\dim M$. 
 
\medskip
 $\dim M=5$: \parbox[t]{26em}{The quantization is a (rational) number, a
weighted sum over homotopy classes of maps $M\to \sX$:}
  \begin{equation}\label{eq:c66}
     F(M) = \sum\limits_{[\phi ]\in \pi _0(\Xm M)}\frac{\#\pi _2(\Xm M,\phi
     )}{\#\pi _1(\Xm M,\phi )} \;=\;
     \frac{\#H^0(M;A)}{\#H^1(M;A)}\;\#H^2(M;A).  
  \end{equation}

\medskip
 $\dim M=4$: \parbox[t]{26em}{The quantization is the vector space of locally
constant complex-valued functions on~$\Xm M$:} 
  \begin{equation}\label{eq:c67}
     F(M) = \Fun\bigl(\pi _0(\Xm M)\bigr) = \Fun\bigl(H^2(M;A)
     \bigr). \phantom{MMMMMMMN} 
  \end{equation}

\medskip
 $\dim M=3$: \parbox[t]{26em}{The quantization is the linear category of flat
vector bundles  (local systems) over~$\Xm M$:}  
  \begin{equation}\label{eq:c68}
     \phantom{M}\begin{aligned} F(M) = \Vect\bigl(\pi _{\le1}(\Xm M) \bigr) &=
      \Vect\bigl(H^2(M;A) \bigr)\times \Rep\bigl(H^1(M;A) \bigr) \\ &\simeq
      \Vect\bigl(H^2(M;A)\times H^1(M;A)\dual \bigr),\end{aligned} 
  \end{equation}
 \hspace{6.5em}\parbox[t]{26em}{where for a finite abelian group~$B$ we denote by $B\dual$ the Pontrjagin dual
group of characters.  (If $M$~is oriented,
there is an isomorphism $H^1(M;A)\dual\cong H^2(M;A\dual)$.)}
  \end{example} 

  \begin{remark}[]\label{thm:c60}
In this example $\sX$~is an infinite loop space---an Eilenberg-MacLane
space---which explains the cohomological translations in
\eqref{eq:c66}--\eqref{eq:c68}.
  \end{remark}

As a further illustration, we describe the quantization of a bordism of top
dimension, which leads to a correspondence diagram of mapping spaces. 

  \begin{example}[$\sX=\BnA2$ redux]\label{thm:c61}
 Suppose $M\:N_0\to N_1$ is a 5-dimensional bordism between closed
4-manifolds $N_0,N_1$.  The restriction maps to incoming and outgoing
boundaries
  \begin{equation}\label{eq:c69}
     \begin{gathered} \xymatrix{&\Xm M\ar[dl]_{p_0} \ar[dr]^{p_1} \\
     \Xm{N_0}&& \Xm{N_1}} \end{gathered} 
  \end{equation}
form a correspondence diagram of mapping spaces.  The quantization
$F(M)\:F(N_0)\to F(N_1)$ maps a function $f\in \Fun\bigl(\pi
_0(\Xm{N_0})\bigr)$ to $(p_1)_*(p_0)^*f$, where the pushforward~$(p_1)_*$~is
the ``weighted finite homotopy sum'' or ``finite path integral''
  \begin{equation}\label{eq:c70}
     \bigl[(p_1)_*g \bigr](\psi ) = \sum\limits_{[\phi ]\in \pi _0(p_1\inv
     \psi )}\frac{\#\pi _2(p_1\inv \psi ,\phi )}{\#\pi _1(p_1\inv \psi ,\phi
     )},\qquad g\in \Fun\bigl(\pi _0(\Xm M)\bigr),\quad \psi \in \Xm{N_1}. 
  \end{equation}
  \end{example}

  \begin{remark}[]\label{thm:c62}
 If $N_0=\emptyset $, then $\Xm{N_0}=*$ and we obtain an element of
$\Hom\bigl(1,F(N_1) \bigr)$, i.e., a vector in the vector space~$F(N_1)$.
  \end{remark}

In terms of the paradigm at the beginning of this subsection, the example given
has trivial~ $\tF$.  We now give an example in which $\tF$~is a nontrivial
invertible theory.  The data that defines it is a pair~$(\sX,\lambda )$
consisting of a $\pi $-finite space~$\sX$ and a cocycle~$\lambda $ on~$\sX$.
Typically we need a generalized orientation to integrate~$\lambda $, depending
on the generalized cohomology theory in which $\lambda $~is a cocycle.

  \begin{example}[twisted $\sX=\BnA2$]\label{thm:c63}
 We continue with $\sX=\BnA2$, and now specialize to $A=\zt$ the cyclic group
of order~2.  Then\footnote{Let $\iota \in H^2(B^2\zt;\zt)$ be the
tautological class.  Then $\iota \smile \Sq^1\!\iota \in H^5(B^2\zt;\zt)$
becomes the nonzero class after extending coefficients $\zt\to \Cx$.}
$H^5(\sX;\Cx)\cong H^6(\sX;\ZZ)$ is cyclic of order~2.  Let $\lambda $~be a
cocycle that represents this class.  The quantizations in
Example~\ref{thm:c59} are altered as follows.  For $\dim M=5$ weight the sum
in~\eqref{eq:c66} by $\langle \phi ^*\lambda ,[M] \rangle$, where $[M]$~is the
fundamental class.\footnote{Since $\lambda $~is induced from a mod~2 class,
orientations are not necessary---we can proceed in mod~2 cohomology.}  For
$\dim M=4$ the transgression of~$\lambda $ to~$\Xm M$ induces a flat complex
line bundle (of order~2) $L\to \Xm M$; now \eqref{eq:c67}~becomes the space of
flat sections of $L\to \Xm M$.  Similarly, for $\dim M=3$ the
cocycle~$\lambda $ transgresses to a twisting of $K$-theory, and the
quantization is a category of twisted vector bundles.  The quantization in
Example~\ref{thm:c61} is also altered using transgressions of~$\lambda $: they
produce line bundles $L_0\to \sX^{N_0}$ and $L_1\to \sX^{N_1}$ as well as an
isomorphism $p_0^*(L_0)\xrightarrow{\;\cong \;}p_1^*(L_1)$.  Then
$(p_1)_*(p_0)^*$~maps sections of $L_0\to \sX^{N_0}$ to sections of $L_1\to
\sX^{N_1}$. 
  \end{example}

  \subsection*{Domain walls and boundaries in finite homotopy theories}

  \subsection{Semiclassical domain walls}\label{subsec:2.10}

 Fix $m\in \ZZ^{\ge1}$ and let $\Xln1$, $\Xln2$ be pairs of $\pi $-finite
spaces and degree~$m$ cocycles.  In the following we use trivializations of
cocycles.  In a model with cochain complexes, a trivialization of a
degree~$m$ cocycle~$\lambda $ is a cochain of degree~$m-1$ whose differential
is~$\lambda $.  In a model in which $\lambda $~is a map to a space in a
spectrum, then a trivialization is a null homotopy of the map, i.e., a
homotopy to the constant map with value the basepoint.

  \begin{definition}[]\label{thm:c69}
 A \emph{semiclassical domain wall} from $\Xln1$ to~$\Xln2$ is a
pair~$(\sY,\mu )$ consisting of a $\pi $-finite space~$\sY$ equipped with a
correspondence  
  \begin{equation}\label{eq:c78}
     \begin{gathered} \xymatrix{&\sY\ar[dl]_{f_1} \ar[dr]^{f_2} \\
     \sX_1&&\sX_2} \end{gathered}
  \end{equation}
and a trivialization~$\mu $ of $f_2^*\lambda _2 - f_1^*\lambda _1$. 
  \end{definition}

  \begin{remark}[]\label{thm:c70}
 \ 
 \begin{enumerate}[label=\textnormal{(\arabic*)}]

 \item We have written~\eqref{eq:c78} to conform to standard practice for a
correspondence from~$\sX_1$ to~$\sX_2$, but to fit our right/left
conventions, as illustrated in Figure~\ref{fig:c2}, we might have
swapped~$\sX_1$ and~$\sX_2$.

 \item If $(\sY',\mu ')$ is a $\pi $-finite space and a degree~$m-1$ cocycle,
then there is a new semiclassical domain wall 
  \begin{equation}\label{eq:c79}
     \begin{gathered} \xymatrix@C-1pc{&(\sY\times \sY',\mu +\mu
     ')\ar[dl]\ar[dr]\\      \Xln1&&\Xln2} \end{gathered} 
  \end{equation}
This corresponds to tensoring with the $(m-1)$-dimensional theory~$(\sY',\mu
')$ on the domain wall.

 \end{enumerate}
  \end{remark}

  \subsection{Quantization of a semiclassical domain wall}\label{subsec:2.11}

To quantize a semiclassical domain wall, we use~\eqref{eq:c78} to construct a
mapping space.  Let $M$~be a closed manifold presented as a union 
  \begin{equation}\label{eq:c80}
     M=M_1\cup \mstrut _ZM_2 
  \end{equation}
of manifolds with boundary along the common boundary~$Z$; then $Z\subset M$ is
a codimension~1 cooriented submanifold.  Form the mapping space
  \begin{equation}\label{eq:c81}
     \sM=\bigl\{ (\phi _1,\phi _2,\psi ):\phi _i\:M_i\to \sX_i,\; \psi \:Z\to
     \sY,\; f_i\circ \psi =\phi _i\res Z \bigr\}. 
  \end{equation}
Now quantize~$\sM$ as illustrated in Example~\ref{thm:c59}.

  \subsection{Composition}\label{subsec:2.12}

Composition of semiclassical domain walls proceeds by homotopy fiber
product.  Suppose $\Xln1$, $\Xln2$, $\Xln3$ are $\pi $-finite spaces and
degree~$m$ cocycles, and let 
  \begin{equation}\label{eq:c82}
     \begin{aligned} (\sY',\mu ')\:\Xln1\longrightarrow \Xln2 \\ (\sY'',\mu
      '')\:\Xln2\longrightarrow \Xln3 \\ \end{aligned} 
  \end{equation}
be semiclassical domain walls.  Their composition 
  \begin{equation}\label{eq:c83}
     (\sY,\mu )\:\Xln1\longrightarrow \Xln3 
  \end{equation}
is constructed via the homotopy fiber product 
  \begin{equation}\label{eq:c84}
     \begin{gathered} \xymatrix{&&\sY\ar@{-->}[dl]\ar@{-->}[dr]\\
     &\sY'\ar[dl]\ar[dr] &&\sY''\ar[dl]\ar[dr] \\ \sX_1&&\sX_2&&\sX_3}
     \end{gathered} 
  \end{equation}
that is the composition of correspondence diagrams (in the homotopy
category); the trivialization~$\mu $ of~$\lambda _3-\lambda _1$ is the sum
$\mu _1+\mu _2$.  (For ease of reading, we omitted pullbacks in the previous
clause.)  We write~\eqref{eq:c84} with cocycles and trivializations as
follows:
  \begin{equation}\label{eq:c85}
     \begin{gathered} \xymatrix@C-1pc{&&(\sY,\mu '+\mu
     '')\ar@{-->}[dl]\ar@{-->}[dr]\\ 
     &(\sY',\mu ')\ar[dl]\ar[dr] &&(\sY'',\mu '')\ar[dl]\ar[dr] \\
     \Xln1&&\Xln2&&\Xln3}  \end{gathered}
  \end{equation}

  \subsection{Boundaries}\label{subsec:2.13}

 As in~\ref{subsec:1.15} we specialize domain walls to boundary theories,
here in the semiclassical world of finite homotopy theories.

  \begin{definition}[]\label{thm:c72}
 Let $\sX$~be a $\pi $-finite space and suppose $\lambda $~is a cocycle of
degree~$m$ on~$\sX$.

 \begin{enumerate}[label=\textnormal{(\arabic*)}]

 \item A \emph{right semiclassical boundary theory} of~$\Xl$ is a
pair~$(\sY,\mu )$ consisting of a $\pi $-finite space~$\sY$, a map $f\:\sY\to
\sX$, and a trivialization~$\mu $ of~$-f^*\lambda $. 

 \item A \emph{left semiclassical boundary theory} of~$\Xl$ is a
pair~$(\sY,\mu )$ consisting of a $\pi $-finite space~$\sY$, a map $f\:\sY\to
\sX$, and a trivialization~$\mu $ of~$f^*\lambda $. 

 \end{enumerate} 
  \end{definition}

\noindent
 The mapping spaces used for quantization specialize~\eqref{eq:c81}. 
 
In this finite homotopy context there is a special form for a regular
boundary theory.

  \begin{definition}[]\label{thm:c73}
 Let $\sX$~be a path connected $\pi $-finite space and suppose $\lambda $~is a
cocycle of degree~$m$ on~$\sX$.  A \emph{semiclassical right regular boundary
theory} of~$\Xl$ is a basepoint $f\:*\to \sX$ and a trivialization~$\mu $
of~$-f^*\lambda $.
  \end{definition}

\noindent
 In terms of Definition~\ref{thm:c72}, the semiclassical right regular
boundary theory is~$(*,\mu )$.  If the cocycle has positive degree in an
ordinary (Eilenberg-MacLane) cohomology theory, then it vanishes on a point
so we can and do take~$\mu =0$.

  \begin{remark}[]\label{thm:c75}
 The regular boundary theory amounts to an extra semiclassical (fluctuating)
field on the boundary, which is a trivialization of the bulk field (map
to~$\sX$).
  \end{remark}

  \begin{exercise}[]\label{thm:27}
 Define a domain wall between boundary theories.  For any $(m,\sX,0)$
consider the right semiclassical boundary theory~ D given by a basepoint $*\to
\sX$ and the right semiclassical boundary theory~ N given by the identity map
$\id_{\sX}\:\sX\to \sX$.  Construct a distinguished domain wall
$\textnormal{D}\to \textnormal{N}$ and a distinguished domain wall
$\textnormal{N}\to \textnormal{D}$.  What is their composition (in both
orders)?
  \end{exercise}

  \begin{example}[]\label{thm:c74}
 Let~$m=2$.  Fix a finite group~$G$ and let $\sX=BG$ with basepoint $*\to
BG$.  The quantization of the interval depicted in Figure~\ref{fig:c5} is the
quantization of the restriction map to the right endpoint 
  \begin{equation}\label{eq:c86}
     \Map\bigl(([0,1],\{0\}),(BG,*) \bigr)\longrightarrow \Map\bigl(\{*\},BG
     \bigr), 
  \end{equation}
which up to homotopy is the map $*\to BG$.  Choose the codomain $\sC=\Cat$ so
that, as in~\eqref{eq:c68}, the quantization of~$\Map(*,BG)$ is the category
$\Vect(BG)\simeq \Rep(G)$.  Then the quantization of the map $*\to BG$, or
better of the correspondence 
  \begin{equation}\label{eq:c87}
     \begin{gathered} \xymatrix{&\ast\ar[dl]\ar[dr] \\\ast&&BG}
     \end{gathered} 
  \end{equation}
of mapping spaces derived from Figure~\ref{fig:c5}, is the pushforward of the
trivial bundle over~$*$ with fiber~$\CC$ (the tensor unit).  This is the
regular representation of~$G$ in~$\Rep(G)$.  If, instead, we choose
$\sC=\Alg(\Vect)$, then $BG$~quantizes to the group algebra~$\GA$ and $*\to
BG$ quantizes to the right regular module.   

  \begin{exercise}[]\label{thm:22}
 Incorporate a nonzero cocycle in the form of a central extension
  \begin{equation}\label{eq:c88}
     1\longrightarrow \Cx\longrightarrow G^\tau \longrightarrow
     G\longrightarrow 1 
  \end{equation}
  \end{exercise}
  \end{example} 

  \subsection{A special sandwich}\label{subsec:2.14}

Let $\Xl$ be given and suppose $(\sY',\mu ')$ and $(\sY'',\mu '')$ are right
and left semiclassical boundary theories for~$\Xl$.  Then, as a special case
of the composition~\eqref{eq:c85}, the $(m-1)$-dimensional semiclassical
sandwich of~$\Xl$ between $(\sY',\mu ')$ and $(\sY'',\mu '')$ has as its
semiclassical data the pair $(\sY'\overset h{\times}\mstrut _{\sX}\sY'',\mu
'+\mu '')$, where $\sY'\overset h{\times}\mstrut _{\sX}\sY''$ is the
homotopy fiber product; observe that $\mu '+\mu ''$ is a cocycle of
degree~$m-1$.

  \subsection*{Defects}

Domain walls and boundaries are special cases of the general notion of a
\emph{defect} in a field theory.  Our discussion here is specifically for
\emph{topological} theories, though with modifications it applies more
generally (see Remark~\ref{thm:c8}(7) below).
 
  \subsection{Preliminary: The category $\Hom(1,x)$}\label{subsec:2.9}

We will encounter the expression `$\Hom(1,x)$' in a higher symmetric monoidal
category, so we begin by elucidating its meaning.  Suppose $\Vect$~is the
symmetric monoidal category of vector spaces, and $V\in \Vect$ is an object,
i.e., a vector space.  The tensor unit~$1$ in $\Vect$ is the vector
space~$\CC$ of scalars.  We usually identify a linear map $T\in \Hom(\CC,V)$
with $T(1)\in V$, so in this case $\Hom(1,V)$~is the space of vectors
in~$V$.  Similarly, if $C\in \Cat$ is a category, then $\Hom(1,C)$ can be
identified with the objects in~$C$.  On the other hand, for the Morita
2-category $\Alg(\Vect)$ of complex algebras, the tensor unit~$1$ is the
algebra~$\CC$ and for any algebra~$A$, the 1-category $\Hom(1,A)$ is the
category of left $A$-modules.   
 
Let $\sC$~be a symmetric monoidal $n$-category.  If $x\in \sC$, then
$\Hom(1,x)$~is an $(n-1)$-category.  It is possible that it is empty, or,
rather, in our usual linear situation that it only contains the zero object.
We also use `$\Hom(1,x)$' when $x\in \Omega ^\ell \sC$ is in some looping
of~$\sC$.  Then the homs are taken in $\Omega ^\ell \sC$.

  \subsection{Definition of a defect in a topological theory}\label{subsec:2.16}

Suppose $m$~is a positive integer, $\sF$~is a collection of background
fields, and 
  \begin{equation}\label{eq:c14}
     \sigma \:\Bord_m(\sF)\longrightarrow \sC 
  \end{equation}
is a topological field theory with values in a symmetric monoidal
$m$-category~$\sC$.  We describe defects of \emph{codimension}~$\ell $ in a
$k$-dimensional manifold~$M$, where $k\in \{0,1,\dots ,m\}$, $\ell \in
\{1,\dots ,m\}$, and $\ell \le k$.  Let $Z\subset M$ be a submanifold of
codimension~$\ell $, and let $\nu \subset M$ be an open tubular neighborhood
of~$Z\subset M$; assume the closure~$\bn$ is the total space of a fiber
bundle $\bn\to Z$ with fiber the closed $\ell $-dimensional disk.  The fiber
over~$p\in Z$ is denoted~$\bn_p$; its boundary~$\partial \bn_p$ is
diffeomorphic to the $\ell $-dimensional sphere~$S^{\ell -1}$.  It is the
\emph{link} of~$Z\subset M$ at~$p$; see Figure~\ref{fig:c4}.

  \begin{figure}[ht]
  \centering
  \includegraphics[scale=.35]{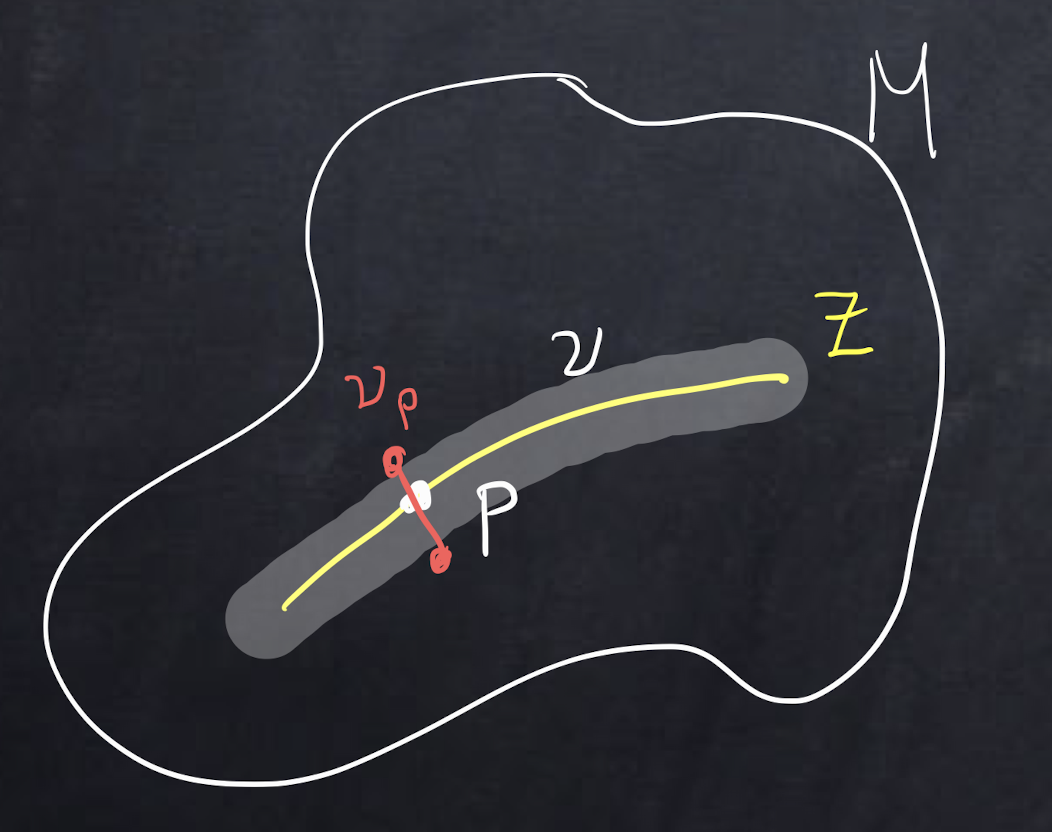}
  \vskip -.5pc
  \caption{The tubular neighborhood and link of a submanifold}\label{fig:c4}
  \end{figure}
 
  \begin{definition}[]\label{thm:c7}
 Assume that $M$~is a closed manifold and $Z\subset M$ is a closed
submanifold. 

 \begin{enumerate}[label=\textnormal{(\arabic*)}]

 \item A \emph{local defect} at~$p\in Z$ is an element 
  \begin{equation}\label{eq:c15}
     \delta _p\in \Hom\bigl(1,\sigma (\partial \bn_p) \bigr). 
  \end{equation}

 \item The \emph{transparent \textnormal{(}local\textnormal{)} defect} is
$\delta _p=\sigma (\bn_p)$.

 \item A \emph{global defect} on~$Z$ is a vector 
  \begin{equation}\label{eq:c16}
     \delta _Z\in \Hom\bigl(1,\sigma (\partial \bn) \bigr). 
  \end{equation}

 \item The \emph{transparent \textnormal{(}global\textnormal{)} defect} is
$\delta _Z=\sigma (\bn)$. 

 \end{enumerate} 
  \end{definition}

\noindent
 The transparent defects can safely be erased.

  \begin{remark}[]\label{thm:c8}
 We make several comments about this definition.

 \begin{enumerate}[label=\textnormal{(\arabic*)}]

 \item $M\setminus \nu $~is a compact manifold with boundary~$\partial \bn$.
Define the bordism $W\:\partial \bn\to \emptyset $ by letting the boundary be
incoming.  If $\delta _Z$~is a global defect, we evaluate the theory
on~$(M,Z,\delta _Z)$ as $\sigma (W)(\delta _Z)$.  This is of the same type
as the value~$\sigma (M)$ on the closed manifold~$M$: a complex number if
$\dim M=m$, a complex vector space if $\dim M=n-1$, etc.

 \item A normal framing of~$Z\subset W$ identifies each link~$\partial
\bn_p$, $p\in Z$, with the standard sphere~$S^{\ell -1}$.  Assuming enough
finiteness, $\sigma (S^{\ell -1})\in \Omega ^{\ell -1}\sC$
defines\footnote{The looping~$\Omega \sC$ of the symmetric monoidal
$n$-category~$\sC$ is the symmetric monoidal $(n-1)$-category $\Hom(1,1)$ of
endomorphisms of the tensor unit.  We can iterate this construction.  The
theory~$\sigma ^{\ell -1}$ takes values in~$\Omega ^{\ell -1}\sC$.  The
cobordism hypothesis, both with and without singularities, is used to define
the theory and its boundary theory.} an $(m-\ell +1)$-dimensional field
theory~$\sigma ^{(\ell -1)}$---the dimensional reduction of~$\sigma $
along~$S^{\ell -1}$---and a local defect~ $\delta _p$~determines a left
boundary theory~$\delta ^{(\ell -1)}$ for~$\sigma ^{(\ell -1)}$, again
assuming sufficient finiteness.  Using the normal framing it makes sense to
assign a single local defect
  \begin{equation}\label{eq:c18}
     \delta \in \Hom\bigl(1,\sigma (S^{\ell -1}) \bigr) 
  \end{equation}
to~$Z$.  The cobordism hypothesis computes an associated global defect
  \begin{equation}\label{eq:c17}
     \delta _Z \in \,\Hom\bigl(1, \sigma ^{(\ell -1)}(Z)\bigr) =
     \Hom\bigl(1,\sigma (Z\times S^{\ell -1}) \bigr) 
  \end{equation}
that we can use to evaluate the theory~$\sigma $ on~$(M,Z,\delta )$ as
in~(1).  Absent the normal framing, we can use a flat family of local defects
and some twisted dimensional reduction.  We can also use objects~$\delta $
that are invariant under a subgroup of the orthogonal group.  For example,
see~\S8.1 of~\footnote{\label{FTi}Daniel~S. Freed and Constantin Teleman,
\emph{Topological dualities in the
  Ising model}, \href{http://arxiv.org/abs/arXiv:1806.00008}{{\tt
  arXiv:1806.00008}}.  } for a discussion of topological defects using
orientation in place of normal framing.

 \item A defect on~$Z$ may be tensored with a standalone field theory on~$Z$
to obtain a new defect.  This corresponds to composing with an element
of~$\Hom(1,1)$ in~\eqref{eq:c15} or~\eqref{eq:c16}.  We illustrated that in a
special case in Remark~\ref{thm:c70}(2).

 \item The sheaf of background fields on a defect need not agree with the
sheaf of background fields in the bulk; the former need only map to the
latter.  Thus we can have a spin defect in an theory of oriented manifolds.

 \item Defects are also defined for manifolds~$M$ with boundaries and
corners, and we also allow boundaries, corners, and singularities in~$Z$.  In
short, we allow~$Z$ to be a stratified manifold.  Then different strata
of~$Z$ have different links, and we compute them and assign (local) defects
working from the lowest codimension to the highest.  We give several
illustrations in these lectures.

 \item If $M$~is closed with no boundary or corners, then $V=\sigma (M)$~is a
vector space and $\sigma \bigl([0,1]\times M \bigr)$ is the identity
map~$\id_V$.  A defect supported in the interior of $[0,1]\times M$ evaluates
under~$\sigma $ to a linear operator on~$V$.  In this situation the terms
`operator' and `observable' are often used in place of with `defect'.

 \item There are also (nontopological) defects in nontopological theories,
but then unless we are in maximal codimension an extension beyond a two-tier
theory is perhaps implicit.  In nontopological theories local defects take
values in a limit as the radius of the linking sphere shrinks to zero.  For
$\dim M=m$ and $\dim Z=0$ the resulting \emph{point defects} are often called
`local operators'.  For $\dim Z=1$ they are \emph{line defects}.\footnote{And
for $\dim Z=2$ they are called \emph{surface defects}.  The progression
point-line-surface is an uncomfortable mishmash of point-line-plane (affine
geometry) and point-curve-surface (differential geometry).}  It is
increasingly common in the physics literature to consider a \emph{1-category}
of line defects; higher dimensional defects and higher categories of defects
also appear.

 \end{enumerate}
  \end{remark}

  \subsection{Composition law on local defects}\label{subsec:2.17}

The value of a topological field theory~$\sigma $ on~$S^{\ell -1}$ is an
$E_{\ell }$-algebra.  By~\ref{subsec:2.5}, this leads to a composition law on
defects, either for local defects \eqref{eq:c15},~\eqref{eq:c18} or global
defects \eqref{eq:c16},~\eqref{eq:c17}.  If $Z$~is normally framed, one can
consider two parallel copies~$Z',Z''$, and then a normal slice of the
complement of open tubular neighborhoods of~$Z',Z''$ inside a closed tubular
neighborhood of~$Z$ is the ``pair of pants'' that defines the composition
law of the $E_\ell $-structure.  The composition law on topological point
defects is a topological version of the usual operator product expansion.
The composition law gives rise to the dichotomy between \emph{invertible
defects} and \emph{noninvertible defects}.

  \subsection*{Semiclassical defects in finite homotopy theories}

Let $F$~be an $m$-dimensional finite homotopy theory based on a $\pi $-finite
space~$\sX$ and a cocycle~$\lambda $ of degree~$m$ on~$\sX$.

  \subsection{Preliminary on transgression}\label{subsec:2.23}

Fix $\ell \in \{1,\dots ,m\}$.  Define 
  \begin{equation}\label{eq:c73}
     \LlX=\Map(S^{\ell -1},\sX).
  \end{equation}
Consider the diagram 
  \begin{equation}\label{eq:11}
     \begin{gathered} \xymatrix{\LlX \times S^{\ell -1}
     \ar[r]^<<<<<e\ar[d]_{\pi _1}
     & \sX \\\LlX} \end{gathered} 
  \end{equation}
in which $e$~is evaluation and $\pi _1$~is projection onto the first factor.
The composition $(\pi _1)_*(e)^*\lambda $ is a cocycle of degree $m-\ell +1$
on~$\LlX$ called the \emph{transgression} of~$\lambda $.  Note that the
stable framing of the sphere is used to execute the pushforward.

  \subsection{Semiclassical local defects}\label{subsec:2.21}

For a defect on a submanifold of codimension~$\ell \in \ZZ^{\ge1}$, the link
is~$S^{\ell -1}$---canonically if the normal bundle is framed---and so the
mapping space on the link is $\LlX$, the mapping space defined
in~\eqref{eq:c73}.  By~\ref{subsec:2.23} the cocycle~ $\lambda $ transgresses
to a cocycle $\tau ^{\ell -1}\lambda $ on~$\LlX$ with a drop of degree by~$\ell
-1$.  Recall the definition of a local defect in Definition~\ref{thm:c7}(1) .

  \begin{definition}[]\label{thm:c68}
 Fix $m,\ell \in \ZZ^{\ge1}$ with $\ell \le m$.  Let $\sX$~be a $\pi $-finite
space and suppose $\lambda $~is a cocycle of degree~$m$ on~$\sX$.  A
\emph{semiclassical local defect} of codimension~$\ell $ for~$\Xl$ is a $\pi
$-finite map
  \begin{equation}\label{eq:c74}
     \delta \:\sY\longrightarrow \LlX 
  \end{equation}
and a trivialization~$\mu $ of~$\delta ^*(\tau ^{\ell -1}\lambda )$. 
  \end{definition}

\noindent
 Since $\LlX$~is $\pi $-finite, \eqref{eq:c74}~ amounts to a $\pi $-finite
space~$\sY$ and a continuous map~$\delta $.  The local quantum defect in
$\Hom\bigl(1,F(S^{\ell -1}) \bigr)$ is the quantization of the
map~\eqref{eq:c74}; see Remark~\ref{thm:c62} for an analogous quantization.

  \subsection{Semiclassical global defects}\label{subsec:2.22}

To pass from local to global we use a tangential structure.  As an example,
if $M$~is a closed manifold and $Z\subset M$ is a \emph{normally framed}
codimension~$\ell $ submanifold on which the defect~\eqref{eq:c74} is placed,
the value of the theory~$F$ on~$M$ with the defect on~$Z$ is the quantization
of the mapping space 
  \begin{equation}\label{eq:c77}
     \Map\bigl((M,Z),(\sX,\sY) \bigr) 
  \end{equation}
consisting of  pairs of maps  $\phi \:M\to \sX$  and $\psi \:Z\to  \sY$ that
satisfy a compatibility condition:  if $Z\times S^{\ell -1}\hookrightarrow M$
is the inclusion  of the boundary of a tubular  neighborhood of~$Z\subset M$,
and $\phi '\:Z\to \LlX$ is the transpose of the composition
  \begin{equation}\label{eq:c75}
     Z\times S^{\ell -1}\longhookrightarrow M\xrightarrow{\;\;\phi
     \;\;}\sX, 
  \end{equation}
then the diagram 
  \begin{equation}\label{eq:c76}
     \begin{gathered} \xymatrix@C+1pc@R+1pc{&\sY\ar[d]^{\delta }\\ Z\ar[r]^{\phi
     '}\ar[ur]^{\psi }&\LlX} \end{gathered}
  \end{equation}
is required to commute. 

  \begin{remark}[]\label{thm:c66}
 \ 
 \begin{enumerate}[label=\textnormal{(\arabic*)}]

 \item One should use instead a mapping space of triples $(\phi ,\psi ,\gamma
)$ where instead of demanding that~\eqref{eq:c76} commute we specify a
homotopy $\gamma \:\delta \circ \psi \to \phi '$.  This sort of derived
mapping space should in principle replace all of the strict mapping spaces we
write throughout the paper.  However, the homotopy can be incorporated into a
tubular neighborhood of~$Z$, so in fact nothing is lost by using the strict
mapping space. 

 \item There are many variations of this basic scenario.  The defect may have
support on a manifold with boundary or corners, or more generally on a
stratified manifold.  Such is the case for the $\rho $-defects in
Definition~\ref{thm:c33} below; a further example is in Figure~\ref{fig:c27}.

 \end{enumerate}
  \end{remark}

   \subsection{Composition of semiclassical local defects}\label{subsec:c5.A}

 The general composition law on local defects is constructed using the higher
dimensional pair of pants or, in the case of $\rho $-defects as in
Figure~\ref{fig:c17}, using the higher dimensional pair of chaps.  Here we
state the semiclassical version of the first. 
 
Resume the setup of Definition~\ref{thm:c68}: $m,\ell \in \ZZ^{\ge1}$ are
integers with $\ell \le m$, and $\Xl$~is the finite homotopy data for an
$m$-dimensional theory~$F$.  Let $P$~be the $\ell $-dimensional pair of
pants: as a manifold with boundary,
  \begin{equation}\label{eq:c89}
     P=D^\ell \;\setminus\; B^\ell \amalg B^\ell , 
  \end{equation}
where $B^\ell \amalg B^\ell $ are embedded balls in the interior of~$D^\ell
$.  As a bordism, 
  \begin{equation}\label{eq:c90}
     P\:S^{\ell -1}\amalg S^{\ell -1}\longrightarrow S^{\ell -1}, 
  \end{equation}
where the domain spheres are the inner boundaries of~$P$ and the codomain
sphere is the outer boundary.  The cocycle~$\lambda $ on~$\sX$ transgresses
to an isomorphism
  \begin{equation}\label{eq:c91}
     \mu \:\pi _1^*(\tll) + \pi _2^*(\tll)\longrightarrow \tll 
  \end{equation}
of cocycles on~$\sX^P$.  Here $\pi _i\:\LlX\times \LlX\to \LlX$ is projection
onto the $i^{\textnormal{th}}$~factor, and in~\eqref{eq:c91} we omit pullbacks
under the source and target maps in the correspondence~\eqref{eq:c92} below.
Then the composition law on~$F(S^{\ell -1})$ is the quantization of the
correspondence
  \begin{equation}\label{eq:c92}
     \begin{gathered} \xymatrix@!@R-13pc@C-13pc{&\bigl(\sX^P\!,\,\mu
     \bigr)\ar[dl]_{r_0}\ar[dr]^{r_1} \\ 
     \bigl(\LlX\times \LlX,\,\pi _1^*(\tll) + \pi _2^*(\tll)\bigr) &&
     \bigl(\LlX,\,\tll\bigr)} \end{gathered} 
  \end{equation}
The composition law on~$F(S^{\ell -1})$ induces the composition law---the
fusion product---on $\Hom\bigl(1,F(S^{\ell -1}) \bigr)$, the higher category
of local codimension~$\ell $ defects.  Suppose given $(\sY_1,\mu _1)$ and
$(\sY_2,\mu _2)$ semiclassical local defects of codimension~$\ell $, as in
Definition~\ref{thm:c68}.  Then their product in $\Hom\bigl(1,F(S^{\ell -1})
\bigr)$ is the quantization of the composition~$r_1\circ g$ in the homotopy
fiber product 
  \begin{equation}\label{eq:c93}
     \begin{gathered} \xymatrix@!@R-12pc@C-12pc{&\bigl(\sY,\,\pi _1^*\mu _1 +
     \pi _2^*\mu _2+\mu \bigr) \ar@{-->}[dl]\ar@{-->}[d]^{g}\\
     \bigl(\sY_1\times \sY_2,\,\pi ^*_1\mu _1 + \pi _2^*\mu _2
     \bigr)\ar[d]&\bigl(\sX^P\!,\,\mu \bigr)\ar[dl]_{r_0}\ar[dr]^{r_1} \\
     \bigl(\LlX\times \LlX,\,\pi _1^*(\tll) + \pi _2^*(\tll)\bigr) &&
     \bigl(\LlX,\,\tll\bigr)} \end{gathered} 
  \end{equation}
This diagram is the general semiclassical composition law on semiclassical
defects. 

  \begin{remark}[]\label{thm:c76}
 The identity object---the tensor unit---in $\Hom\bigl(1,F(S^{\ell
-1})\bigr)$ is the quantization of the semiclassical defect
  \begin{equation}\label{eq:c94}
     \sX^{D^\ell }\longrightarrow \LlX 
  \end{equation}
given by the restriction from maps out of~$D^\ell $ to maps out of its
boundary~$S^{\ell -1}$.  
  \end{remark}

  \subsection*{Defects and symmetry}

Now, finally, we return to the setup for finite symmetry in field theory, as
in Definitions~\ref{thm:c9} and ~\ref{thm:c11}.

  \subsection{$\sr$-defects}\label{subsec:2.18}

Fix a positive integer~$n$.  Suppose $\sr$~is $n$-dimensional symmetry data.
 
  \begin{definition}[]\label{thm:c33}
 A \emph{$\sr$-defect} is a topological defect in the topological field
theory~$\sr$.  We call it a \emph{$\rho $-defect} if its support lies
entirely in a $\rho $-colored boundary.
  \end{definition}

  \begin{figure}[ht]
  \centering
  \includegraphics[scale=.55]{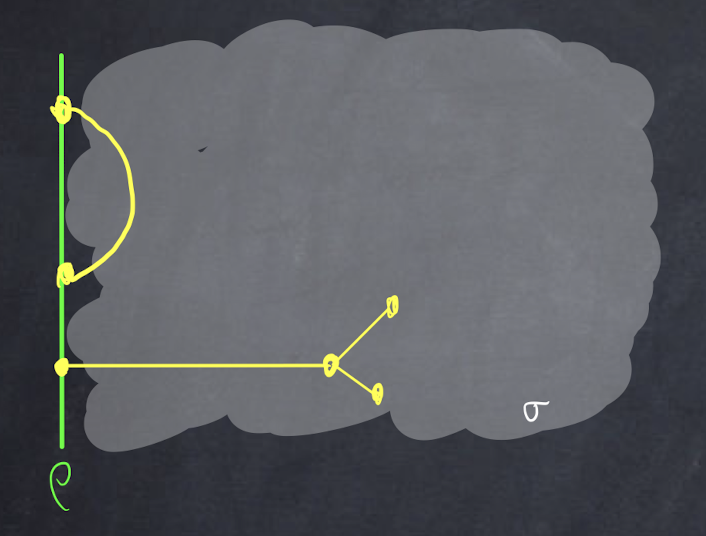}
  \vskip -.5pc
  \caption{$\sr$-defects}\label{fig:10}
  \end{figure}

\noindent
 Figure~\ref{fig:10} depicts some $\sr$-defects.  These are defects in the
abstract symmetry theory.  If $F$~is a quantum field theory equipped with an
$\sr$-module structure $\tFt$, then a $\sr$-defect induces a defect in the
theory~$\srtF$, and then $\theta $~maps it to a defect in the theory~$F$.
Since the defect in the sandwich picture is supported away from $\tF$-colored
boundaries, it is a \emph{topological} defect in the theory~$F$.

  \begin{remark}[]\label{thm:c34}
 Computations with $\sr$-defects, such as compositions, are carried out in
the topological field theory~$\sr$.  They apply to the induced defects in any
$\sr$-module.  
  \end{remark}

  \begin{remark}[]\label{thm:c35}
 In higher dimensions, pictures such as Figure~\ref{fig:10} are interpreted
as a schematic for a tubular neighborhood of the support~$Z\subset M$ of a
defect on a manifold~$M$ (and its Cartesian product with~$[0,1]$).  Also,
unless otherwise stated, for ease of exposition we often implicitly assume a
normal framing to ~$Z$ so that its link may be identified with a standard
sphere.
  \end{remark} 

The image in~$F$ of a defect in the $\srtF$-theory may not be apparent; this
is a significant advantage of the sandwich picture of~$F$.

  \begin{example}[]\label{thm:c40}
 Let $n=3$ and consider a 3-dimensional quantum field theory~$F$ on~$S^3$,
and assume $F$~has an $\sr$-module structure.  In the corresponding
$\srtF$-theory we can contemplate a defect supported on a 2-disk~$D$ in
$[0,1)\times S^3$ whose boundary $K=\partial D\subset \{0\}\times S^3$ is a
knot in the Dirichlet boundary.  (Such a knot is termed `slice'.)  It is
possible that $K$~does not bound a disk in~$S^3$---its Seifert genus may be
positive.  In this case the projection of the slice disk~$D$ to a defect in
the theory~$F$ on~$S^3$ is at best an immersed disk with boundary~$K$, and it
appears that such a topological defect is difficult to describe directly
in the theory~$F$. 
  \end{example} 
 
  \begin{figure}[ht]
  \centering
  \includegraphics[scale=.375]{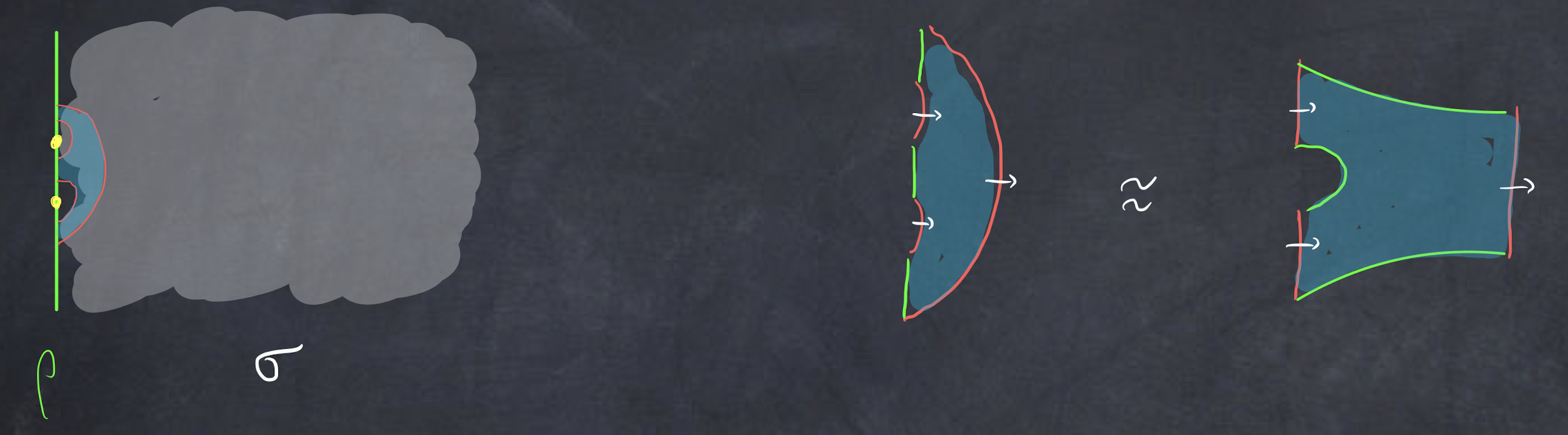}
  \vskip -.5pc
  \caption{The composition law by evaluation on a pair of chaps}\label{fig:c17}
  \end{figure}

  \subsection{Composition of $\rho $-defects}\label{subsec:2.19}

(What we say here also applies to more general $\sr$-defects.)  Since the
labels on $\rho$-defects come from the topological field theory~$\sr$
evaluated on the links, we compute the composition law by applying~ $\sr$ to
a bordism whose boundary consists of links.  For concreteness, we again take
up the quantum mechanical Example~\ref{thm:c14} from Lecture~\ref{sec:1}.
The theory~$\sigma =\sXd2{BG}$ is the 2-dimensional finite gauge theory with
gauge group~$G$ a finite group that acts as symmetries of a quantum
mechanical system~$(\sH,H)$.  Figure~\ref{fig:c17} depicts the composition of
two point $\rho $-defects.  The link of such a defect evaluates under~$\sr$
to the vector space that underlies the group algebra~$A=\GA$.  The
composition law on point $\rho$-defects is computed by evaluating
the\footnote{This particular bordism is also known as
\emph{Gumby}:\raisebox{-2pc}{\includegraphics[scale=.15,trim=0 0 0
2pc]{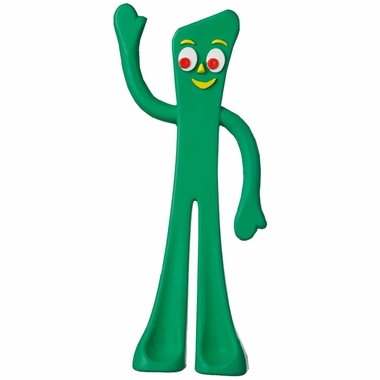}}} ``pair of chaps'' on the right in
Figure~\ref{fig:c17}.  Picture vertical cross sections of this bordism as
links of the two points as the move together and merge into a single point.
The $\sr$-value of the pair of chaps works out to be the multiplication map
$A\otimes A\to A$ of the group algebra.  In particular, on ``classical
labels'' in~$G\subset A$ it restricts to the group product $G\times G\to G$.

  \begin{exercise}[]\label{thm:28}
 Evaluate all of the bordisms in the previous paragraph using the finite path
integral, as described in~\ref{subsec:2.22}.  
  \end{exercise}

  \begin{remark}[]\label{thm:c37}
 We make several observations that apply far beyond this particular example. 

 \begin{enumerate}[label=\textnormal{(\arabic*)}]

 \item This is a hint for Exercise~\ref{thm:28}!  The mapping space of the
link of a point $\rho $-defect is
  \begin{equation}\label{eq:c34}
     \Map\bigl(([0,1],\{0,1\})\,,\,(BG,*) \bigr)\simeq \Omega BG\simeq G.
  \end{equation}
The mapping space of the pair of chaps~$C$ in Figure~\ref{fig:c17} fits into
the correspondence diagram
  \begin{equation}\label{eq:c36}
     \begin{gathered} \xymatrix{&\Map\bigl((C,\partial C_\rho ),(BG,*)
     \bigr)\ar[dl]\ar[dr]\\ \Omega BG\times \Omega BG&&\Omega BG}
     \end{gathered} 
  \end{equation}
that encodes restriction to the incoming and outgoing boundaries.  Here
$\partial C_\rho $~is the $\rho $-colored portion of~$\partial C$.  The left
arrow in~\eqref{eq:c36} is a homotopy equivalence and the right arrow is
composition of loops.

 \item The computation in~\eqref{eq:c36} generalizes to any pointed $\pi
$-finite space $(\sX,*)$ in place of $(BG,*)$.  Then the correspondence is
multiplication on the group~$\OX$, and the quantization is pushforward under
multiplication, i.e., a convolution product.  If the codomain of~$\sigma $
has the form~$\Alg(\sC')$, then compute~$\sigma (\pt)$ as follows:
(1)~quantize~$\OX$ to an object in~$\sC'$, and (2)~induce the algebra
structure from pushforward under multiplication $\OX\times \OX\to \OX$.

 \item Even if we begin with a group symmetry, as in this example, there are
noninvertible topological $\sr$-defects.  In this example, elements of the
group algebra~$\GA$ label point defects on the $\rho $-colored boundary, and
the algebra~$\GA$ contains noninvertible elements.  This fits general quantum
theory, which produces algebras rather than groups.

 \item $\sr$-defects give rise to structure in any $\sr$-module: linear
operators on vector spaces of point defects and on state spaces, endofunctors
on categories of line defects and categories of superselection sectors, etc.
These can be used to explore dynamics.

 \end{enumerate}
  \end{remark}

  \begin{figure}[ht]
  \centering
  \includegraphics[scale=.25]{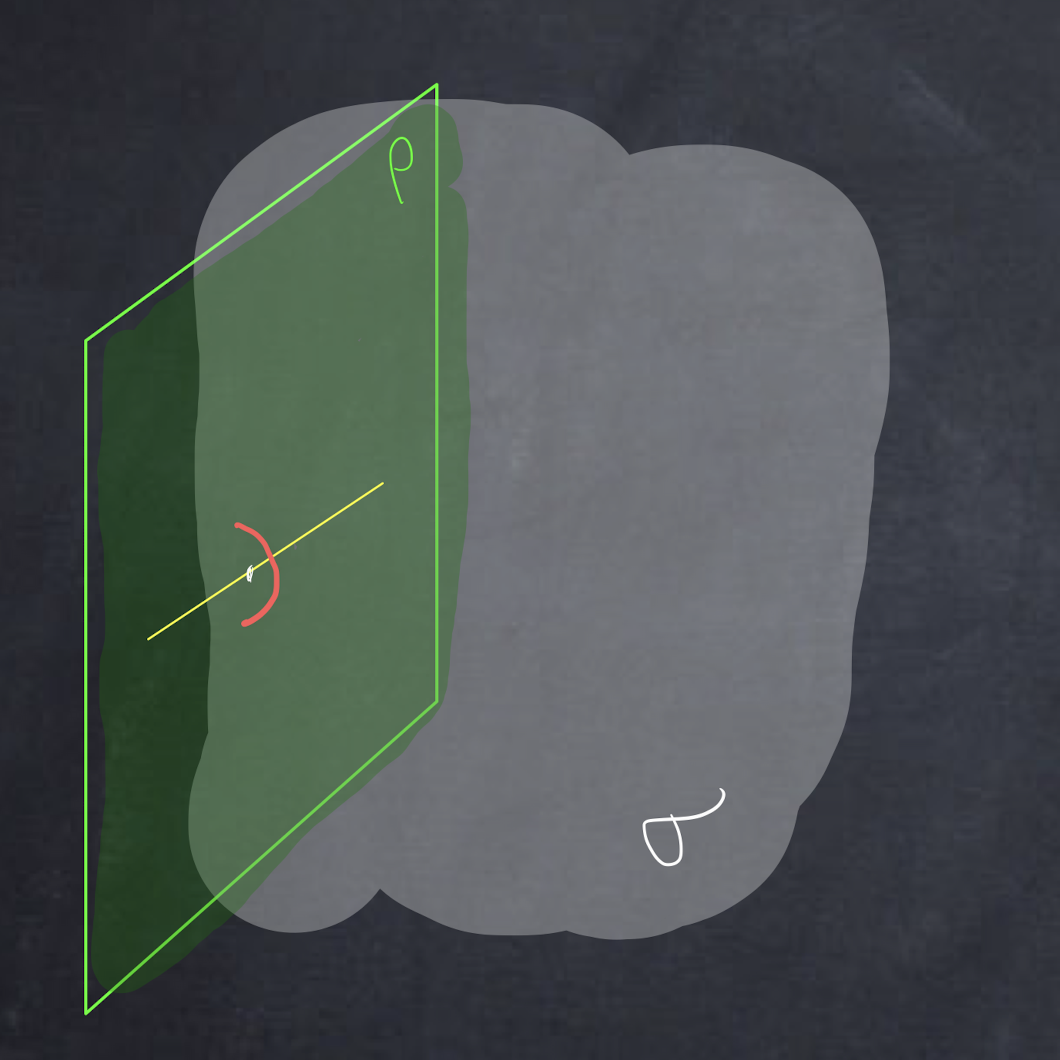}
  \vskip -.5pc
  \caption{A line defect supported on the $\rho $-colored boundary}\label{fig:c19}
  \end{figure}

  \subsection{2-dimensional theories with finite symmetry}\label{subsec:c4.3}

Let $G$~be a finite group and let $\sigma =\sXd3{BG}$~be finite pure
3-dimensional $G$-gauge theory.  As an extended field theory, $\sigma $~can
take values in~$\Alg(\Cat)$, a suitable 3-category of tensor categories, in
which case $\sigma (\pt)$~is the fusion category $\sA=\Vect[G]$ introduced
in~\ref{subsec:c1.5}.  The right regular boundary theory~$\rho $ is
constructed using the right regular module~$\sA_{\sA}$.  Alternatively, in
terms of Definition~\ref{thm:c73}, choose a basepoint in~$BG$.  There are no
background fields for~$\sigma $ or~$\rho $: $\sr$~is an unoriented theory.

  \begin{figure}[ht]
  \centering
  \includegraphics[scale=.25]{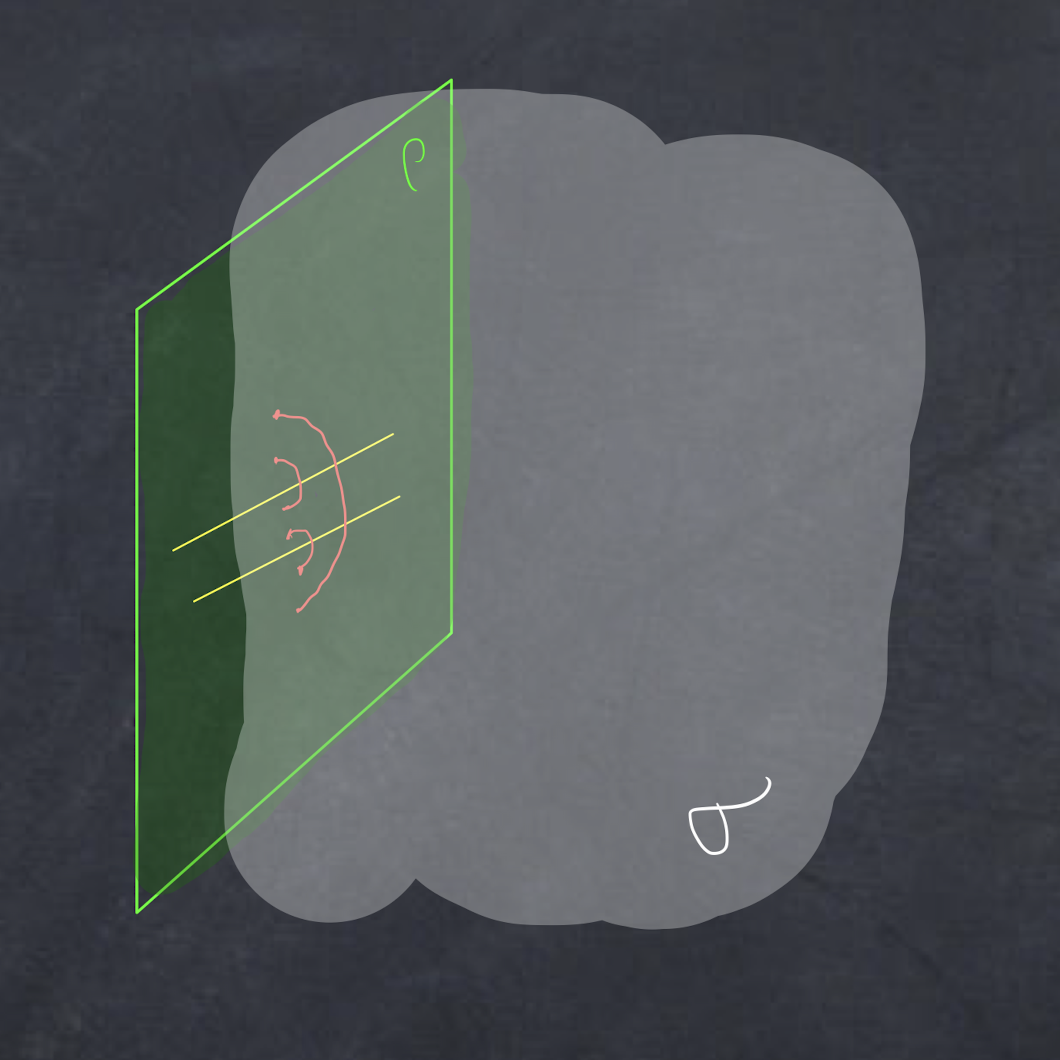}
  \vskip -.5pc
  \caption{Fusion of line defects}\label{fig:c20}
  \end{figure} 

The most familiar $\sr$-defects are the codimension~1 defects supported on
the $\rho $-colored boundary, as depicted in Figure~\ref{fig:c19}.  The link
maps under~$\sr$ to the quantization of the mapping space~\eqref{eq:c34}.
(It is the same mapping space for the link of a codimension~1 defect in
finite gauge theory of \emph{any} dimension.)  That quantization in this
dimension is a linear category, the category~$\Vect(G)$ of vector bundles
over~$G$; it is the linear category that underlies the fusion
category~$\sA$.  The fusion product---computed from the link in
Figure~\ref{fig:c20}, which is the same as the link in
Figure~\ref{fig:c17}---is derived from the correspondence~\eqref{eq:c36} and
is the fusion product of~$\sA$; see Remark~\ref{thm:c37}(2).  Each~$g\in G$
gives rise to an \emph{invertible} defect, labeled by the vector bundle
over~$G$ whose fiber is~$\CC$ at~$g$ and is the zero vector space away
from~$g$.

  \begin{figure}[ht]
  \centering
  \includegraphics[scale=.25]{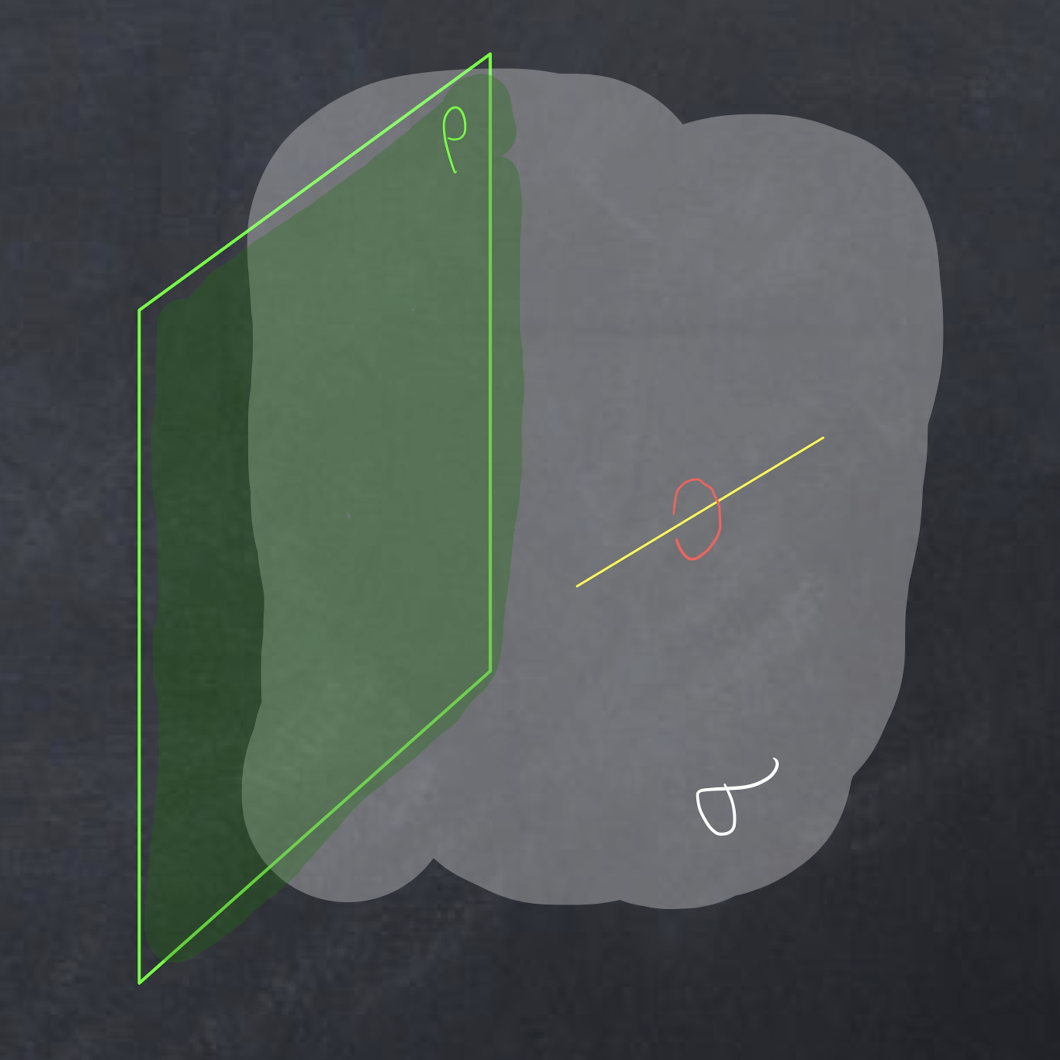}
  \vskip -.5pc
  \caption{A line defect supported in the bulk}\label{fig:c21}
  \end{figure}

Now consider a line defect supported in the bulk, as in Figure~\ref{fig:c21}.
The link is a circle, and so a local defect is an object in the category
$\sigma (\cir)=\Vect_G(G)$ of $G$-equivariant vector bundles over~$G$.  (Here
$G$~acts on itself via conjugation.)  This is the (Drinfeld) center of~$\sA$.
Note that unlike the quantum mechanical situation in Figure~\ref{fig:5}, the
center here is ``larger'' than the algebra~$\sA$.  The simple objects of the
center are labeled by a pair consisting of a conjugacy class and an
irreducible representation of the centralizer of an element in the conjugacy
class.  The corresponding defect is invertible iff the representation is
1-dimensional.  Among these defects are the Wilson and 't~Hooft lines of the
3-dimensional $G$-gauge theory.  There is a rich set of topological defects
that goes beyond those labeled by group elements. 

  \begin{remark}[]\label{thm:30}
 So, even if we begin with the invertible $G$-symmetry, we are inexorably led
to ``non-invertible symmetries''. 
  \end{remark}

  \begin{exercise}[]\label{thm:c42}
 Verify that there are no non-transparent point defects, either on the $\rho
$-colored boundary or in the bulk.
  \end{exercise} 

  \begin{exercise}[]\label{thm:29}
 Evaluate $\sigma (\cir)$ using the finite path integral. 
  \end{exercise}

  \begin{figure}[ht]
  \centering
  \includegraphics[scale=.35]{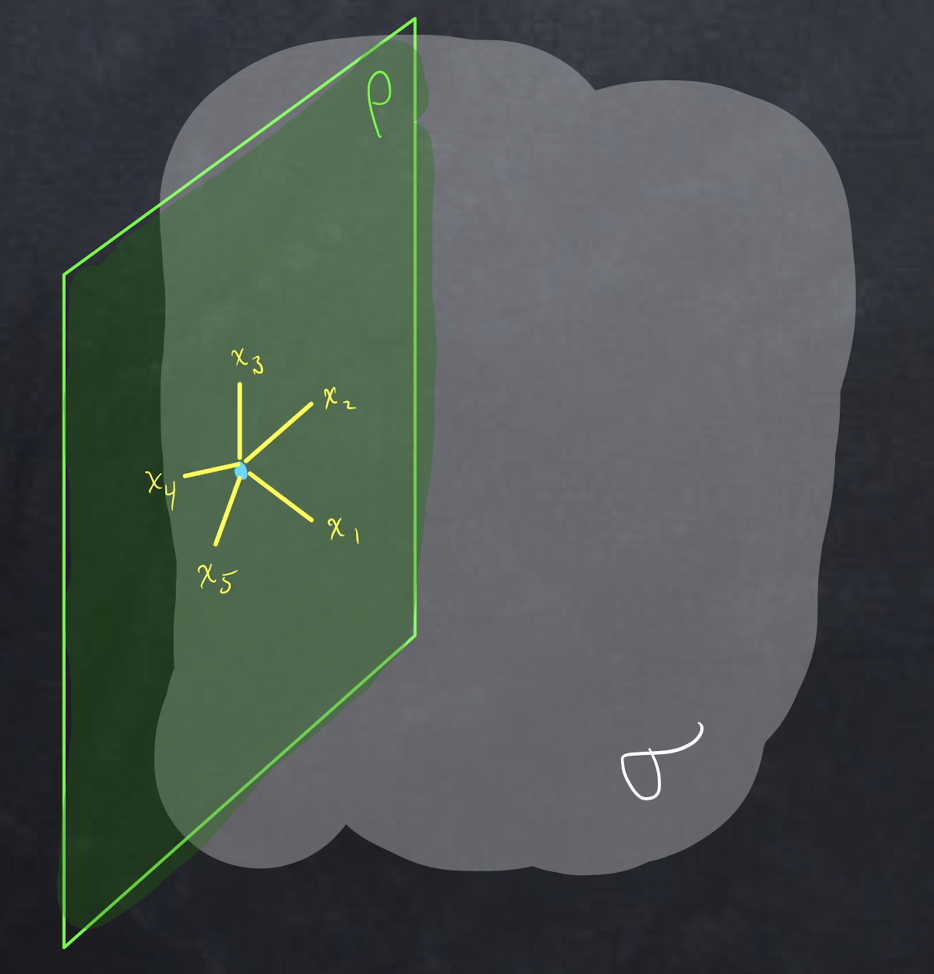}
  \vskip -.5pc
  \caption{A stratified $\rho $-defect}\label{fig:c27}
  \end{figure}

  \subsection{Turaev-Viro theories}\label{subsec:2.20}

The example of finite $G$-gauge theory generalizes to arbitrary Turaev-Viro
theories.  Let $\Phi $~be a spherical fusion category, let $\sigma $ be the
induced 3-dimensional topological field theory (of oriented bordisms) with
$\sigma (\pt)=\Phi $, and define the regular boundary theory~$\rho $ via the
right regular module~$\Phi _\Phi $.  The category of point $\rho $-defects is
the linear category that underlies~$\Phi $, so a defect is labeled (locally)
by an object of~$\Phi $.  We can also have nontrivial stratified $\rho
$-defects, such as illustrated in Figure~\ref{fig:c27}.  In the figure
the~$x_i$ are objects of~$\Phi $ and the label at the central point is a
vector in $\Hom\mstrut _{\Phi }(1,x_1\otimes \cdots\otimes x_5)$.

  \subsection*{Some additional problems}
 
	\problem 
 In~\ref{subsec:c4.3} we computed the local line $\rho $-defects in
3-dimensional pure gauge theory~$\sigma =\sXd3{BG}$ with gauge group a finite
group~$G$.  Recall that $\sigma $~is defined on all (unoriented) manifolds.
Compute the category of line $\rho $-defects supported on $\RP^1\subset
\RP^2$.  (Work on $[0,1)\times \RP^2$ with $\{0\}\times \RP^2$ $\rho
$-colored.)  Is there a composition law for these defects?
	\endproblem

  \begin{figure}[ht]
  \centering
  \includegraphics[scale=.6]{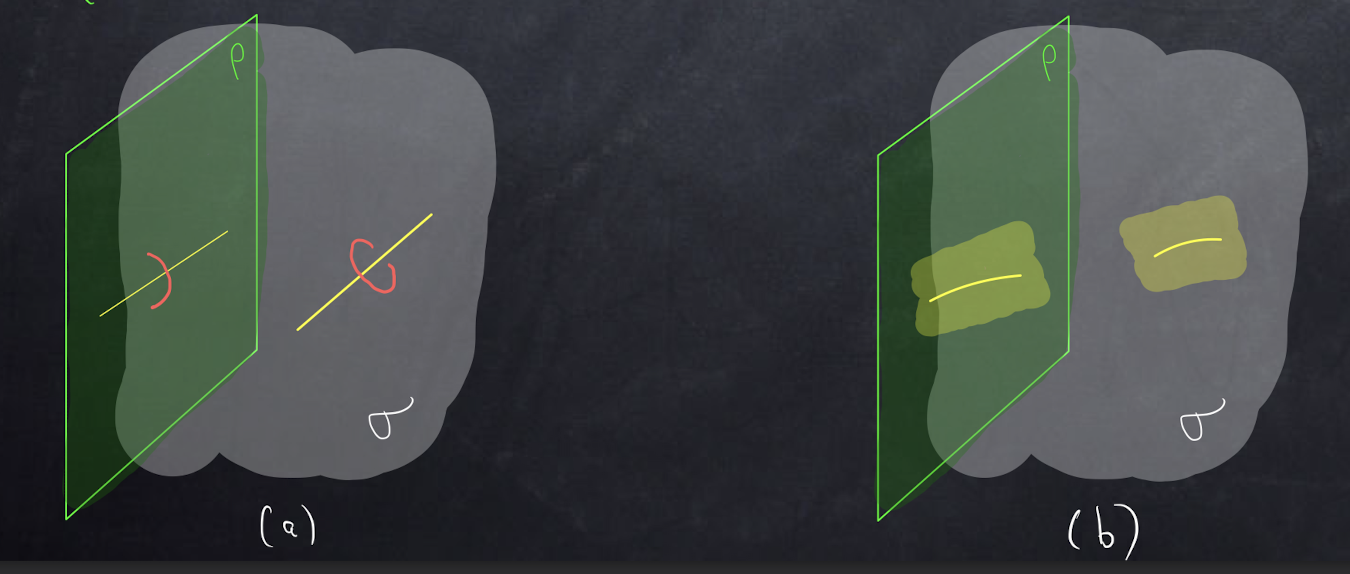}
  \vskip -.5pc
  \caption{(a)~A $\rho $-defect, a $\sigma $-defect, and their
  links; (b)~line defects embedded in surface defects}\label{fig:p3} 
  \end{figure}

	 \probsec\probsecc
 Let $G$~be a finite group, let $A$~be a finite abelian group, and let
$\sX$~be a 2-finite path connected topological space that is an extension 
  $$ B^2A\longrightarrow \sX\longrightarrow BG $$
(This is the Postnikov tower of~$\sX$, which in this case is the classifying
space of a 2-group.)  Fix a basepoint $*\to \sX$ of~$\sX$.  Let $\sigma =\sXd
m{\sX}$ be the corresponding finite homotopy theory in dimension~$m$, and let
$\rho $~be the right regular boundary theory defined by the basepoint.  The
pair~$\sr$ is abstract symmetry data for field theories of
dimension~$n=m-1$. 
 \sec Set $n=3$.  Compute the category of local line $\rho $-defects, i.e.,
of local line defects supported on the $\rho $-colored boundary.  Compute the
category of local line defects in the theory~$\sigma $.  See
Figure~\ref{fig:p3}(a). 
 \sec Repeat for surface defects of both types.  Your answer may be a linear
2-category, a tensor category, \dots 
 \sec Repeat for a line defect in a surface defect of both types; see
Figure~\ref{fig:p3}(b). 
 \sec Repeat the problem for $n=4$.
	\endprobsec

	\probsec\probsecc
A \emph{crossed module} is the following data: a pair of groups~$(H,K)$, a
homomorphism $d\:H\to K$, and an action of~$K$ on~$H$ such that for all $h,
h',h''\in H$ and $k\in K$ we have 
  $$ \begin{aligned} d(h')\cdot h''&= h'h''(h')\inv \\ d(k\cdot h)&=
      k\,d(h)\,k\inv \end{aligned} $$
Define $A=\ker d$ and $G=\coker d$.  (First verify that $d(H)$~is normal
in~$K$.)  Assume that $A$~and $G$~are finite groups. 
 \sec Prove that $A$~is a subgroup of the center of~$H$.  In particular,
$A$~is abelian.
 \sec Construct an action of~$G$ on~$A$. 
 \sec Construct a cohomology class in $H^3(G;A)$. 
 \sec Show that a crossed module is a group object in the category of
groupoids. 
 \sec The data heretofore produces a 2-finite path connected pointed
space~$\sX$ with $\pi _1\sX=G$ and $\pi _2\sX=A$.  In other terms, this is
the classifying space of a 2-group.  How, from a 2-finite path connected
pointed space~$\sX$, do you extract the data ($G$, $A$, action of $G$ on~$A$,
cohomology class)? 
	\endprobsec

	\probsec\probsecc
Suppose $H$~is a Lie group, $A\subset H$ is a finite subgroup of the
center of~$H$, and $\hH$~is a Lie group which contains~$H$ as a normal
subgroup that is a union of components of~$\hH$.  Define $G=\hH/H$.  Assume
that $G$~is a finite group.
 \sec Prove that $A\subset \hH$ is normal.  Set $K=\hH/A$. 
 \sec Construct a crossed module $d\:H\to K$.  Let $\sX$~be the corresponding
2-finite path connected pointed space (defined up to homotopy).
 \sec Construct a fibration $BH\to BK\to \sX$.  In the world of simplicial
sheaves on $\Man$, construct a fibration $B\mstrut _{\nabla }H\to B\mstrut
_{\nabla }G\to \sX$.  
 \sec Find examples of (the result of) this construction in {\tt
arXiv:2204.06564}.  In other words, find examples of the groups $H,\hH,A$.
Can you find some of the higher categories of defects you computed in
a previous problem in that paper?
 \sec For any Lie group~$H$, the automorphism group~$\Aut(H)$ is a Lie group.
Construct a crossed module $d\:H\to \Aut(H)$.  What is the data of the
associated homotopy type?  How/When does this 2-group act on $H$-gauge theory?
	\endprobsec
\endgroup\endgroup\endgroup

\newpage
\setcounter{section}{2}
   \section{Lecture 3: Quotients and projectivity}\label{sec:3}
% \setcounter{footnote}{25}
% \setcounter{figure}{25}
% lastsubsec@ 21

This lecture has two main subjects: quotients and projectivity.  We already
treated these topics in the context of groups and algebras:
see~\ref{subsec:1.8} to the end of Lecture~\ref{sec:1}.  Here we take up
quotients by a symmetry of a field theory.  Working in the ``sandwich''
picture, which separates out the abstract topological symmetry from the
potentially nontopological field theory on which it acts, the quotient is
effected by replacing the right regular boundary theory (Dirichlet) with an
augmentation (Neumann).  We also introduce \emph{quotient defects}, which
amounts to executing the quotient construction on a submanifold.  In the
literature these are often called ``condensation defects'', and the
quotienting process is called ``gauging''.  Quantum theory takes place in
projective geometry, not linear geometry, and so in the second half of the
lecture we take up projectivity in the context of field theory.  It is
expressed via invertible field theories.  The projectivity of an
$n$-dimensional field theory is an $(n+1)$-dimensional invertible field
theory---its \emph{anomaly (theory)}---and trivializations of the anomaly
form a torsor over the group of $n$-dimensional invertible field theories.
In other words, the group of $n$-dimensional invertible field theories acts
on the space of all $n$-dimensional field theories, and theories in the same
orbit share many of the same properties. A symmetry may only act
projectively, in which case it is said to enjoy an \emph{'t~Hooft anomaly},
and that obstructs the existence of an augmentation, so too obstructs the
existence of a quotient.  We conclude with an example of quotients and
twisted quotients for theories with a $BA$-symmetry for a finite abelian
group~$A$, namely line operators in 4-dimensional gauge theory.

  \subsection*{Preliminary remarks}\label{subsec:3.5}

  \subsection{Invertibility}\label{subsec:3.6}

 Anytime we have a (higher categorical) monoid---a set with an associative
composition law~$*$ and unit~$1$---then we have a notion of invertibility: an
element/object~$x$ is \emph{invertible} if there exists~$y$ such that $x*y =
1$ (or $x*y\cong 1$ in a categorical context).  Note that invertibility is a
condition, not data.  This applies to two situations in these lectures:
(1)~the composition law (``stacking'') of field theories, which leads to the
notion of an \emph{invertible field theory}; and (2)~the composition law
(``fusion'') of local topological defects, which leads to the notion of an
\emph{invertible topological defect}.

  \begin{figure}[ht]
  \centering
  \includegraphics[scale=.6]{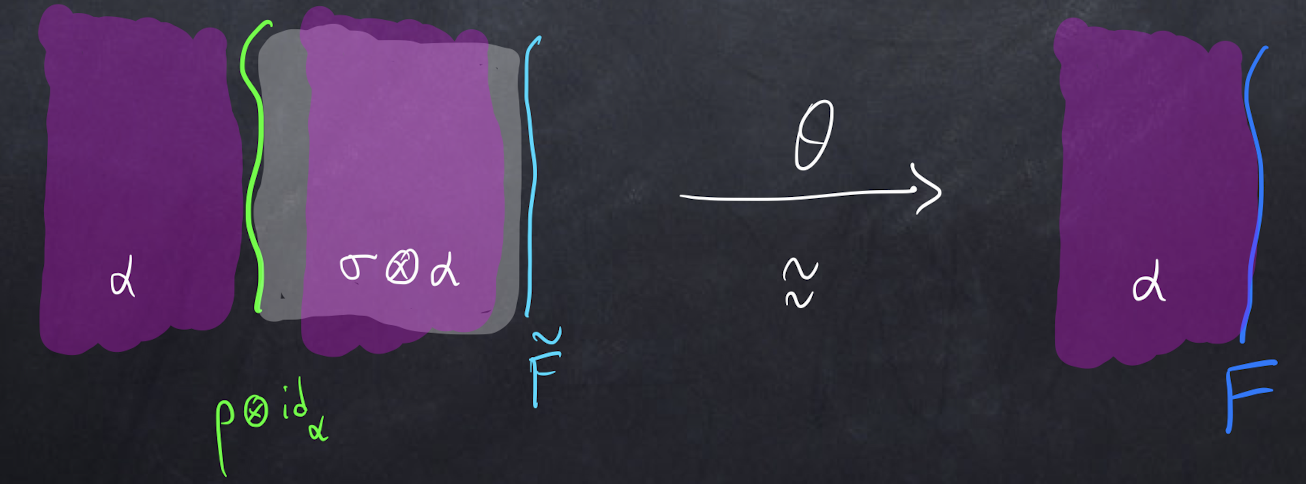}
  \vskip -.5pc
  \caption{The action of~$\sr$ on the boundary theory~$F$ of~$\alpha $}\label{fig:12}
  \end{figure}

  \subsection{Symmetries of a boundary theory}\label{subsec:3.7}

Not every left boundary theory~$F$ of a field theory~$\alpha $ indicates that
$\alpha $~acts as \emph{symmetries} on~$F$.  In these lectures we define
\emph{finite} symmetry in terms of a pair~$\sr$; simply having a left
boundary theory~$F$ for a theory~$\alpha $ is not an action of symmetry.
Furthermore, in this paragraph we do not require that $\alpha $~be
topological.\footnote{Nor do we require that $\alpha $~be invertible; if it
is, then we say $F$~is \emph{anomalous} with \emph{anomaly theory}~$\alpha
$.}  On the other hand, if $\sr$~is finite symmetry data, then there is a
notion of~$\sr$ acting by symmetries on a theory~$F$ that is a left boundary
theory for~$\alpha $.  Namely, the \emph{left module structure} data~$\tFt$
is a left module~$\tF$ for~$\sigma \otimes \alpha $ and an isomorphism as
indicated in Figure~\ref{fig:12}.

  \subsection*{Quotients by a symmetry in field theory}

One should think that the quotients in this section are ``derived'' or
``homotopical'', though we do not deploy those modifiers.  (See
Exercise~\ref{thm:14}.)

  \subsection{Augmentations in higher categories}\label{subsec:3.1}

  \begin{definition}[]\label{thm:c19}
 Let $\sC'$ be a symmetric monoidal $n$-category, and set $\sC=\Alg(\sC')$.
Suppose $A\in \sC$ is an algebra object in~$\sC'$.  Then an
\emph{augmentation} $\eA\:A\to 1$ is an algebra homomorphism from~$A$ to the
tensor unit~$1\in \sC$. 
  \end{definition}

\noindent
 Thus $\eA$~is a 1-morphism in~$\sC'$ equipped with data that exhibits the
structure of an algebra homomorphism.  Augmentations may not exist.  

  \begin{remark}[]\label{thm:c20}
 A general 1-morphism $A\to 1$ in~$\sC$ is an object of~$\sC'$ equipped with
a right $A$-module structure.  An augmentation is a right $A$-module
structure on the tensor unit~$1\in \sC'$. 
  \end{remark}

  \subsection{Augmentations in field theory}\label{subsec:3.2}

  \begin{definition}[]\label{thm:c21}
 Let $\sC'$ be a symmetric monoidal $n$-category, and set $\sC=\Alg(\sC')$.
Let $\sF$~be a collection of $(n+1)$-dimensional fields, and suppose $\sigma
\:\Bord_{n+1}(\sF)\to \sC$ is a topological field theory.  A right boundary
theory~$\epsilon $ for~$\sigma $ is an \emph{augmentation} of~$\sigma $ if
$\epsilon (\pt)$~is an augmentation of~$\sigma (\pt)$ in the sense of
Definition~\ref{thm:c19}.  
  \end{definition}

\noindent
 An augmentation in this sense is often called a \emph{Neumann boundary
theory}.\footnote{There is a more general notion of a Neumann boundary theory
associated to a given Dirichlet boundary theory, but we do not pursue that
here.} 

  \subsection{The quotient theory}\label{subsec:3.3}

We use notations in Definition~\ref{thm:c9} and Definition~\ref{thm:c11} in the
following.  

  \begin{definition}[]\label{thm:c25}
 Suppose given finite symmetry data~$\sr$ and a $\sr$-module structure $\tFt$
on a quantum field theory~$F$.  Suppose $\epsilon $~is an augmentation
of~$\sigma $.  Then the \emph{quotient} of~$F$ by the symmetry~$\sigma $ with
respect to the augmentation~$\epsilon $ is
  \begin{equation}\label{eq:c24}
     F\emod\sigma =\epsilon  \otimes _\sigma \tF. 
  \end{equation}
  \end{definition}

  \begin{figure}[ht]
  \centering
  \includegraphics[scale=.6]{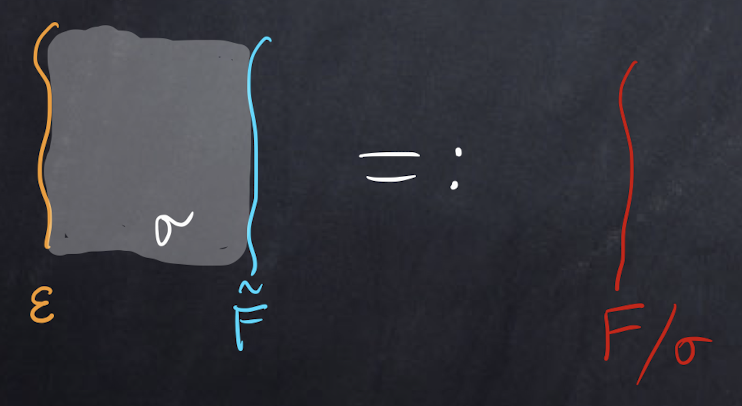}
  \vskip -.5pc
  \caption{The quotient theory}\label{fig:14}
  \end{figure}

\noindent
 We simply write `$F\bs $' if the augmentation~$\epsilon $ is
understood from context.  The right hand side of~\eqref{eq:c24} is the
sandwich in Figure~\ref{fig:14}.

  \subsection{Quotients in finite homotopy theories}\label{subsec:3.4}

Recall the definition of a semiclassical boundary theory
in~\ref{subsec:2.13}.  We now tell what an augmentation is in this context. 

  \begin{definition}[]\label{thm:32}
  Let $\sX$~be a $\pi $-finite space and suppose $\lambda $~is a cocycle of
degree~$m$ on~$\sX$.  A \emph{semiclassical right augmentation} of~$\Xl$ is a
trivialization~$\mu $ of~$-\lambda $. 
  \end{definition}

\noindent
 Observe that if $\lambda =0$, then $\mu $~is a cocycle of degree~$m$.  Also,
there is a canonical choice of~$\mu $ in this instance: $\mu =0$.

  \begin{remark}[]\label{thm:33}
 The cocycle~$\lambda $ encodes an \emph{'t Hooft anomaly} in a finite
homotopy type theory; it is the projectivity of the symmetry.  If $\lambda
=0$, then a cocycle~$\mu $ encodes a twist of the boundary theory, and it
goes by various names: `discrete torsion', `$\theta $-angles', etc.,
depending on the context. 
  \end{remark}

  \begin{example}[]\label{thm:c26}
 Let $G$~be a finite group, and let $\sigma =\sXd{n+1}{BG}$ be the associated
finite gauge theory.  Use the canonical boundary theory $\id_{\sX}\:\sX\to
\sX$.  In the semiclassical picture this corresponds to summing over all
principal $G$-bundles with no additional fields on the $\epsilon $-colored
boundaries.  This is the usual quotienting operation, oft called `gauging'.
  \end{example}

  \subsection{The Dirichlet-Neumann and Neumann-Dirichlet domain walls}\label{subsec:3.8}

  \begin{lemma}[]\label{thm:34}
 Let $\sigma $ be an $(n+1)$-dimensional topological field theory with
codomain $\sC=\Alg(\sC')$, and suppose $\rho $~is the right regular boundary
theory of~$\sigma $ and $\epsilon $~is an augmentation of~$\sigma $.  Then
the category of domain walls from~$\rho $ to~$\epsilon $ is a free module of
rank one over the symmetric monoidal category of theories of dimension~$n$,
as is the category of domain walls from~$\epsilon $ to~$\rho $.
  \end{lemma}

  \begin{figure}[ht]
  \centering
  \includegraphics[scale=.6]{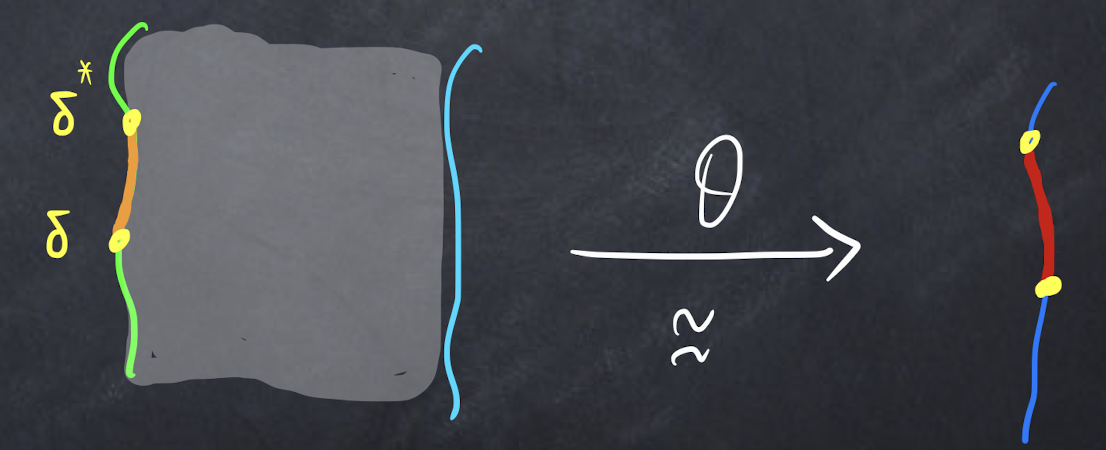}
  \vskip -.5pc
  \caption{The domain walls~$\delta $ and $\delta ^*$}\label{fig:18}
  \end{figure}

\noindent
 Roughly: Use the homomorphism $\epsilon (\pt)\:A\to 1$ to make~$1$ into a
left $A$-module, where $A=\sigma (\pt)$, and so construct a dual \emph{left}
boundary theory~$\epsilon ^L$, the left adjoint.  Then the sandwich $\rho
\otimes _\sigma \epsilon ^L$ is the trivial theory: use the cobordism
hypothesis to compute its value on a point as $A\otimes _A1\cong 1$.  Let
  \begin{equation}\label{eq:12}
     \begin{aligned} \delta \: \rho &\longrightarrow \epsilon \\ \delta
      ^*\:\epsilon &\longrightarrow \rho \end{aligned} 
  \end{equation}
be generating domain walls.

  \subsection{The composition}\label{subsec:3.9}

\begin{figure}[ht]
  \centering
  \includegraphics[scale=.6]{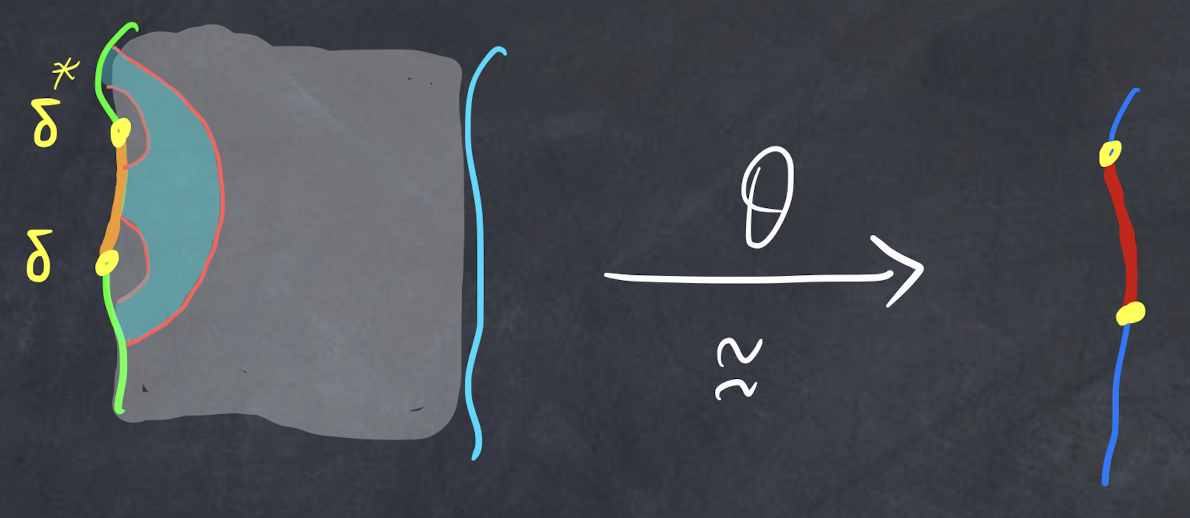}
  \vskip -.5pc
  \caption{Computation of $\delta ^*\circ \delta $}\label{fig:15}
  \end{figure}

\begin{figure}[ht]
  \centering
  \includegraphics[scale=.5]{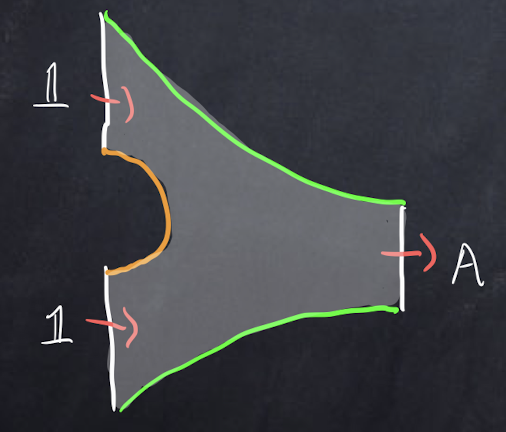}
  \vskip -.5pc
  \caption{The pair of chaps: note the coloring of the boundary components}\label{fig:16}
  \end{figure}

Our task is to compute the composition 
  \begin{equation}\label{eq:13}
     \delta ^*\circ \delta \:\rho \longrightarrow \rho , 
  \end{equation}
which is a self-domain wall of the boundary theory~$\rho $.  (The reverse
composition $\delta \circ \delta ^*$ is similar.)  As always, that computation
is done by tracking the links as the points come together, and we obtain the
pair of chaps extracted in Figure~\ref{fig:15} and isolated in
Figure~\ref{fig:16}.  In that figure we have
%\noindent 
 labeled the incoming boundary components in accordance with
Lemma~\ref{thm:34}.  For the outgoing boundary component we are
assuming~$\sigma (\pt)=A$ is an algebra in $\sC=\Alg(\sC')$; the label~$A$ in
the figure is the underlying object of~$\sC'$.  This evaluates to an object in
$\Hom_{\sC'}(1,A)$.  We evaluate it in two cases.

  \begin{example}[Turaev-Viro symmetry]\label{thm:c52}
 Suppose~$n=2$ and the 3-dimensional theory~$\sigma $ is of Turaev-Viro type
with $\sigma (\pt)=\sA $ a fusion category.  Assume $\rho $~is given by the
right regular module~$\sA _\sA $ and $\epsilon $~is given by a fiber
functor $\eP\:\sA \to \Vect$.  Then the codimension~1
quotient defect has local label the object $\xreg\in \sA $ defined as 
  \begin{equation}\label{eq:c63}
     \xreg=\sum\limits_{x}\eP (x)^*\otimes x, 
  \end{equation}
where the sum is over a representative set of simple objects~$x$.  See
Proposition~8.9 in footnote~\footref{FTi} for a very similar computation.
  \end{example}

  \begin{example}[finite homotopy theories]\label{thm:35}
 Let $\sX$~be a $\pi $-finite space.  Then the composition is the homotopy
fiber product
  \begin{equation}\label{eq:c62}
     \begin{gathered} \xymatrix@R-.5pc@C-.5pc{&&\OX\ar@{-->}[dl]\ar@{-->}[dr]\\
     &\ast\ar[dl]\ar[dr]&&\ast\ar[dl]\ar[dr]\\
     \ast\ar[ddrr]&&\sX\ar[dd]&&\ast\ar[ddll]\\ \\ &&\sX} \end{gathered} 
  \end{equation}
which is then the domain wall 
  \begin{equation}\label{eq:14}
     \begin{gathered} \xymatrix{&\OX\ar[dl]\ar[dr] \\
     \ast\ar[dr]&&\ast\ar[dl] \\&\sX }
     \end{gathered}
  \end{equation}
Note that the points~$*$ in the second row of~\eqref{eq:c62} are obtained as
fiber products from the parts of the diagram that lie below it, and the fact
that they are single points (contractible spaces) proves Lemma~\ref{thm:34}
in the finite homotopy theory case.
  \end{example}

  \begin{figure}[ht]
  \centering
  \includegraphics[scale=.55]{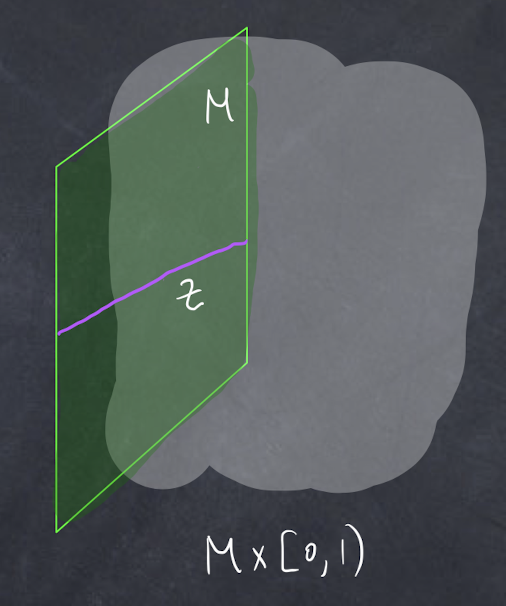}
  \vskip -.5pc
  \caption{The defect $\delta ^*\circ \delta $ on~$Z$}\label{fig:17}
  \end{figure}

  \begin{example}[special case]\label{thm:36}
 Now take~$\sX=BG$ for $G$~a finite group.  Then $\OX\simeq G$, so the
composition $\delta ^*\circ \delta $ sums over maps to~$G$.  Suppose $M$~is a
bordism on which we evaluate~$F$, and suppose $Z\subset M$ is a cooriented
codimension submanifold on which we place the defect~$\delta ^*\circ \delta $.
(As usual, we do not make background fields explicit.)  Form the sandwich
$[0,1)\times M$ with $\{0\}\times (M\setminus Z)$ colored with~$\rho $ and
$\{0\}\times Z$ colored with~$\delta ^*\circ \delta $.  Then the theory sums
over principal $G$-bundles together with a trivialization on the $\rho
$-colored boundary $\{0\}\times (M\setminus Z)$.  For each~$g\in G$ there is
a defect~$\eta (g)$ (of ``'t~Hooft type'') that constrains the jump in the
trivializations across~$Z$ to be~$g$.  Then quantization using~\eqref{eq:14}
shows  
  \begin{equation}\label{eq:15}
     \delta ^*\circ \delta =\sum\limits_{g\in G}\eta (g). 
  \end{equation}
This equation appears in several recent physics papers.  A similar equation
holds for~$\sX$ in general, and in particular for $\sX=B^{q}A$ for a finite
abelian group~$A$, only now automorphisms appear in the quantization.   
  \end{example}

  \subsection*{Quotient defects: quotienting on a submanifold}

The following discussion is inspired by the
paper~\footnote{\label{RSS}Konstantinos Roumpedakis, Sahand Seifnashri, and
Shu-Heng Shao, \emph{Higher
  Gauging and Non-invertible Condensation Defects},
  \href{http://arxiv.org/abs/arXiv:2204.02407}{{\tt arXiv:2204.02407}}.}.
The basic idea is to execute the quotient construction on a submanifold, not
necessarily to take the quotient of the entire theory. 
 
Fix a positive integer~$n$ and finite $n$-dimensional symmetry data~$\sr$.
Suppose $\epsilon $~is an augmentation of~$\sigma $, as in
Definition~\ref{thm:c21}.  As explained in Definition~\ref{thm:c25}, if
$\tFt$ is a $\sr$-module structure on an $n$-dimensional quantum field
theory~$F$, then dimensional reduction of~$\sigma $ depicted in
Figure~\ref{fig:14}, which is the sandwich $\epsilon \otimes _\sigma \tF$, is
the quotient~$F\bs$ of~$F$ by the symmetry.  This can be interpreted as
placing the topological defect~$\epsilon $ on the entire theory.
 
There is a generalization that places the defect on a submanifold. Suppose
$M$~is a bordism on which we evaluate~$F$, and suppose $Z\subset M$ is a
submanifold on which we place the defect.  (As usual, we do not make
background fields explicit.)  Form the sandwich $[0,1)\times M$ with
$\{0\}\times M$ colored with~$\rho $.  Let $\nu \subset M$ be an open tubular
neighborhood of $Z\subset M$ with projection $\pi \:\nu \to Z$, and arrange
that the closure $\bn$ of~$\nu $ is the total space of a disk bundle $\bn\to
Z$.

  \begin{figure}[ht]
  \centering
  \includegraphics[scale=.4]{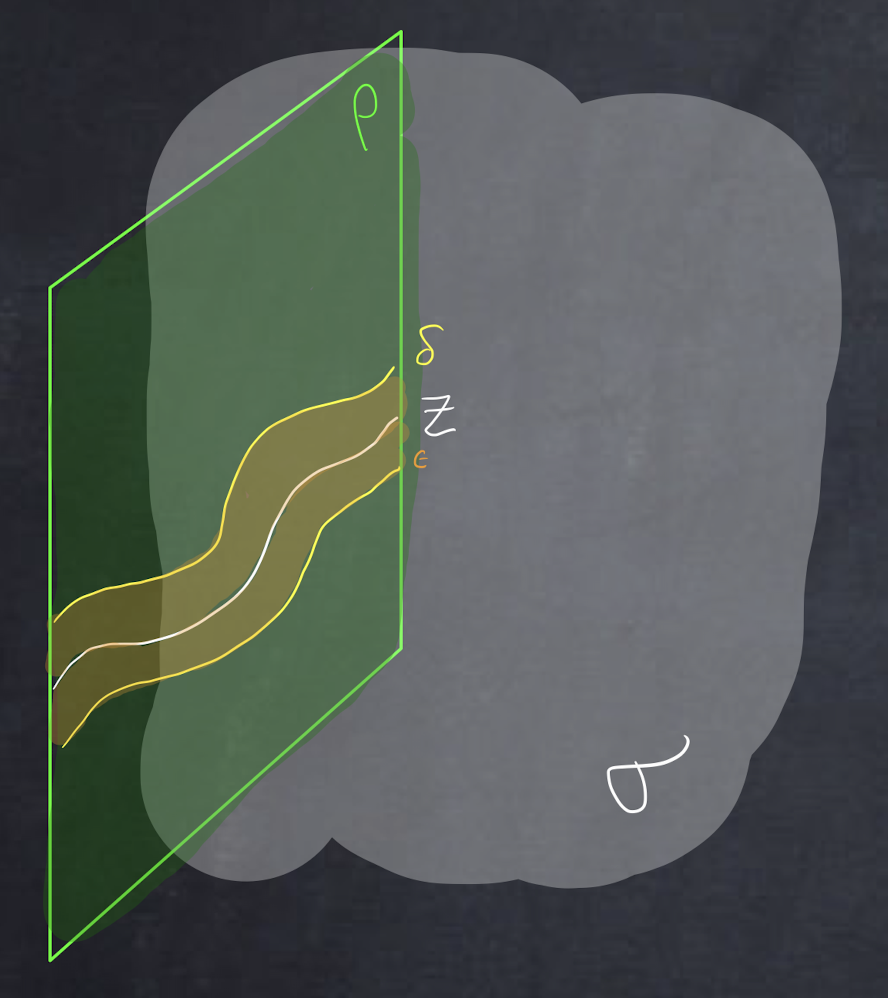}
  \vskip -.5pc
  \caption{The quotient defect~$\epsilon (Z)$}\label{fig:c28}
  \end{figure}

  \begin{definition}[]\label{thm:c45}
 The \emph{quotient defect}~$\epsilon (Z)$ is the $\rho $-defect
supported on $\{0\}\times \bn$ with $\{0\}\times \nu $ colored with~$\epsilon $
and $\{0\}\times \partial \bn$ colored with~$\delta $.  
  \end{definition}

\noindent
 This defect is depicted in Figure~\ref{fig:c28}. 

\begin{figure}[ht]
  \centering
  \includegraphics[scale=.4]{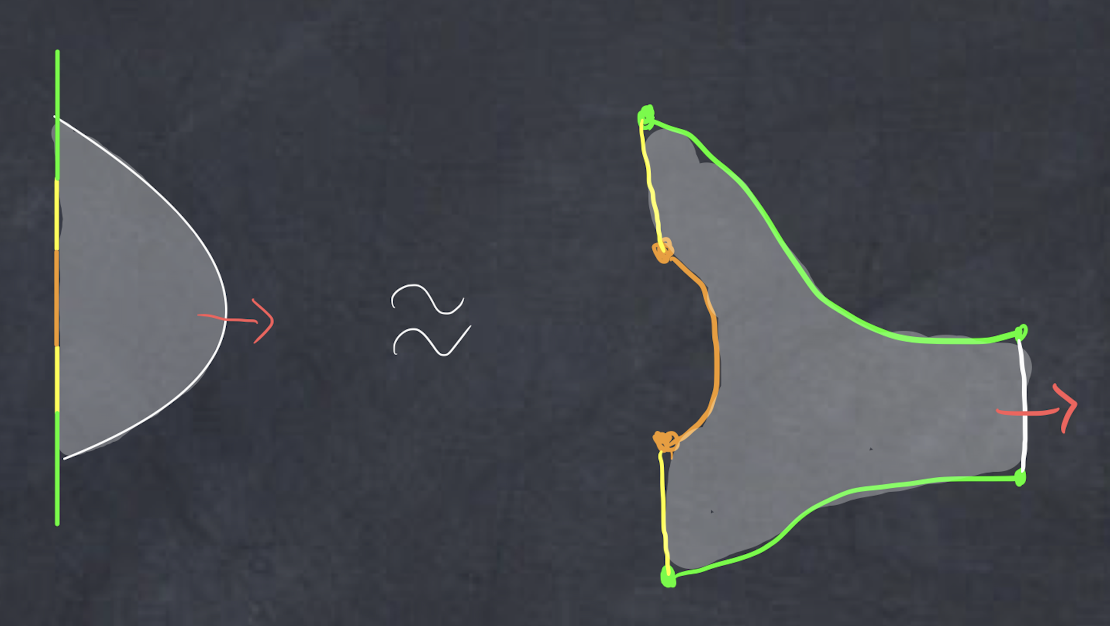}
  \vskip -.5pc
  \caption{The local label of~$\epsilon (Z)$ in codimension~1}\label{fig:c29}
  \end{figure}

Next we compute the local label of the quotient defect~$\epsilon (Z)$, as
in Definition~\ref{thm:c7}(1), and so express~$\epsilon (Z)$ as a defect
supported on~$Z$.  Consider a somewhat larger tubular neighborhood, now of
$\{0\}\times Z\subset \zoh\times M$.  Let $\ell =\codim\mstrut _{\hneg
M}\!Z$.  The tubular neighborhood for~$\ell =1$ is depicted in
Figure~\ref{fig:c29}.  It is a pair of chaps, two of whose incoming boundary
components are $\delta $-colored.  Its value in the topological
theory~$\sigma $---with boundaries and defects $\rho ,\epsilon ,\delta $---is
an object in $\Hom\bigl(1,\sigma (D^1,S^0_\delta ) \bigr)$.  (If
$\sC=\Alg(\sC')$ is the codomain of~$\sigma$, and $\sigma (\pt)=A$ is an algebra
object in~$\sC'$, then $\sigma (D^1,S^0_\delta )=A$ as an object of~$\sC'$.)

  \begin{remark}[]\label{thm:c47}
 The pair of chaps picture makes clear that the defect~$\epsilon (Z)$
for~$\ell =1$ can be interpreted as follows, assuming $Z\subset M$ has
trivialized normal bundle.  Let $Z_1,Z_2$ be parallel normal translates
of~$Z$, color the region in between $\{0\}\times Z_1$ and $\{0\}\times Z_2$
with~$\epsilon $, color the remainder of~$\{0\}\times M$ with~$\rho $, and
use the domain wall~$\delta $ at $\{0\}\times Z_1$ and $\delta ^*$ at
$\{0\}\times Z_2$.  Then $\epsilon (Z)$~is the composition $\delta
^*(Z_2)*\delta (Z_1)$.  If a quantum field theory~$F$ has a $\sr$-module
structure, then $\delta (Z_1)$~is a domain wall from~$F$ to~$F\bs$ and
$\delta ^*(Z_2)$ is a domain wall from~$F\bs$ to~$F$; the
composition~$\epsilon (Z)$ is a self domain wall of~$F$, precisely the one
computed in~\ref{subsec:3.9}.
  \end{remark}

  \begin{figure}[ht]
  \centering
  \includegraphics[scale=.3]{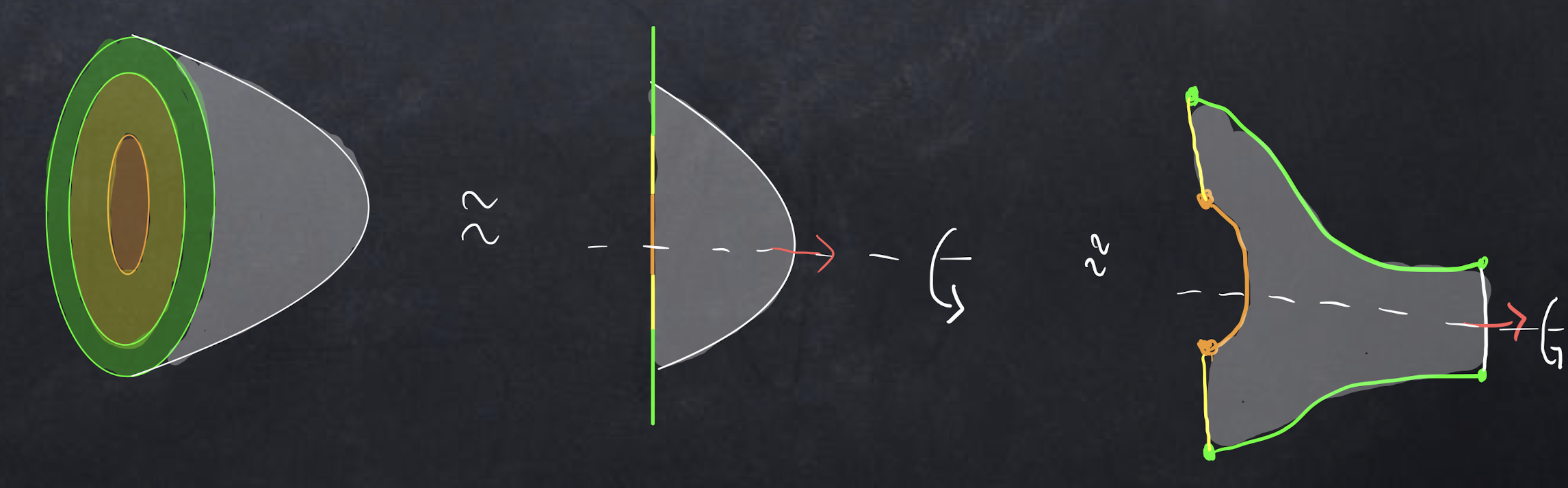}
  \vskip -.5pc
  \caption{The local label of~$\epsilon (Z)$ in codimension~2}\label{fig:c31}
  \end{figure}

The tubular neighborhood of~$\{0\}\times Z\subset \zoh\times M$ for
codimension~$\ell =2$ is the 3-dimensional bordism obtained from the pair of
chaps by revolution in 3-space, as illustrated in Figure~\ref{fig:c31}.  For
general~$\ell >1$, the bordism is the $(\ell +1)$-disk~$D^{\ell +1}$ with
boundary~$S^\ell $ partitioned as 
  \begin{equation}\label{eq:c52}
     \partial D^{\ell +1} = D^\ell _\epsilon \cup A^\ell _\rho \cup D^\ell
     _{\relax} 
  \end{equation}
into disks~$D^\ell $ and an annulus~$A^\ell $ with the domain wall~$\delta $
at the intersection of the~$\epsilon $ and~$\rho $-colored regions.  (In
Figure~\ref{fig:c31} that domain wall is thickened from a sphere~$S^{\ell -1}$
to an annulus~$A^\ell $.) 

  \begin{remark}[]\label{thm:c48}
 These are the local defects.  The global defects are a section of a bundle
(local system) of local defects over the submanifold~$Z\subset M$.
  \end{remark}

  \begin{example}[finite homotopy theory]\label{thm:c49}
 Let $\sigma =\sXd{n+1}\sX$ be the finite homotopy theory built from a $\pi
$-finite space~$\sX$.  Then we can use the semiclassical calculus for $\pi
$-finite spaces to compute semiclassical spaces of defects.  Suppose $\rho
$~is specified by a basepoint $*\to \sX$ and $\epsilon $~is specified by the
identity map $\sX\xrightarrow{\;\id\;}\sX$.  Then $\delta $~is specified by
the homotopy fiber product
  \begin{equation}\label{eq:c53}
     \begin{gathered} \xymatrix{&\ast\ar@{-->}[dl]\ar@{-->}[dr] \\
     \ast\ar[dr]&&\sX\ar[dl]^>>>>>>>{\id}\\ &\sX} \end{gathered} 
  \end{equation}
which is a point.  This is the manifestation of the uniqueness of~$\delta $
(Lemma~\ref{thm:34}), as already remarked after~\eqref{eq:14}.
 
The semiclassical space of local $\rho $-defects of codimension~$\ell $ is
  \begin{equation}\label{eq:c54}
     \Map\bigl((D^\ell ,S^{\ell -1}),(\sX,*) \bigr)=\Omega ^\ell \sX. 
  \end{equation}
Set 
  \begin{equation}\label{eq:c55}
     N^\ell =\bigl(D^{\ell +1},D^\ell \cup A^\ell \cup D^\ell \bigr), 
  \end{equation}
with boundary as in~\eqref{eq:c52}; see Figure~\ref{fig:c31}.  The
semiclassical local label of the defect~$\epsilon (Z)$ is 
  \begin{equation}\label{eq:c56}
     \Map(N^\ell ,\sX)\longrightarrow \OlX ,
  \end{equation}
the map induced by restriction to~$\OlX$. 

  \begin{lemma}[]\label{thm:c50}
 There is a homotopy equivalence $\Map(N^\ell ,\sX)\simeq \OlX$ under which
\eqref{eq:c56}~is the identity map. 
  \end{lemma}

  \begin{proof}
 Use the technique in Example~0.8 of~\footnote{Allen Hatcher, \emph{Algebraic topology}. available at
  \url{https://pi.math.cornell.edu/~hatcher/AT/ATpage.html}.  }.  First,
deformation retract~$A^\ell $ to~$S^{\ell -1}$, and so define
  \begin{equation}\label{eq:c57}
     \bNl = \bigl(D^{\ell +1},D^\ell \cup S^{\ell -1}\cup D^\ell \bigr). 
  \end{equation}
Choose a basepoint for~$\bNl$ on~$S^{\ell -1}$.  Form the correspondence of
pointed spaces  
  \begin{equation}\label{eq:c58}
     \begin{gathered} \xymatrix{&\bNl\cup D^\ell
     \ar[dl]\ar[dr]\\\bNl\!\bigm/\!S^{\ell -1}&& D^\ell \!\bigm/\!S^{\ell
     -1}} \end{gathered} 
  \end{equation}
in which $D^\ell $~is attached to $S^{\ell -1}\subset \bNl$, the left map
collapses this new~$D^\ell $, and the right map collapses~$D^{\ell +1}$.
Since $D^\ell ,D^{\ell +1}$~are contractible, each of these arrows is a
homotopy equivalence.  Now take the pointed mapping spaces into~$\sX$. 
  \end{proof}

The quantization of $\id\:\OlX\to \OlX$ is typically a noninvertible object.
For example, if the quantization is a vector space, then the vector space
is\footnote{The homotopy group $\pi _{\ell }\sX=\pi _\ell (\sX,*)$ uses the
basepoint~$*\in \sX$.} $\Fun(\pi _0\OlX)=\Fun(\pi _\ell \sX)$; the local
label is the constant function~1.  If the quantization is a linear category,
then it is the category $\Vect(\OlX)$ of flat vector bundles over~$\OlX$,
i.e., vector bundles on the fundamental groupoid $\pi _{\le1}\OlX$; the local
label is the trivial bundle with fiber~$\CC$.

  \begin{equation}\label{eq:c59}
     \id\:\Map(Z^\nu ,\sX)\longrightarrow \Map(Z^\nu ,\sX), 
  \end{equation}
where $Z^\nu $~is the Thom space of the normal bundle.  As an example,
suppose~$\ell =1$ and assume that the normal bundle $\nu \to Z$ has been
trivialized.  (This amounts to a coorientation of the codimension~1
submanifold~$Z\subset M$---a direction for the domain wall.)  Then 
  \begin{equation}\label{eq:c60}
     \Map(Z^\nu ,\sX)\simeq \Map(Z,\Omega \sX). 
  \end{equation}
For example, if $A$~is a finite abelian group and $\sX=\BtA$---so $\sigma
$~encodes a $BA$-symmetry---then $\Map(Z^\nu ,\BtA)\simeq \Map(Z,BA)$ is the
``space'' of principal $A$-bundles $P\to Z$.  One should, rather, treat it as
a groupoid, the groupoid $\Bun_A(Z)$ of principal $A$-bundles over~$Z$ and
isomorphisms between them.  A point $*\to \Bun_A(Z)$ is a principal
$A$-bundle $P\to Z$, and this map quantizes to a defect~$\eta (P)$ on~$Z$.
The quantization of $\id\:\Bun_A(Z)\to \Bun_A(Z)$ is a sum of the
quantizations of $*\!\!\bigm/\!\!\Aut P\to \Bun_A(Z)$ over isomorphism
classes of principal $A$-bundles $P\to Z$.  Informally, we might write this
as a sum of
  \begin{equation}\label{eq:c61}
     \frac{1}{\Aut P}\,\eta (P) = \frac{1}{H^0(Z;A)}\,\eta (P). 
  \end{equation}
This sort of expression appears in~\footnote{Yichul Choi, Clay C\'ordova, Po-Shen Hsin, Ho~Tat Lam, and Shu-Heng Shao,
  \emph{Non-Invertible Duality Defects in 3+1 Dimensions},
  \href{http://arxiv.org/abs/arXiv:2111.01139}{{\tt arXiv:2111.01139}}.
}, for example; compare Example~\ref{thm:36}.
 
The $\rho $-defect~$\eta (P)$ has a geometric semiclassical interpretation.
Without the defect one is summing over $A$-gerbes on $\zoh\times M$ that are
trivialized on~$\{0\}\times M$.  The defect~$\eta (P)$ on~$\{0\}\times Z$
tells to only trivialize the $A$-gerbe on $\bigl(\{0\}\times
M\bigr)\;\setminus \;\bigl(\{0\}\times Z\bigr)$ and to demand---relative to
the coorientation of~$Z$---that the trivialization jump by the $A$-bundle
$P\to Z$. 
  \end{example}

  \begin{remark}[]\label{thm:c51}
 \ 
 If the $\pi $-finite space~$\sX$ is equipped with a cocycle~$\lambda $ that
represents a cohomology class $[\lambda ]\in h^n(\sX)$ for some cohomology
theory~$h$, then a codimension~$\ell $ quotient defect has semiclassical
label space~$\OlX$ with transgressed cocycle and its cohomology class $[\tau
^\ell \lambda ]\in h^{n-\ell }(\OlX)$.  A nonzero cohomology class obstructs
the existence of the quotient.  However, as observed in footnote~\footref{RSS} it is
possible that $[\lambda ]\neq 0$ but $[\tau ^\ell \lambda ]=0$ for some~$\ell
$, which means that the quotient of the entire theory by~$\sigma $ does not
exist, but quotient defects of sufficiently high codimension do exist.
  \end{remark}

  \subsection*{Projectivity}

We begin with some ruminations on projective symmetry in quantum theory, in
part to make contact with C\'ordova's \hyperref[cordova]{lecture series}.  A review
of~\ref{subsec:1.11} at this point is warranted.

  \subsection{Linear and projective geometry}\label{subsec:3.10}

Let $V$~be a \emph{linear} space, say finite dimensional and complex.  The
automorphism group~$\Aut V$ consists of invertible linear maps $T\:V\to V$;
after a choice of basis it is isomorphic to the group of invertible square
complex matrices of size equal to~$\dim V$.  The \emph{projective} space~$\PP
V$ is the space of lines (1-dimensional subspaces) of~$V$.  A linear
automorphism $T\in \Aut V$ induces an automorphism $\bT$ of~$\PP V$; the
linear map~$T$ takes lines to lines.  A homothety (scalar multiplication)
of~$V$ induces the identity map of~$\PP V$.  More precisely, there is a group
extension
  \begin{equation}\label{eq:16}
     1\longrightarrow \Cx\longrightarrow \Aut V\longrightarrow \Aut \PP
     V\longrightarrow 1 
  \end{equation}
that serves to define~$\Aut \PP V$.  This extension is \emph{central}; the
kernel~$\Cx$ lies in the center of~$\Aut V$ (and equals the center).  Each
projective transformation in~$\Aut\PV$ has a $\Cx$-torsor of linear lifts.
 
The projective action of a group~$G$ on~$\PP V$ is a group homomorphism $G\to
\Aut\PV$.  One can use it to pull back the central extension~\eqref{eq:16} to
a central extension 
  \begin{equation}\label{eq:17}
     1\longrightarrow \Cx\longrightarrow G^\tau \longrightarrow
     G\longrightarrow 1 
  \end{equation}
of~$G$.  The group~$G^\tau $ acts linearly on~$V$ through a group
homomorphism $G^\tau \to \Aut V$ and, as our starting point, the group~$G$
acts projectively on~$\PV$.  The central extension~\eqref{eq:17} is a measure
of the projectivity of this projective action.  If $G$~is a discrete group,
say a finite group, then the equivalence class of the central extension is an
element of the cohomology group~$H^2(G;\Cx)$.  If this class is zero, then
there exist splittings of~\eqref{eq:17}.  Furthermore, the splittings form a
torsor over the group of characters of~$G$, i.e., over the cohomology group
$H^1(G;\Cx)$: given a splitting~$s$ any other splitting is the product of~$s$
with a character $\chi \:G\to \Cx$.  In summary:
  \begin{equation}\label{eq:18}
     \begin{aligned} \textnormal{existence: }&H^2(G;\Cx) \\
      \textnormal{uniqueness: }&H^1(G;\Cx) \\ \end{aligned} 
  \end{equation}

  \subsection{Quantum theory is projective, not linear}\label{subsec:3.11}

The space of pure states of a quantum system is a projective space~$\scP$, at
first with no topology.  Instead there is a function 
  \begin{equation}\label{eq:19}
     \scP\times \scP\longrightarrow [0,1] 
  \end{equation}
that maps an ordered pair of states to the probability of transitioning from
one state to the other.\footnote{If we write $\scP=\PP\sH$ for a Hilbert
space~$\sH$, and $\psi _1,\psi _2\in \sH$ are nonzero vectors, then the value
of~\eqref{eq:19} on the pair of lines generated is 
  \begin{equation}\label{eq:20}
     \frac{|\langle \psi _1,\psi _2 \rangle|^2}{\|\psi _1\|^2\,\|\psi
     _2\|^2}. 
  \end{equation}
}  Symmetries of the quantum system are automorphisms of~$\scP$ that
preserve the function~\eqref{eq:19}. 
 
A fundamental theory due to Wigner (and von Neumann) states that for
$\scP=\PP\sH$ any symmetry of the quantum system lifts either to a unitary
automorphism of~$\sH$ or an antiunitary automorphism of~$\sH$.
(See~\footnote{Daniel~S. Freed, \emph{On Wigner's theorem}, Proceedings of
the Freedman Fest
  (Vyacheslav~Krushkal Rob~Kirby and Zhenghan Wang, eds.), Geometry \& Topology
  Monographs, vol.~18, Mathematical Sciences Publishers, 2012, pp.~83--89.
  \href{http://arxiv.org/abs/arXiv:1211.2133}{{\tt arXiv:1211.2133}}.  } for
geometric proofs.)  In fact, up to a simple transformation, the transition
function~ \eqref{eq:19} equals the distance function for the Fubini-Study
metric on~$\PP\sH$, and so Wigner's theorem becomes a theorem about its
isometries.  It follows that the projectivity of a group~$G$ of symmetries of
quantum theory is measured by a $\zt$-graded central extension of~$G$, where
the $\zt$-grading keeps track of the unitary vs.~antiunitary dichotomy.  (See
\footnote{Daniel~S. Freed and Gregory~W. Moore, \emph{Twisted equivariant
matter},
  \href{http://dx.doi.org/10.1007/s00023-013-0236-x}{Ann.~Henri Poincar{\'e}
  \textbf{14} (2013)}, no.~8, 1927--2023,
  \href{http://arxiv.org/abs/arXiv:1208.5055}{{\tt arXiv:1208.5055}}.  I
apologize for the self-referential tendencies I learned from Gilderoy
Lockhart.  The paper on which these summer school notes are based will not be
so disgustingly provincial.} for a fuller discussion of symmetry in quantum
mechanics.)

  \subsection{Projectivity in quantum field theory: anomalies}\label{subsec:3.12}

Here the metaphor~\eqref{eq:c95} comes in handy.  A field theory
$F\:\Bord_n(\sF)\to \sC$ is a linear representation of $\Bord_n(\sF)$, and
since quantum theory is projective we expect this representation to be
projective, and furthermore its projectivity should be measured by some kind
of ``cocycle'' on~$\BnF$.  Indeed, the right kind of cocycle in this context
is an \emph{invertible field theory}.  In this context the projectivity is
called an \emph{anomaly (theory)};\footnote{Note that the anomaly is a theory
over the background fields~$\sF$, which may or may not have to do directly
with symmetry.} it is an invertible field theory over~$\sF$ of
dimension~$n+1$.  Observe that in the case of a finite group~$G$ acting on a
quantum mechanical system ($n=1$), this matches the cocycle of a group
extension~\eqref{eq:17}, which has degree~2.
 
Anomalies are not a sickness of a theory; to the contrary, they are useful
tools for investigating its behavior.  They are only potentially a sickness
when we want to ``integrate out'' some fields, i.e., turn some background
fields into fluctuating fields.  Roughly speaking, that is because we can
integrate functions valued in a \emph{linear space}, not functions valued in
a \emph{projective space}.  So to carry out the integration, we must lift the
projective geometry (field theory) to linear geometry.  In this situation we
have a fiber bundle of fields, such as
  \begin{equation}\label{eq:21}
     \begin{gathered} \xymatrix@C-2pc{\sF \ar[d]_{\pi }&= 
     \{\textnormal{Riemannian metric,
     orientation, $H$-connection}\}  \\ \overline{\sF} &=
     \{\textnormal{Riemannian metric, orientation}\}
     \phantom{\textnormal{
     $H$-connection}}} \end{gathered}  
  \end{equation}
The total space is the space of background fields in the starting theory, the
base is the space of background fields in the pushforward theory, and the
fibers are the background fields in the original theory that we have promoted
to fluctuating fields.  Suppose the original theory over~$\sF$ has an
anomaly~$\alpha $: an invertible $(n+1)$-dimensional theory over~$\sF$.  Then
to integrate over the fibers of~$\pi $---to push forward under~$\pi $---we
need to provide \emph{descent data} for~$\alpha $.  In other words, we need
to provide an $(n+1)$-dimensional field theory~$\ba$ over~$\overline{\sF}$
and an isomorphism $\alpha \xrightarrow{\;\cong \;}\pi ^*\ba$.  This is the
formal part---the main work is the analysis required to integrate over an
infinite dimensional space---but if this can be done, we obtain a pushforward
theory over~$\overline{\sF}$ with anomaly~$\ba$.  This descent problem has
existence and uniqueness aspects, analogous to~\eqref{eq:18}, only now with
invertible field theories.

  \begin{remark}[]\label{thm:37}
 Changing descent data by tensoring with an $n$-dimensional invertible theory
is sometimes called ``changing the scheme'', and pushforwards that differ in
this way share many physical properties. 
  \end{remark}

  \subsection{'t Hooft anomalies in finite homotopy theories}\label{subsec:3.13}

Recall~\ref{subsec:2.7}, which I recommend you review at this point.  In
terms of the discussion above, a ``cocycle'' on a $\pi $-finite space~$\sX$
defines an invertible field theory in which the map to~$\sX$ remains a
background field.  Once we sum---the finite path integral---over these maps,
with the cocycle as a weight, then we obtain a typically noninvertible
theory~$\sigma $.  When equipped with a semiclassical regular boundary
theory~$\rho $, the quantization of data in~\ref{subsec:2.13}, then
$\sr$~represents a symmetry with an \emph{'t~Hooft} anomaly represented by
the cocycle.

  \subsection{Twisted boundary theories}\label{subsec:3.14}

At this point recall Definition~\ref{thm:c72}.  Note that if $\sX$ is a
$\pi $-finite space, and we take the zero cocycle, then a (right or left)
semiclassical boundary theory is a map $\sY\to \sX$ of $\pi $-finite spaces
together with a \emph{cocycle}~$\mu $ on~$\sY$.  If we are working with an
$m=(n+1)$-dimensional theory, then $\mu $~has degree~$m$.  The cocycle~$\mu $
is used to weight/twist the quantization of the boundary.

  \subsection*{Example: $BA$-symmetry in 4~dimensions and line defects}

The following discussion is inspired by~\footnote{\label{AST}
Ofer Aharony, Nathan Seiberg, and Yuji Tachikawa, \emph{{Reading between the
  lines of four-dimensional gauge theories}},
  \href{http://dx.doi.org/10.1007/JHEP08(2013)115}{JHEP \textbf{08} (2013)},
  115,
\href{http://arxiv.org/abs/1305.0318}{{\tt arXiv:1305.0318 [hep-th]}}; see
also Section~2.3 of 
Davide Gaiotto, Gregory~W. Moore, and Andrew Neitzke, \emph{{Framed BPS
  States}}, \href{http://dx.doi.org/10.4310/ATMP.2013.v17.n2.a1}{Adv. Theor.
  Math. Phys. \textbf{17} (2013)}, no.~2, 241--397,
  \href{http://arxiv.org/abs/1006.0146}{{\tt arXiv:1006.0146 [hep-th]}}.
This example is discussed in~\footref{F1} as well.}.

  \subsection{Symmetry data}\label{subsec:3.15}

Let $A$~be a finite abelian group, set $\sX=\BnA2$, and fix a basepoint $*\to
\BnA2$.  This defines the semiclassical data of a $BA$-symmetry, what is
often referred to as a ``1-form $A$-symmetry''.  We set~$\lambda =0$: there
is no 't~Hooft anomaly.  For definiteness set~$n=4$, so we use the
5-dimensional finite homotopy theory $\sigma =\sXd5{\BnA2}$.  The basepoint
gives a regular right boundary theory, and we study the pair~$\sr$ as
abstract 4-dimensional symmetry data.

  \subsection{The left $\sr$-module: 4-dimensional gauge theory}\label{subsec:3.16}

\begin{figure}[ht]
  \centering
  \includegraphics[scale=.6]{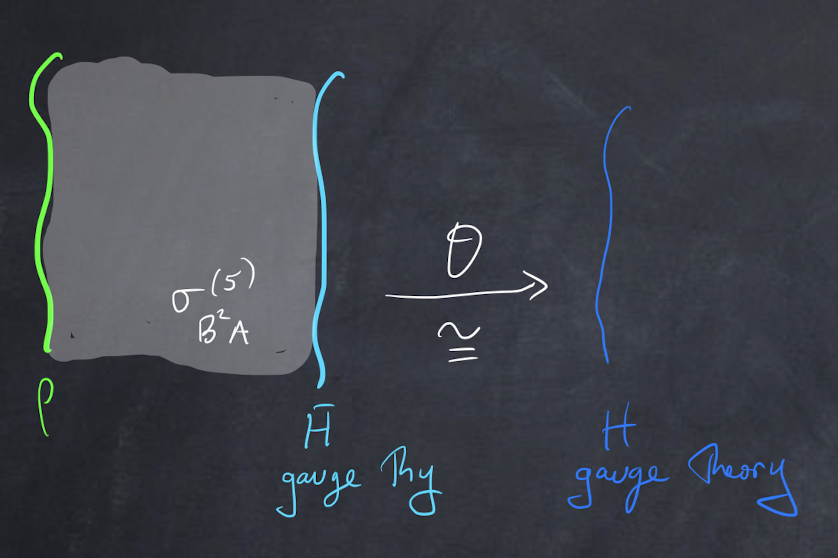}
  \vskip -.5pc
  \caption{The $BA$-action on $H$-gauge theory}\label{fig:19}
  \end{figure}

Let $H$~be a Lie group, and suppose $A\subset H$ is a subgroup of its center.
Set $\bH=H/A$.  From the exact sequence 
  \begin{equation}\label{eq:22}
     A\longrightarrow H\longrightarrow \bH 
  \end{equation}
of Lie groups we obtain a sequence of fiberings 
  \begin{equation}\label{eq:23}
     BA\longrightarrow BH\longrightarrow B\bH\longrightarrow \BnA2 
  \end{equation}
In fact, we can promote~\eqref{eq:23} to a fibering of classifying
spaces\footnote{These are simplicial sheaves on~$\Man$; see footnote~\footref{FHop}.}
of connections: 
  \begin{equation}\label{eq:24}
     B\mstrut _{\nabla }H\longrightarrow B\mstrut _{\nabla
     }\bH\longrightarrow \BnA2  
  \end{equation}
The fibering of $B\mstrut _\nabla H$ over~$\BnA2$ encodes the action of~$BA$ on
$B\mstrut _\nabla H$.  (Given a principal $H$-bundle with connection and a
principal $A$-bundle, use the \emph{homomorphism} $A\times H\to H$ to construct
a new principal $H$-bundle with connection.)  If $H$~is a finite group, and
therefore $\bH$~too is a finite group, then the map $B\mstrut _{\nabla
}\bH=B\bH\to \BnA2$ would be a semiclassical left boundary theory.  In general,
of course, there is no finiteness nor is $\bH$-gauge theory a topological field
theory.  Nonetheless, we set $\tF$~to be 4-dimensional $\bH$-gauge theory, and
so obtain the $\sr$-module exhibited in Figure~\ref{fig:19}.

  \begin{remark}[]\label{thm:38}
 We have used the phrase ``$H$-gauge theory'' without specifying which one.
All we need here that it is a 4-dimensional gauge theory with $BA$-symmetry.
Other details do not enter this discussion. 
  \end{remark}

  \subsection{Topological right $\sr$-modules}\label{subsec:3.17}

\begin{figure}[ht]
  \centering
  \includegraphics[scale=.6]{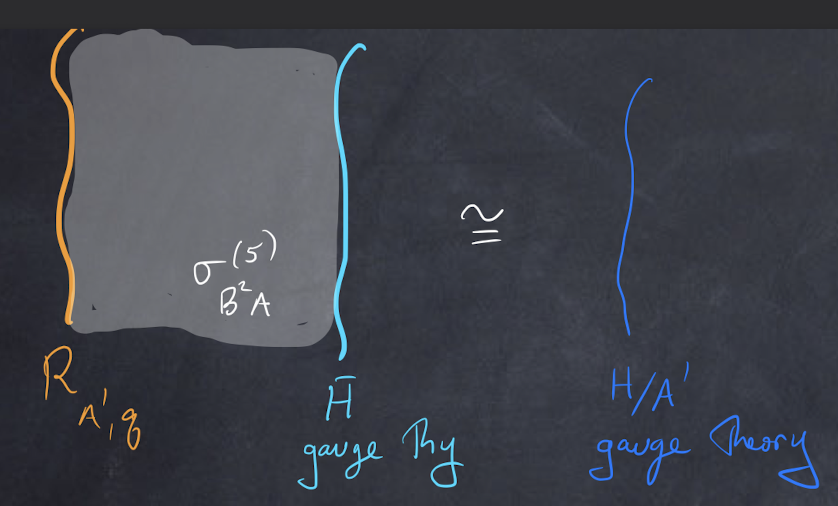}
  \vskip -.5pc
  \caption{A quotient of $H$-gauge theory}\label{fig:20}
  \end{figure}

For any subgroup $A'\subset A$ there is a map of $\pi $-finite spaces 
  \begin{equation}\label{eq:25}
     \BnA2'\longrightarrow \BnA2 
  \end{equation}
We use this as semiclassical right boundary data, but we allow a twisting as
in~\ref{subsec:3.14}, i.e., a degree~4 cocycle~$\mu $ on~$\BnA2'$.  In this
case we use ordinary cohomology (so assume an orientation is among the
background fields), and then the class of~$\mu $ lives in the cohomology
group $H^4(\BnA2';\Cx)$.  A theorem of Eilenberg-MacLane computes 
  \begin{equation}\label{eq:26}
     H^4(\BnA2';\Cx) \cong \{\textnormal{quadratic functions
     }q\:A'\longrightarrow \Cx\} 
  \end{equation}
Thus pairs~$(A',q)$ determine a right topological boundary theory~$\RAq$.
The sandwich $\RAq\otimes _{\sigma }\tF$, depicted in Figure~\ref{fig:20}, is
a twisted form of $H/A'$-gauge theory.

  \begin{remark}[]\label{thm:39}
 Under the isomorphism~\eqref{eq:26}, the quadratic form~$q$ gives rise to
the \emph{Pontrjagin square} cohomology operation 
  \begin{equation}\label{eq:27}
     \scaP_q\:H^2(X;A')\longrightarrow H^4(X;\Cx) 
  \end{equation}
on any space~$X$.  It enters the formula for the partition function in the
theory $\RAq\otimes _{\sigma }\tF$.
  \end{remark}

  \begin{figure}[ht]
  \centering
  \includegraphics[scale=.6]{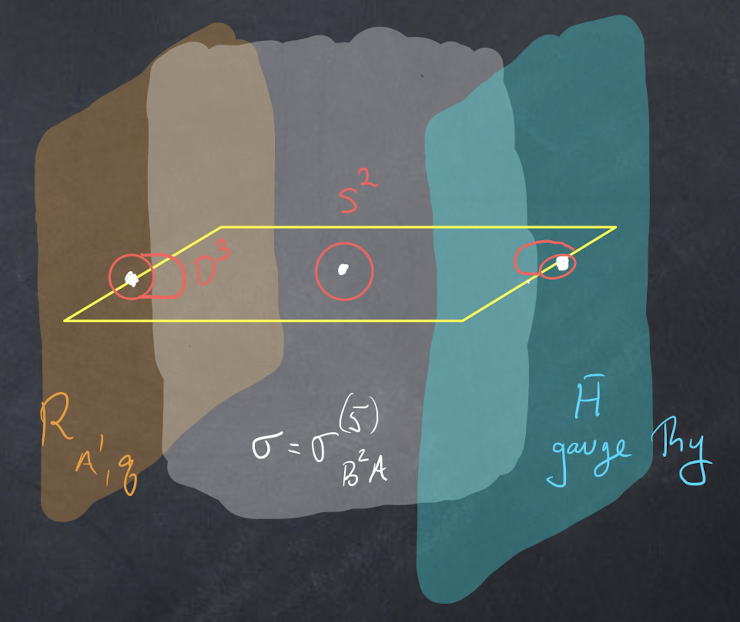}
  \vskip -.5pc
  \caption{Line defects in twisted $H/A'$-gauge theory}\label{fig:21}
  \end{figure}

  \subsection{Local line defects: interior label}\label{subsec:3.18}
 
\begin{figure}[ht]
  \centering
  \includegraphics[scale=.6]{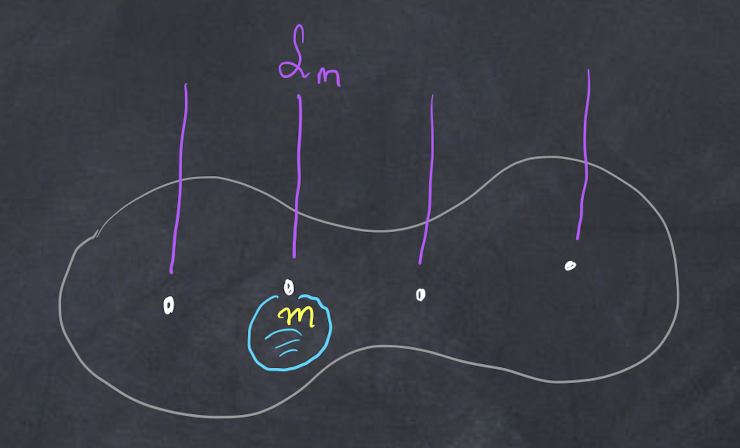}
  \vskip -.5pc
  \caption{The 2-category $\Cat\bigl((\BnA2)^{S^2}\bigr)$}\label{fig:22}
  \end{figure}

We study local line defects in the twisted $H/A'$-gauge theory using the
sandwich picture in Figure~\ref{fig:21}.  So we have the surface~$[0,1]\times
C$ built over a curve~$C$ (in some manifold~$M$), with topological
$\RAq$-boundary at $\{0\}\times C$ and with nontopological $\tF$-boundary at
$\{1\}\times C$.  See Figure~\ref{fig:6} for an analogous figure in quantum
mechanics for point defects.  We find the relevant labels by working down in
dimension of strata, starting at the top.  Thus we begin with the link of an
interior point, which is~$S^2$.  Its quantization~$\sigma (S^2)$ is, say, a
linear 2-category---one step beyond our illustration in
Example~\ref{thm:c59}.  Observe that for the mapping space $\Map(S^2,\BnA2)$
we have
  \begin{equation}\label{eq:28}
     \begin{aligned} \pi _0\bigl(\Map(S^2,\BnA2) \bigr)&= H^2(S^2;A)\cong A
      \\ \pi _1\bigl(\Map(S^2,\BnA2) \bigr)&= H^1(S^2;A)=0 \\ \pi
      _2\bigl(\Map(S^2,\BnA2) \bigr)&= H^0(S^2;A)\cong A \\ \end{aligned} 
  \end{equation}
The quantization consists of flat bundles (local systems) of linear
categories over the 2-groupoid with these homotopy groups, depicted in
Figure~\ref{fig:22}.  In other words, for each 
  \begin{equation}\label{eq:29}
     m\in H^2(S^2;A)\cong A 
  \end{equation}
there is a linear category~$\sL_m$.  Furthermore, $\pi _2$~based at~$m$ acts
on~$\sL_m$ as automorphisms of the identity functor.  Under suitable
assumptions, then, we can decompose according to the characters 
  \begin{equation}\label{eq:30}
     e\in H^0(S^2;A)\dual\cong A\dual 
  \end{equation}
to obtain 
  \begin{equation}\label{eq:31}
     \sL_m = \bigoplus\limits_{e}\sL_{m,e}\cdot e
  \end{equation}
In summary, then, an object of~ $\sigma (S^2)$ is a collection of linear
categories~$\sL_{m,e}$ labeled by $m\in A$ and~$e\in A\dual$.  

  \subsection{Local line defects: collating the labels}\label{subsec:3.19}

Now we proceed to the lower dimensional strata, so determine the labels on
the boundary strata of $\zo\times C$, depicted with their links in
Figure~\ref{fig:21}.  A label on the $\RAq$-colored boundary is an object~
$\sL_0$ in this category---part of the topological field theory~$\sr$---and a
label of the $\tF$-colored boundary is another object~$\sL_1$.  The image
under~$\theta $ of this configuration is the sum over~$m,e$ of
$\Hom\bigl((\sL_0)_{m,e},(\sL_1)_{m,e} \bigr)$, which is the category of
defects in the twisted $H/A'$-gauge theory.  What we will now compute is that
$\sL_0$~is supported at a subset of pairs~$(m,e)$ determined by the
subgroup~$A'\subset A$ and the quadratic function $q\:A'\to \Cx$.  This is
the information gained from the symmetry; it matches the examples
in footnote~\footref{AST}.  Note that the information is from the topological part of
the sandwich; the particulars of the analytic part (the $\bH$-gauge theory)
do not enter.

  \subsection{Higher Gauss law}\label{subsec:3.20}

We begin with the lower Gauss law.  Suppose $\sG$~is a finite 1-groupoid, and
$L\to \sG$ is a complex line bundle over~$\sG$.  We want to compute the
global sections (which is a limit; the colimit is equivalent).  At a point
$m\in \sG$ the group $\pi _1(\sG,m)$~acts by a character on the
fiber~$L_m$.  The value of a global section at~$m$ must be a fixed point of
this action, so it lies in the invariant subspace of~$L_m$, that is, the section
vanishes unless the character vanishes.
 
Now suppose $\sK\to \sG$ is a bundle of invertible complex linear categories
over a 2-groupoid~$\sG$, the higher analog of a complex line bundle.  (We
call the fibers \emph{$\VV$-lines}, where $\VV=\Vect$ is the category of
complex vector spaces, regarded as the categorification of a ring in this
context.)  Then at each~$m\in \sG$ there are two layers to consider, and we
need fixed point data for both.  If $f\in \pi _1(\sG, m)$ and $x\in \sK_m$,
then the first piece of fixed point data is an isomorphism
  \begin{equation}\label{eq:32}
      \eta _f(x)\:x\longrightarrow  f(x) 
  \end{equation}
Then if $a\:f\to g$ is a 2-morphism in~$\sG$ at~$m$, we also need an
isomorphism~$\lambda _a(x)$ in
  \begin{equation}\label{eq:33}
     \begin{gathered} \xymatrix@R-.3pc@C+1pc{&f(x)\ar[dd]^{\lambda _a(x)}\\
     x\ar[ur]^{\eta 
     _f(x)} \ar[dr]_{\eta _g(x)}\\&g(x)} \end{gathered} 
  \end{equation}
that makes the diagram commute.  Now specialize to $f=g=\id_x$.  Then
\eqref{eq:33}~becomes  
  \begin{equation}\label{eq:34}
     \begin{gathered} \xymatrix@R-.3pc@C+1pc{&x\ar[dd]^{\lambda _a(x)}\\
     x\ar[ur]^{\eta } \ar[dr]_{\eta }\\&x} \end{gathered} 
  \end{equation}
for some automorphism~$\eta $.  This commutes only if $\lambda _a(x)=\id$.
Hence, if $\pi _1\sG=0$ then the global sections are only nonzero on the
components on which $\pi _2\sG$~acts trivially.  This is the higher Gauss
law.  

  \begin{exercise}[]\label{thm:52}
 Make this paragraph precise using the theory of semiadditive categories.
  \end{exercise}

  \begin{figure}[ht]
  \centering
  \includegraphics[scale=.7]{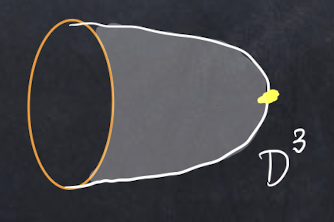}
  \vskip -.5pc
  \caption{The link in Figure~\ref{fig:21} at the $\RAq$-colored boundary}\label{fig:23}
  \end{figure}

  \subsection{Local line defects: the missing link}\label{subsec:3.21}

First, associated to the quadratic function $q\:A'\to \Cx$ is a
bihomomorphism 
  \begin{equation}\label{eq:35}
     b\:A'\times A'\longrightarrow \Cx 
  \end{equation}
It induces a Pontrjagin-Poincare duality 
  \begin{equation}\label{eq:38}
     H^2(S^2;A')\times H^0(S^2;A')\longrightarrow \Cx 
  \end{equation}
and so an isomorphism 
  \begin{equation}\label{eq:39}
     e'\:H^2(S^2;A')\longrightarrow H^0(S^2;A')\dual 
  \end{equation}

We proceed to quantize the link depicted in Figure~\ref{fig:23}, which is a
3-disk $D^3$ with boundary colored by~$\RAq$.  Label the center
point~$(m,e)$, which means the tensor unit category~$\Vect$ sitting over~ $m$
with $\pi _2$~acting via the character~$e$; see~\eqref{eq:31}.  (Thus
$\sL_{m,e}=\Vect$ and the linear categories labeled by other pairs~$(m',e')$
are zero.)  There is a semiclassical description of this defect~$(m,e)$ in
terms of Definition~\ref{thm:c68}: the space~$\sY=\BnA2$ is equipped with a
2-cocycle that represents the class in $H^2(\BnA2;\Cx)\cong A\dual$ given by
the character~$e$; the map to $\Map(S^2,\BnA2)$ is the identity onto the
component indexed by~$m$.  At the $\RAq$-colored boundary, in semiclassical
terms the boundary theory is given as the map
  \begin{equation}\label{eq:36}
     \bigl(\Map(S^2,\BnA2'),\tau ^2(\mu _q )\bigr)\longrightarrow
     \Map(S^2,\BnA2), 
  \end{equation}
where $\tau ^2\mu _q$~is the transgression of a cocycle that represents the
class of the quadratic function in~\eqref{eq:26}.  Altogether we have a
diagram 
  \begin{equation}\label{eq:37}
     \begin{gathered} \xymatrix@C-3pc{\bigl(\Map(S^2,\BnA2'),\tau ^2(\mu _q)
     \bigr)\ar[dr] && \Map(D^3\setminus B^3,\BnA2)\ar[dl]\ar[dr] &&
     (B^2A,e)\ar[dl]^{m} \\ 
     &\Map(S^2,\BnA2) && \Map(S^2,\BnA2)} \end{gathered} 
  \end{equation}
Interpret in terms of the bordism obtained by cutting out a ball around the
yellow defect; in the bottom row the first entry is the orange boundary and
the second the link of the yellow defect.  Take the homotopy limit to
conclude that the quantization is supported on pairs~$(m,e)$ which satisfy
  \begin{equation}\label{eq:40}
     \begin{aligned} m&\in A' \\ e &= e'(m)\inv \end{aligned} 
  \end{equation}
This selection rule is the one in footnote~\footref{AST}. 

  \begin{exercise}[]\label{thm:40}
 Check this last statement in some examples from footnote~\footref{AST}. 
  \end{exercise}

  \subsection*{Some additional problems}
 
  \begin{figure}[ht]
  \centering
  \includegraphics[scale=.5]{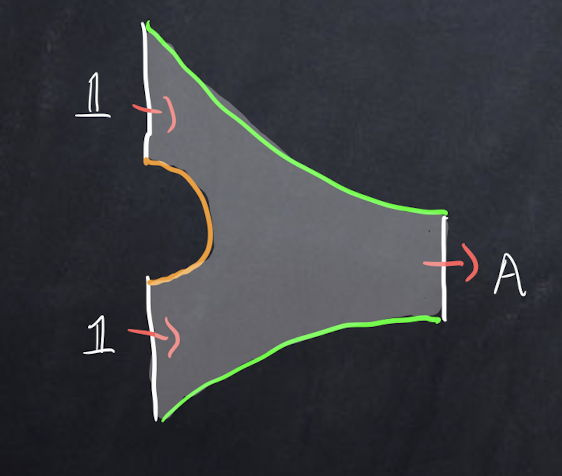}
  \vskip -.5pc
  \caption{An element in $\Hom(1,A)$}\label{fig:p4}
  \end{figure}

	\problem 
 Let $G$~be a finite group, let $\chi \:G\to \Cx$ be a character (it could be
the trivial character that sends each $g\in G$ to $1\in \Cx$), and let
$\sigma =\sXd2{BG}$~be the 2-dimensional finite $G$-gauge theory with values
in $\sC=\Alg(\Vect)$ the Morita 2-category of algebras.  Then $\sigma
(\pt)=A=\GA$.  The regular boundary theory (constructed from the regular
module $A_A$) is indicated in Figure~\ref{fig:p4} in green, and the
augmentation boundary theory in orange.  The depicted bordism maps to an
element of $\Hom(1,A)$, where this is Hom in $\Omega
\sC=\Hom_{\sC}(1,1)=\Vect$.  (Why?)  So it can be identified with an element
of $A=\GA$.  Compute that element.
	\endproblem

\begin{figure}[ht]
  \centering
  \includegraphics[scale=.6]{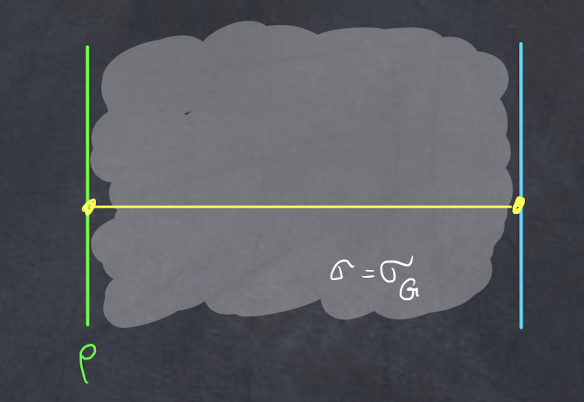}
  \vskip -.5pc
  \caption{Point defects in quantum mechanics}\label{fig:p5}
  \end{figure}

	\problem 
 Let $(\sH,H)$ define a quantum mechanical theory~$F$ that is invariant under
the action of a finite group~$G$.  Let $\sigma $~be the 2-dimensional
$G$-gauge theory and consider Figure~\ref{fig:p5}, which shows~$F$ in the
``sandwich'' picture.  There a general point defect is depicted.  Point
defects in~$F$ form a vector space.  What extra structure is encoded in this
sandwich picture?  Now replace the right regular boundary~$\rho $ with a
quotient $\epsilon _{G'}$ for a subgroup~$G'\subset G$: it is associated to
the right module $\CC\langle G'\backslash G  \rangle$ for the group algebra
$A=\GA$.  What theory is obtained?  What are the point defects in terms of
the point defects for~$F$?  What happens if you twist by a character of~$G'$? 
	\endproblem

\newpage
\setcounter{section}{3}
   \section{Lecture 4: Duality defects and applications}\label{sec:4}
%\setcounter{footnote}{36}
%\setcounter{figure}{41}
% lastsubsec@ 21

So far we have emphasized the structure of symmetry in field theory at the
expense of concrete examples.  In many ways this lecture is no different: its
focus is on abelian duality in many forms and then on a particular defect
related to duality.  But at least we have a few definite applications to
honest quantum field theories!  They come from two
papers,\footnote{\label{CK}Yichul Choi, Clay C\'ordova, Po-Shen Hsin, Ho~Tat
Lam, and Shu-Heng Shao,
  \emph{Non-Invertible Duality Defects in 3+1 Dimensions},
  \href{http://arxiv.org/abs/arXiv:2111.01139}{{\tt arXiv:2111.01139}}\\
\indent Justin Kaidi, Kantaro Ohmori, and Yunqin Zheng, \emph{{Kramers-Wannier-like
  Duality Defects in (3+1)D Gauge Theories}},
  \href{http://dx.doi.org/10.1103/PhysRevLett.128.111601}{Phys. Rev. Lett.
  \textbf{128} (2022)}, no.~11, 111601,
  \href{http://arxiv.org/abs/2111.01141}{{\tt arXiv:2111.01141 [hep-th]}}.
}
whose authors include two of the other lecturers at this school.  This
lecture is to a large extent an exposition of these papers. 
 
We begin with a broad outline of the dynamical question addressed and how
symmetry plays a role in solving it.  It emerges that the question addressed
is whether a topological field theory has an augmentation in the sense
of~\ref{subsec:3.2}.  Then we turn to \emph{abelian duality}: for finite
abelian groups; for spectra; and finally for special finite homotopy
theories, where it is a form of \emph{electromagnetic duality}.  This duality
induces a map on boundary theories, and one interest here is that the
transform of a theory under electromagnetic duality is the quotient theory by
the symmetry (Corollary~\ref{thm:53}).  We also discuss \emph{self-duality}:
topological theories that are self-electromagnetic dual.  In the last part
of the lecture we consider situations in which the quotient~$F\bs$ of a
theory by a symmetry is isomorphic to the original theory~$F$.  Composition
with the topological defect~$\delta $ in~\ref{subsec:3.8} is a self-domain
wall of the theory~$F$, the so-called \emph{duality defect}.  At the end of
the lecture we return to the applications in footnote~\footref{CK}.  The main argument
is in~\ref{subsec:4.19}.

  \subsection*{Trivially gapped theories with finite symmetry}

In quantum field theory one is often interested in the low energy, or quantum
relativistically equivalent\footnote{In terms of L=length, T=time, and M=mass,
energy has units~$\rM\rL^2/\rT^2$.  The fundamental constants have units
$[c]=\rL/\rT$ and $[\hbar]=\rM\rL^2/\rT$.  Use~$\hbar$ to relate energy and
inverse time, and then use~$c$ to get inverse length.} long distance, behavior
of a system.  As with classical physics, posed in terms of differential
equations, a quantum field theory is formulated at short distance/time and the
interesting dynamics---the ``answers'' to the short range
``questions''---happen at long distance/time.  The term \emph{ultraviolet}~(UV)
is used for short range, and the term \emph{infrared}~(IR) is used for long
range.\footnote{These derive from wavelength~$\lambda $, which is related to
energy in a relativistic quantum theory by $E=2\pi \hbar c/\lambda $.}  The
\emph{renormalization group flow} is meant to be\footnote{The cautious language
reflects the lack of rigorous constructions in many situations.} a flow on a
space of theories which has trajectories that limit at negative infinite time
to an ultraviolet theory and limit at positive infinite time to a corresponding
infrared theory.

  \subsection{Possible infrared behaviors}\label{subsec:4.1}

First, there is a basic dichotomy in quantum systems: \emph{gapped}
vs.~\emph{gapless}.  A quantum mechanical system is \emph{gapped} if its
minimum energy is an eigenvalue of finite multiplicity of the Hamiltonian,
assumed bounded below, and is an isolated point of the spectrum.  This notion
generalizes to a relativistic quantum field theory if we understand `spectrum'
to mean the spectrum of representations of the translation group of Minkowski
spacetime.  In the infrared, a gapless system is typically thought to be
well-approximated by a conformal field theory.  The low energy effective theory
of a gapped system is typically thought to be well-approximated by a
topological field theory, roughly up to tensoring by an invertible field theory
(which need not be topological).\footnote{These statements are hardly
universal; much more thought is needed here.  For example, fracton models in
condensed matter physics do not have long range field theory approximations in
the traditional sense.}  In the gapped case there is a further dichotomy: the
low energy approximation can be invertible or not.  In case it is invertible,
the term `trivially gapped' is sometimes used.  One can view this invertibility
condition as a nondegeneracy condition on the vacuum, namely that there be a
unique vacuum state on each space.  That condition typically holds for quantum
mechanical systems, but is often violated in supersymmetric field theories;
noninvertible low energy topological field theories also occur for
nonsupersymmetric systems.

  \subsection{Persistence of symmetry}\label{subsec:4.2}

Let $\sr$~be symmetry data that acts on an ultraviolet theory~$\FUV$.  Under
renormalization group flow, one imagines that the symmetry persists: 
  \begin{equation}\label{eq:42}
     \begin{tikzcd} \sr \lact\FUV \arrow[d, rightsquigarrow, shift
     left=3.8ex, "\textnormal{ RG flow}"] \\\sr\lact\FIR \end{tikzcd}  
  \end{equation}
In other words, if there is $\sr$-symmetry in the ultraviolet, then there
should be $\sr$-symmetry in the infrared as well. 

  \begin{remark}[]\label{thm:41}
 In general, the symmetry in the infrared could be bigger: an \emph{emergent
symmetry} may occur.  As well, there may be symmetries in the~UV that act
trivially in the~IR.  The assertion is that there is a homomorphism from
UV-symmetry to IR-symmetry.  (We have not discussed ``homomorphisms'' of
symmetry in these lectures.) 
  \end{remark}

  \subsection{Trivially gapped theories with symmetry}\label{subsec:4.3}

The question, then, of whether a quantum field theory with symmetry~$\sr$ can
be trivially gapped is whether there exists a left $\sigma $-module~$\tl$
such that
  \begin{equation}\label{eq:43}
     \lambda :=\rho \otimes _\sigma \tl 
  \end{equation}
is an \emph{invertible} field theory.  If we are in the situation of
Definition~\ref{thm:c13}, if $\tl$~is topological, and if $A=\sigma (\pt)$ is
an algebra, then this is equivalent\footnote{Evaluate~\eqref{eq:43} on a point
to see that $\lambda (\pt)=A\otimes _A\tl(\pt)$, and so $\tl(\pt)$~is an
invertible $A$-module.}  to asking for an augmentation $A\to 1$ of~$A$; see
Definition~\ref{thm:c19}.  One can envision obstructions to the existence of an
augmentation without knowing anything specific about~$\FUV$ or the left
$\sr$-module structure on it.  Indeed, that is precisely how we argue at the
end of the lecture.  (See~\ref{subsec:4.18} to be sure which theory does not
have an augmentation.)

  \subsection*{Abelian duality}

  \subsection{Finite abelian groups}\label{subsec:4.4}

This is the classical version of \emph{Pontrjagin duality}. 

  \begin{definition}[]\label{thm:42}
 Let $A,A'$~be finite abelian groups.  A bihomomorphism 
  \begin{equation}\label{eq:44}
     b\:A\times A'\longrightarrow \Cx 
  \end{equation}
is a \emph{duality pairing} if $b$~is nondegenerate in the sense that (1)~if
$a\in A$ satisfies $b(a,a')=1$ for all~$a'\in A'$, then $a=1$; and (1)~if
$a'\in A'$ satisfies $b(a,a')=1$ for all~$a\in A$, then $a'=1$. 
  \end{definition}

\noindent
 The bihomomorphism~$b$ induces homomorphisms $A\to (A')\dual$ and $A'\to
A\dual$; the nondegeneracy condition states that these are isomorphisms.
Here $A\dual$~is the Pontrjagin dual abelian group of homomorphisms $A\to
\Cx$.  The duality relation is symmetric: $(A\dual)\dual\cong A$
(canonically).    

  \begin{remark}[]\label{thm:45}
 For any finite abelian group~$A$ there is a canonical duality pairing
(evaluation) 
  \begin{equation}\label{eq:48}
     A\dual\times A\to \Cx
  \end{equation}
  \end{remark}

If $A,A'$ are in Pontrjagin duality, then there is a Fourier transform
isomorphism
  \begin{equation}\label{eq:45}
     \Fun(A)\xrightarrow{\;\cong \;}\Fun(A') 
  \end{equation}
on the vector spaces of complex functions.

  \subsection{Example}\label{subsec:4.5}

For $\ell \in \ZZ^{>0}$, the group~$\bmul$ is the subgroup of~$\Cx$
consisting of $\ell ^{\textnormal{th}}$~roots of unity.  It is a cyclic group
of order~$\ell $; there are several generators but none canonically picked
out (Galois symmetry).  The group~$\zmod\ell $ of classes of integers under
the mod~$\ell $ equivalence is also cyclic of order~$\ell $, again not
canonically so.  These two groups are not canonically isomorphic, but they
are canonically dual via the pairing 
  \begin{equation}\label{eq:46}
     \begin{aligned} b\:\bmul\times \zmod\ell &\longrightarrow \Cx \\ \lambda
      \;\;\;,\quad k\;\;\;&\longmapsto \lambda ^k\end{aligned} 
  \end{equation}

  \begin{remark}[]\label{thm:43}
 For~$\ell =2$ these are cyclic groups of order~2, and any two cyclic groups
of order~2 are canonically isomorphic.  Perhaps it is worth pointing out that
the center of~$\SU_N$ is canonically~$\bmu {\!N}$: think of the scalar
matrices that comprise the center. 
  \end{remark}

  \begin{remark}[]\label{thm:44}
 Classical Pontrjagin duality extends to topological abelian groups that are
locally compact.  In some sense the $\ell \to\infty $ limit of~\eqref{eq:46}
is the duality pairing 
  \begin{equation}\label{eq:47}
     \begin{aligned} \TT\times \ZZ&\longrightarrow \Cx \\ \lambda
      \;\;,\; k&\longmapsto \lambda ^k\end{aligned} 
  \end{equation}
The corresponding isomorphism~\eqref{eq:45} of infinite dimensional vector
spaces (of $L^2$~functions) is the usual Fourier transform. 
  \end{remark}

  \subsection{Pontrjagin self-duality}\label{subsec:4.8}

If $A=A'$ in Definition~\ref{thm:42}, then we say that $b$~is a
\emph{Pontrjagin self-duality pairing}; it identifies~$A$ as its own
Pontrjagin dual.  We have already said that $\bmut$~has a canonical such
Pontrjagin self-duality.  We emphasize that Pontrjagin self-duality is
\emph{data}, not merely a condition.

  \subsection{$\pi $-finite spectra}\label{subsec:4.6}

Recall Definition~\ref{thm:c53}(3) of a $\pi $-finite spectrum~$\sT$.  There is
an extension of duality to such spectra; the role of~$\Cx$ is played by an
analog of the \emph{Brown-Comenetz} dual,\footnote{They use~$\QQ/\ZZ$ in place
of~$\Cx$, as could we.} the spectrum~$I\Cx$, which is a ``character dual'' to
the sphere characterized by the universal property
  \begin{equation}\label{eq:49}
     [\sT,I\Cx]\cong (\pi _0\sT)\dual 
  \end{equation}
for any spectrum~$\sT$.  (Here $[\sT,\sT']$~denotes the abelian group of
homotopy classes of spectrum maps $\sT\to \sT$.)  The \emph{spectrum} of maps
$\sT\to I\Cx$ is the character dual~$\sTd$ to the $\pi $-finite
spectrum~$\sT$; it is also $\pi $-finite.

  \begin{example}[]\label{thm:46}
 Let $A$~be a finite abelian group and $HA$~the corresponding
Eilenberg-MacLane spectrum.  There is a duality pairing 
  \begin{equation}\label{eq:50}
     HA\dual\!\wedge HA\longrightarrow H\Cx\longrightarrow I\Cx 
  \end{equation}
that identifies~$HA\dual$ as the character dual to~$HA$. 
  \end{example}

  \subsection*{Finite electromagnetic duality}

  \subsection{The electromagnetic dual theory}\label{subsec:4.7}

Let $\sT$~be a $\pi $-finite spectrum, and denote by~$\XT$ its 0-space, which
is an infinite loop space.  By the definition of a spectrum, there is a
basepoint $*\to \XT$.  For any dimension~$n\in \ZZ^{>0}$, the semiclassical
data~$(\XT,*)$ quantizes to symmetry data~$\sr$ in which $\sigma
=\sXd{n+1}{\sT}$~is an $(n+1)$-dimensional topological field theory.  The
duality pairing
  \begin{equation}\label{eq:c25}
     \STd\wedge  \sT\longrightarrow \Sigma ^nI\Cx 
  \end{equation}
determines a cohomology class on $\XTd\wedge \XT$; let $\mu $ be a cocycle
representative.\footnote{In many cases of interest the pairing~\eqref{eq:c25}
factors through a simpler cohomology theory than $I\Cx$.  For example, if
$\sT=\Sigma ^sHA$ is a shifted Eilenberg-MacLane spectrum of a finite abelian
group, then \eqref{eq:c25}~factors through~$\Sigma ^nH\Cx$ and we can
choose~$\mu $ to be a singular cocycle with coefficients in~$\sT$.  Recall
that we use the word `cocycle' for any geometric representative of a
generalized cohomology class.}

  \begin{definition}[]\label{thm:c29}
 \ 

 \begin{enumerate}[label=\textnormal{(\arabic*)}]

 \item The \emph{dual symmetry data}~$\sdrd$ to~$\sr$ is the finite homotopy
theory $\sd=\sXd{n+1}{\XTd}$ with Dirichlet boundary theory~$\rd$ from the
basepoint $*\to \XTd$.

 \item The \emph{canonical domain wall}~$\zeta \:\sigma \to \sd$---i.e., a
$(\sd,\sigma )$-bimodule---is the finite homotopy theory constructed from the
correspondence of $\pi $-finite spaces
  \begin{equation}\label{eq:c26}
     \begin{gathered} \xymatrix{&(\XTd\times \XT,\mu )\ar[dl]\ar[dr] \\
     \XTd&&\XT} \end{gathered} 
  \end{equation}
in which the maps are projections onto the factors in the Cartesian product.
There is a similar canonical domain wall $\zeta \dual\:\sigma \dual\to
\sigma $. 

 \item The \emph{canonical Neumann boundary theories}~ $\epsilon ,\ed$ are
the finite homotopy theories induced from the identity maps on~$\XT,\XTd$,
respectively. 

 \end{enumerate}  
  \end{definition}

\noindent
 Our formulation emphasizes the role of~$\sigma $ as a symmetry for another
quantum field theory.  But $\sigma $~is a perfectly good $(n+1)$-dimensional
field theory in its own right.  From that perspective $\sd$~is the
$(n+1)$-dimensional \emph{electromagnetic dual} theory.
See~\footnote{Yu~Leon Liu, \emph{Abelian duality in topological field
theory},
  \href{http://arxiv.org/abs/arXiv:2112.02199}{{\tt arXiv:2112.02199}}.
} for more about electromagnetic duality in this context.

  \begin{remark}[]\label{thm:c30}  
\ 
 \begin{enumerate}[label=\textnormal{(\arabic*)}]

 \item In this finite version of electromagnetic duality there is a shift
from what one expects based on usual electromagnetic duality of $\U_1$-gauge
fields.  So, for example, if $A$~is a finite abelian group, then in
3~dimensions the finite gauge theory that counts principal $A$-bundles has
as its electromagnetic dual the gauge theory that counts principal
$A\dual$-bundles.

 \item  As usual, we have not made explicit the background fields for~$\sigma $,
$\sd$, and~$\zeta $, $\zeta \dual$.  In fact, the theories $\sigma $~and
$\sd$~are defined on bordisms unadorned by background fields: they are
``unoriented theories''.  For~$\zeta $ we need a set of (topological)
background fields that orient manifolds sufficiently to integrate~$\mu $.
For example, if $\mu $~is a singular cocycle with coefficients in~$\TT$, then
we need a usual orientation.

 \end{enumerate}
  \end{remark}

  \subsection{Electromagnetic self-duality}\label{subsec:4.9}
 
Analogous to Pontrjagin self-duality of abelian groups~\ref{subsec:4.8},
there is the possibility of an isomorphism $\sigma \xrightarrow{\;\cong
\;}\sd$, which is a self-duality of finite abelian gauge theories.  It is
based on a self-duality pairing 
  \begin{equation}\label{eq:53}
     \sT\wedge \sT\longrightarrow \Sigma ^nI\Cx 
  \end{equation}
(Keep in mind that the theory~$\sigma $ is in dimension~$n+1$.) 

  \begin{example}[Finite version of ``$p$-form gauge fields'']\label{thm:48}
 Suppose that $A$~is a finite abelian group equipped with a Pontrjagin
self-duality, and for~$p\in \ZZ^{>0}$ set $\sT=\Sigma ^{p}HA$.  Then in odd
dimension $2p+1$ there is an electromagnetic self-duality of
$\sigma =\sXd{2p+1}{\sT}$.  The cases~$p=1$ and~$p=2$ are of particular interest
later in this lecture.   
  \end{example} 

  \begin{example}[A non-Eilenberg-MacLane spectrum]\label{thm:49}
 Consider the spectrum~$\sT$ that is an extension (Postnikov tower) 
  \begin{equation}\label{eq:51}
     \Sigma ^3H\zt\longrightarrow \sT\longrightarrow \Sigma ^2H\zt 
  \end{equation}
with extension class ($k$-invariant or Postnikov invariant) 
  \begin{equation}\label{eq:52}
     \Sq^2\:\Sigma ^2H\zt\longrightarrow \Sigma ^4H\zt 
  \end{equation}
Then there is an electromagnetic self-duality of~$\sXd6{\sT}$. 
  \end{example}

  \begin{remark}[]\label{thm:47}
 There is a field theory built on a ``self-dual'' field, which is very
interesting; it is a kind of square root of~$\sXd{n+1}\sT $, and it requires
a quadratic refinement of~\eqref{eq:53} to define.  
  \end{remark}

  \subsection*{Duality and quotients}

  \subsection{Duality swaps Dirichlet and Neumann}\label{subsec:4.10}

We keep the notation of~\ref{subsec:4.7}, including
Definition~\ref{thm:c29}.  The topological field theories~$\sigma $ and~$\sd$
have dimension~$n+1$.

  \begin{proposition}[]\label{thm:c31}
 There is an isomorphism of right $\sigma $-modules 
  \begin{equation}\label{eq:c27}
     \psi \:\rd\otimes _{\sd}\zeta \xrightarrow{\;\;\cong \;\;}\epsilon 
  \end{equation}
  \end{proposition}

  \begin{figure}[ht]
  \centering
  \includegraphics[scale=1.3]{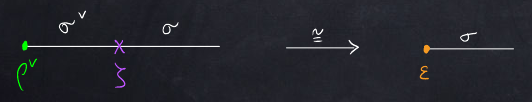}
  \vskip -.5pc
  \caption{An isomorphism of right $\sigma $-modules}\label{fig:c10}
  \end{figure}

\noindent
 This isomorphism is depicted in Figure~\ref{fig:c10}. 

  \begin{proof}
 We use the calculus of $\pi $-finite spectra, as described
in~\ref{subsec:2.12}---see especially the composition
law~\eqref{eq:c85}.  The theory~$\sigma $ is induced from~$\XT$, the
theory~$\sd$ from~$\XTd$, the boundary theory~$\rd$ from $*\to \XTd$, and the
domain wall~$\zeta $ from the correspondence diagram~\eqref{eq:c26}.  Hence
$\rd\otimes _{\sd}\zeta $ is induced from the homotopy fiber product:
  \begin{equation}\label{eq:c28}
     \begin{gathered}
     \xymatrix{\ast\ar[dr]&(\XT,0)\ar@{-->}[l]\ar@{-->}[r]&(\XTd\times \XT,\mu
     )\ar[dl]\ar[dr] \\ &\XTd&&\XT} \end{gathered} 
  \end{equation}
Here we use that the restriction of~$\mu $ to~$*\times \XT$ is zero.   So the
sandwich is the right $\sigma $-module induced from the composition 
  \begin{equation}\label{eq:c29}
     \begin{gathered} \qquad \qquad \xymatrix{(\XT,0)\ar@{-->}[r] &(\XTd\times
     \XT,\mu )\ar[dr] \\&&\XT} \end{gathered} 
  \end{equation}
which is~$\id_{\XT}$.  That theory is the augmentation boundary theory~$\epsilon
$.  
  \end{proof}

  \begin{figure}[ht]
  \centering
  \includegraphics[scale=1.25]{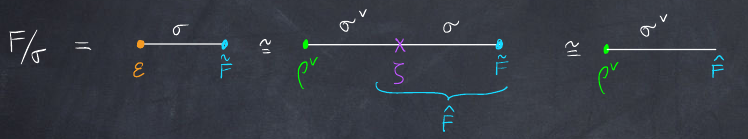}
  \vskip -.5pc
  \caption{The dual symmetry on the quotient~$F\bs$}\label{fig:c11}
  \end{figure}

  \subsection{The effect on left $\sigma $-modules}\label{subsec:4.11}

The duality domain wall~$\zeta $ maps left $\sigma $-modules to left
$\sd$-modules; the domain wall~$\zeta \dual$ induces a map in the opposite
direction.  The following theorem, which follows immediately from
Proposition~\ref{thm:c31}, illuminates this involutive correspondence.

  \begin{corollary}[]\label{thm:c32}
 Let $F$~be a quantum field theory equipped with a $\sr$-module structure.
Then the quotient~$F\bs$ carries a canonical $\sdrd$-module structure.  
  \end{corollary}

  \begin{corollary}[]\label{thm:53}
 The transform of~$F$ under electromagnetic duality is the
quotient~$F\bs$.\qed
  \end{corollary}

  \begin{proof}[Proof of Corollary~\ref{thm:c32}]
 The proof is contained in Figure~\ref{fig:c11}.  In words: Let $\tFt$ be the
$\sr$-module data, as in Definition~\ref{thm:c11}.  Define the left
$\sd$-module 
  \begin{equation}\label{eq:c30}
     \hF=\zeta \otimes _\sigma \tF 
  \end{equation}
and the isomorphism 
  \begin{equation}\label{eq:c31}
     \hth\:\rd\otimes _{\sd}\hF = \rd\otimes _{\sd}\zeta \otimes _{\sigma
     }\tF\xrightarrow{\;\;\psi \,\otimes\, \theta \;\;} \epsilon \otimes _\sigma
     \tF = F\bs
  \end{equation}
Then $(\hF,\hth)$ is the desired $\sdrd$-module structure. 
  \end{proof}

  \subsection*{The 2-dimensional Ising model}

Here we briefly indicate an important example of duality: the Ising model.
We take it as a \emph{stat mech model}, which is a discrete quantum system,
rather than a continuous quantum field theory.  (There is another basic class
of discrete quantum models---\emph{lattice systems}---that are discussed in
Mike Hopkins' lectures.)  The same basic structure
applies.  The paper footnote~\footref{FTi} contains much more information.  The story
there is for an arbitrary finite group; here for simplicity we take the
finite abelian group~$A=\bmut$.

  \begin{figure}[ht]
  \centering
  \includegraphics[scale=1.2]{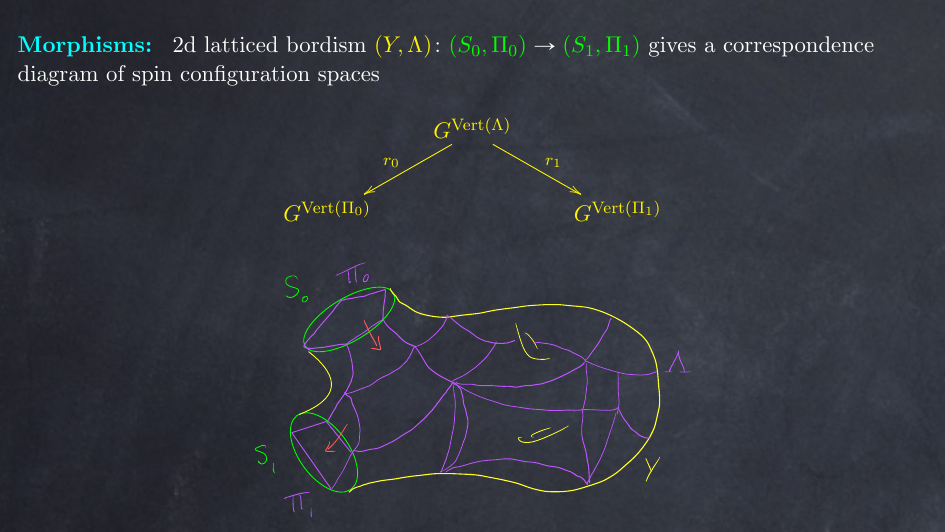}
  \vskip -.5pc
  \caption{A bordism $(Y,\Lambda )\:(S_0,\Pi _0)\to (S_1,\Pi _1)$ with lattices}\label{fig:24}
  \end{figure}

  \subsection{The basic model}\label{subsec:4.12}
 
Fix a positive real number~$\beta \in \RR^{>0}$.  There is a 2-dimensional
field theory~$F_\beta $ of ``latticed'' 1-~and 2-manifolds; a typical bordism
in the bordism category of such is depicted in Figure~\ref{fig:24}.  (We
defer to footnote~\footref{FTi} for details, such as the definition of the lattices.)
The ``fluctuating field'' over which we sum---the ``spin''---is a function
from the vertices of the lattice to the group~$A$, and as usual a bordism
gives rise to a correspondence of fluctuating fields:
  \begin{equation}\label{eq:54}
     \begin{gathered} \xymatrix{&A^{\Vertices(\Lambda )}\ar[dl]_{r_0}
     \ar[dr]^{r_1} \\ A^{\Vertices(\Pi _0)}&& A^{\Vertices(\Pi _1)}}
     \end{gathered} 
  \end{equation}
Linearize by push-pull
  $$F_\beta (Y,\Lambda ) = (r_1)_*\circ K_\beta \circ (r_0)^*\:F_\beta (S_0,\Pi
  _0)\longrightarrow 
     F_\beta (S_1,\Pi _1)$$
where we include an ``integral kernel''~$K_\beta $.  The vector space of
functions $F_\beta (S_i,\Pi _i)=\Fun\bigl(A^{\Vertices(\Lambda )} \bigr)$ is
independent of~$\beta $.  The function~$K_\beta $ is defined in terms
of a weight on the group:
  \begin{equation}\label{eq:55}
     \begin{aligned} \theta _\beta \:\bmut&\longrightarrow \RR^{>0} \\
      +1\;&\longmapsto \;1 \\ -1\;&\longmapsto \;e^{-2\beta }\end{aligned} 
  \end{equation}
Physically, $\beta $~is the inverse temperature.  Then
for~$s\:\Vertices(\Lambda )\to A=\bmut$ we define
  $$ K_\beta (s)=\prod\limits_{e}\theta  \bigl(g(s;e)
     \bigr),\qquad \textnormal{$e$ incoming or interior}, $$
where for an edge~$e\in \Edges(\Lambda )$ we let $g(s;e)\in A$ be the ratio
of the two spins~$s\res{\partial e}$ on the endpoints of~$e$: it is~$+1$ if
they agree and $-1$~if they disagree.

  \begin{exercise}[]\label{thm:50}
 How does this model behave as~$\beta \to 0$?  As $\beta \to \infty $? 
  \end{exercise}

  \subsection{Symmetry}\label{subsec:4.13}

The group~$A$ acts as a symmetry that exchanges spin~$+1$ with spin~$-1$,
vertex by vertex.  It is straightforward to construct a model~$\tF_\beta $
that couples the theory to a background principal $A$-bundle $Q\to Y$: the
spin~$s$ becomes a section of this bundle over the vertices, and the ratio
function~$g(s;e)$ is still well-defined using parallel transport along the
edge~$e$.  Then, summing over principal $A$-bundles, we see that $\tF_\beta
$~is a left $\sigma$-module, where $\sigma =\sXd3{BA}$ is the 3-dimensional
finite $A$-gauge theory.  Let $\rho $~be the usual regular boundary theory
for~$\sigma $.

  \subsection{Kramers-Wannier duality}\label{subsec:4.14}

Under electromagnetic duality $\sr\to \sdrd$, we obtain a left
$\sd$-module~$\hF_\beta $ and a corresponding dual theory
$F^{\textnormal{dual}}_\beta =\rd\otimes _{\sd}\hF_{\beta }$.  The
isomorphism~\eqref{eq:c31} is $F^{\textnormal{dual}}_\beta =F_\beta
\!\bigm/\!\hneg \sigma$.  Now we need an input which, of course, depends on
the details of the Ising model and which we do not elaborate here.  Namely,
there is a \emph{Kramers-Wannier} isomorphism 
  \begin{equation}\label{eq:56}
     \phi _\beta \:F_\beta\!\bigm/\!\hneg \sigma\xrightarrow{\;\;\cong \;\;}
     F_{\beta \dual}, 
  \end{equation}
where the dual inverse temperature~$\beta \dual$ is defined by the condition 
  \begin{equation}\label{eq:57}
     \sinh(2\beta )\sinh(2\beta \dual)=1. 
  \end{equation}
Furthermore, this isomorphism is equivariant for the symmetry action: it is
induced from an isomorphism of left $\sd$-modules.  The map $\beta
\leftrightarrow\beta \dual$ is an involution on~$\RR^{>0}$, and it has a
unique fixed point~$\beta _c$.  At this self-dual point there is an
isomorphism
  \begin{equation}\label{eq:58}
     \phi =\phi _{\beta _c}\:F\bs\xrightarrow{\;\;\cong \;\;} F,
  \end{equation}
where $F=F_{\beta _c}$.  Use the electromagnetic self-duality to conclude
that \eqref{eq:58}~is an isomorphism of left $\sr$-modules.

  \subsection*{The duality defect and an obstruction to trivial gappedness}

  \subsection{The defect}\label{subsec:4.16}

We resume our general setup: $\sr$~is symmetry data in $n$~dimensions, with
$\rho $~a regular right $\sigma $-module, and $F$~is a quantum field theory
equipped with a $\sr$-module structure $\srtF$.  Assume further that $\sigma
$~has an augmentation and $F$~is equipped with an isomorphism
  \begin{equation}\label{eq:59}
     \phi \:F\bs\xrightarrow{\;\;\cong \;\;} F.
  \end{equation}

  \begin{remark}[]\label{thm:54}
 One could ask for some compatibility of~\eqref{eq:59} with symmetry.  By
Corollary~\ref{thm:c32} the quotient~$F\bs$ is a $\sdrd$-module, so what we
would need is self-duality data for~$\sr$; then we could ask that
\eqref{eq:59}~be compatible with the $\sr$-module structure.  I do not see
where that enters the arguments below, but it would be a more satisfactory
story and would make the duality defect part of \emph{self}-duality.  More
likely, is that there is a more specialized story with self-duality, and it
would be nice to track down the extra information one can glean.
  \end{remark}

\begin{figure}[ht]
  \centering
  \includegraphics[scale=.5]{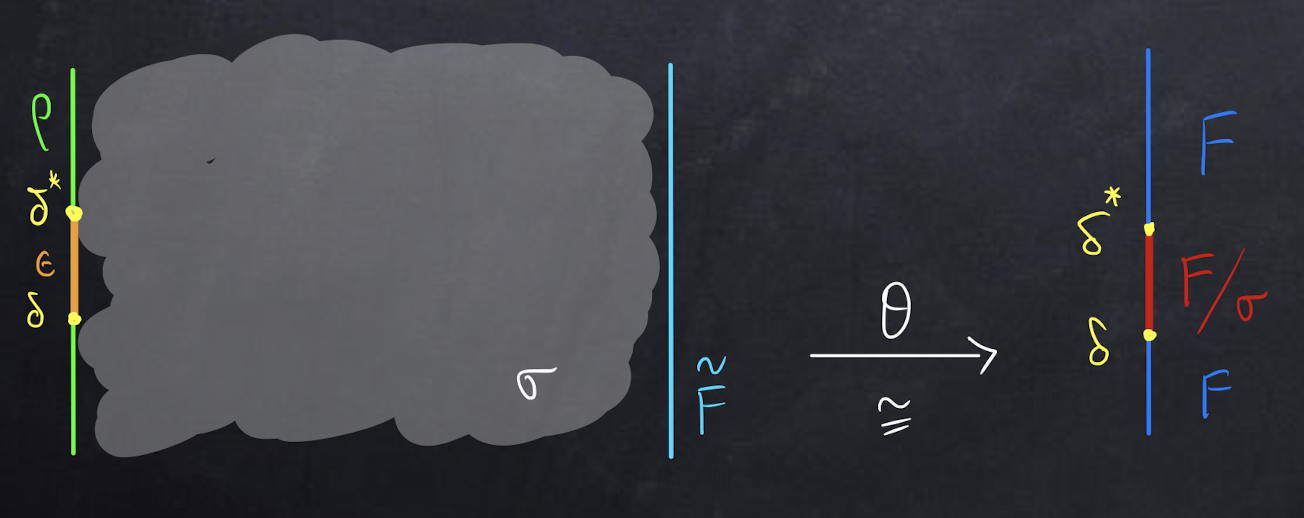}
  \vskip -.5pc
  \caption{Figure~\ref{fig:18} redux}\label{fig:25}
  \end{figure}

Figure~\ref{fig:25} recalls the domain walls $\delta \:F\to F\bs$ and
$\delta^*\:F\bs\to F$ that are defined in~\ref{subsec:3.8}.

  \begin{definition}[]\label{thm:51}
 The \emph{duality defect} is the self-domain wall
  \begin{equation}\label{eq:60}
     \Delta =\phi \circ \delta \:F\longrightarrow F. 
  \end{equation}
  \end{definition}

\noindent 
 Since $\delta $~is a topological defect, and $\phi $~is simply an
isomorphism of theories, the composition~$\Delta $ is also a topological
defect.

  \subsection{The composition~$\Delta ^*\circ \Delta $}\label{subsec:4.17}

As a preliminary we observe the following.  View~$\phi $ as a domain wall
from~$F$ to~$F\bs$, and furthermore imagine that there is a 2-category of
theories, domain walls, and domain walls between domain walls.  Then we can
contemplate the adjoint~$\phi ^*$.  It is a general fact that if $\phi $~is
invertible, then its adjoint equals its inverse: $\phi ^*=\phi \inv $. 

  \begin{exercise}[]\label{thm:55}
 Formulate and prove this result in 2-categories. 
  \end{exercise}

Using this we compute 
  \begin{equation}\label{eq:61}
     \Delta ^*\circ \Delta  =(\phi \delta )^*(\phi \delta )=\delta ^*\phi
     ^*\phi \delta =\delta ^*\phi\inv 
     \phi \delta =\delta ^*\circ \delta 
  \end{equation}
The composition $\delta ^*\circ \delta $ is discussed in~\ref{subsec:3.9}.
We use those computations below.

  \subsection{A larger symmetry}\label{subsec:4.18}

The theory~$F$ has a $(\hat\sigma ,\hat\rho )$-module structure for a
topological field theory~$\hat\sigma $ that is~$\sigma $ with $\Delta
$~adjoined.  I have not thought through in detail how to define~$(\hat\sigma
,\hat\rho )$.  The equation~\eqref{eq:61} holds in~$\hat\sigma $.  One can
also work out compositions of~$\Delta $ with other defects in~$\sr$.

It is this larger theory which we will prove in some instances does not admit
an augmentation.

  \subsection{The obstruction to infrared triviality}\label{subsec:4.19}

Here `triviality' means `trivially gapped' in the sense of~\ref{subsec:4.1}:
the low energy approximation to a gapped theory is invertible.\footnote{It is
not necessarily topological, just invertible.}  Continuing in this situation
with the duality defect, suppose as in~\ref{subsec:4.3} that there exists a
left $\sigma $-module~$\tl$ such that $\Delta $~also acts, $\Delta
$~satisfies~\eqref{eq:61}, and the sandwich 
  \begin{equation}\label{eq:62}
     \lambda :=\rho \otimes _\sigma \tl 
  \end{equation}
is invertible.  (In terms of~\ref{subsec:4.18}, $\tl$~ is a left $\hat\sigma
$-module.)  Because $\lambda $~is invertible, self-domain walls act as
multiplication by an $(n-1)$-dimensional field theory; they do not couple
to~$\lambda $.  (Compare: an endomorphism of a line is multiplication by a
complex number, an endomorphism of a $\VV$-line is tensoring by a vector space,
etc.)  So $\delta ^*\circ \delta $~acts as multiplication by an
$(n-1)$-dimensional \emph{topological} field theory, and so too does $\Delta
$~act as multiplication by a topological field theory.  Those theories
satisfy~\eqref{eq:61}: $\Delta $~is a kind of square root of~$\delta ^*\circ
\delta $.  But in some situations no such square root exists, as we can prove
using the well-developed principles of topological field theory.  The
conclusion in that case is that no sandwich~\eqref{eq:62} exists with $\lambda
$~invertible.

We emphasize that the argument does not use any left $\sr$-module; it just
uses~$\sr$ and the additional defect~$\Delta $.  So it applies to any quantum
field theory with a $\sr$-module structure.

  \subsection*{Examples}

  \subsection{A 2-dimensional example}\label{subsec:4.20}

Let $\sigma =\sXd3{B\!\bmut}$ be 3-dimensional $\bmut$-gauge theory with its
usual regular right boundary theory.  Adjoin a codimension one defect~$\Delta
$ that satisfies~\eqref{eq:61}.  This acts on 2-dimensional theories~$F$
equipped with an isomorphism $\phi \:F\bs\xrightarrow{\;\cong \;}F$.  Now the
composition~$\delta ^*\circ \delta $ was computed in Example~\ref{thm:36},
and it acts on an invertible 2-dimensional theory as multiplication by the
1-dimensional topological field theory which is the $\sigma $-model
into~$\bmut$.  In particular, the vector space attached to a point has
dimension~2.  Hence there is no square root: $\Delta $~acts as multiplication
by a 1-dimensional topological field theory, as does~$\Delta ^*$, and the
vector space attached to a point has the same dimension in both.  Since
$\sqrt2$~is not an integer, this cannot happen. 
 
We can an alternative proof in this case that directly shows that
$\hat\sigma $~in \ref{subsec:4.18} does not have an augmentation.  Namely,
$\hat\sigma (\pt)$ is the fusion category $\Vect[\bmut]$ with an additional
simple object~$x$ adjoined.  Let $\bmut=\{1,g\}$, so that the set of simple
objects in~$\hat\sigma (\pt)$ is $\{1,g,x\}$.  Then the relations are 
  \begin{equation}\label{eq:63}
     \begin{aligned} g^2&=1 \\ gx&=x \\ x^2&=1+g\end{aligned} 
  \end{equation}
One can see directly that no fiber functor exists, for example by the same
dimension argument applied to the last relation.  The fusion
category~$\hat\sigma (\pt)$ was introduced by Tambara-Yamagami. 
 
An example of a theory with this symmetry, including~$\Delta $, is the Ising
model at the critical temperature; see~\eqref{eq:58}.  But that theory is not
gapped, so learning that it is not trivially gapped is not a particular
victory.

  \subsection{A 4-dimensional example}\label{subsec:4.21}

Now take $\sigma =\sXd5{B^2\!\bmut}$ to be the $\bmut$-gerbe theory in
5~dimensions; it acts on 4-dimensional theories with $B\!\bmut$-symmetry.  We
give examples below.  Assume that $\Delta $~is adjoined and that
\eqref{eq:61}~is satisfied.  The composition~$\delta ^*\circ \delta $ is
computed in Example~\ref{thm:35}; from~\eqref{eq:14} we see that it acts on
an invertible 4-dimensional theory as multiplication by 3-dimensional
$\bmut$-gauge theory~$\Gamma =\sXd3{B\!\bmut}$.  We claim there is no
3-dimensional topological field theory~$T$ such that $T^*\circ T=\Gamma $.
If so, evaluate on a point to obtain fusion categories~$T(\pt)$ and $\Gamma
(\pt)=\Vect[\bmut]$.  The number of simple objects in~$\Vect[\bmut]$ is~2,
which is not a perfect square.  The number of simple objects in
$T^*(\pt)\otimes T(\pt)$ is a perfect square.  This contradiction proves
that there is no invertible left $\sigma $-module on which $\Delta $~acts. 

  \begin{remark}[]\label{thm:56}
 We have not identified the 5-dimensional topological field
theory~$\hat\sigma $.  It would be interesting to do so and to argue directly
that $\hat\sigma (\pt)$ has no augmentation.  (Perhaps the argument is a
rewording of the one given.) 
  \end{remark}

We conclude with two 4-dimensional gauge theories to which the argument
applies to rule out trivial gappedness; see footnote~\footref{CK} for details. 
 
The first is $\U_1$~Yang-Mills theory with coupling constant~$\tau $.  The
map~$\phi $ is S-duality, which maps $\tau \mapsto -1/\tau $.  The quotient
by the $B\!\bmut$-symmetry maps $\tau \mapsto \tau /4$, so the fixed value is
$\tau =2\sqrt{-1}$.  This is the theory to which the argument applies. 
 
The second example is $\SO_3$~Yang Mills theory with $\theta $-angle $\theta
=\pi $.

\bigskip\bigskip\bigskip
}

%\newpage
%\newyear{\Large Les Diablerets Summer School}
%{\Large September 3--8, 2023}

%\newpage

%{\clearpage\import{constantin/}{constantin.tex}}

%\newpage
%{\clearpage\import{appendix/}{main.tex}}

\bibliographystyle{JHEP}
\bibliography{bibli.bib}

\end{document}